\documentclass[a4paper,11pt,twoside]{book}

\usepackage{fancyhdr}
\usepackage{titlesec}
\usepackage[dvips]{graphicx}
\usepackage{epsfig}
\usepackage{subfigure}
\usepackage{amssymb}
\usepackage{amsmath}
\usepackage{amsfonts}
\usepackage[english]{babel}
\usepackage[section]{placeins}

\begin{document}
{
\pagestyle{fancy}
\setlength{\headheight}{14pt}
\renewcommand{\sectionmark}[1]{\markright{\thesection.\ #1}}
\renewcommand{\chaptermark}[1]{\markboth{\thechapter.\ #1}{}}
\fancyhead[LE,RO]{\textbf\thepage} %
\fancyhead[RE]{\textbf{\textsl{\rightmark}}} %
\fancyhead[LO]{\textbf{\textsl{\leftmark}}}
\fancyfoot{} %
\titleformat{\chapter}[display]{\Huge\bf}%
{%
\addvspace{-4ex}%
{\Large\chaptertitlename}\ \Huge\thechapter}%
{2ex}{}
\newcommand{\boh}[2]{\underline{#1}\marginpar{{\bf$\leftarrow$~?}%
{ \small #2}}}
\newcommand{\margriga}[3]{\marginpar{%
    \makebox[0pt][r]{%
        \raisebox{0pt}[0pt][0pt]{%
            \rule[#1]{0.25ex}{#2}\ %
        }%
    }#3}}
\voffset=0in
\hoffset=.0in
\topmargin=0pt
\leftmargin=0pt
\headheight=0pt
\headsep=0pt
\begin{figure}[p!]
\centerline{\includegraphics[angle=0]{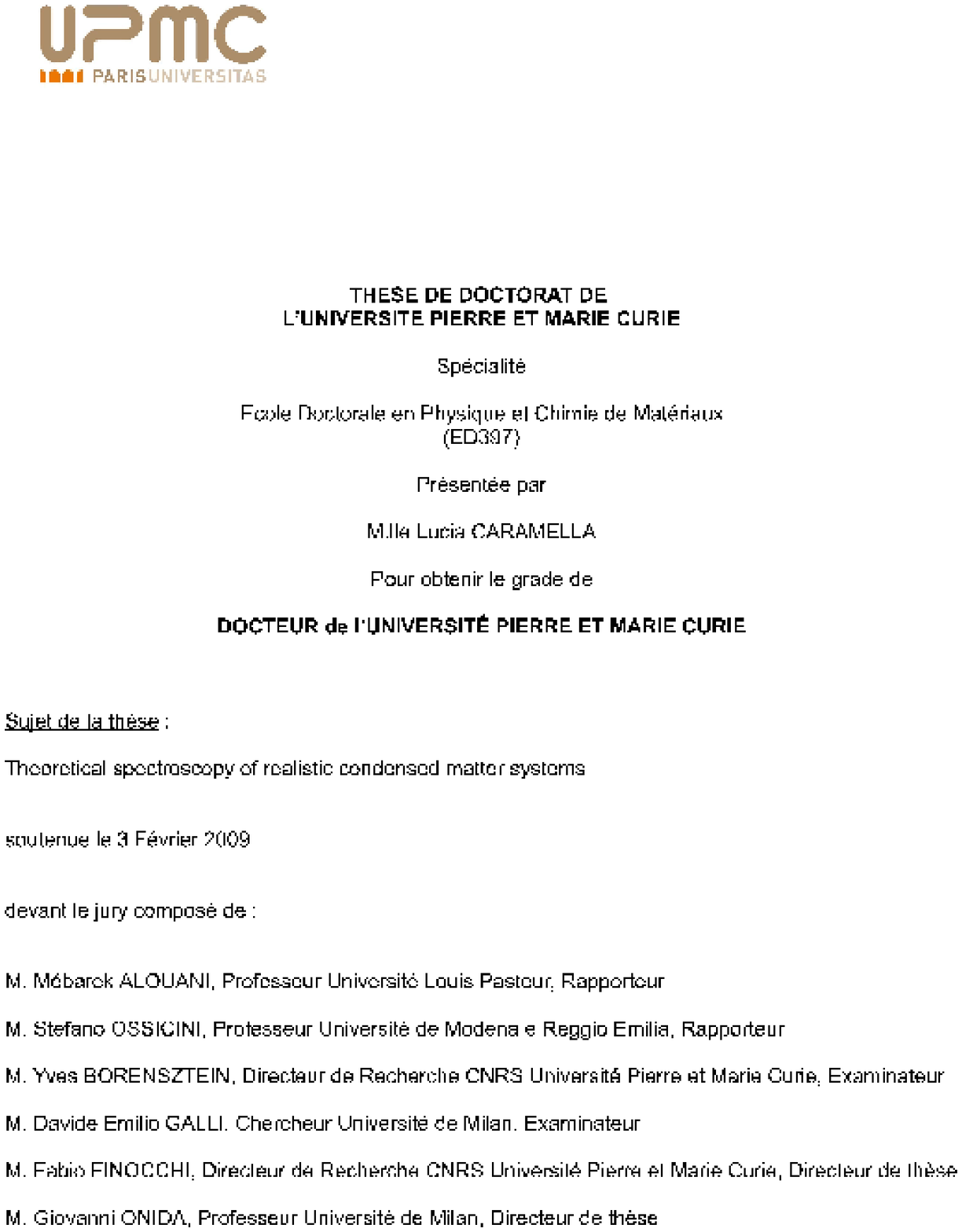}}
\end{figure}
\cleardoublepage
}

\frontmatter
\thispagestyle{empty}
%\documentclass[11pt,italian,a4paper,titlepage]{article}
%\usepackage[italian]{babel}
%
%
%%\pagestyle{fancy} \addtolength{\headwidth}{\marginparsep}
%%\addtolength{\headwidth}{\marginparwidth}
%%
%%\lhead[\fancyplain{}{\thepage}]{\fancyplain{}{\textit{\rightmark}}}
%%\rhead[\fancyplain{}{\textit{\leftmark}}]{\fancyplain{}{\thepage}}
%%\cfoot{}
%
%\begin{document}
%
\begin{titlepage}

\begin{center}
{\large{{Universit\`{a} degli Studi di Milano\\
Dottorato di ricerca in Fisica, Astrofisica e Fisica Applicata\\
and\\
Universit\'e Pierre et Marie Curie\\
Ecole Doctorale Physique et Chimie de Mat\'eriaux\\}}} 
\vspace{\stretch{12}} 
{\LARGE{\textbf{Theoretical spectroscopy of \\realistic condensed matter systems}}\\}
\vspace{\stretch{1.0}} s.s.d. FIS/03
\end{center}
\vspace{\stretch{10.5}}
\begin{flushleft}
{\large 
Director of the Doctoral School (I): Prof. Gianpaolo BELLINI\\[0.15cm]
Director of the Doctoral School (F): Prof. Jean--Pierre JOLIVET\\[0.55cm]
Thesis Director: Prof. Giovanni ONIDA\\[0.15cm]
Thesis Director: Dr. Fabio FINOCCHI\\[0.15cm]
}
\end{flushleft}
\vspace{\stretch{6.5}}
\begin{flushright}
{\large PhD Thesis of:\\[0.15cm]
Lucia CARAMELLA \\[0.15cm]
Ciclo XXI\\[0.15cm] }
\end{flushright}
\centering \vspace{\stretch{15.5}} {\large{ Academic year: 2008/2009 }}
\end{titlepage}
%
%
%\end{document}

\clearpage
\thispagestyle{empty}

%\title{\bf Theoretical spectroscopy of realistic condensed matter systems}
%\author{L. Caramella}

  \newcommand{\ea}{ \textit{et al.}}
  \newcommand{\ptwo}{$p(2\times2)$}
  \newcommand{\pone}{$p(2\times1)$}
  \newcommand{\ctwo}{$c(4\times2)$} 
  \renewcommand{\[}{\left[}
  \renewcommand{\]}{\right]}
  \renewcommand{\(}{\left(}
  \renewcommand{\)}{\right)}
  \newcommand{\ra}{\rangle}
  \newcommand{\la}{\langle}
  \newcommand{\bk}{\mathbf{k}}
  \newcommand{\br}{\mathbf{r}}
  \newcommand{\Eref}[1]{Eq.~(\ref{#1})}

%\begin{document}
%\maketitle
\tableofcontents
\markboth{Contents}{Contents}
\listoffigures
\markboth{List of figures}{List of figures}
%\addcontentsline{toc}{chapter}{List of figures}
\listoftables
\markboth{List of tables}{List of tables}
%\addcontentsline{toc}{chapter}{List of tables}

\mainmatter
%%%%%%%%%%%%%%%%%%%%%%%%%%%%%%%%%%%%%%%%%%%%%%%%%%%%%%%%%%%%%%%%%%%%%%%%%%%
%                              TEXT                                       %
%%%%%%%%%%%%%%%%%%%%%%%%%%%%%%%%%%%%%%%%%%%%%%%%%%%%%%%%%%%%%%%%%%%%%%%%%%%
\chapter*{Introduction}\label{intro:sec}
\markboth{Introduction}{Introduction}
\addcontentsline{toc}{chapter}{Introduction}
 %%
%  1 - Importanza interazione radiazione-materia
%%
A deep understanding of the interaction between matter and radiation
(including electrons and light) is a key issue in order to describe
the physical nature and the properties of materials.
This can be achieved with a joint effort of numerical simulations
and experiments. 
In fact, the physical origin of the experimental spectral features 
can often be understood unambiguously 
with the help of numerical simulations. \\
%
% 2 - Grandi successi di calcoli ab initio ground state per sistemi realistici
%     negli ultimi anni (superfici ricostruite in particolare)
%  
\indent Nowadays numerical computation of ground state properties 
of condensed matter systems can be successfully treated within 
the density functional theory (DFT).
In this context, the problem of solving the Schr\"odinger equation
for the ground state of a many body system can be exactly recast
into the variational problem of minimizing an energy functional
with respect to the charge density.
The success of this approach has been shown during the last
years by ab initio calculations describing the ground state properties 
of realistic systems, in particular in case of reconstructed surfaces.
%
% 3 - Ground state non basta piu' molti esperimenti fanno eccitazioni
% 
However, ground state properties are not enough to describe those
experiments involving excitations of the electronic
system. In most cases, an external probe modifies
the charge distribution of the sample, producing excited states 
and a dynamical rearrangement of the density. \\
\indent In the last years, excited state theories
providing an overcome of limits of DFT have been proposed.
A particularly fruitful attempt to go beyond the ground state theory 
is offered by time dependent density functional theory (TDDFT),
providing an exact reformulation of quantum mechanics in terms 
of time evolving density. 
Within this theory, the complexity of the problem is confined 
to the exchange--correlation potential $V_{\textrm{xc}}$, 
whose analytic form is unknown but several approximations 
are available in literature,
giving correct predictions in many realistic cases.
However, in many interesting applications, the simpliest approximation
(as independent particle RPA) are successful. 
This is the case, for example, of the simulation of the surface optical
spectroscopy, such as reflectivity anisotropy (RA) or differential spectroscopy (SDR). \\
\indent On the other hand, new experiments require
more complete theories including, for example, 
spin degree of freedom in order to treat magnetic systems, or
local field effects in order to describe strong anisotropic systems, 
or the inclusion of semicore and core levels in order 
to obtain information about core and semicore spectroscopies. 
It is hence important to implement these more complete
theories in ab initio computational codes in order to 
describe more realistic condensed matter systems.
In particular, these implementations are fundamental 
to be able to treat systems with explicit inclusion 
of surfaces, or isolates sytems.
Moreover, new more efficient algorithms are essential
and the improvement of existing codes is required in order to 
extend the range of numerical simulation applicability 
to systems with a larger size (in terms of number of atoms). \\
\indent  Theoretical spectroscopy is a successful 
combination of these quantum theories and computer 
simulation intended to describe the fundamental mechanisms 
of interaction between materials and perturbing external fields. \\ 
The present work is an example of what is possible to 
obtain with the theoretical and numerical tools we have 
just mentioned.
In particular, we will discuss problems of numerical efficiency
and the inclusion of some aspects neglected up to now, such as the 
inclusion of spin variable and semicore levels. \\
\indent This manuscript contains different parts: 
a thread can be drawn from the technical 
development of methods, to simulations of a variety of physical systems.
Moreover, the study of a large variety of complex physical applications
helps to point out the limits and the advantages of the theories adopted. \\
After a brief review of the theoretical background presented 
in chapter~\ref{ch:DFT}, in chapter~\ref{ch:surface}
we describe in details the methods used to simulate surface 
spectroscopies and the main experimental techniques.\\
From the following chapter, we approach the core of the work 
developed in this thesis, in paticular in chapter~\ref{ch:hilbert} 
we focus on the dynamical response function $\chi(\br,\br',\omega)$ 
and we present the developement of an Hilbert transform (HT) 
based method to evaluate the independent particle 
response. 
The time scaling analysis of the HT--method on a model
shows that it is convenient for large systems. As an application,
we studied the crystal local field effects on the optical spectra of the 
Si(100)-(2$\times$2) surface, weakly oxidized. 
In chapters~\ref{ch:a-si100} and~\ref{ch:oxi}, we focused 
RA and EEL spectra of clean and oxidized Si(100) surfaces.
Chapter~\ref{ch:a-si100} is devoted to the clean surface, 
for which we discuss the spectra calculated on three reconstructions: 
p(2$\times$1), p(2$\times$2) and c(4$\times$2). 
The oxidation process of this surface is then analyzed 
in chapter~\ref{ch:oxi}, where several oxygen adsorption sites 
are studied through geometric optimization and calculation of EEL spectra. \\
The following part of this manuscript is devoted to spin polarized systems.
The limits of DFT and TDDFT-LDA approach are highlighted in 
the case of BeH, a simple molecule with unpaired number of electrons 
(see chapter~\ref{ch:a-molecules}). 
The successful calculation of the optical conductivity 
for bulk iron is then presented in chapter~\ref{ch:a-iron}
and spin resolved electronic properties of this system are provided
including the $3s$ and $3p$ semicore states. \\
In the last part of the manuscript, chapter~\ref{a:sulfur}, it is summarized 
the interesting electronic and magnetic properties of iron, cobalt and nickel 
pyrites.
These compounds are complex spin polarized systems that could have 
stimulating applications in the new field of spin electronics.
%

%
%%%%%%%%%%%%%%%%%
  \part{Theory}
%%%%%%%%%%%%%%%%%
%
\chapter{Theoretical background} \label{ch:DFT}
 The problem of finding the electronic ground state of a condensed matter system
is equivalent to solve the fundamental equation of quantum mechanics, i.e. 
the Schr\"odinger equation for a set of interacting electrons immersed 
in an external potential.
The Density Functional Theory (DFT) provides a successful tool to treat 
the problem giving a variational reformulation of the equations in terms 
of the electronic density.
In this chapter we briefly review this theory.
The fundamental theorems and its Time Dependent generalization (TDDFT)
is reviewed with a presentation of the methods used in this thesis for
the numerical simulation of realistic systems.
% 
%In conclusion an accent to the % its nature as a minimization problem and
%role of the spin variable is presented in order to 
%evidence the highlights and the limits in using the local spin density approximation 
%to describe magnetic properties of the matter.
% 

%In order to describe ground state properties and excitations 
%of a many body system, a theory able to treat 
%a N$\times$M--dimensional problem is required.
%A many body system is here defined a set of N electrons and M nuclei 
%interacting and coupled. 
%The number of degrees of freedom and interactions involved 
%make the problem intractable a part of few ideal cases.
%
%Density Functional Theory (DFT) and his Time Dependent generalization
%(TDDFT) answer to these requirement and 
%nowadays is a well-established tool succesfully 
%employed in \emph{ab initio} calculations of ground state properties
%of atoms, molecules and solids.
%
%In the following sections we will summarize the
%the framework of Density Functional Theory (Sec. \ref{sec:th-dft})
%and his Time Dependent generalization (Sec. \ref{sec:th-tddft}).
%Moreover we will conclude this Chapter underlying the
%introducing the dielectric function and the optical conductivity,
%the quantities that are directly related to experiments (Sec. \ref{sec:th-lf}).

 \section{The Schr\"odinger equation for condensed matter systems}
  The non-relativistic time-indipendent Schr\"odinger equation
of a system constisting of N interacting electrons in an external potential
generated by M atomic nuclei is given by:
\begin{equation}
  \widehat{H} \Psi(\br_1, ... , \br_N) = E \Psi(\br_1, ... , \br_N)
 \label{eq:H1}
\end{equation}
where $ \Psi(\br_1, ... , \br_N) $ represents the wave function of the N-electron
many body system. The hamiltonian in eq.~(\ref{eq:H1}) is the sum of four operators:
\begin{eqnarray}
 \widehat{H} &=& \widehat{T}_e + \widehat{V}_{ee} + \widehat{V}_{ei} + \widehat{T}_i + \widehat{V}_{ii}  \\
             &=&  \sum^N_{i=1} \[ -\frac{1}{2}\nabla^2_i + \frac{1}{2}\sum_{i \ne j} \frac{1}{|\br_i -\br_j|}   \] 
                - \sum^M_{\alpha = 1} \frac{Z_{\alpha}}{|\br_i-\mathbf{R}_{\alpha}|} 
                + \sum^M_{\alpha = 1} \[-\frac{1}{2}\nabla^2_{\alpha}
                + \frac{1}{2}\sum^M_{\beta \ne \alpha}\frac{Z_{\alpha}Z_{\beta}}{R_{\alpha\beta}}\] \nonumber
 \label{eq:H}
\end{eqnarray}
where we assumed atomi units $\hbar=m=e=a_0=1$.
The terms in eq.~(\ref{eq:H}) are associated to the kinetic energy and the 
Coulomb interaction of electrons ($T_e$ and $V_{ee}$ respectively) 
and nuclei ($T_i$ and $V_{ii}$ respectively),
and to the potential energy of the electrons in the field of 
the nuclei $V_{ei}$. Z$_{\alpha(\beta)}$ are the atomic numbers of the elements 
involved and R$_{\alpha\beta}$ are the nuclei distances.\\
Several approaches can be adopted in order to solve eq.~(\ref{eq:H}).\\ 
First, the \emph{pseudopotental approach} assumed in this thesis,
simplifies the treatment of the problem reducing the number of the
active electrons just to the valence ones and describing for each atom 
the joint effect of the nucleus and the core electrons with a suitable 
potential. For this reason in the following we will refer to ions 
instead of nuclei in the previous treatment. \\
\indent Secondly, ions and electrons masses are extremely differents
(M$_i >>$ m$_e$) determining different time scale motions.
Assuming that ions are allowed to move adiabatically in the 
field of the electrons ground state, the problem can be treated
perturbatively within the Born Oppenheimer approximation.
Hence, writing the wavefunction as:
\begin{eqnarray}
 \Psi(\{r\},\{R\}) &=& \psi_{\{R\}}(\{r\}) \phi(\{R\}) \\
 \textrm{where}\,\,\,\,\{r\} &=& \{\br_1,...,\br_N\} \nonumber\\
 \textrm{and} \,\,\,\, \{R\} &=& \{\mathbf{R}_1,...,\mathbf{R}_M\} \nonumber
\end{eqnarray}
it is possible to decouple the hamiltonian separating the ionic 
and electronic part:
\begin{eqnarray}
 \[T_e + V_{ee} + V_{ei}\] \psi_{\{R\}}(\{r\}) &=& E^{n}_e(\{R\})\psi_{\{R\}}(\{r\}) \label{eq:He} \\
 \[T_i + V_{ii} + E^{n}_e(\{R\}) \]\phi(\{R\})  &=& E_{tot}\phi(\{R\})  \label{eq:Hi}
\end{eqnarray}
where label $i$ refers to ions and $E^{n}_e(\{R\})$ 
in eq.~(\ref{eq:He}) represents the electronic 
contribution to the potential energy, i.e. the glue for the nuclei, in fact 
without this attractive term, the system would not be bonded.
Conversely in eq.~(\ref{eq:Hi}) ions contribute to the potential $V_{ei}$
and are seen by the electrons as fixed point charges. \\
\indent In conclusion, taking in account the approximations assumed, 
the Eq.~\ref{eq:H} reduce to the eigenvalues problem of the operator:
\begin{equation}
 H = - \sum^N_{i=1} \frac{1}{2}\nabla^2_i
     + \sum_{i<j} v(\br_i,\br_j)
     - \sum^N_{i=1} v(\br_i) 
\end{equation}
where the first term is the kinetic energy of electrons, the third 
is the external potential due to the ions in which the electons are immersed, 
the second term represents the complexity of the problem, it describes the 
interaction between electrons and prevents the decoupling of the equation in 
N one particle equations.

 \section{The variational formulation}
  If we consider a time--independent Hamiltonian, as described
in the previous section, and we assume that periodic boundary conditions
are applied, the spectrum of eigenvalues and eigenfunctions is
discrete. For an arbitrary function $\Psi$ with non vanishing norm, 
we can define the quantity:
\begin{equation}
 E\[\Psi\] = \frac{\la \Psi|\widehat{H}|\Psi\ra}{\la\Psi|\Psi\ra}
\label{eq:var}
\end{equation}
It is then possible to prove that the Schr\"odinger 
equation $\widehat{H}|\Psi\ra=E|\Psi\ra$ is equivalent 
to the variational principle $\delta E\[\Psi\] = 0$. \\
In fact, taking the variation of eq.~(\ref{eq:var}) we obtain:
\begin{eqnarray}
 \delta(E\[\Psi\]\la\Psi|\Psi\ra) &=& \delta E\[\Psi\]\la\Psi|\Psi\ra + E\[\Psi\]\la\delta \Psi|\Psi\ra + E\[\Psi\]\la\Psi|\delta \Psi\ra \nonumber\\
                              &=& \la\delta \Psi|\widehat{H}|\Psi\ra  + \la\Psi|\widehat{H}|\delta \Psi\ra.
\end{eqnarray}
and considering that $\widehat{H}$ is hermitian we can conclude that: 
\begin{eqnarray}
 \delta E\[\Psi\] = 0 &\Leftrightarrow& \la\delta \Psi|\widehat{H}-E|\Psi\ra + \la\Psi|\widehat{H}-E|\delta \Psi\ra= 0  \nonumber\\
                      &\Leftrightarrow& \(\widehat{H}-E\[\Psi\]\)|\Psi\ra= 0 
\end{eqnarray}
The importance of the functional defined in eq.~(\ref{eq:var}) can be seen 
expanding $\Psi$ over the wavefunctions $\{\psi_n\}$:
\begin{eqnarray}
 \Psi = \sum_n c_n\psi_n &\Rightarrow& E\[\Psi\] = \frac{\sum_n |c_n|^2E_n}{\sum_n|c_n|^2} \nonumber \\
                         &\Rightarrow& E\[\Psi\]-E_0 = \frac{\sum_n|c_n|^2(E_n-E_0)}{\sum_n|c_n|^2}
\end{eqnarray}
where $E_0 = \textrm{min}\{E_n\}$ is the ground state energy. \\
Hence we can conclude that:
\begin{equation}
 E\[\Psi\] \ge E_0 \,\,\, ; \quad \Psi =\alpha \Psi_0\,\, \Leftrightarrow \,\, E\[\Psi\] = E_0
 \label{eq:min}
\end{equation}
i.e. the ground state energy $E_0$ is implicitly 
defined by the minimization~(\ref{eq:min}).

 \section{Density functional theory}
  Density Functional Theory (DFT) is a successful tool
largely used in order to study ground state properties of
condensed matter systems.
Within this theory the problem of solving the Schr\"odinger
equation for the ground state can be exactly recast into the 
variational problem of minimizing a functional
with rispect to the charge density.
The complexity of the problem is reduced in principle from having to deal
with a function of 3N variable to one, the density, 
that depends only on the 3 spatial coordinates.
In fact, the key quantity of the theory is the electronic
density $\rho(\br)$ that, respect to the many--body 
wavefunction $\Psi(\br_1,...,\br_N)$, is a real quantity 
and has an intuitive physical interpretation.
A review of the topic can be found in literature~\cite{DFT,DFT2,ORR02,pastori_book},
in this section we will review briefly the most important milestones.
\label{sec:th-dft}
   \subsection{Hohenberg-Kohn theorem}
    The essential role that is played by the charge density in the
search for the electronic ground state was pointed out 
for the first time by Hohenberg and Kohn~\cite{DFT2}.\\
Let us consider a system of N interacting
electrons immersed in an external potential $V_{\textrm{ext}}$
with hamiltonian:
\begin{equation}
 \widehat{H} = \widehat{H}_{\textrm{int}} + \widehat{V}_{\textrm{ext}}
\end{equation}
in particular $H_{\textrm{int}}=T_e+V_{ee}$ and the external potential 
$V_{\textrm{ext}}$ is due to the interaction, for example, 
between electrons and ions.
Assuming that the ground state is not degenerate, 
the first part of the Hohenberg--Kohn theorem asserts that for every density $\rho(\br)$
V-representable\footnote{A density is V-representable
if it is positive defined, normalized to a number N, and such that there
exists an external potential $V(\br)$ for which there is a non-degenerate
ground state corresponding to that density.}, 
the external potential $V_{\textrm{ext}}$ is a functional of the charge density 
$V_{\textrm{ext}}=V_{\textrm{ext}}[\rho(\br)]$, 
within an additive constant.\\
\indent Let us assume, \emph{ad absurdum}, that there exists a different
potential $V^{\prime}_{\textrm{ext}}$ with a ground state $\Psi^{\prime}$ 
corresponding to the same ground state density $\rho(\br)$.
If E and E$^{\prime}$ are the respective ground state energies we can 
write:
\begin{eqnarray}
 E < \langle \Psi^{\prime}|\widehat{H}| \Psi^{\prime} \rangle &=&
     \langle \Psi^{\prime}|\widehat{H}^{\prime}| \Psi^{\prime} \rangle + 
     \langle \Psi^{\prime}|\widehat{H}-\widehat{H}^{\prime}|\Psi^{\prime}\rangle \nonumber \\
   &=& E^{\prime} + \int\rho(\br) [V_{\textrm{ext}}(\br) -V^{\prime}_{\textrm{ext}}(\br)]d\br
 \label{eq:eq1}
\end{eqnarray}
A similar equation can be written in case of 
$\langle \Psi|\widehat{H}^{\prime}| \Psi \rangle$, in fact:
\begin{eqnarray}
 E^{\prime} < \langle \Psi|\widehat{H}^{\prime}| \Psi \rangle &=&
     \langle \Psi|\widehat{H}| \Psi \rangle + 
     \langle \Psi|\widehat{H}^{\prime}-\widehat{H}|\Psi \rangle \nonumber \\
   &=& E - \int\rho(\br) [V_{\textrm{ext}}(\br) -V^{\prime}_{\textrm{ext}}(\br)]d\br
 \label{eq:eq2}
\end{eqnarray}
and now adding eq.~(\ref{eq:eq1}) to eq.~(\ref{eq:eq2}):
\begin{eqnarray}
 E + E^{\prime} < E + E^{\prime}
\end{eqnarray}
that is the \emph{absurdum}. \\
Therefore, the theorem establishes the legitimacy of the
charge density as the fundamental variable in the electronic 
problem, demonstrating a one-to-one correspondence between the
density and the external potential $\rho = \rho[V_{ext}]$.
Hence, this relation is invertible so the external potential 
can be viewed as a functional of the density $V_{ext} = V_{ext}[\rho]$. \\
Moreover, since the ground state energy 
is a function of the external potential $E_0 = E_0[V_{ext}]$,
it is now possible to write it as a function 
of the charge density (HK functional):
\begin{eqnarray}
 E^{HK}[\rho(\br)] = T[\rho(\br)] + E_{H}[\rho(\br)] + \int V_{\textrm{ext}} \rho(\br) d\br 
\end{eqnarray}
where $E_{H}[\rho(\br)]$ is the Hartree energy given by:
\begin{equation}
 E_{H}[\rho(\br)] = \frac{1}{2} \int d\br \int d\br' \frac{\rho(\br)\rho(\br')}{|\br -\br'|}
\end{equation}
Once that the existance of the HK functional is established 
the second part of the theorem affirms that the minimum of the 
functional $E^{HK}[\rho(\br)]$ is obtained when the 
charge density $\rho$ is exactly the ground state density 
(\emph{energy variational principle}). \\
In conclusion, the total energy of an N interacting particles system
can be written as a functional of the density. This functional exists, is universal 
and non depending on the form of the external potential, however its analytical 
form is unknown.

   \subsection{The Kohn-Sham equations}
    The Hohenberg-Kohn theorem provides the theoretical justification to 
reformulate the search for the many--body ground state as a varational 
problem on the charge density.
Although the analytical form of the HK functional is unknown, the 
minimization procedure lead to a set on N associated differential 
equations:
\begin{eqnarray}
&&  \delta E^{HK}[\rho] = 
     \delta \left[ T[\rho] + V_{H}[\rho] + \int V_{ext}\rho(\br) d\br 
                - \lambda \( \int \rho(\br) d\br - N\) \right] = 0 \nonumber \\
&& \frac{\delta T[\rho]}{\delta \rho(\br)} + \frac{\delta V_{H}[\rho]}{\delta \rho(\br)} + V_{ext} = \lambda 
\end{eqnarray}
where $V_H$ is the Hatree potential and $\lambda$ are Lagrange multipliers 
required by the normalization constraint. \\ 
The Kohn and Sham approach is based on the introduction of an 
auxiliary non interacting system of N electrons, having the same density of
the real interacting system in a suitable external potential. \\
We can write the ground state density of the interacting system 
expanded on a basis of N independent orthonormals orbitals:
\begin{equation}
 \rho(\br) = \sum_i f_i \phi^*_i(\br)\phi_i(\br) \label{eq:den1}
\end{equation}
where $f_i$ represents the occupation factor of the orbital $i$. \\
Hence, the HK functional can be written in terms of the kinetic energy of the 
non interacting system:
\begin{equation}
 E^{HK}_{V_{ext}}[\rho] = T_0 + E_{H}[\rho] + \int V_{ext} \rho(\br) d\br + E_{xc}[\rho] \label{eq:hk-ks}
\end{equation}
where $T_0$ is the kinetic energy of the non--interacting system:
\begin{eqnarray}
   T_0[\rho] &=& T_0[{\phi_i}] 
              = \sum_i \int \phi^*_i(\br) \[-\frac{\nabla^2}{2}\phi_i(\br) \]d\br 
\end{eqnarray}
$E_H$ is the Hartree energy and $E_{xc}$ is the only unknown term:
\begin{eqnarray}
  E_{xc}[\rho]  &=& E - T_0 - E_{\textrm{H}} - V_{\textrm{ext}}.   
 \label{eq:xc-ks}
\end{eqnarray}
In particular, the $E_{xc}$ term contains the contributions 
given by the difference in the kinetic energy of the interacting 
and non interacting system, i.e. $\Delta T[\rho] = T[\rho] - T_0[\rho]$, 
the exchange effects (Fermi correlation) and the correlation effects (Coulomb correlation). \\
\indent Now, we can calculate the minimum of the HK functional 
(\ref{eq:hk-ks}) for the KS non interacting system in a fixed external 
potential $V_{\textrm{ext}}$ and with the N$\times$N constraints due to the orthonormality 
of the orbitals:
\begin{eqnarray}
  \frac{\delta}{\delta\phi^*}\left[E - \sum_{n,m} \lambda^N_{m,n} 
             \left( \int \phi^*_m\phi_n -\delta_{m,n} \right) \right] = 0 \label{eq:min-ks}
\end{eqnarray}
where:
\begin{eqnarray} 
\frac{\delta}{\delta\phi^*_i} 
      = \frac{\delta \rho}{\delta \phi^*_i} \frac{\delta}{\delta\rho} 
      = \phi_i\frac{\delta}{\delta\rho}
\end{eqnarray}
From eq.~(\ref{eq:min-ks}) we can obtain the following set of equations:
\begin{equation}
  \left[ -\frac{1}{2}\nabla^2 + V^{\textrm{eff}}_{\textrm{DFT}} \right] \phi_i = \lambda_i \phi_i
 \label{eq:reqks}
\end{equation}
where $H^{\textrm{KS}} = -\frac{1}{2}\nabla^2 + V^{\textrm{eff}}_{\textrm{DFT}}$ 
is the Kohn-Sham Hamiltonian and $V^{\textrm{eff}}_{\textrm{DFT}}$ is the sum of 
three contributions:
\begin{equation}
  V^{\textrm{eff}}_{\textrm{DFT}} = V_{\textrm{H}} + V_{\textrm{ext}} +  V_{\textrm{xc}}
\end{equation}
the Hartree potential $V_{\textrm{H}}$, the external potential $V_{\textrm{ext}}$ 
and the exchange-correlation potential given by:
\begin{eqnarray}
  V_{\textrm{xc}} = \frac{\delta E_{xc}[\rho]}{\delta\rho(\br)}
\end{eqnarray}
Now, if we rewrite the density eq.~(\ref{eq:den1}) expanded
over the N occupied orbitals:
\begin{equation}
 \rho(\br) = \sum_i^N |\phi_i(\br)|^2 \label{eq:den2}
\end{equation}
we must solve the set of N one particle equations.
Now, if we assume that $\left\{\phi_i\right\}$ diagonalize 
the N$\times$N hermitian matrix H$^{KS}$:
\begin{equation}
 \lambda_{mn} = \langle \phi_m | H^{KS} | \phi_n \rangle
\end{equation}
we can write N one-particle equations:
\begin{eqnarray}
 \left\{ -\frac{\nabla^2}{2} + V_{\textrm{ext}}(\br) 
    + \int \frac{\rho(\br')}{|\br - \br'|}d\br'  
    + V_{\textrm{xc}}(\br) \right\} \phi_i(\br) = \lambda_i \phi_i(\br)
 \label{eqa:ks}
\end{eqnarray}
where $\lambda_i$ are now interpreted as the KS energies $\epsilon^{KS}_i$. \\
Since the last two terms of the hamiltonian depend on the eigenvectors throught
eq.~(\ref{eq:den2}), the eigenvalues and eigenvectors can be determined 
self consistently. \\
The equations~(\ref{eq:den2}) and~(\ref{eqa:ks}) are called 
\emph{Kohn and Sham equations} and provide a procedure to calculate 
the total ground--state energy of the system:
\begin{equation}
 E = E_{V_{\textrm{ext}}}[\rho_0] = \sum_{i=1}^{N} \epsilon^{KS}_i 
              - E_{H}[\rho_0] + E_{xc}[\rho_0] 
              - \int \rho_0(\br) V_{\textrm{xc}}(\br)d\br
\end{equation}
In conclusion it is worth mentioning that the KS eigenvalues do not have any 
physical meaning, as, for instance, Hartree--Fock eigenvalues that are 
related to real orbital energies via the Koopmans theorem.
However, there exists a number of approximations of the exchange-correlation potential 
(see sections~\ref{sec:dft-LDA},~\ref{sec:dft-LSDA} and~\ref{sec:dft-GGA}) providing good agreements with experimental results 
in many applications. This justifies the practical usefullness of the KS scheme.

 \section{Methods}
   \subsection{Local density approximation} \label{sec:dft-LDA}
    Once the Kohn-Sham scheme is defined there 
still exists the problem of the missing 
analytical representation of the 
exchange--correlation energy. \\
A commonly used approximation is offered by the Local Density
Approximation (LDA)~\cite{LDA}, which makes DFT practically 
applicable to a wide variety of systems and provides 
a correct description of systems in which the 
density varies slowly in space. 
The form of $E_{xc}$ is given by:
\begin{eqnarray}
    E_{xc}^{LDA}[\rho(\br)] = \int \epsilon^{heg}_{xc}
                (\rho(\br))\rho(\br)d\br 
\end{eqnarray}
where the local dependence of $E_{xc}$ on the 
density $\epsilon^{heg}_{xc}(\rho(\br))$ is given
in terms of the exchange--correlation energy 
of the homogeneous electron gas of constant 
density $\rho=\rho(\br)$.
Hence the systems is locally approximated to an homogeneous 
electrons system.
The function $\epsilon^{heg}_{xc}$ can be separated
in an exchange part:
\begin{equation}
 \epsilon_x(\rho) = -\frac{3}{4}
   \left(\frac{3}{\pi}\right)^{1/3} \rho^{1/3}
\end{equation}
and a correlation term $\epsilon_c(\rho)$, a function
that can be obtained 
by Quantum Monte Carlo simulations (QMC), the most popular form has been given for different densities by 
Ceperley and Adler~\cite{ceperley_prl_v45_p566_y1980}. \\
Moreover the exchange--correlation potential can be written as:
\begin{eqnarray} 
    V_{xc}^{LDA}(\br) &=& 
        \frac{\delta E^{LDA}_{xc}}{\delta n(\br)} \nonumber \\
       &=& \epsilon_{xc}(\rho(\br)) + \rho(\br)
        \frac{d \epsilon_{xc}}{d\rho}
  \label{eq:vxc-lda}
\end{eqnarray}
The domain of applicability of the LDA has proved to be 
valid for a large amount of systems, even not homogeneous ones.
However, its results are not appropriate for the case of 
few electron systems (see chapter~\ref{ch:a-molecules}).
For localized systems self--interaction corrections 
(SIC~\cite{perdew_prb_y1981_v23_5048}) are usually used.

   \subsection{Local spin density approximation} \label{sec:dft-LSDA}
    The Local Spin Density Approximation (LSDA) provides a 
generalization of the LDA to the case of spin polarized
calculations.
%The magnetic polarization of the system is defined by:
%\begin{equation}
% \mu(\br) = \mu_B \left(\rho_{\uparrow}(\br) - \rho_{\downarrow}(\br)\right) 
%\end{equation}
%
Let us define the spin polarization parameter $\zeta$:
\begin{equation}
 \zeta = \frac{\rho_{\uparrow}-\rho_{\downarrow}}
              {\rho_{\uparrow}+\rho_{\downarrow}} \,\,\,\,\,\,\,\,\, 0\le \zeta \le 1
\end{equation}
In the limiting case where $\zeta = 0$, $\rho_{\uparrow}=\rho_{\downarrow}$
and we will recover the LDA for unpolarized systems (U), 
conversely, if $\zeta = 1$ the system is completely
spin polarized (P) and it is possible to write 
the following parametrizations (see Ref.~\cite{perdew_prb_y1981_v23_5048}):
\begin{eqnarray}
  \epsilon^{P}_{x}(\rho) &=& 2^{1/3} \epsilon^{U}_{x}(\rho) \\
  \epsilon^{P}_{c}(\rho) &=& \frac{1}{2} \epsilon^{U}_{c}(2^{4/9}\rho) 
\end{eqnarray}
for the exchange and correlation part, respectively.
In the intermediate cases the parametrization for 
$\epsilon_{xc}$ is given by:
\begin{equation}
 \epsilon_{xc}(\rho) = \epsilon^{U}_{xc}(\rho) 
        + \mathit{f}(\zeta)\left[\epsilon^P_{xc}(\rho) - \epsilon^U_{xc}(\rho) \right]
\end{equation}
where the smooth interpolation function 
$\mathit{f}(\zeta)$ is defined by:
\begin{equation}
  \mathit{f}(\zeta)= \frac{(1+\zeta)^{4/3}+(1-\zeta)^{4/3}-2}{2^{4/3}-2}
\end{equation}

   \subsection{Generalized gradient approximation} \label{sec:dft-GGA}
    A natural way to improve the LDA in order to account 
for the inhomogeneities of the density
is to make a gradient expansion of the
exchange-correlation energy with respect
to the density.
In this way $\epsilon_{xc}$ results to 
be dependent on the local derivative
of the density:
\begin{equation}
 E^{GGA}_{xc}[\rho_{\uparrow},\rho_{\downarrow}] =
  \int \mathit{f}(\rho_{\uparrow},\rho_{\downarrow}, 
                               \nabla\rho_{\uparrow},
                               \nabla\rho_{\downarrow} ) d\br  
\end{equation}
This is the so called Generalized Gradient 
Approximation (GGA), often used in terms of the 
Perdew-Burke-Ernzerhof (PBE)~\cite{perdew_prb_54_1996,PBE}
parametrization. \\
The GGA improves the LDA with respect 
to some applications (molecules or systems with
strongly inhomogeneous density distribution) 
but it does not offer a systematic advance in the 
DFT calculation tools.
 
   \subsection{Brief review of pseudopotential method}
    Pseudopotential approach treats an all-electron
variational calculation of ground state properties in terms of the only valence
wavefunctions immersed in a modified potential.
In this way, core states, being the most localized and expensives to be represented,
are not directly included in the calculations:
their effect on valence electrons is described by a suitable \emph{pseudo}potential.
A review of the topic can be found in the literature,
ranging from the most influential works~\cite{phillips_pr_v116_y1959_p287, HSC79,K80,BHS82,kleinman_prl_v48_y1982_p1425,vanderbildt_prb_v41_y1990_p7892,blochl_prb_v41_y1990_p5414,gonze_prb_v44_y1991_p8503} 
to other important but less fundamental ones~\cite{kleinman_prb_v21_y1980_p2630,louie_prb_v26_y1982_p1738,
troullier_prb_v43_y1991_p1993,hamann_prb_v40_y1989_p2980}. \\
In the following we briefly summarize the most important steps of the method.
All-electron valence orbitals can be represented as a linear combination
of core orbitals $|\psi_c \rangle$ and a smooth function $|\phi^P_v\ra$:
\begin{equation}
 |\psi_v\ra = |\phi^P_v\ra + \sum_c \alpha_{cv}|\psi_c\ra
 \label{eq:psiv}
\end{equation}
where $\alpha_{cv}=\la\psi_{c'}|\phi^P_v\ra$
are coefficients that guarantee the core-valence orthogonality.\\
By inverting eq.~(\ref{eq:psiv}) with respect to $|\phi^P_v\ra$, 
it is possible to write valence pseudo wavefunction in terms of 
all-electron core and valence states.
Then, applying the Hamiltonian to $|\phi^P_v\ra$ it is possible to show that
they are eigenstates of a modified hamiltonian with the same eigenstates of the
all-electron wavefunctions:
\begin{equation}
    \left[\widehat{H}+\sum_c(\epsilon_v - \epsilon_c)|\psi_c\ra\la\psi_c|)\right]
      |\phi^P_v\ra = \epsilon_v|\phi^P_v\ra
 \label{eq:Hmod}
\end{equation}
The projector defined in eq.~(\ref{eq:Hmod}) by 
$\widehat{\mathcal{P}}=\sum_c(\epsilon_v -\epsilon_c)|\psi_c\ra\la\psi_c|$
is not local. Moreover, because $\la \phi_v|\widehat{\mathcal{P}}|\phi_v\ra$
is positive defined it represents a repulsive and short range potential,
as it should be to correctly describe core orbitals. \\
\indent In the general scheme, norm conserving pseudopotentials are derived
from an atomic reference state requiring that \emph{pseudo} and
all-electron valence eigenstates have the same
energies and density outside a chosen core cutoff radius.
Normalization of the pseudo orbitals guarantees that
they include the same amount of charge in the core region.
Futhermore \emph{pseudo} and all-electron logarithmic
derivatives agree, at the reference energies, beyond
the cutoff radius. Finally, norm conservation ensures
that the \emph{pseudo} and all-electron logarithmic derivatives
agree also around each reference level to first
order in the energy. \\% 
\indent In this way a pseudopotential exhibits the same scattering
properties as an all--electron potential in a neighboorhood of the
atomic eigenvalues~\cite{FS99}.
This property provides a measure of the transferability of
the pseudopotential.

 \section{Time dependent density functional theory}
    Density functional theory is a successful tool for a large
range of applications, however some limits can be underlined.\\
First, DFT is a ground state theory and it is not obvious how to 
generalize the KS eigenvalues in order to represent the quasi particle 
energies\footnote{For instance, the direct interpretation of the 
KS eigenvalues as the quasiparticle energies of the system leads 
to the understimation of the bandgap of semiconductors.}. 
Secondly, DFT is a theory dealing with stationary states, hence it is
not possible to apply it to the case of time--dependent hamiltonians. \\
\indent Part of those limits are overcomed by the Time Dependent Density Functional Theory (TDDFT) 
that is an exact reformulation of time dependent quantum mechanics 
where the fundamental variable is the time--dependent electronic
density $\rho(\br,t)$ instead of the many--body function of the system.
The first milestone is the Runge--Gross theorem~\cite{TDDFT}
that provide a generalization of the Hohenberg--Kohn theorem
to time dependent densities. The theorem states that there exists 
a one--to--one correspondence between the time dependent external 
potential $W(t)$ and the time dependent density of an evolving 
system at a fixed initial state:
\begin{equation}
    W(t) \leftrightarrow \rho(\br,t)
\end{equation}
If we consider the Hamiltonian describing an N-electrons system
given by:
\begin{equation}
 \widehat{H}(t) = \widehat{T} + \widehat{V} + \widehat{W}(t)
\end{equation}
where, beyond the kinetic and the coulombian term ($T$ and $V$ respectively), 
a time--dependent external potential 
$W(t) = \sum_i V_{\textrm{ext}}(\br,t)$ appears 
that can be expandend around an initial time $t_0$ 
such as $V_{\textrm{ext}}(\br,t_0) = V_{\textrm{ext}}(\br)$.\\
The time evolution of the system is described by the 
Schr\"odinger equation:
\begin{eqnarray}
 H(t)\psi(t) = i\frac{\partial}{\partial t}\psi(t)
\end{eqnarray}
Two time dependent densities $\rho(\br,t)$ and $\rho'(\br,t)$,
having a commun initial state $\psi(t_0)=\psi_0$ 
and influenced by two different external potentials 
$V_{\textrm{ext}}$ and $V'_{\textrm{ext}}$, expandable
around $t_0$ and such as $V'_{\textrm{ext}} \ne V_{\textrm{ext}} + c(t)$,
are always different.
Hence $\rho(\br,t)$ determines the external potential but
for a time dependent function $c(t)$. Conversely the potential 
fixes the density but for a time dependent phase:
\begin{equation}
 \psi(t) = e^{-i\alpha(t)}\psi\[\rho,\psi_0\](t)
\end{equation}
Hence, for every time--dependent observable $\widehat{O}(t)$ 
that is not depending on time derivative or time integral, 
is a functional of the density:
\begin{equation}
 \la \psi(t) |\widehat{O}(t)|\psi(t) \ra = O\[\rho\](t)
\end{equation}
From the Runge--Kohn theorem it is straightforward 
to build the Kohn--Sham scheme for the time dependent case 
(see Refs.~\cite{TDDFT2} and~\cite{tddft_book}).
We can write the action:
\begin{equation}
  A[\psi] = \int^{t_1}_{t_0} dt \la \psi(t)| 
              i\frac{\partial}{\partial t} - \widehat{H}(t)|
            \psi(t) \ra 
   \label{eq:action}
\end{equation}
where $\psi$ is the many--body wavefunction with initial 
condition $\psi(t_0) = \psi_0$.
The time--dependent Schr\"odinger equation corresponds to 
a stationary point of $A[\psi]$ similarly 
to classical mechanics, where the trajectory is a stationary point
of the action $\mathit{A} = \int^{t_1}_{t_0}\mathit{L}(t)dt$ 
with $\mathit{L}$ the Lagrangian of the system.
The action in eq.~(\ref{eq:action}) is a functional of the 
density and has a stationary point corresponding
to the correct $\rho(\br,t)$, i.e. solving the Euler equations:
\begin{equation}
 \frac{\delta A[\rho]}{\delta \rho(\br,t)} = 0
\end{equation}
it is possible to recover the density. \\
Similarly to the static case, we can write:
\begin{equation}
  A[\rho] = B[\rho] - \int^{t_1}_{t_0} dt \int d\br \rho(\br,t)V_{ext}(\br,t)
 \label{eq:aro}
\end{equation}
where B is a universal functional given by:
\begin{equation}
  B[\rho] = \int^{t_1}_{t_0} dt \la \psi(t)|
              i\frac{\partial}{\partial t} - \widehat{T} - \widehat{V}| \psi(t) \ra
   \label{eq:action}
\end{equation}
Now, an auxiliary non interacting system can be associated to the 
interacting one in a similar way than to the Kohn--Sham scheme.
The stationary condition can be applied to eq.~(\ref{eq:aro}) with
the condition $\rho(\br,t) = \sum_i|\phi_i(\br,t)|^2$ in order to 
obtain the time dependent KS equations:
\begin{equation}
 \[-\frac{1}{2}\nabla^2 + V_{ext}(\br,t) + 
    \int V(\br,\br')\rho(\br',t)d\br' + 
    V_{xc}(\br,t) \] \phi_i(\br,t) = i\frac{\partial}{\partial t}\phi_i(\br,t)
 \label{eq:vedd}
\end{equation}
In eq.~(\ref{eq:vedd}) it is possible to recognize three contributions
to the effective potential:
\begin{equation}
  V_{\textrm{eff}}(\br,t)=V_{\textrm{H}}(\br,t)+V_{\textrm{ext}}(\br,t)+V_{\textrm{xc}}(\br,t)
\end{equation}
the Hartree and the external potential ($V_{\textrm{H}}$ and $V_{\textrm{ext}}$
respectively) and the exchange--correlation potential defined by
the functional derivative:
\begin{equation}
   V_{\textrm{xc}}(\br,t) = \frac{\delta A_{\textrm{xc}} }{\delta \rho(\br,t)}
  \label{eq:vx}
\end{equation}
where $A_{\textrm{xc}}$ is the exchange--correlation part 
of the action~(\ref{eq:action}).

 \section{Linear response}
    Within the TDDFT framework we can calculate the linear response 
of an N particle system to an external time dependent perturbation.
The response will be related the excited states of the system and can be 
defined as the variation of the density with respect to 
the variation of the time dependent external potential causing the 
perturbation:
\begin{equation}
 \chi(\br,t , \br',t') = 
   \frac{\delta \rho(\br,t)}{\delta V_{\textrm{ext}}(\br',t') }
    \vert_{V_{\textrm{ext}=0}}
 \label{eq:chi1}
\end{equation}
Similarly, the linear response in the case of the auxiliary non interacting 
KS system can be expressed by:
\begin{equation}
 \chi^{0}(\br , t , \br', t') = 
   \frac{\delta \rho(\br,t)}{\delta V_{\textrm{eff}}(\br',t')}|_{V_{\textrm{eff}=0}}
 \label{eq:chi2}
\end{equation}
where the functional derivatives are calculated at the first order in $V_{\textrm{ext}}$
in eq.~(\ref{eq:chi1}) and in $V_{\textrm{eff}} $ in eq.~(\ref{eq:chi2}). \\
Now, using the following relation:
\begin{equation}
 \frac{\delta \rho}{\delta V_{\textrm{ext}}} = 
   \frac{\delta \rho}{\delta V_{\textrm{eff}}}\frac{\delta V_{\textrm{eff}}}{\delta V_{\textrm{ext}}} =
   \chi^0\frac{\delta V_{\textrm{eff}}}{\delta V_{\textrm{ext}}} 
 \label{eq:az1}
\end{equation}
we can write:
\begin{equation}
 \frac{\delta V_{\textrm{eff}}(\br,t)}{\delta V_{\textrm{ext}}(\br',t')} = \delta(\br-\br')\delta(t-t') + 
       \int \[ \frac{\delta(t-t'')}{|\br - \br''|} + f_{xc}(\br,t,\br'',t'') \] \chi(\br'',t'',r',t')d\br''dt''
 \label{eq:az2}
\end{equation}
where:
\begin{equation}
 f_{xc}(\br,t,\br',t') = \frac{\delta V_{xc}[\rho(\br,t)]}{\rho(\br',t')}|_{V_{\textrm{ext}} = 0}
\end{equation}
is the exchange--correlation kernel, the quantity that contains the core of the 
complexity of the problem.
Combining eq.~(\ref{eq:az1}) and~(\ref{eq:az2}) 
it is possible to write a Dyson equation for $\chi$ and $\chi^0$:
\begin{equation}
 \chi(\br,\br',\omega) =  \chi^0(\br,\br',\omega) + \int d\br_1 d\br_2 \chi^0(\br,\br_1,\omega) 
    \[\frac{1}{|\br_1 - \br_2|} + f_{xc}(\br_1,\br_2,\omega) \]\chi(\br_2,\br',\omega)
\end{equation}
The analytical form of the exchange--correlation kernel is unknown, 
for this reason, the solution of this integral equation is not trivial.\\
In the case of $f_{xc} = 0$ the approximation is called the independent particle
random phase approximation (IP-RPA) that is equivalent to the 
Hartree theory but with the addition of time dependency.
In this scheme the density fluctuation at the first order is written as:
\begin{equation}
 \rho(\omega) = \int \chi^0(\br,\br',\omega)V_{\textrm{eff}}(\br',\omega)d\br' 
              = \int \chi(\br,\br',\omega)V_{\textrm{ext}}(\br',\omega)d\br'
\end{equation} 
where $\chi^0$ is built using the KS eigenvalues and eigenvectors
calculated with an approximation for the exchange--correlation
potential $V_{\textrm{xc}}$ in the KS hamiltonian. \\
\indent The problem of the efficient evaluation of the response function will be
discussed extensively in chapter~\ref{ch:hilbert}, where a new method for the 
calculation of $\chi^{(0)}$ will be also presented.
For this reason, we postpone to that chapter the details on the analytical form 
of this quantity.

 \subsection{Adiabatic (spin) local density approximation}
    The adiabatic local density approximation (ALDA) furnishes 
a way to compute the excitation energies of a system within the
TDDFT.
Within this approximation the exchange--correlation potential
defined in eq.~(\ref{eq:vx}) is written as:
\begin{equation}
 V_{xc}(\br,t) \simeq 
   \frac{\delta E_{xc}[\rho] }
         {\delta \rho_t(\br) } 
\end{equation}
where the functional derivative is taken respect to the 
instantaneous density in such a way that the exchange--correlation 
energy depends just on the density at a
fixed time\footnote{For this reason, in adiabatic
LDA memory effects are neglected.}. 
By consequence, the exchange--correlation kernel becomes:
\begin{equation}
 f_{xc}(\br,t,\br',t') = 
   \frac{\delta V_{xc}(\br,t)}
        {\delta \rho(\br',t')} \simeq 
 \delta(t-t') \frac{\delta V_{xc}(\br,t) }
        {\delta \rho(\br',t') }
\end{equation}
and using local density approximation, see eq.~(\ref{eq:vxc-lda}), 
we can also write:
\begin{equation}
  f^{ALDA}_{xc}(\br, t, \br', t') = \delta(t-t')
     \delta(\br - \br')\( 
 2\frac{d\epsilon^{heg}_{xc}(\rho) }{d\rho} +
     \rho \frac{d^2\epsilon^{heg}_{xc}(\rho)}{d^2\rho} 
          \)
\end{equation}
Moreover, if we want to include the spin variable (ALSDA)
we obtain the following expression:
\begin{eqnarray}
 f^{ALDA}_{xc}(\br, t, \br', t') 
   &=& \delta(t-t')
     \delta(\br - \br') \\ 
   & & \(
   \frac{d\epsilon^{LSDA}_{xc}(\rho,\zeta)}
        {d\rho_{\uparrow}} + 
   \frac{d\epsilon^{LSDA}_{xc}(\rho,\zeta)}        
        {d\rho_{\downarrow}} +
   \rho \frac{d^2\epsilon^{LSDA}_{xc}(\rho,\zeta)}
        {d\rho_{\uparrow} d\rho_{\downarrow}}
         \)  \nonumber
\end{eqnarray}
where the derivation is defined as:
\begin{eqnarray}
 \frac{d\epsilon(\rho,\zeta)}{d\rho_{\uparrow}} 
    &=& \frac{\partial \epsilon(\rho,\zeta)}{\partial \rho} + \frac{\partial \epsilon(\rho,\zeta)}{\partial \zeta} \nonumber \\
 \frac{d\epsilon(\rho,\zeta)}{d\rho_{\downarrow}} 
    &=& \frac{\partial \epsilon(\rho,\zeta)}{\partial \rho} - \frac{\partial \epsilon(\rho,\zeta)}{\partial \zeta}   \nonumber 
\end{eqnarray}

 \section{Dielectric function}
    The key quantity connecting the theories presented in the 
previous sections and the experimental spectra is the dynamical 
dielectric function $\varepsilon(\br,\br',\omega)$.
When an external perturbing field is applied to the sample, the 
charge density rearrages and an additional potential 
$V_{\textrm{ind}}$ is induced by the polarization of the system.
The total potential, or \emph{screened} potential,
is due to the contribution of the external
and the induced potential:
\begin{equation}
 V_{tot} = V_{ext} + V_{ind}
\end{equation}
and it can be also written in terms
of the dielectric function:
\begin{equation}
 V_{tot} = \int \varepsilon^{-1}(\br,\br')V_{\textrm{ext}}(\br') d\br'
 \label{eq:xeps}
\end{equation}
The dynamical dielectric function can be recovered as:
\begin{equation}
  \varepsilon(\br,\br',\omega) = \delta(\br-\br') 
      - \int d\br'' v(\br-\br') \chi(\br'',\br',\omega)
\end{equation}
where $ v(\br-\br') $ is the bare coulomb interaction.
The dynamical dielectric function $\varepsilon(\omega)$
takes in account the rearrangement of the charge density
presenting hole and charge accumulation
due to the perturbation.
However, the screened potential usually 
is calculated from the external potential inducing 
the polarizabilty of the system (see eq.~(\ref{eq:xeps})).
%\begin{equation}
% V_{\textrm{tot}} = \varepsilon^{-1}V_{\textrm{ext}}
%\end{equation}
Hence, the important microscopic quantity is the 
inverse of the dielectric function that can be written as:
\begin{equation}
  \varepsilon^{-1}(\br,\br',\omega) = \delta(\br-\br') 
      + \int d\br'' v(\br-\br'') \chi(\br'',\br',\omega)
\end{equation}
%In the case of periodic systems it can be useful
%to represent the dielectric function in the reciprocal space
%where $\varepsilon$ is represented as a {\bf G}{\bf G}'-matrix.
%The macroscopic dielectric function is then recovered by the 
%relation:
\indent In conclusion, the response function $\chi$ and the dynamical
dielectric function $\epsilon(\br,\br',\omega)$ represent the 
key ingredients for theoretical spectroscopy.
In the next chapter we will present the connections between $\epsilon$
and the three class of experimental spectroscopies considered in the present
thesis: energy loss, reflectivity and absorption spectra.

\chapter{Surface spectroscopies} \label{ch:surface}
   Every real solid is surrounded by surfaces. 
Moreover the miniaturization of the technological devices requires a better
understanding of mechanisms at the atomistic scale at which surface
effects become important\footnote{In effect if we look at the number of surface
atoms ($N_A(S)$) with respect to the bulk ($N_A(B)$) 
in a 1~cm$^3$ volume cube we can say that
surface effects are negligible because: 
$\frac{N_A(S)}{N_A(V)}=\frac{10^{15}}{10^{23}}=10^{-8}$.
On the contrary in the case of a 100~\AA\ length cube we 
write: $\frac{N_A(S)}{N_A(V)}=\frac{10^{4}}{10^{6}}=10^{-2}$,
hence the surface signals are not negligible anymore.}. \\
From the experimental point of view,
surface atoms are only visible in sensitive techniques
or by studying processes involving
atoms at the surface (crystal growth, adsorption, oxidation, etching, ...). \\
Under normal conditions (atmospheric pressure and room temperature) 
a real surface of a solid is different from an ideal truncated bulk because 
of a reordering of the surface atomic bonds and because prepared surfaces 
are normally very reactive to atoms and molecules in the environment.
From chemisorption to physisorption, all kinds of particle adsorption 
gives rise to an adlayer on the topmost atomic layers of the solid.

Because of this complexity, first principles calculations can be very helpful 
to better understand the physics of such a system.
In this section we will briefly review the significant experiments and 
the theoretical tools devoted to describe surface physics.

 \section{Experimental issues}
   Spectroscopy is a useful tool to get information 
 about the physical nature or geometrical reconstruction of surfaces. 
 Many  high level experimental technologies  has been developed in the 
 last decades in order to create and analyse the surfaces 
 of materials, an exhaustive review of the topic can be found 
 in the literature~\cite{luth_book,zangwill_book,bechstedt_book}.\\
 Here we summarize the highlights to introduce our results presented
 in the followings chapters.

\subsection{Preparation and structural properties}
  In order to get spectroscopic information, a well defined surface 
  has to be prepared on a particular solid, using a special preparation process and 
  under well defined external conditions.\\
  There are several ways to prepare a surface from a crystalline material and they
  can be grouped into three categories:
   (i) \emph{cleavage} (limitated to cleavage planes), 
   (ii) treatment of imperfect and contaminated surfaces by \emph{ion bombardment and 
    thermal annealing} and 
   (iii) \emph{epitaxial growth} of a crystal layer by means of evaporation 
  or molecular beam epitaxy (MBA). 
   In all cases Ultra High Vacuum (UHV), i.e pressure conditions lower than $10^{-8}$Pa
   ($10^{-10}$ torr) are required. \\
  However, despite the great care in preparing surfaces, irregular deviations 
  from perfect smoothness and purity are always present 
  (steps, terraces or in general \emph{surface roughness}) 
  making real surfaces far from the ideal ones.
 
  Surface atoms rearrange with respect to the bulk crystal positions
  because forces acting on the on top atoms differ from interactions
  between atoms inside the volume, and as a results this difference can be enhanced 
  depending on the bounding behaviour of the material.\\
  However, the deviation of atom positions from that of an infinite crystal 
  decreases with increasing distance from the surface.\\
  Hence in our theoretical models we will assume with confidence that positions
  of atoms deep inside the bulk are the same as those in an infinite crystal.
  On the contrary, the distortions of the atomic configuration due to the termination 
  of the crystal, are important close to the surface. \\
  In the case of silicon, the main element considered in this work, when
  a surface is created tetrahedral bonds are broken, and a 
  non negligible atomic rearrangement is expected to destroy the translational
  symmetry of an ideal bulk truncated surface.
  Moreover dangling bonds are usually unstable because rebonding
  lowers the total energy pushing surface atoms closer to form pairs (dimers).
  For this reason we can expect that a silicon surface is a good example of a
  \emph{reconstruction} process.\\
  On the contrary, in the case of materials where chemical bonds are less
  directional (as the case of metals), 
  surfaces are created by \emph{relaxation} of the topmost layers
  along the direction perpendicular to the surface plane. In this case the 
  changes may conserve the translational-symmetry of the bulk.

\begin{figure}
  \centerline{\epsfig{file=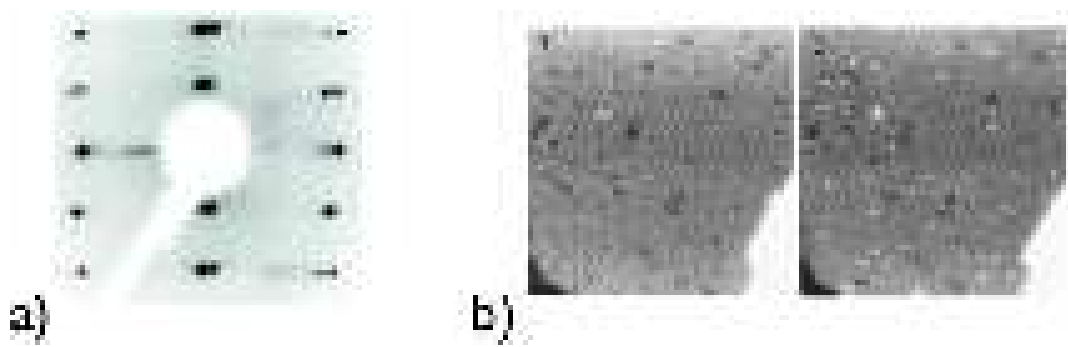,clip=,angle=0,width=14cm}}
  \caption[LEED, STM (experiments)]{
      Examples of experimental images obtained by LEED (left) for a vicinal surface, explaining the 
      double spots in the image, and STM (right). In particular in a) LEED patterns of a Si(001)-($1\times2$) 
      are reproduced from~\cite{power_v67_p115315_y2003} in b) an STM image 
      of the Si(100)-(2$\times$1) surface at 61K, filled (a) and empty (b) states, 
      from~\cite{hemeryck_jcp_v126_p114707_y2007}.
  \label{fig:exp-leed-stm}
 }
\end{figure}

   The main experimental techniques used in the study of surface structure
   exploit the diffraction of neutral atomic beams or electrons.
   With Low-Energy Electron Diffraction (LEED) the surface periodicity and 
   a reconstruction are observed directly 
   via the diffraction pattern, which give an image of the reciprocal lattice.
   The size and shape of the spots contain information about the extention of 
   domains and the presence of surface defects.
   The atomic positions inside the unit cell and the relaxation can be studied
   through the intensity profiles of the diffracted beam, 
   i.e., by plotting the measured intensity of each diffraction spot as a function 
   of the energy of the incoming electron.
   The information of the atomic position is obtained by comparing the experimental
   LEED profiles with those obtained by a theoretical simulation of the electron 
   diffraction in the crystal, where the atomic positions are the input data.
    An example of LEED patterns is reported in Fig.~\ref{fig:exp-leed-stm}.\\
   Reflection High Energy Electron Diffraction (RHEED) 
   is also used, principally to monitor the thin film growth.
   In this technique incident energies of $10-100$keV 
   and incident angles of about $3-5$ degrees are used.

   The structural analysis of LEED is often performed togheter with 
   Auger Electron Spectroscopy (AES) to control the chemical composition of the sample. 
   In AES a beam of electrons with energies beyond 1keV strikes the surface and
   the number of electron backscattered N(E) is analysed as 
   a function of the energy. 
   The $\frac{dN}{dE}$ gives the signature of the elements present in the sample.

   Moreover we mention the light-ion Rutherford backscattering (RBS), 
   where beams of H$^+$ or He$^{2+}$ ions are used with energies of 
   hundreds of keV (LEIS up to 20~keV, MEIS from 20~eV to 200~keV and HEIS to 2~MeV).
   The ions are scattered by the nuclei of the crystal following 
   the dynamics of classical Rutherford scattering and lose energy 
   along straight trajectories through interaction with the electrons. 
   Information on the composition and atomic displacements in the surface layers can be obtained 
   thanks to this method detecting the number of ions as a function of the energy
   and the outcoming directions of ions diffused backward. 

   Another important technique employed to investigate the structural properties
   of semiconductors is Scanning Tunnel Microscopy (STM).
   Due to Binnig and Rohrer (1982),
   this technique gives the local density of occupied and empty
   states integrated over a given energy range around the Fermi level.
   %information about the electronic states in the lattice and
%   providing an image of the corrugation of the last atomic plane of the crystal.
   This technique does not need UHV.
   An example of STM image is reported in Fig.~\ref{fig:exp-leed-stm}.
 
   From all these techniques we can get structural information about
   the system but a comprehensive interpretation of the data must usually be 
   supported by a theoretical description which involves the knowledge of the 
   electronic structure~\cite{note_superfici,bertoni_book}.

\subsection{Electronic properties}
\begin{figure}
      \centerline{\epsfig{file=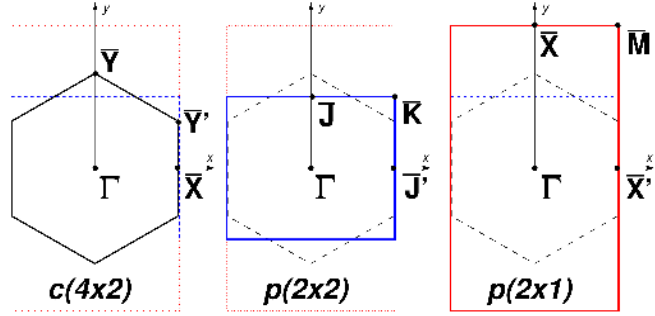,clip=,angle=270,width=8.5cm}}
      \caption[Surface Brillouin Zone]{
              Surface Brillouin Zone for the three surface
              reconstructions that we considered in the present thesis: a c-rectangular, 
              a p-square and a p-rectangular.
        \label{fig:th-SBZ}
         }       
   \end{figure}

   In translationally invariant systems the wave vector ${\bf k}$ defines a set of
   good quantum numbers for each type of elementary excitation.
   In the case of an ordered surface of a crystal, such a wavevector $\bar{{\bf{k}}}$,
   is restricted to two dimensions (parallel to the surface) because in the third direction
   the system is not translationally invariant anymore.
   The Surface Brillouin Zone (SBZ) becomes 2 dimensional 
   and is defined as the smallest polygon
   in the 2D reciprocal space situated symmetrically 
   with respect to a given lattice point (the origin) 
   and bounded by points $\bar{{\bf{k}}}$, satisfying the equation:
    \begin{equation}
      \bar{{\bf k}} \cdot {\bf g} = \frac{1}{2}|{\bf g}|^2
     \end{equation}
   where ${\bf g}$ is a surface reciprocal lattice vector.
   Figure~\ref{fig:th-SBZ} represents three of the five SBZ, referring 
   to the ones we considered in the next chapters of this thesis:
   p-rectangular, c-rectangular and square. Further details on surface 
   theory can be found in~\cite{bechstedt_book}.
%%%%%%%%%%%%%
%  FIGURA SBZ theory
%%%%%%%%%%%%%
   Among the experimental techniques which allow to inspect directly 
   the band structure and the electronic structure 
   in the 2 dimensional BZ we mention the most fundamental: 
   Photoemission (PES), Angle--Resolved Photoemission (ARPES) 
   and Inverse Photoemission (IPES).

   PES is the most important technique able to give a picture of 
   the Density of States (DOS) at the upper atomic planes.
   The physics behind the PES technique is an application of 
   Einstein's photoelectric effect.
   The sample is exposed to a beam of light inducing photoelectric ionization;
   synchrotron radiation is the ideal isochromatic radiation source.
   The energies of the emitted photoelectrons are characteristic of their original 
   electronic states.
   For solids, photoelectrons can escape only from a depth 
   of the order of nanometers, so that it is the surface layer which is 
   mostly analyzed. 
   PES can be performed with X-ray (XES), E$\simeq$20--150~eV, where it is possible to
   see transitions between surface bands below the edge of the bulk bandgap 
   (the small cross section can be improved using grazing angles)~\cite{zangwill_book}.
%   In the ultraviolet range (UPS), E$>$150eV, where 
%   excitations involve localized core levels and surface empty bands 
%   (large cross section)~\cite{zangwill_book}.
%    According to the range energy of the radiation involved this techique 
%    is called XPS (involving X-rays) or
%    UPS (using Ultra Violet radiation)
%    The energy distribution curve obtained by choosing
%    a photon energy so as to produce photoelectrons in the energy range 
%    where the escape 
%    depth has a minimum value (few~\AA). As bulk and surface states  contribute to it. 
%    The surface features can be identified by removing or modifying them through the absrption 
%    of different atoms at the surface.

   ARPES gives information about the k-dispersion of bands 
   and allows to separate the contributions from bulk and surface states.
   The former connect states with the same 3D ${\bf k}$-vector, the latter involve photoemission
   processes conserving only the component parallel to the surface ${\bf k}_{||}$.
    The detected 2D vector connecting surface states and the continuum 
   can be written as: 
 \begin{eqnarray}
   k^{out}_{||} &=& k^{in}_{||} + {\bf g}   \,\,\,\,\,\,\,\,\,\,\, \forall k_{\perp}
 \end{eqnarray}
   where ${\bf g}$ is a vector of the surface reciprocal lattice.
   Hence, plotting the electron energy as a function of the emission
   angle $\theta$ by:
\begin{eqnarray}
   E_{kin} &=& \frac{\hbar^2}{2m}(k^2_{||}+k^2_{\perp}) \nonumber \\
   k_{||}(\theta)  &=& k_{||} \textrm{sin}\theta = \frac{\sqrt{2mE_{||}}}{\hbar} \textrm{sin}\theta 
\end{eqnarray}
   we can say that the peaks in the energy distribution curve represent
   the initial state of the solid labelled by $k_{||}$.
    
   Inverse photoemission (IPES) allows to detect the energy of the photon emitted 
   when an electron of an external beam of given E and ${\bf k}$ falls into an empty 
   conduction surface band or an image state.
   Even optical adsorption is a useful method  to 
   study the occupied and unoccupied states that in a first 
   approximation can be described by the Joint Density of States (JDOS)
   defined by:
\begin{equation}
  JDOS(\omega) = \int \rho(E) \theta(E_F -E) \rho(E-\hbar\omega)\theta(E+\hbar\omega - E_F)dE
\end{equation}
   Finally STM is used to describe electronic states of the surfaces 
   by introducing a potential difference between the tip and the sample.% (see Ref.~\cite{?}).
   In this way it is possible to have a spatial map of the wave function at different energies 
   for both empty and filled states (see Fig.~\ref{fig:exp-leed-stm}).\\
   In conclusion it is worth mentioning that spectroscopies which study 
   the electronic structure are also an indirect test of the surface atomic structure.
 
\subsection{Reflectivity anisotropy experimental spectroscopy}
   Optical spectroscopy is an important tool to probe surfaces since 
   they allow for \emph{in situ}, non--destructive and real--time 
   measurements. Moreover, material damage or contamination associated with 
   charged particle beams are avoided.\\
   However, since the light penetration and wavelength are much larger than typical surface 
   thicknesses (few~\AA), optical spectroscopy is less sensitive to the surface.

   Nevertheless, a trick can be used in order to resolve the surface signal.
   This is the case of Reflectivity Anisotropy Spectroscopy (RAS)
   and Surface Differential Reflectivity (SDR), optical techinques of great importance
   for detecting transitions between surface states.
   Surface sensitivity is greatly enhanced
   with the use of appropriate conditions which enhance
   the contribution of interband transitions involving surface states~\cite{rds_book}.
   
\begin{figure}
     \begin{center}
    \includegraphics[width=6.5cm,clip=,angle=90]{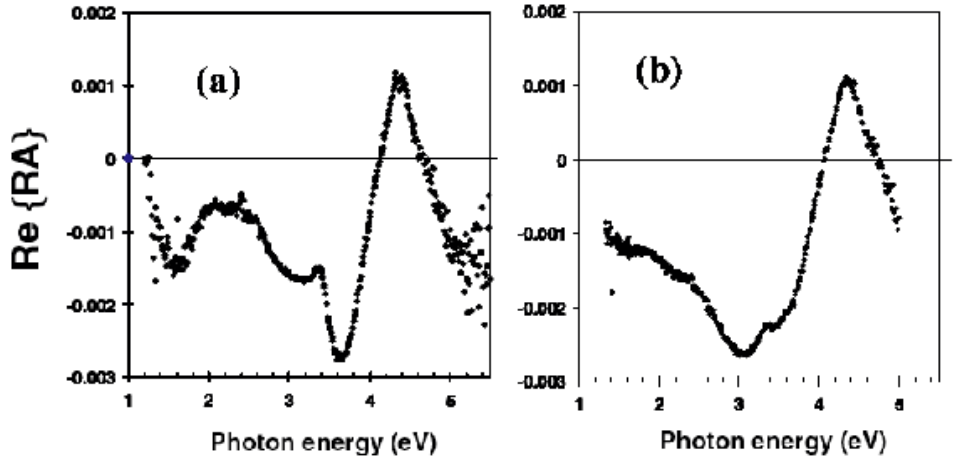}
      \end{center}
   \caption[RAS (experiments)]{
        Experimental RA spectrum of the clean Si(100). We report as an example, the results
        taken from Ref.~\cite{pluchery_pss_y2005_v242_2696} to show how the choice of the 
        surface can affect a RA spectra: Nominal (a) and vicinal (b) surfaces are considered, 
        and it is evident how the low energy spectral features are enhanced in the case of
        the nominal surface.
    \label{fig:ras-exp}
           }
 \end{figure}

   RAS is defined as the difference between the normalized reflectivities 
   measured at normal incidence, for two orthogonal polarizations of light
   belonging to the surface plane:
   \begin{eqnarray} 
     RAS &=& 2\frac{R_y - R_x}{R_y + R_x} \nonumber\\
         &=& 2\frac{(R_0+\Delta R_y) - (R_0+\Delta R_x)}{2R_0+\Delta R_y+\Delta R_x}
   \label{eq:ras1} 
  \end{eqnarray}
   where R$_0$ is the isotropic Fresnel reflectivity.
   Since the bulk of a cubic material is optically isotropic, any reflectivity
   anisotropy must be related to the reduced symmetry of the surface or to 
   another symmetry breaking perturbation, for instance an electric field.
   In the case of $\frac{\Delta R_{\alpha}}{R_0}<<1$ we can write:
   \begin{equation}
      RAS \simeq \frac{\Delta R_y - \Delta R_x}{R_0}
   \end{equation}
    An example of measured RAS is represented in Fig.~\ref{fig:ras-exp} and 
    geometry scattering is shown in Fig.~\ref{fig:ras-geo}.

  On the contrary an SDR spectrum is defined by the difference 
  in the reflectivity measured on a clean surface before and after passivation
  (e.g. by adsorbing atoms or molecules on the surface). Passivation
  (oxidation, the case we discuss in chapter~\ref{ch:oxi}), removes surface
  states but does not affect bulk contributions. One hence obtains, for
  the optical response specific to the surface:
    \begin{equation}
      SDR = \frac{\Delta R_{clean} - \Delta R_{pass}}{R_{clean}}
      \label{eq:SDR}
    \end{equation}
 
  In conclusion, it is important to mention that all these techinques 
   can be appreciably sensitive to the experimental definition of the surface 
   in the sense that in some cases (e.g. Si(100) and Si(100):O,
   as treated in this work: see chapters~\ref{ch:si100,ch:oxi}) 
   the RA signal is modified because of the presence of steps, 
   terrace or different oriented domains.
   The influence of steps on the reflectance spectra has been analysed 
   by Jaloviar\ea\ \cite{jaloviar_prl_v82_p791_y1999}; hence
   Shioda and der Weide~\cite{shioda_prb_y1998_v57_R6823} use highly 
   oriented surfaces (with terraces 1000 times larger than vicinal surfaces) 
   in order to obtain more accurate RA profiles.
   Finally, a comparison of RA spectra obtained by nominal and vicinal surfaces
   is shown as an example in Fig.~\ref{fig:ras-exp} where
   we reproduce data from Ref.~\cite{pluchery_pss_y2005_v242_2696} to 
   illustrate how the use of nominal surfaces improves the
   spectral resolution in the low energy region of the spectrum.

\subsection{Electron energy loss spectroscopy at surfaces}
     A natural complement to optical spectroscopy is 
       Electron Energy Loss Spectroscopy (EELS)
       which, despite some complications in the interpretation 
       of the data, turns out to be surface sensitive probe, 
       particularly in High Resolution EELS (HREELS), 
       which uses low energy incoming beams.\\
\begin{figure}[h!]
\begin{minipage}{5cm}
 \includegraphics[width=7.5cm]{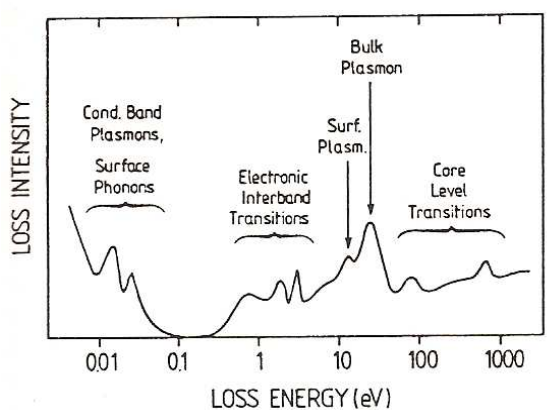}
\end{minipage}
\hspace{4cm}
\begin{minipage}{5cm}
  \caption[EELS excitations]{ Schematic representation of all kind of excitations
            that can be detected by EEL spectroscopy. This figure is reproduced from~\cite{luth_book}
    \label{fig:luth-eels}
    }
\end{minipage}
\end{figure}

     In an EEL experiment a material is exposed to a beam of electrons 
     with a defined narrow range kinetic energy. 
     Some of the electrons undergo inelastic scattering, 
     losing part of their energy and having their paths 
     slightly and randomly deflected from the specular direction. 
     The amount of energy loss can be measured via an electron spectrometer
     and interpreted in terms of excitations of the sample.
     Inelastic interactions include phonon excitations, 
     inter and intra-band transitions,
     plasmon excitations, and inner shell ionizations (see Fig.~\ref{fig:luth-eels}).

     Usually EELS is performed in transmission, but to be surface
     sensitive it must be applied in a reflection geometry (REELS), 
     see Fig.~\ref{fig:reels-geo}, 
     and use relatively low incident energies (around 50--100~eV, versus 1~keV for bulk).
     In HREELS the beam is highly monochromatic and the energies of the electrons range up to 
     few eV. %10~meV is the energy resolution.

 \section{Theoretical surface spectroscopies}
  A real surface is a complex physical system
 whose geometry, i.e. atomic positions, is generally unknown and 
usually involve many degrees of freedom.
This makes calculations very heavy and can create 
serious obstacles when fully treating excited states.
For this reason calculations are usually performed at different
levels of sophistication, involving various simplifications and 
approximations, according to the accuracy required and the numerical
heaviness. \\ 
We assume to be able to calculate the dielectric tensor 
of a general bulk system (as discussed in the previous chapters) 
and we are going to describe how to use it in order to reproduce 
and predict surface spectroscopic experiements.

\subsection{The slab method}
\begin{figure}[h!]
\begin{center}
 \includegraphics[width=10.5cm]{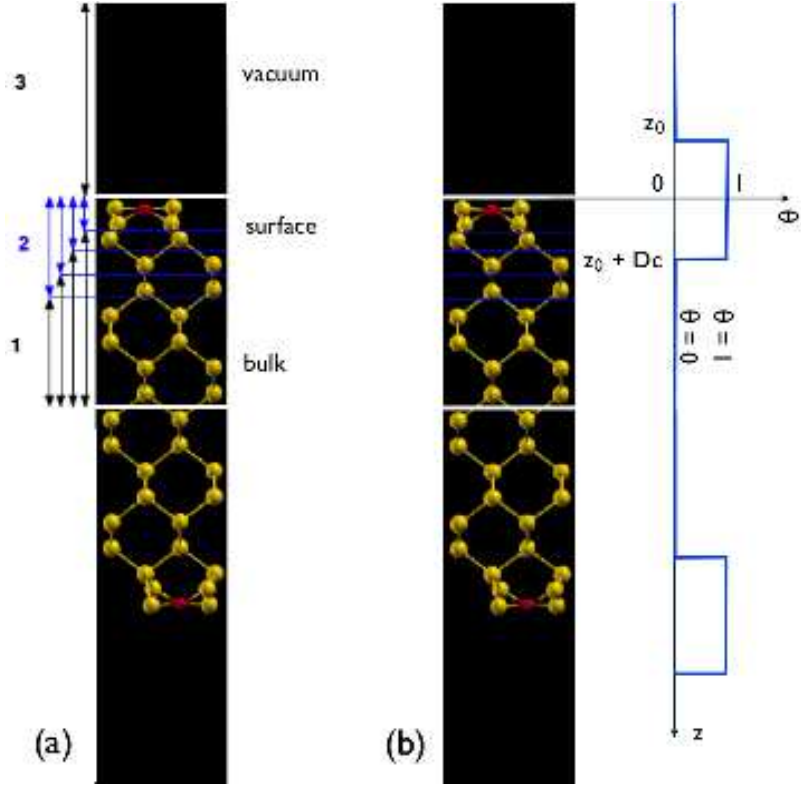}
\end{center}
\caption[Three Layers Model]{\label{fig:3LM}
    Schematic representation of an oxidised silicon surface whithin the supercell 
    approach used in our calculations. 4 oxygen atoms (red circles) and 64 silicon 
    atoms (yellow circles) form the symmetric slab.
    The supercell is the union of the 16 layer slab and the vacuum region.
    In figure (a), the bulk, surface and vacuum regions of the three-layer 
    model are indicated for the upper half of the slab.
    On Fig. (b) blue line represents an example of a cutoff function
    used in the real-space slicing technique.   
}
\end{figure}

The description of the crystal termination is solved here using 
the slab method, i.e. representing the surface by means of an atomic
slab of suitable thickness (usually 20-30~\AA).
Using plane-wave basis sets, the three dimensional periodicity
of the system can be recovered by considering repeated slabs, separated by a
sufficiently large region of empty space.\\
In Figure~\ref{fig:3LM} we illustrate the example of a slab inside 
a supercell and indicate the three layers involved (giving
the name to the three layers model):
a bulk region (composed by the inner atoms of the slab), 
a surface layer with thickness $d$ (top layers of the slab),
and a vacuum volume.
The thickness of the surface layer must be smaller than the wavelength 
$\lambda$ of the light.
The bulk properties are assumed to be described by an isotropic 
dielectric function $\varepsilon_b(\omega)$ and the surface 
is described by a frequency dependent dielectric tensor, 
where the complex diagonal elements are defined by 
$\varepsilon_{xx}(\omega)$, $\varepsilon_{yy}(\omega)$ and $\varepsilon_{zz}(\omega)$.

In practical calculations, the vacuum region is chosen large enough 
to avoid the interaction between the two surfaces of the slab and 
careful convergence tests have to be performed.
Figure~\ref{fig:3LM} shows the case of a symmetric 
slab geometry describing an oxidised silicon surface.
The slab method is general; also non symmetric slabs 
can be used, even if the case is not treated in this thesis.
 
\subsection{Real--Space slicing technique}
Within the description of the three-layers model, an electron
impinging on the surface feels the potential from this
surface layer through its dielectric function $\varepsilon_s$
as well as the potential of the bulk region. However, microscopic
calculations generally output the dielectric function of the 
supercell $\varepsilon_c$. In previous works~\cite{dels-moch-bar-91},
$\varepsilon_s$ was extracted from $\varepsilon_c$ by using 
the expression:
\begin{equation}
 I(\omega) = (N_b - 2N_s) d_l\varepsilon_b(\omega) + 2N_sd_l\varepsilon_s(\omega)
\end{equation}
Here $d_l$ is the interlayer spacing, $N_s$ ($N_b$) is the number
of layers in each surface (bulk) region, and $I$ is the integral 
of the slab RPA dielectric susceptibility $\varepsilon(\omega,z,z')$
over $z$ and $z'$, i.e., $I(\omega)=N_cd_l\varepsilon_c$. However
this approach cannot always guarantee perfect cancellation of the
bulklike layers in the supercell, and may even lead to unphysical
negative loss features.
A more reliable approach is to extract $\varepsilon_s$ directly
using a \emph{real--space slicing} technique, i.e. by projecting out 
the response of a defined surface layer using a \emph{cut off} function
in real space (for a detailed treatment see Ref.~\cite{hogan_prb_y2003_v68_p035405}).\\
This technique is usually reported in terms of the polarizibility
of the half slab $\alpha^{hs}$, which,
in case of symmetric slabs, is obtained dividing by 2 
the full polarizability $\alpha$:
\begin{equation}
 \textrm{Im}[4\pi\alpha^{hs}_{ii}(\omega)]=\frac{4\pi^2e^2}{m^2\omega^2A}\sum_k \sum_{v,c} |P^i_{ v{\bf k},c{\bf k} }|^2 
                                           \delta(E_{c{\bf k}}-E_{v{\bf k}} - \hbar \omega)
\end{equation}
where $P^i_{ v{\bf k},c{\bf k} }$ are the matrix elements of the 
momentum operator\footnote{If the pseudopotential is
non local, $\hat{v}m$ is used instead of the momentum operator}.\\
We introduce now a cutoff function $\theta(z)$ aimed at \emph{projecting out}
the optical transitions related to a certain selected region of the slab.
The function $\theta(z)$ is a sum  of two Heaviside step functions:
 \begin{equation}
   \theta(z) = H(z-z_0)-H(z-(z_0+D_c))
 \end{equation}
where $D_c$ is the thickness of the cutoff function (see Fig.~\ref{fig:3LM}).\\
The cutoff function is introduced into the calculation of the optical 
properties through the use of a modified matrix element
$\tilde{P}_{v{\bf k},c{\bf k}}$, defined by:
\begin{equation}
  \tilde{P}^i_{v{\bf k},c{\bf k}} = -i\hbar \int d{\bf r} \psi^*_{v{\bf k}}({\bf r})
          \theta(z)\frac{\partial}{\partial r_i} \psi_{c{\bf k}}({\bf r}),
\end{equation}
and hence the polarizability of the slice is described by the relation:
\begin{equation}
 \textrm{Im}[4 \pi \alpha^{cut}_{ii} (\omega)] = 
     \frac{8 \pi^2 e^2 }{ m^2 \omega^2 A} \sum_k \sum_{v,c} [P^i_{v{\bf k},c{\bf k}}]^*\tilde{P}^i_{v{\bf k},c{\bf k}}
     \delta(E_{c{\bf k}}-E_{v{\bf k}} -\hbar\omega).
\end{equation}
In the next paragraphs we will show an application of this technique 
to analyse the layer-by-layer contribution of the electron energy loss
spectra.

  \subsection{Theory of RAS}
   By exploiting a reflection geometry and the polarizability 
of the incident wavevector, certain spectroscopies are able 
to resolve the surface contribution to the optical properties
of a material.
\begin{figure}[h!]
   \centerline{\epsfig{file=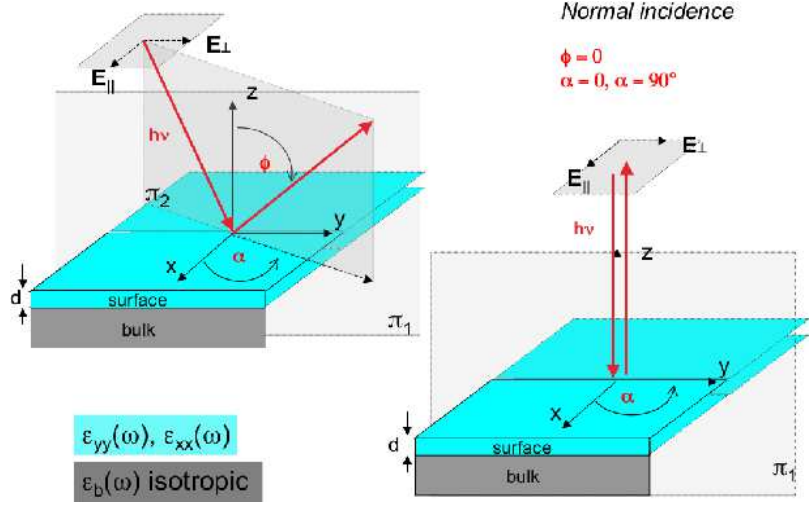,clip=,angle=0,width=10.5cm}}
   \caption[RAS geometry]{
        Schematic illustration of the general SDRS geometry in oblique incidence (left),
        and RAS assuming normal incidence (right). The light propagation direction
        is represented by red lines, and we indicate the polarization of
        light in two orthogonal directions belonging to the surface plane. 
        The bulk is assumed to be isotropic, while the surface is described by a frequency 
        dependent dielectric tensor with eigenvectors parallel to $x$ and $y$.
    \label{fig:ras-geo}
        }
\end{figure}
In the case of normal incidence light travelling from a first medium 
with $\varepsilon = \varepsilon_1$ to a second medium with 
$\varepsilon = \varepsilon_2$ (see Fig.~\ref{fig:ras-geo})
the Fresnel reflectivity is defined by :
\begin{equation}
 R_0 = \left|\frac{N_2-N_1}{N_2+N_1}\right|^2 = 
  \left|\frac{\sqrt{\varepsilon}_2-\sqrt{\varepsilon}_1}
             {\sqrt{\varepsilon}_2+\sqrt{\varepsilon}_1}\right|^2
\end{equation}
where $N_i = \sqrt{\varepsilon_i}$ is the complex refraction index 
$N = n_1 + i n_2$. \\
Accordingly, \emph{reflectance} is defined as a complex number $r$ for which 
$R = |r|^2$.
From the experimental point of view, measurements are usually performed 
with respect to this quantity by:
\begin{equation}
 I(\omega) = \frac{I_{rif}(\omega)}{I_{inc}(\omega)}
           = \left|\frac{E^{rif}_0}{E^{inc}_0} \right|^2
           = |r|^2
\end{equation}
On the contrary, reflectivity is a real number and $R \in [0,1]$
with special cases occurring when $R = 1$ (large difference between $N_1$ and $N_2$), 
$R = 0$ ($N_1 = N_2$) or when $N_i$ for the two media are  both real or imaginary.
Reflectance is a complex quantity linking the electric field amplitudes:
\begin{eqnarray}
 E = E_0 e^{i(\beta {\bf k}{\bf x} - \omega t )} = E_0e^{i\frac{\omega}{c}(Nx - ct)} 
   = E_0e^{i\frac{\omega}{c}(n_1x - ct)}e^{-\frac{\omega}{c}n_2x}   
 \label{eq:E}
\end{eqnarray}
where $\omega = vk$, and hence $k = \frac{\omega}{v}=\frac{\omega}{c}N$.
From Eq. \ref{eq:E} we can write:
\begin{equation}
 |E|^2=|E_0|^2e^{-2\frac{\omega}{c}n_2x}
\end{equation}
from which we deduce that energy decreases exponentially.\\
If now we consider the energy density $\bar{\omega}$ and
the absorption coefficient $\alpha$:
\begin{equation}
 \frac{\partial \bar{\omega}}{\partial x} = -\alpha\bar{\omega}
\end{equation}
we obtain:
\begin{eqnarray}
  \alpha &=& \frac{2\omega}{c}n_2 \\
  \textrm{Re}\varepsilon &=& n_1^2-n_2^2 \\
  \textrm{Im}\varepsilon &=& n_1 n_2 
\end{eqnarray}
and finally
\begin{equation}
  \alpha = \frac{\varepsilon_2}{n_1}\frac{\omega}{c}.
\end{equation}
In the case of an interface with vacuum,
the reflectivity is simplified:
\begin{equation}
 R(\omega) = \left|\frac{N-1}{N+1}\right|^2 =
             \left|\frac{(n_1 -1)^2+n_2}{(n_1+1)^2+n_2}\right|^2
\end{equation}
In the case of non normal incidence (see Fig. \ref{fig:ras-geo})
two contributions are distinguished:
\begin{eqnarray}
R_s&=& \left|\frac{\textrm{cos}\theta - \sqrt{\varepsilon - \textrm{sin}^2\theta }      }
                  {\textrm{cos}\theta + \sqrt{\varepsilon - \textrm{sin}^2\theta }      } \right|^2 \\
R_p&=& \left|\frac{\varepsilon \textrm{cos}\theta - \sqrt{\varepsilon - \textrm{sin}^2\theta }   }
                  {\varepsilon \textrm{cos}\theta + \sqrt{\varepsilon - \textrm{sin}^2\theta }   } \right|^2
\end{eqnarray}
related to $p$ and $s$ waves respectively.

In experiments, the RA spectrum is calculated starting 
from the knowledge of the reflectance and hence the reflectivity
by means of Eq. \ref{eq:ras1}. \\
Theoretical models link reflectivity to the dielectric tensor.
In the particular case where $\frac{\Delta R_i}{R_0}<<1$,
the relative deviation of the reflectivity with
respect to the Fresnel contribution is given in terms of the surface 
and bulk dielectric tensor $\varepsilon_s$, $\varepsilon_s$ by the equation:
\begin{equation}
   \frac{\Delta R_i}{R_0} = \frac{4\omega}{c}\textrm{cos}\theta \textrm{Im}
     \left(\frac{\varepsilon^s_{ii} - \varepsilon_b}{\varepsilon_b -1}\right)
\end{equation}
representing SDRS formula in s-polarization, or, in the case of normal incidence:
\begin{equation}
   \frac{\Delta R_i}{R_0} = \frac{4\omega}{c} \textrm{Im} 
       \left(\frac{4\pi\alpha_{ii}(\omega)}{\varepsilon_b -1}\right)
  \label{eq:ras-alpha}
\end{equation}
In Eq.~\ref{eq:ras-alpha} we used the slab polarizability 
instead of the surface and bulk dielectric function. 
More details of this theory can be found in reference~\cite{rds_book}.

  \subsection{Theory of electron energy loss at surfaces} \label{sec:th-eels}
\begin{figure}[h!]
  \centerline{\epsfig{file=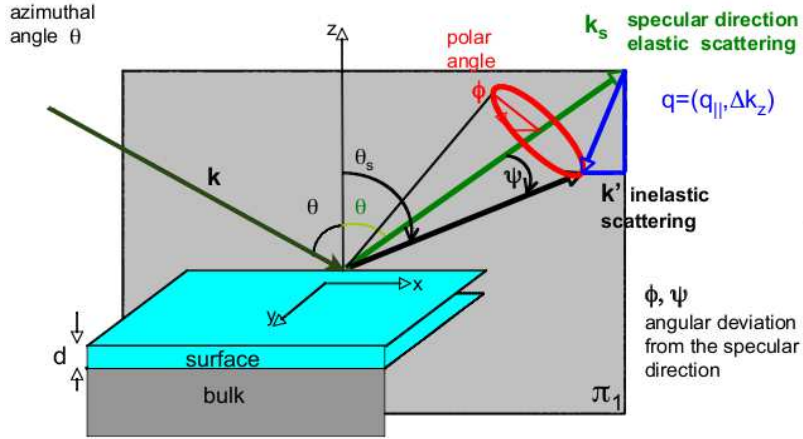,clip=,angle=0,width=10.5cm}}
   \caption[REELS geometry]{
        Schematic representation of the reflection geometry for
        Electron Energy Loss Spectroscopy experiments.
       \label{fig:reels-geo}
           }
 \end{figure}

   We use a semiclassical dipole scattering theory
that accounts for the long-range interaction between the incident
electrons and the medium under study~\cite{luth_book, ibach_book}.
Assuming planar scattering, and taking $yz$ as being the scattering
plane ($z$ is the surface normal, see Fig.~\ref{fig:reels-geo}) 
the scattering probability is defined by:
 \begin{equation}
   \textrm{P}({\bf k},{\bf k'}) = \textrm{A}({\bf k},{\bf k'}) 
       \textrm{ Im } \textrm{g}({\bf q}_{||}, \omega)
   \label{eq:Pkk}
 \end{equation}
where ${\bf k}$ and ${\bf k'}$ are the incident and scattered wavevectors
respectively. The kinematic factor, A$({\bf k},{\bf k'})$:
   \begin{equation}
      \textrm{A}({\bf k},{\bf k'}) = \frac{2}{(e a_0 \pi)^2} \frac{1}{\cos \theta}
                          \frac{k'}{k} \frac{q_{||}}{|q_{||}^2 + q_{\bot}^2|^2}
      \label{eq:Akk}
   \end{equation}
mostly contains the information concerning the scattering geometry 
(see Fig.~\ref{fig:reels-geo}). \\
The angle $\theta$ is the direction of the incident
beam with respect to the normal to the surface plane, and $q_{||}, q_{\bot}$
are the parallel and perpendicular components of the transferred
momentum ${\bf q} = {\bf k} - {\bf k}'$.\\
The loss function is defined by:
   \begin{equation}
       \textrm{g}({\bf q}_{||}, \omega) = - \frac{1}{1 + \varepsilon_\textrm{eff}({\bf q}_{||}, \omega)}
   \label{eq:gkk}
   \end{equation}
and represents the part of Eq.~\ref{eq:Pkk} involving the approximation
of the model and the separation between bulk and surface contributions
to the dielectric function.
If the surface were to be modelled as a semi-infinite truncated bulk,
$\varepsilon_\textrm{eff}$ would be replaced by $\varepsilon_\textrm{b}$
and we would obtain the familiar expression of Mills~\cite{ibach_book}.

\begin{figure}[!h]
  \begin{tabular}{c}
   \centerline{\epsfig{file=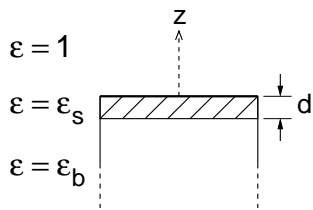,angle=0,clip=,width=4cm}}
  \end{tabular}
 \caption[Three Layers Model]{Schematic representation of the constituent parts 
   of the three-layer model of the surface.
          }
 \label{fig:threelayermodel}
\end{figure}

In this work we adopt an anisotropic three-layer model of the surface as
derived by Selloni and Del Sole~\cite{selloni_ss_y1986_v186_p35,palummo_prb_y2006_v74_p235431}.
The surface is modelled as in Fig.~\ref{fig:threelayermodel}: a semi-infinite
layer of vacuum, a surface layer of thickness $d$, represented by the surface
dielectric \emph{tensor} $\varepsilon_s$, and a semi-infinite layer of bulk
(dielectric function $\varepsilon_b$).
%%%%%%%%%%%%%
The effective dielectric function is defined by:
   \begin{eqnarray}
     &&\varepsilon_\textrm{eff}({\bf q}_{||}, \omega) = 
                 \varepsilon_s({\bf q}, \omega)\times \nonumber \\ 
     &&\times\frac{\varepsilon_s({\bf q}, \omega)  + 
                   \varepsilon_b({\bf q}, \omega)  + 
                   \Delta({\bf q}, \omega)
                  e^{-2{\bf q}_{||}d \varepsilon_\textrm{aux}({\bf q}, \omega)} }
                {\varepsilon_s({\bf q}, \omega) + 
                  \varepsilon_b({\bf q}, \omega) - 
                    \Delta({\bf q}, \omega) 
                e^{-2{\bf q}_{||}d \varepsilon_\textrm{aux}({\bf q}, \omega)} }
   \label{eq:eps_eff}
   \end{eqnarray}
   where $d$ is the thickness of the surface, $\varepsilon_b({\bf q}, \omega)$ and 
   $\varepsilon_s({\bf q}, \omega)$ are the bulk and surface dielectric function 
   and $\Delta({\bf q}, \omega) = \varepsilon_b({\bf q}, \omega)  - \varepsilon_s({\bf q}, \omega)$.\\
   In particular $\varepsilon_s({\bf q}, \omega)$ and the auxiliary function 
   $\varepsilon_{aux}({\bf q}, \omega)$ are written as a function of the $y,z$ 
   components of the dielectric tensor:
   $\varepsilon_s({\bf q}, \omega) = \sqrt{\varepsilon_{s,y}({\bf q}, \omega) 
   \varepsilon_{s,z}({\bf q}, \omega)}$ and $\varepsilon_{\textrm{aux}}({\bf q},\omega) 
   = \sqrt{ \frac{\varepsilon_{s,y}({\bf q}, \omega)}{\varepsilon_{s,z}({\bf q}, \omega)} }$.\\
 Although the dielectric functions appearing in Eq.~\ref{eq:eps_eff} are fully dependent
 on $q_{||}$ and $\omega$, such quantities are not easy to calculate, since $q$ and $\omega$
 are not independent.
 Hence we make the approximation of replacing $\varepsilon_s({\bf q},\omega)$ with the
 optical dielectric function $\varepsilon_s(\omega) \approx \lim_{{\bf q} \to 0} 
 \varepsilon_s({\bf q},\omega)$.
 This appears to be a reasonable assumption since for most of the experiments modelled
 in this work, $q$ is rather small. \\
%(order of XXX BZ).
%
 Surface dielectric functions are calculated according to Ref.~\cite{hogan_prb_y2003_v68_p035405}
 using a cutoff function (as discussed in the previous sections) 
 in order to select the number of terminal layers contributing to the
 surface response.

%
%%%%%%%%%%%%%%%%%%%%
 \part{Development}
%%%%%%%%%%%%%%%%%%%%
%
\chapter{Efficient calculation of the electronic polarizability} \label{ch:hilbert}
In this section we show the application of an efficient numerical scheme 
to obtain the independent--particle dynamic polarizability matrix $\chi^{(0)}(r,r',\omega)$, 
a key quantity in modern \emph{ab initio} excited state calculations. 
The method has been applied to the study of the optical response of 
a realistic oxidized silicon surface, including the effects of crystal local fields. 
The latter are shown to substantially increase the surface optical anisotropy 
in the energy range below the bulk bandgap. Our implementation in a large--scale 
\emph{ab initio} computational code allows us to make a quantitative study of 
the CPU time scaling with respect to the system size, and demonstrates 
the real potential of the method for the study of excited states in large systems. 

\section{Motivations}
   The recent developments of experimental techniques for the non 
   destructive study of solid surfaces call for a simultaneous improvement 
   of the theoretical tools:
   the interpretation and prediction of optical and dielectric
   properties of surfaces require more and more quantitative and
   reliable \emph{ab initio} calculations, possibly including many--body effects.
   Such an improvement of the theoretical description can be achieved, for example,
   by lifting some of the usual approximations adopted in the calculation of 
   the optical response.
   However, making less approximations increases the computational heaviness,
   and is only possible if efficient numerical algorithms can be adopted.
   A good example is given by the calculation of the independent--particles
   dynamical polarizability matrix $\chi^{(0)}({\bf r}, {\bf r}', \omega)$, which is
   often required as the starting point in Time-Dependent Density
   Functional Theory (TDDFT) \cite{TDDFT} and in  Many-Body Perturbation Theory-based 
   calculations, such as in the  GW \cite{H65}, or GW+Bethe-Salpeter 
   schemes (for a review, see e.g. ref. \cite{ORR02}).
   Evaluating the full response matrix for realistic, many-atoms systems
   can be computational challenging, since it requires a computational effort
   growing as the fourth power of the number of atoms, and the availability
   of efficient numerical schemes becomes a key issue. \\
   Recently, schemes allowing to decouple the sum-over-states and the frequency
   dependence have been presented. Miyake and Aryasetiawan \cite{MA99}
   and Shishkin and Kresse \cite{SK06} have shown that methods based on the Hilbert
   transform can substantially reduce the computational cost of frequency-dependent 
   response functions, making it comparable to that of the static case.
   In particular the approach presented in \cite{MA99} has been applied to a
   linear-muffin-tin-orbital (LMTO) calculation of the spectral
   function of bulk copper, while in \cite{SK06}, a work focused on the 
   GW implementation using the Projector Augmented-Wave method (PAW \cite{B94}), 
   a similar approach is used to compute the spectral function of bulk silicon 
   and materials with \emph{d} electrons (GaAs and CdS).
   Another recent work by D. Foerster \cite{DF05} is focused on the same issue
   and demonstrates how the use of a basis of local orbitals can reduce the
   scaling of a susceptibility calculation for an N--atom system from $N^4$ to
   $N^3$ operations for each frequency, but at the cost of disk space. \\
   However, the application of such non-traditional
   methods to large supercells, such as those involved in real
   surface calculations, have not been presented so far. \\
   It may be stressed that for a given application the computational burden is determined
   not only by general scaling law, but also by prefactors.
   In particular, prefactors determine the crossover where one method becomes
   more convenient than the other.
   This crossover has not yet been discussed for the Hilbert transform methods.

   In the present work, we demonstrate the application of a scheme -similar to that introduced 
   in \cite{MA99} and \cite{SK06}- based on the efficient use of the Hilbert transforms,
   by performing the calculation of the optical properties of a realistic,
   reconstructed surface: Si(100)(2$\mathbf{\times}$2):O, covered with 1 monolayer (ML) of oxygen. \\
   We provide a \emph{quantitative} evaluation of the computational gain for this 
   calculation of the full dynamical independent-particle polarizability. The latter is constructed
   from Kohn-Sham eigenvalues and eigenvectors and is then used to compute surface optical spectra, 
   including for the first time the local field (LF) effects on Reflectance Anisotropy (RAS) 
   and Surface Differential Reflectivity (SDR) spectra of this surface.

\subsection{Local filed effects}   
   The impact of local fields on surface optical spectra has been a controversial issue for decades,
   specially concerning the so-called \emph{intrinsic} or bulk--originated effects. 
   The latter have been measured, for energies above the bulk bandgap, 
   since the seminal works by Aspnes and 
   Studna \cite{asp-stu-prl-1985} showing that the normal incidence optical reflectivity of natural  
   Si(110) and Ge(110) surfaces displays an anisotropy of the order of 10$^{-3}$. 
   The effect was called intrinsic since it is not due to the existence of surface states, 
   nor to surface reconstruction. 
   Early model calculations by Mochan and Barrera \cite{moc-bar-prl-1985} 
   performed for a lattice of polarizable entities and exploiting the Clausius-Mossotti relation 
   pointed out that intrinsic anisotropies could be due to LF effects. A subsequent 
   work by Del Sole, Mochan and Barrera \cite{dels-moch-bar-91} based on Tight Binding (TB)
   method has shown that the RAS spectra calculated for Si(110):H within this semiempirical scheme did
   not reproduce well the experimental data, despite the inclusion of surface LF effects.
   However, more recent calculations based on realistic bandstructures   
   (within DFT-LDA with GW-corrected band gap) \cite{schmidt-bernholc-prb-2000}
   have suggested that intrinsic anisotropies at the bulk critical points for the 
   (almost ideally terminated) Si (110):H  surface could arise as a consequence
   of surface perturbation of bulk states, without invoking LF effects.
   Other tight-binding calculations (see, e.g., ref. \cite{onida-delsole-prb-1999})
   suggested the existence of intrinsic surface optical anisotropies not due 
   to surface local fields. \\ 
   \indent A substantial advance in clarifying the role of local fields has been achieved 
   only recently by F. Bechstedt and co-workers, who carried out a calculation of the 
   RAS spectra of Si(110):H \cite{bech-prl-2002} and monohydride Si(100)(2x1) \cite{bech-prb-2003} 
   including self-energy, crystal local fields, and excitonic effects from a 
   fully \emph{ab initio} point of view.
   In both the considered surfaces, which have no surface states within the bulk bandgap, 
   the LF were found to cause a slight decrease of the optical reflectivity; however,
   the effect was found to cancel to a large extent in the RAS spectra,
   being almost identical for the two polarizations of the incident light.
   The situation may be different in the case of extrinsic optical anisotropies, i.e. those 
   directly related to surface states and surface reconstruction, and appearing 
   below the bulk bandgap. In at least one case
   substantial effects due to LF have been reported \cite{rodolfo2}.
   However, further calculations for a wider class of surfaces are 
   necessary in order to assess this point more precisely. \\
   The system we consider here belongs to a widely studied family of surfaces,
   because of their importance in the understanding of silicon-silicon dioxide
   interfaces in semiconductor technology.
   Despite the many experimental 
   \cite{WK98, nakajima-prb-2001, dreiner-prl-2004, pi-surfsci-2001, yoshigoe-surfsci-2003, ikegami-jjap-1996, itoh-surfsci-2001} 
   and theoretical
   \cite{uchiyama-surf-sci-1999,kato-prb-2000, richard-comput-mat-sci-2005, chabal-prb-2002,widjaja-jchemphys-2002,ciacchi-prl-2005} 
   works appeared in recent years,
   the debate on the oxidation mechanism of Si(100) is still open.
   However, the most favorable oxygen adsorption sites in the first stages of (room-temperature) 
   oxidation process have been identified as the dimer-bridge
   position, and a bridge position on the backbond corresponding to the lower
   atom of the dimer.  This remark is supported by STM experiments \cite{itoh-surfsci-2001} and by 
   a first-principles molecular dynamics calculation \cite{ciacchi-prl-2005}.
   From the theoretical point of view, ground and excited state properties of
   Si(100)(2$\mathbf{\times}$2):O at 0.5 and 1ML coverage have been recently studied by some of the 
   authors \cite{incze_prb_y2005_v71_035350}; however, computational limits prevented till now the inclusion
   of the local--field effects in the \emph{ab initio} calculation of optical properties. \\
   \indent In the following paragraphs we brefly summarize 
   the theoretical framework and the expression
   of $\chi^{(0)}$ usually employed in plane--wave based calculations.
   Then we show how the Hilbert transform (HT) technique can be applied,
   as a generalization of the Kramers--Kronig relations, in order to decouple
   the sum-over-states and the frequency dependence in $\chi^{(0)}$. Moreover  
   an estimation of the accuracy and the possible computational gain
   are presented for a model system.
 \label{sec:motivation}
\section{Theoretical framework} 
   The starting point of our work is a DFT-LDA ground state calculation performed with
   the ABINIT code \cite{ABINIT} yielding independent-particle eigenvalues and eigenvectors
   within the Kohn-Sham scheme \cite{DFT,DFT2}.   
   Besides to the occupied ones, empty (conduction) states up to an energy 
   of several eV above the Fermi level are obtained by means of iterative
   diagonalization techniques.\\
   However, in order to study the optical and dielectric response, the level of
   theory must be brought beyond the ground state one, using, e.g., 
   many--body perturbation theory or TDDFT \cite{TDDFT}.
   The latter is particularly suited for the study of neutral excitations,
   as those involved in optical reflectivity and electron energy-loss.
   
  \subsection{Fundamental ingredients}
   Within TDDFT, it is possible to obtain the retarded density-density response function
   $\chi({\bf r},{\bf r}',\omega)$ from its non-interacting Kohn-Sham counterpart
   $\chi^{(0)}({\bf r},{\bf r}',\omega)$ through a Dyson-like equation:
   \begin{equation}
      \chi = \chi^{(0)} + \chi^{(0)}K\chi
     \label{eq:dyson}
   \end{equation}
   where the kernel $K$ contains two terms: the Coulomb potential, $v_{c}$,
   and the exchange-correlation kernel, $f_{xc}({\bf r},{\bf r}',\omega)$.
   An explicit expression for $\chi$ is then given by
   \begin{equation}
      \chi = \chi^{(0)}\left[1 - (v_{c} + f_{xc})\chi^{(0)} \right]^{-1}.
     \label{eq:dyson_solved}
   \end{equation}
   Eq.s (\ref{eq:dyson}) and (\ref{eq:dyson_solved})  are matrix equations,
   involving two-points functions such as $\chi$ and $\chi^{(0)}$.
   In the present case, working within a plane-waves expansion,
   $\chi_{{\bf G}{\bf G}'}({\bf q}, \omega)$ and $\chi^{(0)}_{{\bf G}{\bf G}'}({\bf q}, \omega)$
   are matrices in reciprocal space, and
   $ v_{c}({\bf q} + {\bf G}) = \frac{4\pi}{|{\bf q} + {\bf G}|^2}$   
   is the Coulomb potential.
   The exchange-correlation contribution, $f_{xc}$, is not exactly known.
   It can be included in an approximate form, e.g. using the LDA functional \cite{LDA, perdew_prb_y1981_v23_5048} 
   in the adiabatic approximation (ALDA), or in a more sophisticated approximation
   such as those described in \cite{noi_kernel, noi_kernel_lungo, ADSM03, DSA03,ST04, SK03}.
   In order to compare with optical experiments, the macroscopic dielectric function
   $\varepsilon_{M}(\omega)$ must be calculated. The latter is defined as:
   \begin{equation}
     \varepsilon^{}_{M}(\omega) = \lim_{q \to 0}
     \frac{1}{ \varepsilon^{-1}_{{\bf G}={\bf G}'=0}({\bf q},\omega) }
    \label{eq:eps_M}
   \end{equation}
 where the inverse dielectric function $\varepsilon^{-1}_{{\bf G},{\bf G}'}({\bf q},\omega) $ is
 linked to the response function $\chi$ by:
   \begin{equation}
     \varepsilon^{-1}_{{\bf G},{\bf G}'}({\bf q},\omega) =
      1 + v_{c}({\bf q} + {\bf G})\chi_{{\bf G},{\bf G}'}({\bf q},\omega).                                 
    \label{eq:eps-1}
   \end{equation}
When only $v_{c}$ is included in the kernel $K$ of eq.(\ref{eq:dyson}) exchange
and correlation effects in the response are neglected, while the use of the correct
expression (\ref{eq:eps_M}) still consider the LF effects \cite{Baroni}.
Already at this level the calculations can become time consuming from the 
 computational point of view when the full
 $\chi^{(0)}({\bf r},{\bf r}',\omega)$ matrix has to be obtained.
 In complex systems with large unit cells the only tractable way to proceed is often to neglect
 local fields, by assuming that $\varepsilon_{M}(\omega)$ is well approximated by the average
 of the microscopic dielectric function:
  \begin{equation}
     \varepsilon^{NLF}_{M}(\omega) =\lim_{{\bf q} \to 0}\varepsilon_{{\bf 0},{\bf 0}}({\bf q}, \omega)
    \label{eq:epsNLF}
  \end{equation}
 This corresponds to neglecting the off-diagonal elements
 of $\varepsilon$ in reciprocal space \footnote{In real space, this corresponds to assume a dependence of
  $\varepsilon_{M}({\bf r},{\bf r}')$ only on the difference $({\bf r} - {\bf r}')$}.
 When moreover exchange and correlation effects are neglected,
 (independent quasiparticle approximation or IP-RPA) the imaginary part of the macroscopic
 dielectric function $\varepsilon^{NLF}_M$ takes the simple Ehrenreich and Cohen \cite{EC59} form:
  \begin{equation}
    \textrm{Im}\varepsilon^{NLF}_{M}(\omega) = \frac{16\pi}{\omega^2}\sum_{ij}
            |<\psi_i|{\bf v}|\psi_j>|^2\delta(\epsilon_j -\epsilon_i -\omega)
   \label{eq:EC59}
 \end{equation} 
 where ${\bf v}$ is the velocity operator and i, j stand for occupied 
 and unoccupied states respectively. 
 The substantial simplification obtained in this case explains why most
 of the calculations of the optical properties of real surfaces are done
 within the independent quasiparticle approach, neglecting local--field effects.
 On the other hand, a fast and efficient scheme to compute the
 full matrix $\chi^{(0)}$ represents a key issue in order to be able to go beyond this approximation,
 e.g. by including the local fields, as we do in the present work.
 Moreover, an efficient method giving access to the full $\chi^{(0)}$ is 
 of paramount importance when the screened coulomb interaction $W_{{\bf GG}'}({\bf q})$ is needed, 
 such as in ab-initio GW calculations.
 In the following, we hence concentrate on the expression of $\chi^{(0)}$ itself, i.e.:
 \begin{eqnarray}
     \label{eq:chizero}
      \chi^{(0)}({\bf r},{\bf r}',\omega) &=& 2 \sum_{ij} f_i(1-f_j)               
            \psi_i^*({\bf r})\psi_j({\bf r})\psi_j^*({\bf r}')\psi_i({\bf r}') \times \nonumber \\
                       && \times \[
                               \frac{1}{\omega - ( \epsilon_j - \epsilon_i) + i\eta} -
                               \frac{1}{\omega + (\epsilon_j - \epsilon_i) + i\eta}
                               \]
   \end{eqnarray}
    where $f_{i}$ are occupation numbers (0 or 1 in the present case), $\eta$ is an infinitesimal
    and the factor 2 is due to the spin degeneracy.
    Switching to reciprocal space and focusing on the case of  semiconductors,
    we make valence (v) and conduction (c) bands to appear explicitly,
    and rewrite this equation as:
 \begin{equation}
     \label{eq:chi0heavy}
       \chi^{(0)}_{{\bf G},{\bf G}'}({\bf q}, \omega) = \frac{2}{\Omega_0 N_k}\sum_{{\bf k}}\sum_{c,v}
                                       \left[
               \frac{ \widetilde{\rho}_{vc{\bf k}}({\bf q}+{\bf G})\widetilde{\rho}_{cv{\bf k}}({\bf q}+{\bf G}') }
                    { \omega - (\epsilon_{c{\bf k}} - \epsilon_{v{\bf k}} ) + i\eta } -
               \frac{ \widetilde{\rho}_{cv{\bf k}}({\bf q}+{\bf G})\widetilde{\rho}_{vc{\bf k}}({\bf q}+{\bf G}') }
                    { \omega + (\epsilon_{c{\bf k}} - \epsilon_{v{\bf k}} ) + i\eta   }
                                       \right]
  \end{equation}
  where $\Omega_0$ is the volume of the unitary cell and we have also introduced the notation 
  $\widetilde{\rho}_{vc{\bf k}}({\bf q}+{\bf G}) $ to indicate the
  Fourier transform of $\phi^*_{v{\bf k}+{\bf q}}({\bf r})\phi_{c{\bf k}}({\bf r})$.
  From the numerical point of view the evaluation of these sums for each frequency $\omega$
  can become very heavy. Indeed, for a realistic system the evaluation of eq. (\ref{eq:chi0heavy})
  involves, for each frequency, the summation over a large number of terms, which for a
  system of 50 atoms typically is of the order of $10^{8}$.

\subsection{The Hilbert-transform approach}
 Since we consider the case of the ${\bf q}\to 0$ limit to study optical properties, 
 in the following the label ${\bf q}$ will be omitted to simplify the notation. 
 The generalization to the case of finite {\bf q} is straightforward. Introducing a simplified notation 
 for band and ${\bf k}$-point indexes, we define a single index of transition t to represent 
 the triplet $\{v, c, {\bf k} \}$. 
  In this way, $\omega_{t}$ indicates an (always positive) energy difference,
  $(\varepsilon_{c,{\bf k}} - \varepsilon_{v,{\bf k}})$. 
  We also introduce the  two complex quantities:
 \begin{eqnarray}
     Z_{1,t} & = &   \widetilde{\rho}_{vc{\bf k}}({\bf G})\widetilde{\rho}_{cv{\bf k}}({\bf G}') \\
     Z_{2,t} & = & - \widetilde{\rho}_{cv{\bf k}}({\bf G})\widetilde{\rho}_{vc{\bf k}}({\bf G}')
  \end{eqnarray}
 such that:
 \begin{equation}
      \chi^{(0)}_{{\bf GG}'}(\omega)  =  \sum_{t}\[
                                                  \frac{Z_{1,t}}{\omega - \omega_t + i\eta} +
                                                  \frac{Z_{2,t}}{\omega + \omega_t + i\eta}
                                                  \].
     \label{eq:chi0noi}
  \end{equation}
   When ${\bf G}={\bf G'}$ (diagonal elements) the $Z_{i,t}$ are real, and
   $Z_{1} = -Z_{2}$.
   Using 
   \begin{equation}
     \lim_{\eta \to 0^+}\frac{1}{x \pm i\eta} = \mathcal{P}\(\frac{1}{x}\) \mp i \pi \delta(x)
   \end{equation}
   one can rewrite the $\eta \to 0^+$ limit of equation (\ref{eq:chi0noi})
   as the sum of four terms:
  \begin{eqnarray}
        \chi^{R1}_{{\bf GG}'}(\omega) &=& \sum_{t}\frac{Z_{1,t}}{\omega - \omega_t}  \label{eq:r1}  \\  
       \chi^{R2}_{{\bf GG}'}(\omega) &=& i \pi \sum_{t} Z_{1,t}\delta(\omega - \omega_t)\label{eq:r2} \\
        \chi^{A1}_{{\bf GG}'}(\omega) &=& \sum_{t}\frac{Z_{2,t}}{\omega + \omega_t} \label{eq:a1} \\   
     \chi^{A2}_{{\bf GG}'}(\omega) &=& i \pi \sum_{t} Z_{2,t}\delta(\omega + \omega_t) \label{eq:a2}
  \end{eqnarray}
   R and A label resonant and anti resonant contributions, respectively, 
   and the four terms are general complex quantities.
   In $\chi^{R2}(\omega)$ and $\chi^{A2}(\omega)$ each term $Z_{t}$ 
   contributes to the function $\chi$ only at
   $\omega = \omega_t$, and has no effect elsewhere. 
   By discretizing the frequency axis, the sums over $t$ appearing in $\chi^{R2}$ 
   and $\chi^{A2}$ can hence be performed once and for all, at difference 
   with those labeled by R1 and A1 for which the sums should be calculated
   for each $\omega$.
   Thanks to the linearity of the Hilbert transform, defined as
   \begin{equation}
   H\textrm{\emph{f}} (t)= \frac{1}{\pi}\mathcal{P}
            \int^{+\infty}_{-\infty}\frac{f(x)}{ x - t }dx,
   \label{eq:hilbert}
   \end{equation}
   one can however directly  obtain $\chi^{A1}$ and $\chi^{R1}$ from $\chi^{A2}$ and $\chi^{R2}$: 
   \begin{eqnarray}
     \chi^{A1} &=& H\[\chi^{A2}\]  \label{eq:Ha2}\\
     \chi^{R1} &=& H\[\chi^{R2}\]  \label{eq:Hr2}
   \end{eqnarray}

In such a way\footnote{In the case of real matrix elements, $Z_{i,n} \in \mathbb{R}$, one recovers
  the Kramers-Kronig relations linking real and imaginary parts of the response},
it is possible to recover the complete $\chi_{{\bf G},{\bf G}'}^{(0)}(\omega)$
in the spectral range of interest from the knowledge of a single sum performed over the poles $\omega_t$.
In other words, one can avoid the explicit summation over $t=\{c,v,{\bf k}\}$ to be repeated for each frequency.
The present procedure for the calculation of the frequency-dependent polarizability
matrices is similar to the method of Miyake and Aryasetiawan \cite{MA99},
with the difference that those authors represented $\delta$-functions using Gaussians,
instead of bare rectangular functions as in our case \footnote{Similarly, Shishkin and \ Kresse \cite{SK06} used triangular functions}.

\subsection{Numerical efficiency for a toy system}
Our scheme has been first tested on a model system\footnote{We considered 
bulk silicon Kohn-Sham energies, increasing the number of transitions to build $\chi^{(0)}$,
 and randomly redefining the transition matrix elements} in order to check both the accuracy and the efficiency of the
algorithm. Figure (\ref{fig:model}) shows the results of the test, 
comparing $\chi^{(0)}(\omega)$ (real part) as obtained in the traditional 
way (i.e. by evaluating expression (\ref{eq:chi0heavy})
for several frequencies), and by the Hilbert transform (HT) algorithm.
The results are practically indistinguishable on the scale of the plot.
The same figure shows the growth of the required CPU time as
a function of the number of transitions (number of $\{ v,c,{\bf k} \}$ triplets).
The gain appears to be proportional to the system size.
The possibility to achieve such a large gain, at least in principle and for
a simple system, was also noticed in the previous works describing
efficient algorithms for the calculation of $\chi^{(0)}$ \cite{MA99,SK06}.\\
Alternative approaches for efficient TDDFT calculations have also been suggested. 
In particular, another promising scheme based on a superoperator approach and
allowing to access TDDFT \emph{spectra} in a numerically efficient way has been
recently introduced by Walker and coworkers \cite{WB06}.                              
This approach is however not designed for the calculation of the
whole matrix $\chi^{(0)}$, contrary to the method studied here.                                 
 \begin{figure}[h!]
\begin{minipage}{8cm}  
   \centerline{\epsfig{file=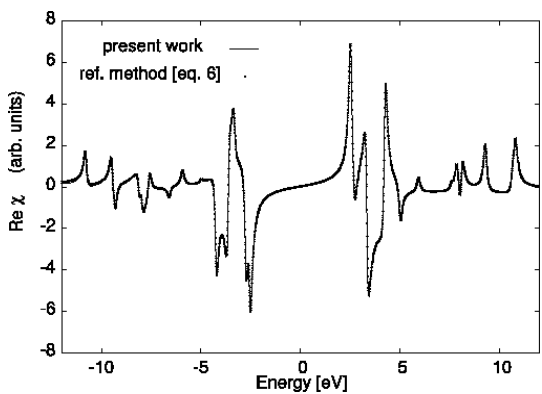,angle=270,clip=, width=7.0cm}} 
\end{minipage}
 \hspace{0.5cm}
 \begin{minipage}{5cm}
 \caption[Accuracy HT method]{
          Accuracy test for the HT-based algorithm, shown for the real part of
          $\chi^{(0)}_{{\bf GG}'}(\omega)$ of a model system (see text).
          The two curves turn out to be indistinguishable on the scale of
          the plot (maximum error less than 0.5\% ).
  \label{fig:model1}}
 \end{minipage}
 \end{figure}
In order to know the actual CPU requirements for the calculation of  $\chi^{(0)}$, and to
explore the possibilities to study complex systems, such as the impurity levels and
band offsets mentioned in ref. \cite{SK06}, in practice one has to keep into account
the time used to compute the matrix elements (numerators in eq. \ref{eq:chi0noi}),
and the time used to perform the Hilbert transforms, which was not explicitly evaluated in previous works.  In
the following, we hence applied our approach, similar in its essence
to that used in \cite{MA99} and \cite{SK06}, to a large system investigating the actual
numerical performances of the algorithm. As it will be shown below, substantial improvements
can actually be achieved in such realistic calculations. \\
  \begin{figure}[h!]
\begin{minipage}{8cm} 
  \centerline{\epsfig{file=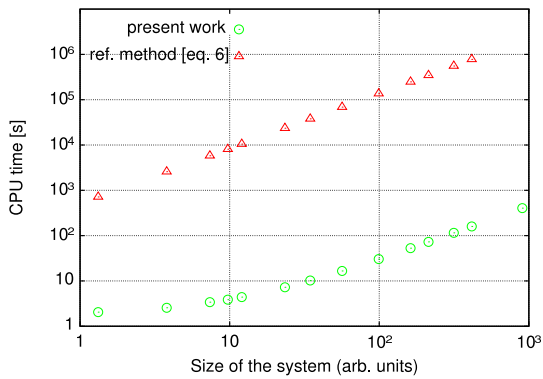,angle=270,clip=,width=7.0cm}}
\end{minipage} 
 \hspace{0.5cm}
\begin{minipage}{5cm} 
    \caption[Time scaling HT method (model system)]{
          Computational load requested to evaluate $\chi^{(0)}_{{\bf GG}'}(\omega)$
          on a model system, as a function of the number of transitions (which scales
          as the size of the system), for both the traditional and the HT-based methods.
  \label{fig:model} }
\end{minipage}
 \end{figure}
  Therefore, in the following section, we use our implementation
  of the HT scheme in the \emph{ab initio} DP code~\cite{DP}
  to study a real reconstructed surface: the oxidized Si(100)-(2$\times$2), 
  for which we present the first calculation of its optical reflectivity spectra 
  (RAS and SDR) with the inclusion of local--field effects.\\
  Finally, we carefully compare the numerical performance of the
  DP code with and without the use of HTs, and we draw our conclusions.
 \label{sec:hilbert}
\section{Optical properties of oxidized Si(100)-(2$\times$2)}
   The HT method has been implemented into the large scale,
   plane--waves \emph{ab initio} TDDFT code named DP, developed
   by the French node of ETSF \cite{etsf}. 
   As mentioned in the previous paragraphs, we used it to calculate
   the optical properties of Si(100)(2$\mathbf{\times}$2):O.
   For this surface we adopt the equilibrium structure for 1ML
   coverage shown in figure  (\ref{fig:str}), which is representative 
   of a situation in which dimer and backbond sites are both occupied 
   by an oxygen atom (structure c3 in ref. \cite{katalin_inpreparation,katalin_submitted}
   and see chapter \ref{ch:oxi} for further considerations).
   The surface is simulated with a slab composed by 6 layers, containing 48 Si
   and 8 oxygen atoms, in a repeated supercell approach.
   Our structural results agree well with those of previous calculations
   \cite{kato-prb-2000, NM00, katalin_inpreparation,katalin_submitted}.
   We use standard norm conserving pseudopotentials of
   the Hamann type \cite{hamann_prb_v40_y1989_p2980},
   and an energy cut off of 30 Ry, yielding 15000 plane waves in our unit cell.
   Eight special  (Monkhorst-Pack, \cite{MP76}) ${\bf k}$-points in the irreducible
   Brillouin zone (IBZ) are used for the self-consistent
   ground state calculation, while a 7$\mathbf{\times}$7 grid
   is used in the evaluation of $\chi^{(0)}_{{\bf GG}'}(\omega)$.
   Kohn-Sham eigenvalues and eigenvectors are obtained for all occupied states (120)
   and for empty states up to 15 eV above the highest occupied state (top valence).
   Optical properties are computed through the evaluation of the macroscopic
   dielectric function with and without the inclusion of local field effects.
%
% FIGURE 3 - EPS
% --------
\begin{figure}[h!]
 \begin{minipage}{6.5cm}
   \centerline{\epsfig{file=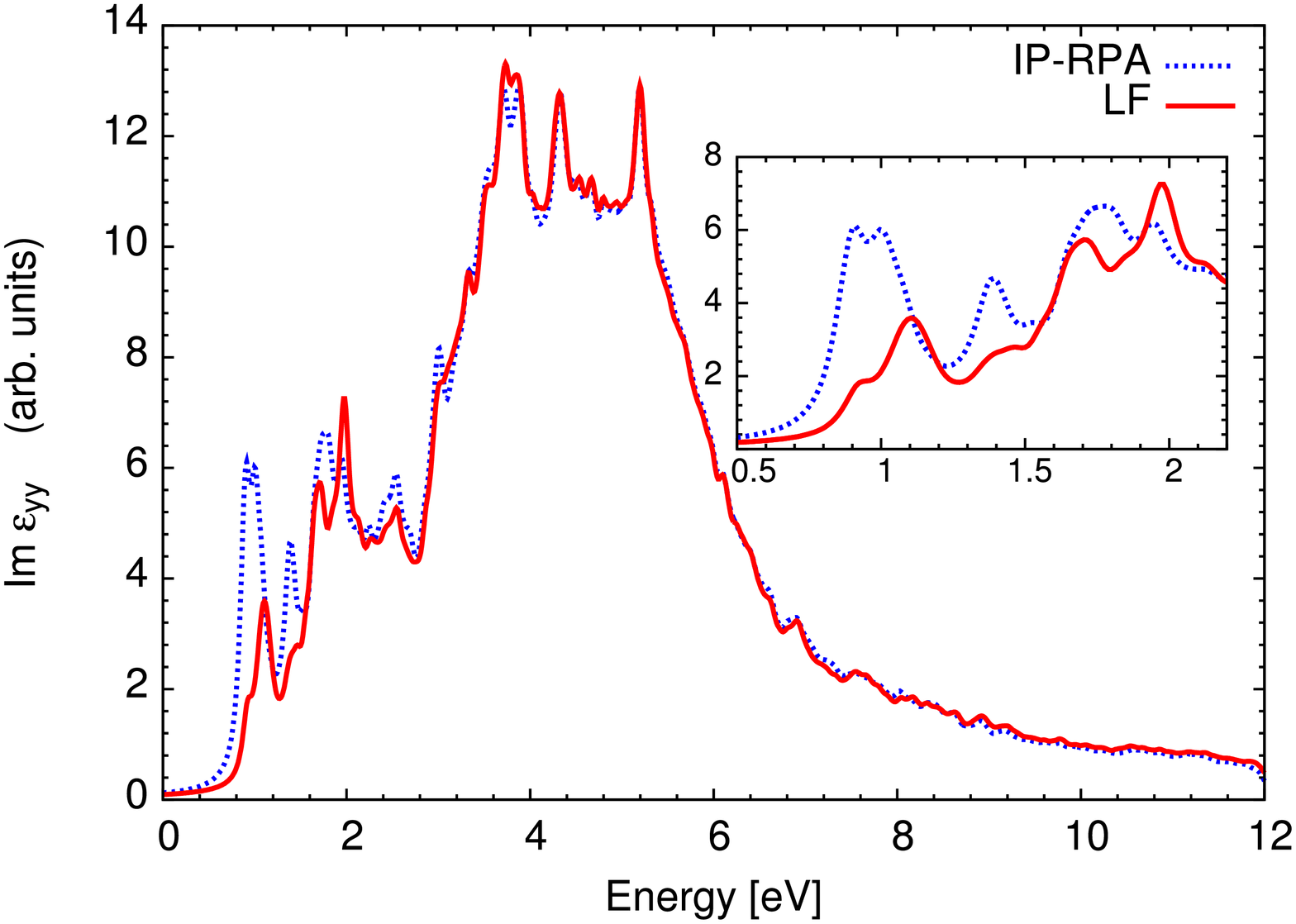,angle=270,clip=,width=6.5cm}}  
 \end{minipage}
\hspace{0.5cm}
 \begin{minipage}{6.5cm}
   \centerline{\epsfig{file=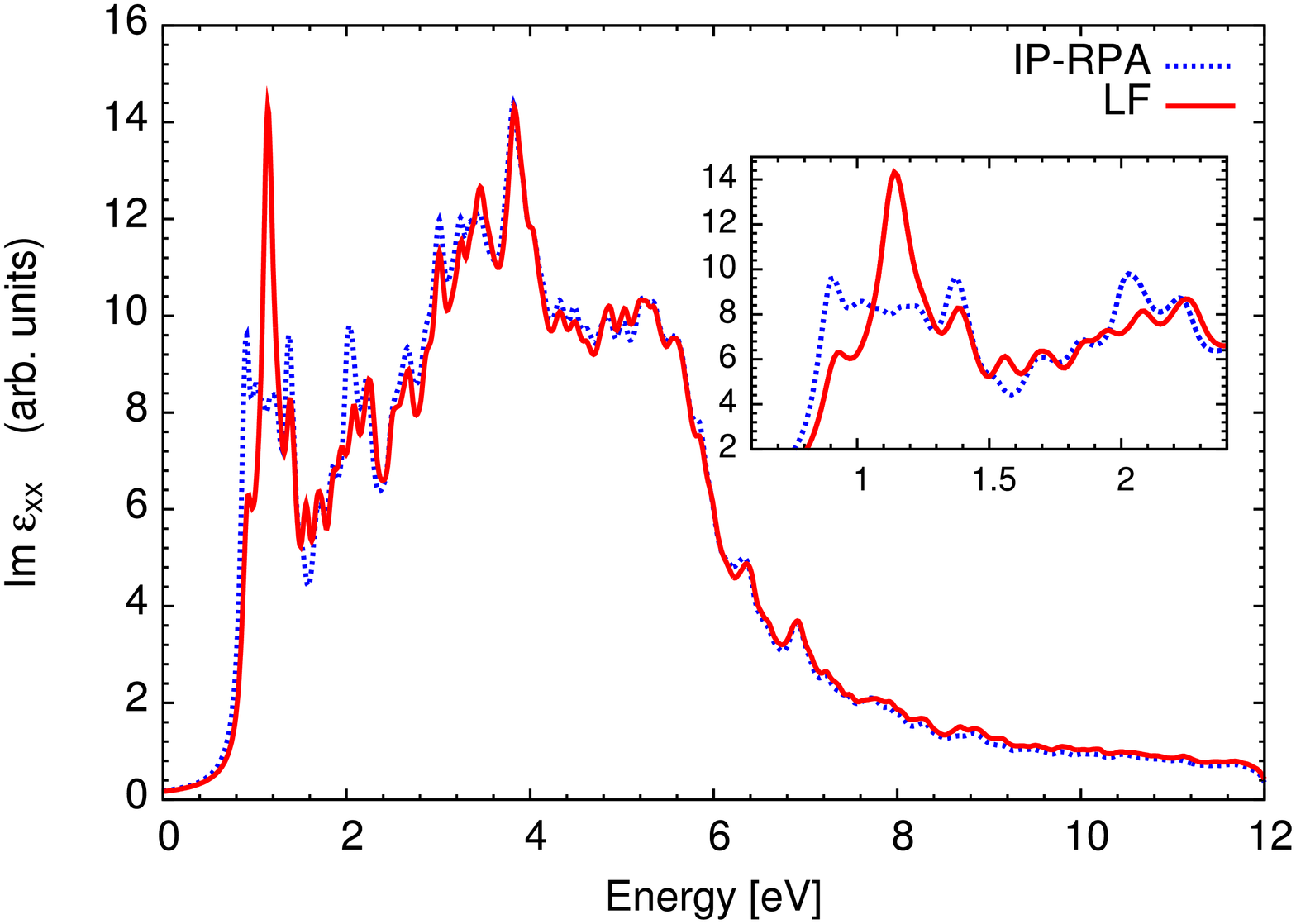,angle=270,clip=,width=6.5cm}}
  \end{minipage}
 \end{figure}
 \begin{figure}[h!]
 \begin{minipage}{6.5cm}
   \centerline{\epsfig{file=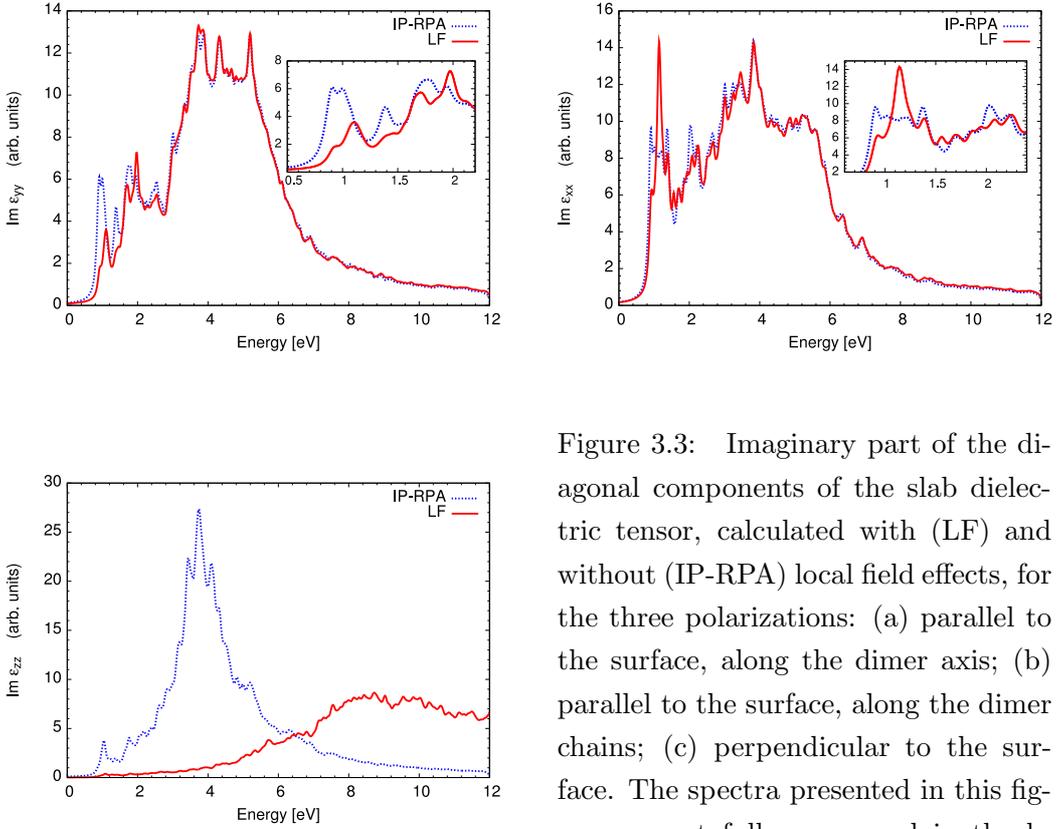,angle=270,clip=,width=6.5cm}}
  \end{minipage}
  \hspace{0.5cm}
 \begin{minipage}{6.5cm}
 \caption[Si(100)-p(2$\times$2):O (1ML) Imaginary part of the dielectric function]{ 
           Imaginary part of the diagonal components of the slab dielectric
           tensor, calculated with (LF) and without (IP-RPA) local field
           effects, for the three polarizations:
           (a) parallel to the surface, along the dimer axis; (b) parallel to the
           surface, along the dimer chains; (c) perpendicular to the surface.
           The spectra  presented in this figure are not fully converged
           in the k-points sampling.
\label{fig:ImE_x-y-z} }
  \end{minipage}
\end{figure}

   Figure \ref{fig:ImE_x-y-z} shows the imaginary part
   of the slab dielectric function as a function of the energy.
   Local--field effects are quite important in
   the low energy region (0-2)eV, enhancing $\varepsilon_M$
   for light polarized along the direction of the dimers
   chains (x direction, see Figure \ref{fig:str}),
   and suppressing it for light polarized
   along the dimers axis (y direction).
   This goes in the direction of
   a better description of the microscopic inhomogeneities of the system.
  In the present case, the extrinsic surface optical anisotropy, as defined in the
   introduction, is hence found to be visibly affected by LF.
   In the case of the third polarization, i.e. the
   one perpendicular to the surface
   (not experimentally relevant in the case of
   normally incident light), local--field effects
   are huge, and introduce a blueshift of the absorption
   edge as large as 5 eV.
   This can be explained by the strong inhomogeneity
   of the charge distribution in passing
   from the slab to the vacuum, leading to a classical
   depolarization effect. Similar behaviors
   have been found for example in GaAs/AlAs superlattices \cite{SB02},
   in graphite \cite{MR02} and nanowires \cite{FS05,FS052}.

   Starting from the slab dielectric function, we computed
   Reflectance Anisotropy (RAS) and Surface
   Differential Reflectivity (SDR) Spectra \cite{YB05}, 
   with and without inclusion of LF effects.
   We used theoretical models (see chapter \ref{ch:surface} or \cite{rds_book}) 
   linking the RAS and SDR spectra to the dielectric 
   functions evaluated for the bulk crystal ($\varepsilon_{b}$)
   and  for the slab ($\varepsilon_{ii}$) through the relation:
    \begin{equation}
         \frac{\Delta R_{i}}{R_{0}} = \frac{4 \omega}{c}
             \textrm{Im} \[ \frac{\varepsilon_{yy}(\omega) - \varepsilon_{xx}(\omega) }{\varepsilon_{b}(\omega) } \]
    \end{equation}
   where $\varepsilon_{xx}$ and $\varepsilon_{yy}$ are the diagonal components
   of the surface dielectric tensor.   We show our results for RAS and SDR
   in figures \ref{fig:RAS} and  \ref{fig:SDR} respectively.
%%%%%%%%%%%%%%%
\begin{figure}[h!]
 \begin{minipage}{7cm}
 \centerline{\epsfig{file=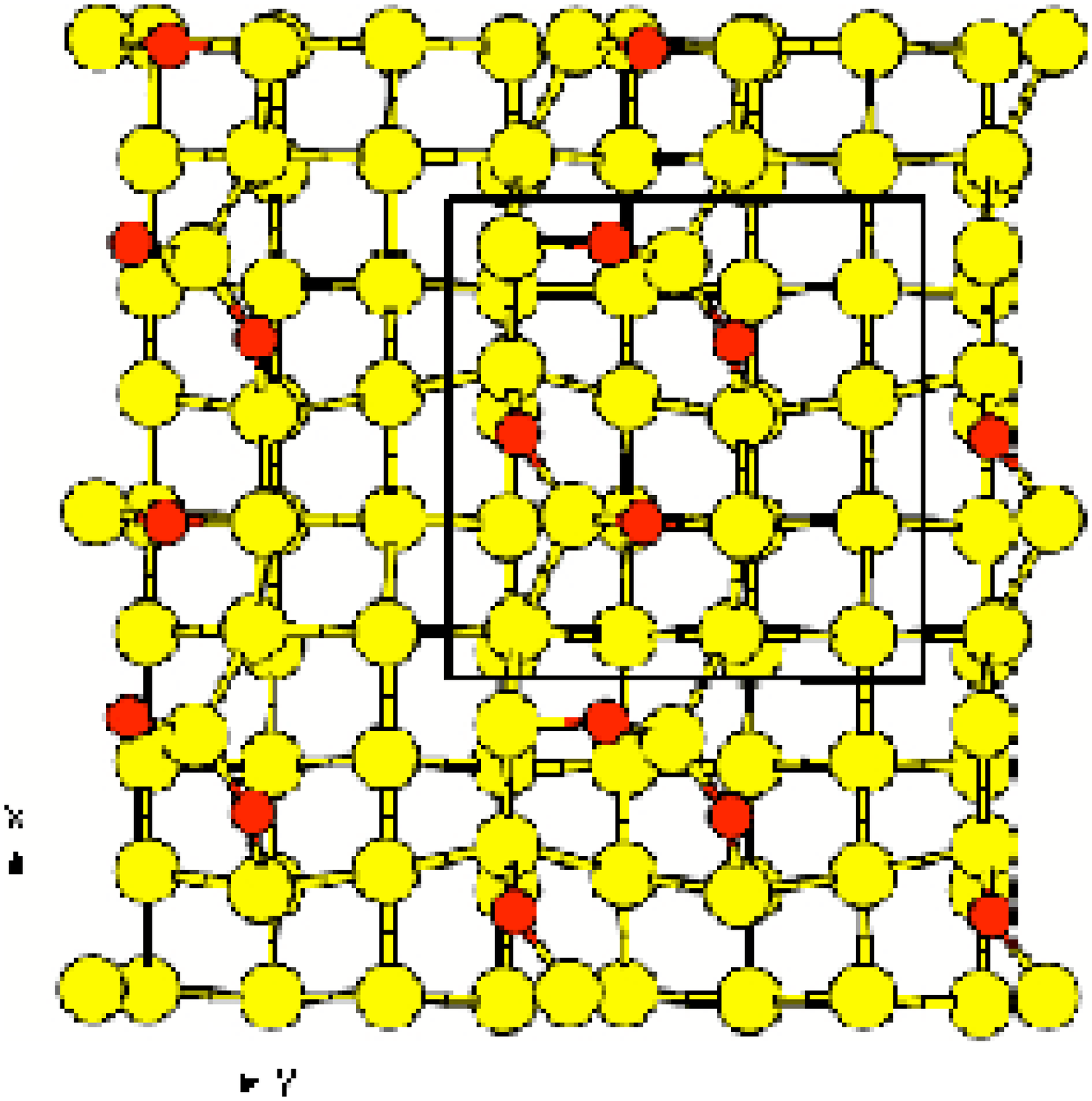,angle=0,clip=, width=6cm}} \hfill
 \centerline{\epsfig{file=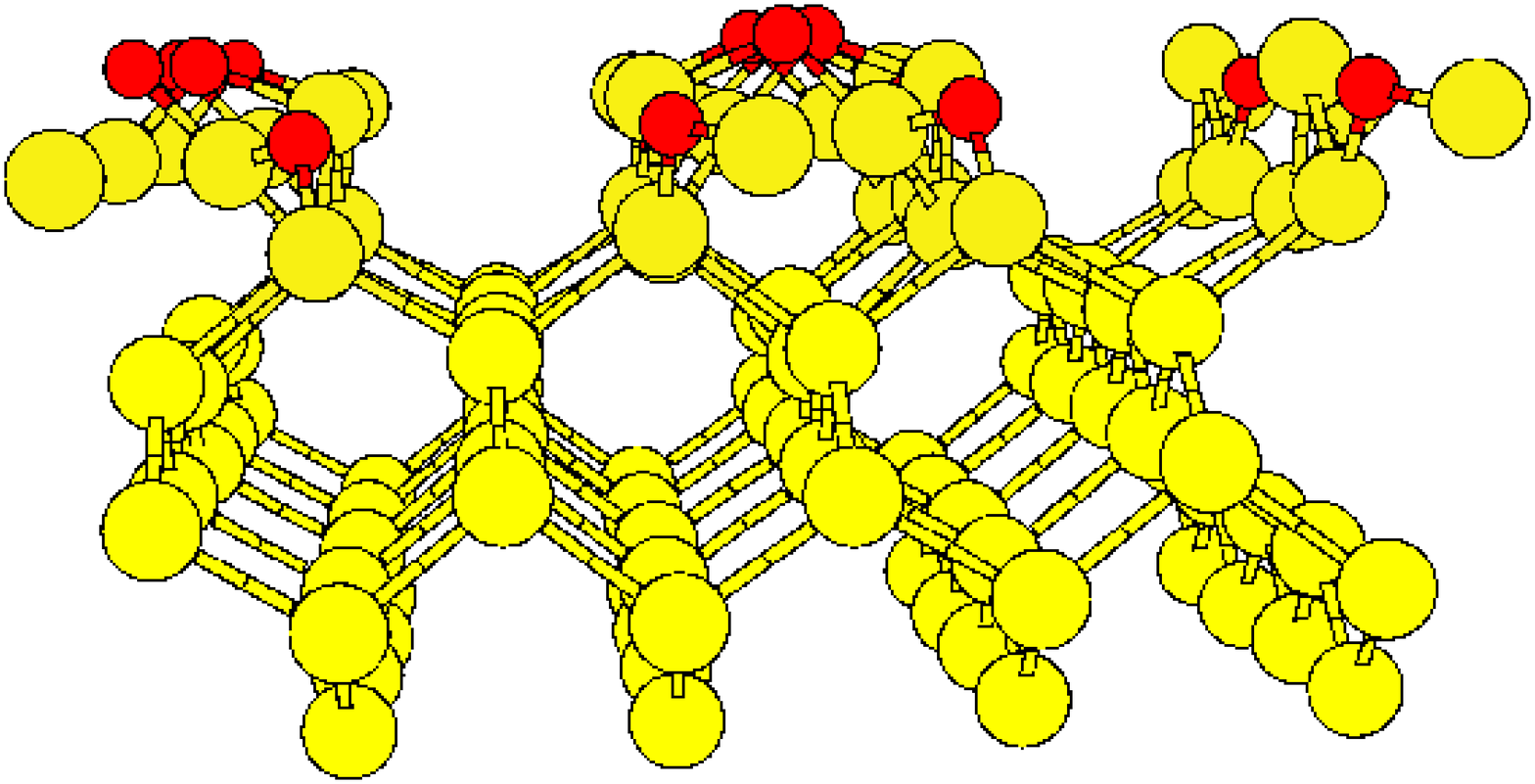,clip=,width=6cm}}\hfill
\end{minipage}
 \hspace{1.5cm}
 \begin{minipage}{5cm}
 \caption[Si(100)-p(2$\times$2):O (1ML)]{
     Surface structure of Si(100)(2x2):O at 1ML coverage
     with oxidation of Si dimers and backbonds.
     oxygen atoms are depicted in dark gray (red), while light gray (yellow)
     circles represent  bulk and surface Si atoms.
     Dimer chains are oriented along the x direction.
     (a): top view of the surface (xy plane), with the surface unit cell;
     (b): lateral view of the half slab (yz plane).
\label{fig:str} }
 \end{minipage}
\end{figure}
%
% FIGURE. 4 - RAS
% ---------
\begin{figure}[h!]
   \centerline{\epsfig{file=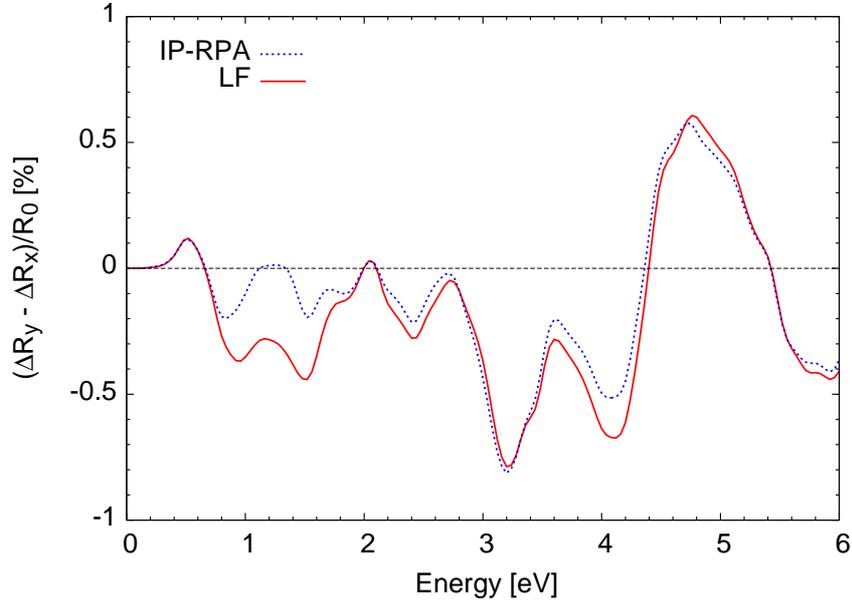,angle=270,clip=,width=12cm}} 
   \caption[Si(100)-p(2$\times$2):O (1ML) - RAS]{ 
           Calculated RAS spectrum of Si(100)(2x2):O at convergence, for the structural model shown in
           figure \ref{fig:str}. Results including (LF) or neglecting (IP-RPA)
           the local field effects are very similar, except for the region between 0.8
           and 1.8 eV, where the LF effects strongly enhance the RAS signal.
           The energy scale has been shifted by 0.6 eV to compensate for the neglect of
           self-energy effects.
           }
\label{fig:RAS}
\end{figure}
%
%
%
% FIGURE 5 - SDR
\begin{figure}[h!]
   \centerline{\epsfig{file=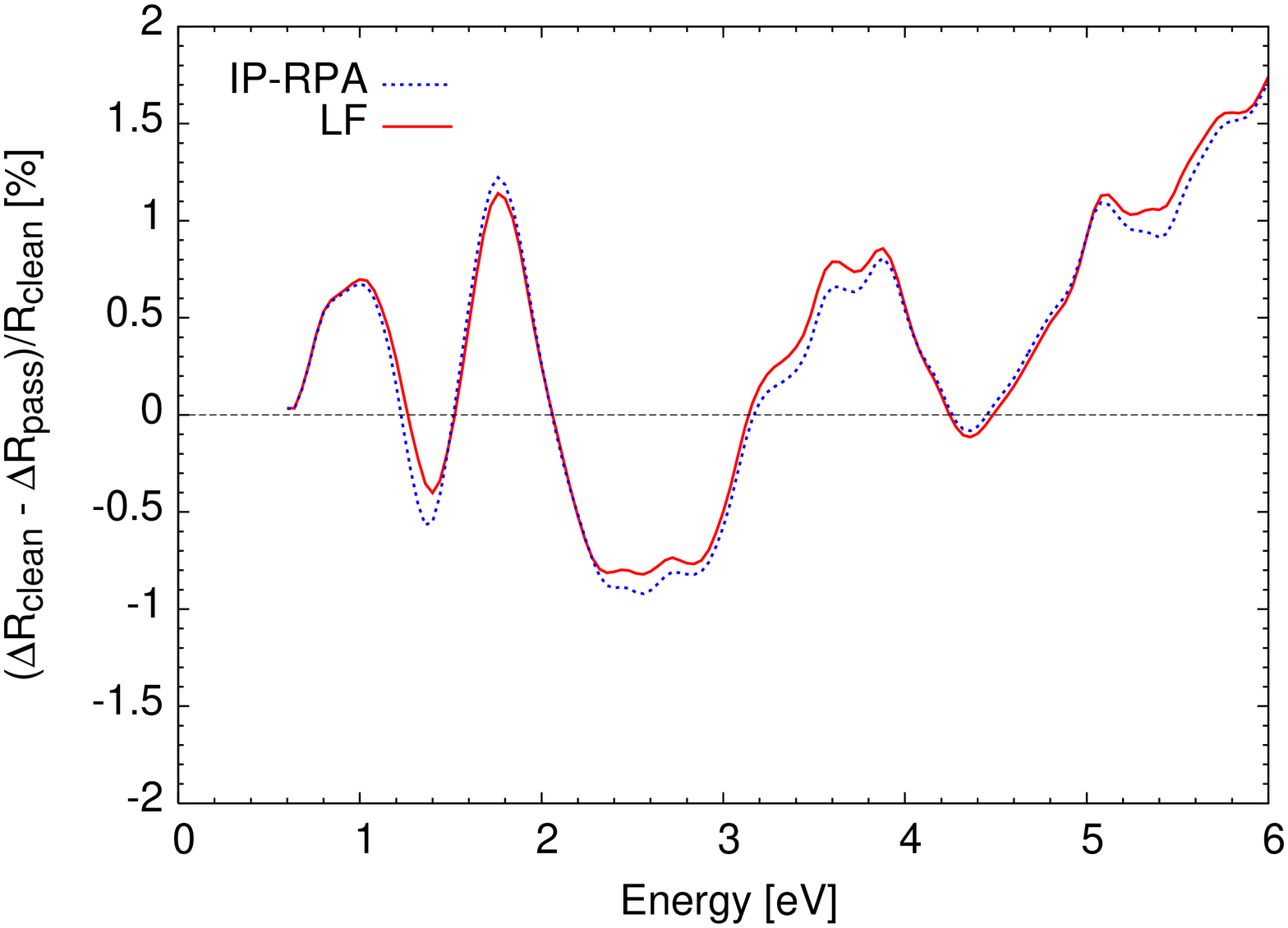,angle=270,clip=,width=12cm}} 
   \caption[Si(100)-p(2$\times$2):O (1ML) - SDR]{ 
           Calculated SDR spectrum (unpolarized light) of Si(100)(2x2):O, for
           the structural model shown in figure \ref{fig:str}.
           Results including (LF) or not (IP-RPA) the local fields are
           almost indistinguishable, showing that the effects visible in the
           low-energy part of figure \ref{fig:ImE_x-y-z} are canceling each other
           in the SDR spectrum. The same energy shift as in figure \ref{fig:RAS}
           has been applied. In \emph{polarized} SDR  the local field effects
           would be of the same size as for the RAS spectra.
           }
\label{fig:SDR}
\end{figure}

   We first discuss the case of RAS. The effects of local fields on the imaginary part
   of the dielectric tensor are most evident in the low-energy region of the spectrum
   (below 2 eV), as shown in the inset of Figures \ref{fig:ImE_x-y-z}a and \ref{fig:ImE_x-y-z}b.
   In particular, LF are found to enhance and sharpen the strong $\varepsilon_{2}$
   peak at about 1.2eV for light polarized along the dimer chains (fig. \ref{fig:ImE_x-y-z}b),
   and to reduce the first three peaks for light polarized along the dimer axis.
   As a result, LF induce a strong enhancement in the surface optical anisotropy (of the order
   of 100\%) in the region between 0.8eV and 2eV, as displayed in fig. \ref{fig:RAS}.
   This low-energy region (below the direct gap of bulk Si) corresponds to surface-localized
   states, which are expected to carry the surface anisotropy. The fact that LF
   evidence this anisotropy is consistent with the fact that dimer chains realize a structure
   which is geometrically strongly inhomogeneous in the direction perpendicular to
   the dimer chains (see fig. \ref{fig:str}).
   At higher energies (above 2eV) bulk contributions dominate $\varepsilon_2$,
   and the resulting RAS is mainly due to surface perturbed bulk states.
   The latter appear to be less affected by local fields than the true surface
   states, and lead to a RAS spectrum which, above 2.0eV, is almost
   insensitive to the inclusion of local field effects.
   This picture is confirmed by the analysis of SDR results.
   The latter are in fact calculated for unpolarized light, i.e. by averaging
   $\varepsilon_{xx}$ and $\varepsilon_{yy}$. Since LF enlarge $\varepsilon_{yy}$
   and reduce  $\varepsilon_{xx}$, their effects almost completely cancel out when the average is
   taken. Our calculated (unpolarized) SDR spectrum, displayed in fig. \ref{fig:SDR},
   appears in fact to be very little affected by the local fields, in the whole
   energy range between 0 and 6~eV.
   However, if a polarized SDR spectrum is computed, then local fields are found to
   influence the low-energy region ($\leq 2~$eV), in a way which is very similar to the
   behavior of the RAS.

   Unfortunately, it is not possible to perform here a comprehensive
   comparison with RAS and SDR experimental data, since this would require the calculation of
   several possible reconstruction and geometries.
   In fact, the oxidation mechanism of Si(100) has been shown to be exceedingly complex,
   with different mechanisms playing their role depending on the oxidation temperature:
   a barrierless oxidation of the first Si layer \cite{ciacchi-prl-2005}, or an ``active oxidation''
   involving etching of the surface and penetration of oxygen in a layer-by-layer manner
   at higher temperature \cite{yasuda-prl-87,yasuda-prb-67,albao-surfsci-2005}.
   Recently, the Si(100)(2x1):O surface optical anisotropy has been shown to be sensitive to
   the structural details of the oxygen adsorption by ab--initio calculations of the atomic
   geometries and optical response of a large number of Si(100):O structures
   \cite{bechstedt-prb-2005, bechstedt-jphyschem-2005}.
   An highly structured potential energy surface has been found, with minima at the
   backbonds of the ``down'' atoms in Si--Si dimers \cite{bechstedt-jphyschem-2005}.
   Moreover, an appreciable amount of disorder is probably present after oxidation
   of the first Si monolayer, and the local strain induced by oxygen adsorption
   is expected to have a sizable impact on the optical anisotropy spectra \cite{bechstedt-prb-2005}.

   However, our findings for LF effects in the single case studied here suggest an important
   general remark about surface optical spectroscopies.
   In fact, it is well known that, due to the large penetration depth of visible
   and UV photons, the surface--specific optical reflectivity signal is very
   small with respect to the bulk contribution.
   For materials with an isotropic bulk, the RAS spectroscopy has
   indeed been developed in order to extract the surface signal, by exploiting its
   anisotropy. A correct evaluation of the latter has hence the highest priority
   in  theoretical calculations of surface optical spectra.
   The fact that crystal local fields are potentially able to alter significantly
   the surface optical anisotropy, at least below the bulk bandgap,
   should hence be kept in mind, particularly when the anisotropy of electronic
   states is associated with a large  structural anisotropy at the surface,
   such as in the case of dimer chains on Si(100)(2$\times$1).

%%%%%%%%%%%%%%%%%%%%%%%%%%%%%%%%%%%%%%%%%%%%%%%%
%%%%%%%%% COMPUTATIONAL SCALING AND PERFORMANCES
%%%%%%%%%%%%%%%%%%%%%%%%%%%%%%%%%%%%%%%%%%%%%%%%
\section{Computational scaling and performances}

 In this section, we present a quantitative analysis
 of the numerical performance of the HT-based approach,
 as implemented in the large scale code DP \cite{DP}, with respect to the traditional
 approach.
 Several calculations have been done
 by varying the three main convergence parameters:
 (i) the number of valence--conduction transitions ($N_{t}=N_{{\bf k}}\times N_{v}\times N_{c}$);
 (ii) the number of frequency intervals considered in the spectrum,
 i.e. the spectral resolution (number of frequencies, $N_{\omega}$);
 (iii) the number of plane--waves considered in the response matrix ($N_{{\bf G}}$). 
 Optical properties usually converge at an $N_{{\bf G}}$  value which can be substantially
 smaller than the total number of plane waves,  $N_{g}$, used to describe the
 wavefunctions.

 The calculation of $\chi^{(0)}$ is expected to scale, in the case of the reference
 approach, as \footnote{$N_t N_{\bf G}^2$ leads to the $ N^4_{at}$ scaling mentioned 
 in the introductory section.}:
 \begin{equation}
  T_{ref} = N_{{\bf k}}\left[ \alpha N_{v}N_{c}N_{g}\textrm{log}N_{g} +
                     \beta N_{v}N_{c}N^{2}_{{\bf G}}N_{\omega}
                         \right] +
                         \textrm{A}
   \label{eq:Tref}
 \end{equation}
 where $N_{{\bf k}}$ is the number of {\bf k}-points, and $\alpha$ and $\beta$ are prefactors
 which are independent on $N_{v},N_{c},N_{{\bf G}},N_{{\bf k}},N_g$, and $N_{\omega}$.
 The first term in equation (\ref{eq:Tref}) is due to the evaluation of
 the numerators $Z_{n}$ in equations (\ref{eq:r2}) and (\ref{eq:a2}) by using FFT,
 and is present both in the reference and Hilbert approach.
 The second term stems from to the evaluation of equation (\ref{eq:chi0noi})
 in the traditional way.
 The remaining term A keeps into account residual parts of the calculation,
 as the matrix inversions, which contribute much less to the CPU time than the first two terms.
 
 The expected scaling in the case of the Hilbert-based scheme is instead:
 \begin{equation}
T_{new} = N_{{\bf k}}\left[ \alpha N_{v}N_{c}N_{g}\textrm{log}N_{g} + \beta ' N_{v}N_{c}N^{2}_{{\bf G}}\right]
                      + \gamma N^{2}_{{\bf G}}N^2_{\omega}
                      + \textrm{A}'
  \label{eq:Tnew}
 \end{equation}
In this case, the second term does not contain the factor $N_{\omega}$
anymore, and its prefactor becomes $\beta '$, due to the calculation
of $\chi^{A2}$ and $\chi^{R2}$ (equations \ref{eq:r2} and \ref{eq:a2}).
The calculation becomes, in this sense, comparable to a static one.

However, the actual evaluation of the Hilbert transforms
(equations \ref{eq:Ha2} and \ref{eq:Hr2}) introduces a new term scaling as
$N^2_{{\bf G}}N^2_{\omega}$. Due to the small prefactor $\gamma$, the latter
term can often be neglected (see, e.g., figure 3 of reference \cite{MA99}).
In the present work, we found that the CPU time spent inside the Hilbert
transform itself can be made negligible by an optimized algorithm\footnote{Exploiting 
the fact that the principal value numerical integration routine has to be 
called $N^2_{{\bf G}}$ times, always on the same energy intervals, a substantial 
gain could be achieved by tabulating the (about $10^6$) required values of the complex logarithm
once and for all at the beginning of the double loop over the number of ${\bf G}$ vectors
in the construction of $\chi^{(0)}$.}.

\begin{figure}[h!]
 \begin{minipage}{7.2cm}
 \includegraphics[width=5.5cm,angle=270]{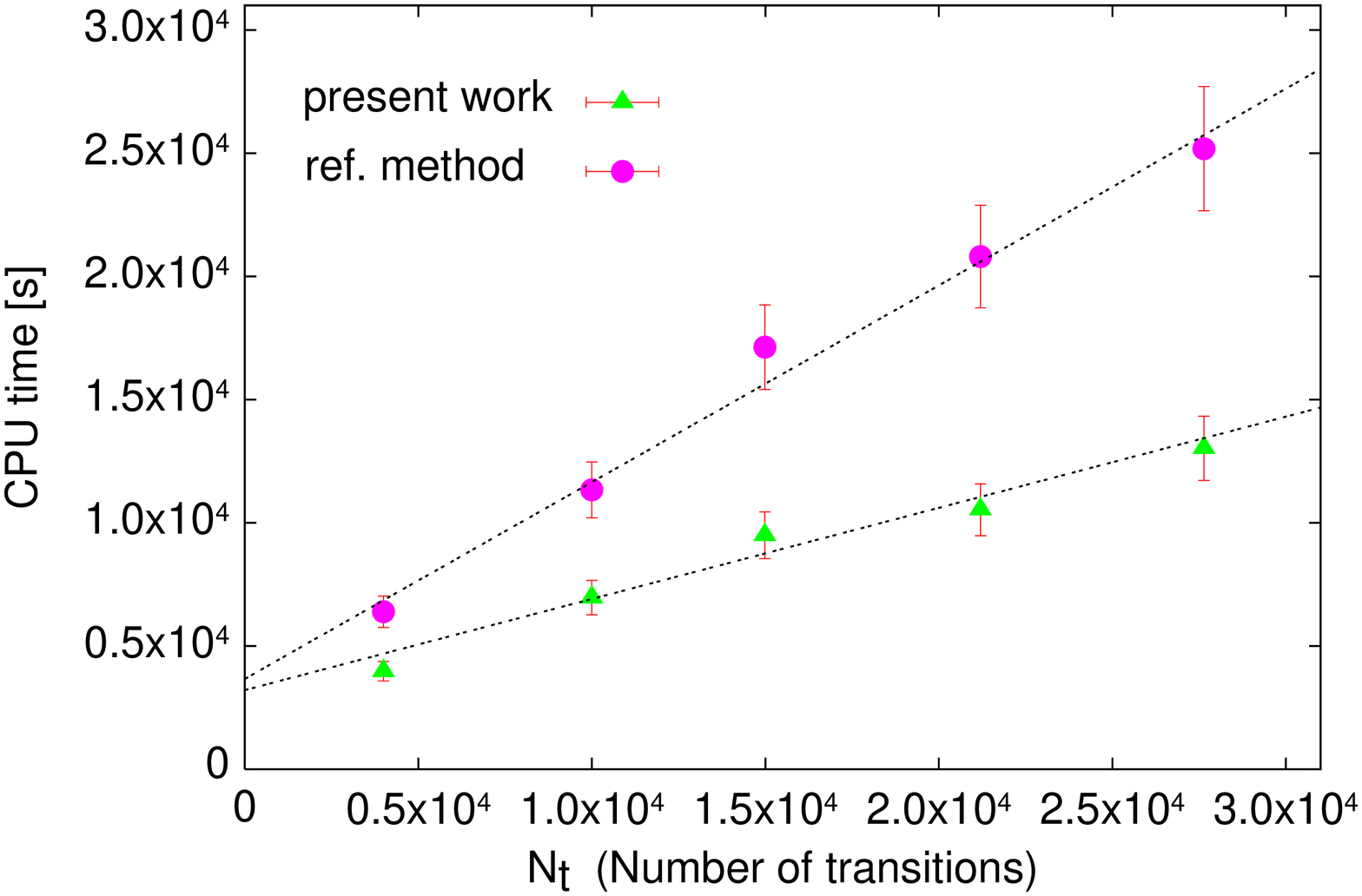}\\
 \includegraphics[width=5.5cm,angle=270]{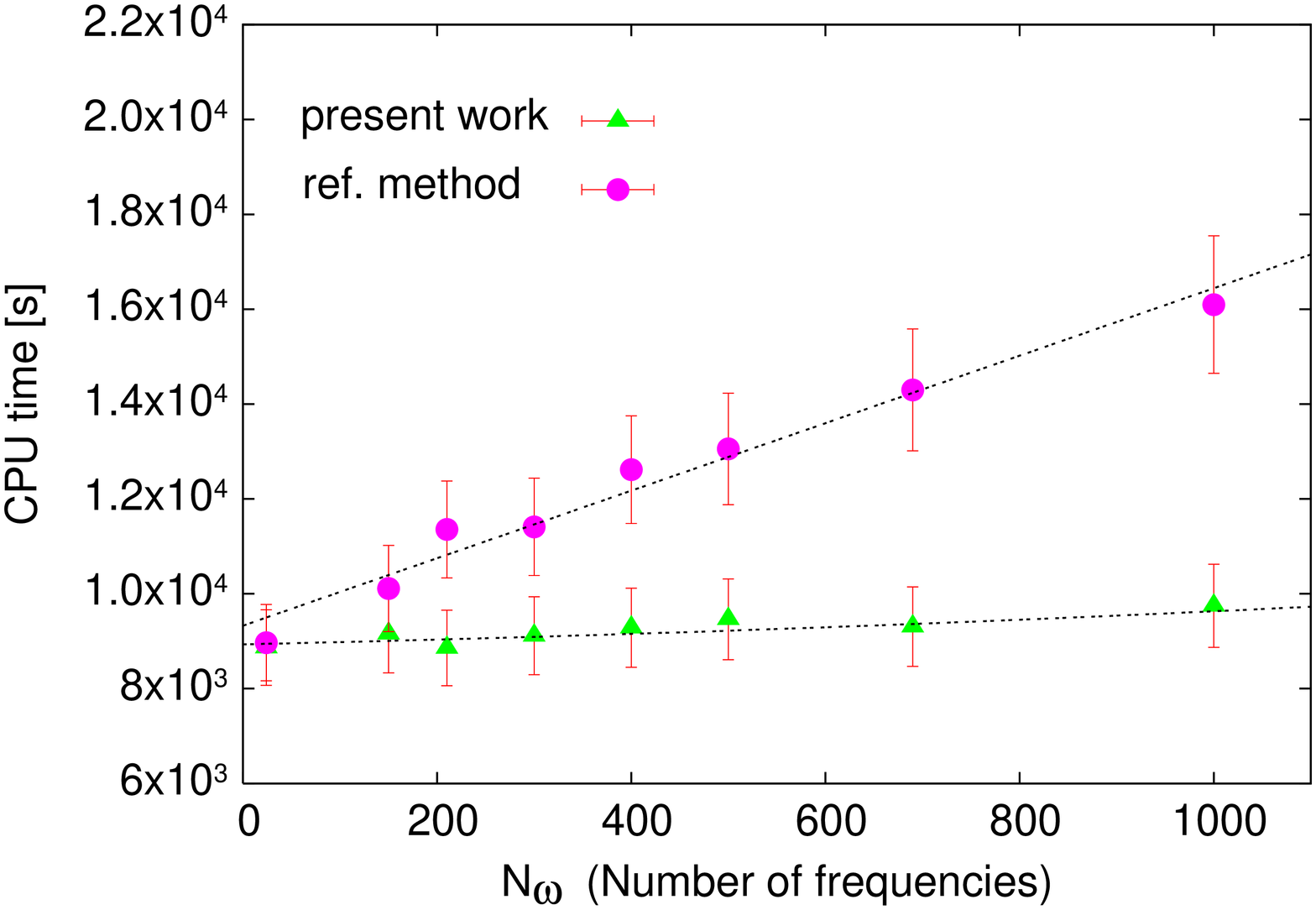}\\
 \includegraphics[width=5.5cm,angle=270]{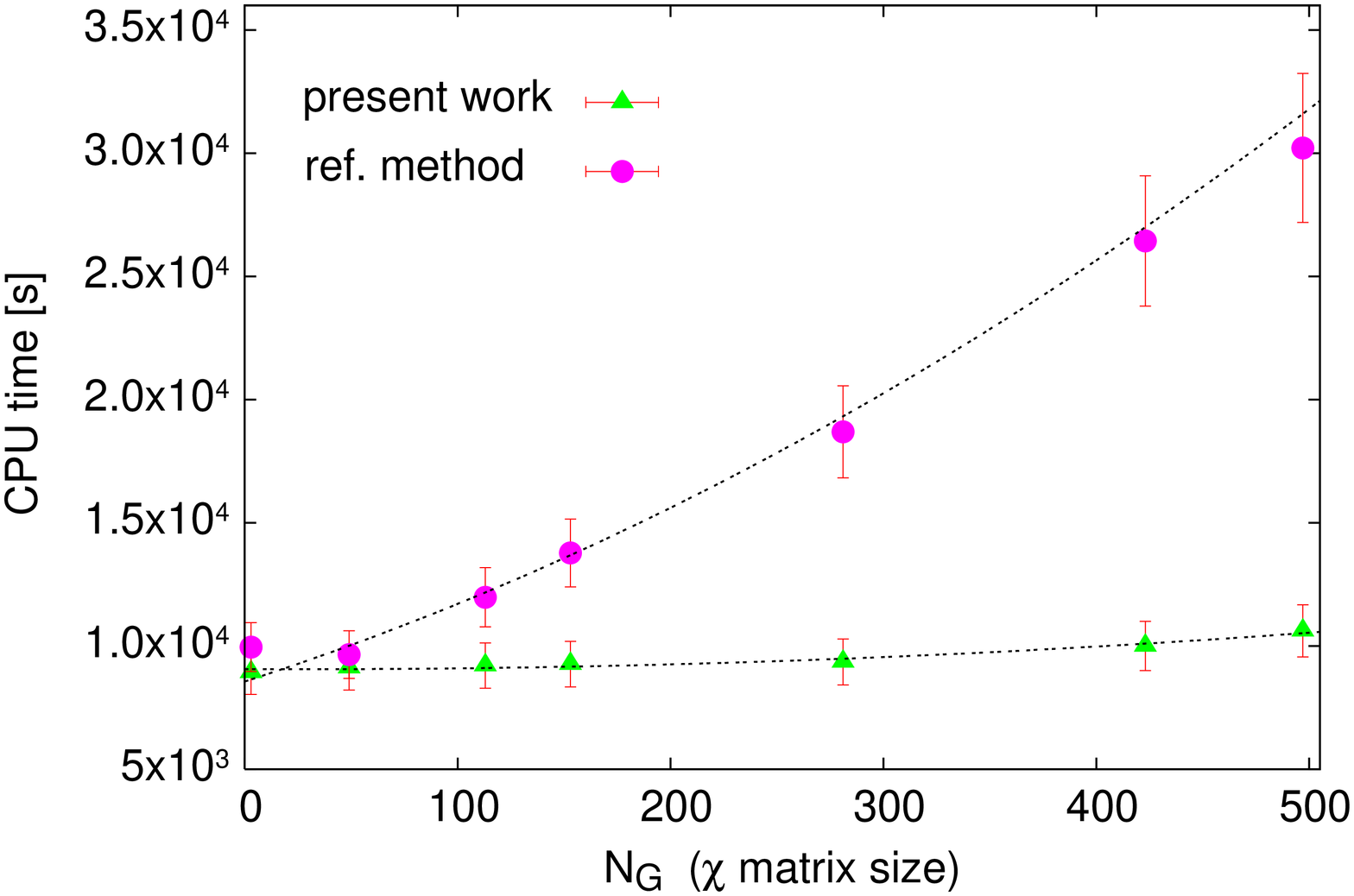}
 \end{minipage}
 \hspace{1.2cm}
\begin{minipage}{5cm}
\caption[Time scaling HT method (real system)]{\label{fig:cpu_realsystem}
     Quantitative study of the computational load required to evaluate the
           $\chi^{(0)}$ matrix for the 56-atoms slab representing a Si(100)(2x2):O surface.
           The effects of the three main parameters determining the numerical convergence
           of the theoretical spectra are studied separately.
           (a): number of valence--conduction transitions,
                determined by the energy cutoff on the empty
                (conduction) bands included in eq. (\ref{eq:chi0heavy}),
                the number of occupied bands being fixed;
           (b): number of frequency intervals taken on the $\omega$ axis,
                determined by the requested spectral resolution;
           (c): size of the $\chi^{(0)}$ matrix in reciprocal space,
                roughly proportional to the system size
                (for a fixed real-space resolution).
}
\end{minipage}
\end{figure}

Considering an N-atoms unit cell, the number of transitions $N_{t}$ is
clearly the parameter growing fastest with the system size,
since it is proportional to N$^2$.
About 22000 transitions per ${\bf k}$-point, corresponding to the inclusion
of about 200 empty bands, are requested to converge the
dielectric tensor of the Si(100)(2x2):O slab up to 12 eV.
The number of frequency intervals, $N_{\omega}$, is instead independent on N, 
but it grows linearly with the required spectral resolution.
In the present case, 300 frequencies have been necessary
in order to achieve a 40 meV resolution over a spectral range of 12 eV.
Finally, $N_{\bf G}$, i.e. the size of $\chi^{(0)}$ in reciprocal
space, depends on the requested real-space resolution needed
in the description of the induced density variations. This means
that larger $N_{\bf G}$ will be necessary to describe systems with
smaller interatomic distances, or with larger polarizability.
The real-space resolution  is  independent on the system size;
however, for a fixed resolution $N_{\bf G}$ will  grow linearly
with the volume of the unit cell in direct space. $N_{\bf G}$ is hence
proportional to the number of atoms (in a bulk system) or to the
volume of the supercell (for a finite or semiinfinite system).
In our slab calculation, converging the spectra with local field effects 
requires to consider at least 113x113 matrices (incidentally, we stress that
the IP-RPA spectra, requiring just the ${\bf G}=0$, ${\bf G}'=0$ matrix element
of $\chi^{(0)}$, do not depend on $N_{\bf G}$ ).

The HT algorithm turns out to be clearly advantageous with respect to
$N_{t}$ and $N_{\bf G}$, as shown in the first and last panels of figure
\ref{fig:cpu_realsystem}.
Concerning $N_{\omega}$,  despite an unfavorable scaling in the limit
of infinite spectral resolution  (the CPU time grows quadratically with
the number of frequency intervals), one must notice that, due to the small
prefactor $\gamma$, the HT method is largely convenient in whole range of
interest ($N_{\omega} \simeq 10^3$).

 \label{sec:scaling}
\section{Conclusions}
   Two classes of conclusions can be drawn from the presented results.
   First, the physics: the study of Si(100)(2x2):O has shown that
   local field effects,  although playing only a minor role on the surface optical
   properties above the bulk bandgap, are able to enhance substantially the
   surface optical anisotropy in the low--energy end of the spectra.
   A similar effect can be expected for other surfaces, when the
   anisotropy of electronic states is associated with a large structural
   anisotropy, such as in the case of the dimer chains on
   Si(100)(2x2).
   Moreover, large local field effects are found for light polarized
   normally to the surface.
   Second, the numerics: the computational gain achievable by
   using the Hilbert transform-based algorithm has been shown to be substantial,
   both in  a model system \emph{and} in a real, physical application.
   A successful implementation of the Hilbert transform method in the large-scale
   plane-waves \emph{ab initio} computer code DP \cite{DP}
   allows us to locate the crossover (starting from which the Hilbert transform algorithm
   becomes convenient) already at medium size systems (less than 50 atoms).

%
%%%%%%%%%%%%%%%%%%%%%%%
  \part{Applications}
%%%%%%%%%%%%%%%%%%%%%%%
%
\chapter{The clean Si(100) surface} \label{ch:a-si100}
 Surfaces are complex physical systems that are very important from
both the fundamental and technological point of view.
One of the most important surfaces of all is the (100) surface of silicon,
as it appears in most electronic devices. Nevertheless,
in spite of its many applications, a completely unambiguous structural and
physical description of Si(100) is lacking.
In this chapter we summarize our results about
first principles calculations of electron energy loss spectra
of the Si(100) clean surface.

\section{Si(100): which reconstruction?}
Due to its enormous technological importance, the Si(100) surface has been the
subject of a wide range of experimental and theoretical studies spanning several decades.
In fact, quality publications continue to appear
regarding the atomic structure and electronic properties of the clean surface.
Following early LEED experiments~\cite{schlier_jcp_y1959_v30_p917}, it was understood that Si(100)
forms a \pone\ reconstruction. The classic explanation of the LEED observation
is that the surface is composed of rows of Si dimers separated by trenches
(Fig.~\ref{fig:structures_clean}), as confirmed by various scanning
tunnelling microscopy studies~\cite{ono_prb_y2003_v67_p201306,sagisaka_prl_y2003_v91_p146103}.
Although some quantum chemistry studies have found that a symmetric dimer
structure (causing a metallic surface) forms the global
minimum~\cite{shoemaker_jcp_y2000_v112_2994},
several total energy calculations based on density functional
theory~\cite{ramstad_prb_y1995_v51_p14504} have found that dimer buckling
induces a small energy gain, such that the dimers adopt an asymmetric configuration
and the surface remains semiconducting~\cite{chadi_prl_y1979_v43_p43}. \\% Cite some STS? \\
\indent Three distinct structures have been proposed for the Si(100) surface: the \pone,
whereby all dimers are buckled the same way (see Fig.~\ref{fig:structures_clean}(a));
the \ptwo\ structure, where alternating dimers in a row are buckled in
opposite directions, and adjacent rows are buckled in phase 
(see Fig.~\ref{fig:structures_clean}(b));
the \ctwo\ phase, being the same as the \ptwo\ but with adjacent rows
buckled out of phase (see Fig.~\ref{fig:structures_clean}(c)).
Total energy calculations have found~\cite{ramstad_prb_y1995_v51_p14504} 
that the \pone\ reconstruction is prohibitively higher in energy 
than the other two (at zero K), and that the \ctwo\ is only slightly 
favoured over the \ptwo.
Which reconstruction occurs on the surface depends critically on the temperature.
LEED studies have shown that an order-disorder phase transition occurs at about
200K~\cite{ihm_prl_y1983_v51_p1872,tabata_ss_y1987_v179_p63}.
Below this critical temperature, a \ctwo\ phase is generally observed; above it,
a \pone\ periodicity is seen.
Direct observation of the surface structure with STM is complicated by two factors however.
Firstly, it is well established now that the experimental measurement itself
can influence the result, and drive \ctwo$\to$\ptwo\ phase 
transitions~\cite{yoshida_prb_y2004_v70_p235411,nakamura_prb_y2005_v71_p113303}.
Charge injection, or electric fields induced by the STM tip, can cause dimers to
flip, according to various 
experimental~\cite{sagisaka_prb_y2005_v71_p245319,nakamura_prb_y2005_v71_p113303}
and theoretical works~\cite{kantorovich_prb_y2006_v73_p245420,seino_prl_y2004_v93_p036101}.
Nevertheless, the general consensus is that the \ctwo\ reconstruction is the more
stable structure below the critical temperature~\cite{nakamura_prb_y2005_v71_p113303,shirasawa_prl_y2005_v94_p195502}. \\
\indent Above 200K, STM images appear to show a symmetric dimer configuration.
However, at these temperatures the dimer rocking mode is activated,
and hence it is believed that the observed symmetric \pone\ structure is merely a time average
of the thermal flip-flop motion of the buckled dimers.
Based on molecular dynamics simulation of the dimer motion, it was suggested that
the surface consists of simultaneous local presence of asymmetric dimers, and
of instantaneously flat symmetric dimers~\cite{shkrebtii_prb_y1995_v51_p11201}.
More recent studies have suggested that the dimers remain short-range correlated
(see refs. 32--35 in Ref.~\cite{kutschera_apa_y2007_v88_p519}).
In particular, a two photon photoemission (2PPE) study found minor difference 
between the surface band dispersion at 90K and at room temperature~\cite{weinelt_apa_y2005_v80_p995}. \\
\indent Theoretical simulations of the reflectance anisotropy (RA) spectra
confirm that the \pone\ reconstruction does not reproduce correctly the
experimental lineshape~\cite{palummo_prb_y1999_v60_2522,fuchs_prb_y2005_v72_p075353,fuchs_thesis_y2004}.
On the other hand, both \ctwo\ and \ptwo\ structures
are quite similar to the experiment, with the \ctwo\ yielding a slightly
better agreement~\cite{palummo_prb_y1999_v60_2522}, in particular predicting
the observed SDR structure below 1~eV.
Simulation of the surface differential reflectance also favours the \ctwo\ 
surface~\cite{palummo_prb_y1999_v60_p2522}. \\
\indent In addition to optical techniques such as RAS or SDR,
electron energy loss spectroscopy (EELS) in the reflection geometry (REELS or RELS)
offers an enhanced surface sensitivity and easy access to a wide energy range
(see chapter~\ref{ch:surface}).
Although the majority of literature considering REELS of Si(100)
has focused on vibrational properties~\cite{ibach_book},
several studies have examined the nature of electronic states
at the clean Si(100) surface~\cite{ibach_prl_y1973_v31_p102}.
Indirect information about surface states was derived from
related studies looking at the changes in the REEL spectrum following 
oxidation~\cite{ibach_prb_y1974_v10_p710,ibach_prb_y1974_v9_p1951,ludeke_prl_y1975_v34_1170}.
High resolution (HREELS) measurements were carried out by Farrell\ea\ \cite{farrell_prb_y1984_v30_p721}
and Gavioli\ea\ \cite{gavioli_ss_y1997_v377_p360}.
In the latter work, tight binding calculations were performed on the \pone\ symmetric
and asymmetric dimer models, and suggested that a mixture of the two structures
was necessary to explain the room temperature EEL spectra.
However, the \ctwo\ structure was not considered in that work, and therefore some
of the conclusions reached are not complete. \\
\indent In the following sections, we present a computational study of high resolution EELS
for the different reconstructions of Si(100): \pone, \ptwo\ and \ctwo\.
We consider the energy range that probes the excitation of
interband, i.e., about 0--6~eV, and hence
connect the experimental observation directly with the atomic structure and
microscopic electronic response.

 \section{First principle scheme}
 We use density-functional theory within the local density approxiation (DFT-LDA),
within a plane-wave and pseudopotential framework.
The ABINIT \cite{gonze_cms_y2002_v25_p478,abinit2} and 
quantum-espresso/PWSCF \cite{pwscf-long}
codes were used for computing the relaxed atomic structures, electronic bandstructures, and
Kohn-Sham eigenvalues and eigenvectors required for the evaluation of the 
optical properties.
However, in order to be consistent and since we found only minor differences between spectra computed with the
two codes, we report only the final results obtained using the output of PWSCF. \\
For all our calculations we used standard norm conserving pseudopotentials
of the Hamann type \cite{hamann_prb_v40_y1989_p2980} for both silicon and oxygen
generated with FHI98PP package \cite{FHI98PP} within DFT-LDA
(Perdew-Zunger parametrization \cite{perdew_prb_y1981_v23_5048}) framework.

\section{Geometric structure}
\begin{figure}
 \begin{center}
  \includegraphics[angle=270,width=9.5cm]{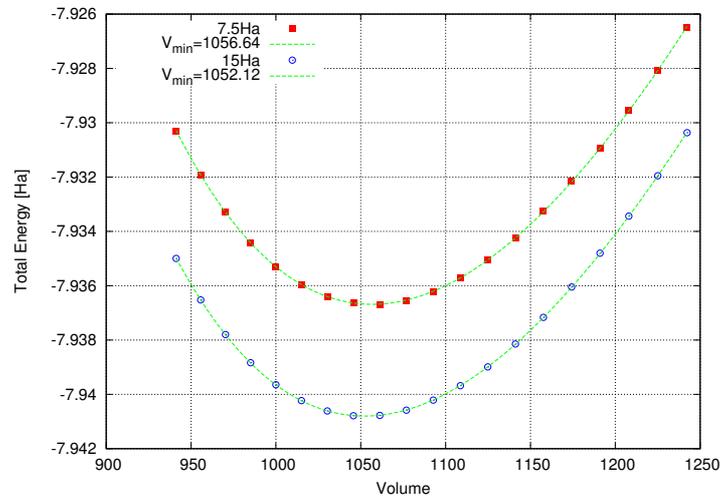}
 \caption[Bulk silicon lattice parameter]{
         Optimization of the lattice parameter of bulk silicon 
         at 7.5~Ha and 15~Ha cutoff energy. \label{fig:si-a}}
 \end{center}
\end{figure}

 Even if a 7.5~Ha kinetic energy cutoff 
is rasonable in order to treat the case of a clean 
silicon surface, we used a 15~Ha cutoff throughout, 
as was previously found to be sufficient for oxidized 
Si(100) with the same pseudopotential~\cite{incze_prb_y2005_v71_035350}. 
The effect of the two cutoff is shown in Fig.~\ref{fig:si-a} where 
it is clear that the minimum of the curve is not appreciably different.

Therefore we used the theoretical lattice constant of 5.393~\AA,
as determined at 15~Ha %10.191 au
A standard repeated-slab and supercell approach was adopted in order to model the surface
structure. \\
\begin{figure}[!tbh] 
 \centerline{\epsfig{file=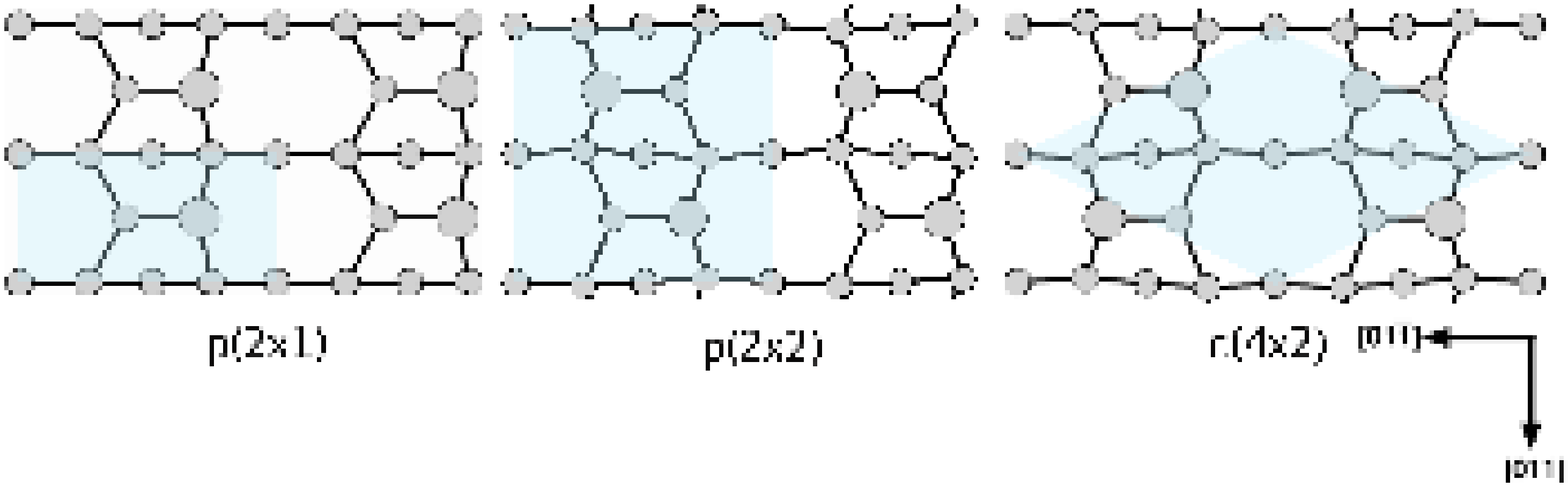,angle=0,clip=,width=13.5cm}}
\caption[Si(100)-\pone, -\ptwo\ and -\ctwo]{ 
        Ball and stick model of \pone, \ptwo\ and \ctwo\ surface reconstructions.
        Large circles indicate ``up'' silicon dimer atoms.
        Unit cells are indicated by shaded regions.\label{fig:structures_clean}
} 
\end{figure}
We use relatively thick slabs (16 atomic layers) separated by 8 layers of vacuum (about 10~\AA).
During the geometry optimization, the central four layers were fixed at the bulk positions and
structures were relaxed until the cartesian force components were less than 20~meV/\AA. % ABINIT
Our obtained structural parameters are similar to those obtained previously for this surface (see 
for example Ref.~\cite{palummo_prb_y1999_v60_2522})
such as a dimer buckling of 0.755~\AA\ and a dimer length of 2.33~\AA.
\newline

\section{Computational aspect}
 Optical spectra and energy loss spectra were calculated using the 
YAMBO code \cite{yambo-long}, taking Kohn-Sham eigenvalues and 
eigenvectors from a non-self-consistent run of the PWSCF code \cite{pwscf-long}.
We carried out convergence tests on the optical and energy loss properties with
respect to the number of bands and the number of $\bk$-points.
For the calculation of energy loss we %above the surface plasmon feature ($~\sim 12$~eV) we
included up to 320 unoccupied bands, corresponding to minimum $e$--$h$ transition energy
of $\sim 13.5$~eV.%, and which yields well-converged surface plasmon peaks.
  \begin{figure}[h!]
   \centerline{\epsfig{file=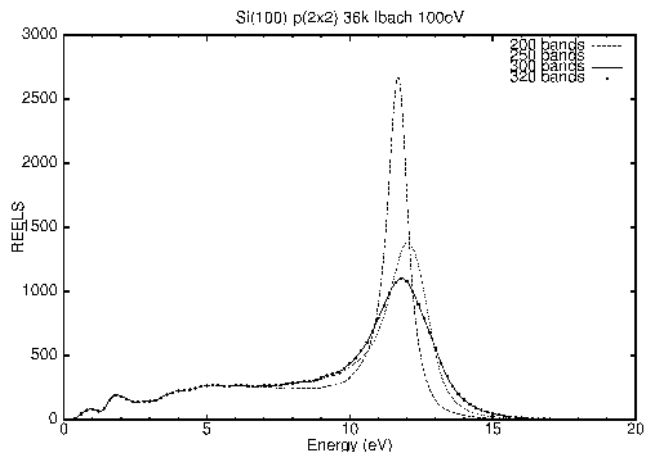,angle=270,width=8.5cm}}
     \caption[EELS: band convergence]{
         EEL spectrum of the clean Si(100)-p(2$\times$2) surface
         calculated for incident energy E$_i$=100eV and $\theta=42$.
         The height of the surface plasmon peak is mostly affected by the
         convergence respect to the number of bands.\label{fig:conv}
            }
  \end{figure}
Further convergence tests were carried out
regarding $\bk$-point sampling. For the most converged calculations, we used
roughly equivalent density sets for the three reconstructions:
for the \ctwo, 72 $\bk$-points in the irreducible Brillouin zone (IBZ), 
which is equivalent to 1152 points in the $(1\times1)$ BZ 
(for both clean and oxidized cells, see next chapter);
for the \ptwo, 64 $\bk$-points in the IBZ 
(equivalent to 1024 points in the $(1\times1)$ BZ);
for the \pone, 100 $\bk$-points in the IBZ 
(equivalent to 800 points in the $(1\times1)$ BZ). \\
%These parameters gave us a precision in peak energy of XXX, 
%and in peak intensity of XXX\%.
All spectra reported in this work were obtained using the non-interacting particle (RPA) level
of theory. Many-body effects, including local field and excitonic effects, were compensated for
by applying a scissors operator of $+0.5$~eV to the unoccupied states, following the recipe of
Del Sole and Girlanda~\cite{dels-moch-bar-91}.
In this way we account for the well-known underestimation of
the DFT band-gap, and partially include self-energy and excitonic shifts in energy.
The value of $+0.5$~eV was determined in other works on Si(100) as giving best agreement with the
experimental RAS \cite{incze_prb_y2005_v71_035350}. \\
In the following subsection we present some discussion of major technical
tools used in order to compute REELS.

 \subsection{Broadening}
 Inspection of Eq.~\ref{eq:Akk} reveals that the cross section goes as $1/q^3$ as $\omega \to 0$.
 Hence any features appearing in the loss function Im~$g(q_\parallel,\omega)$
 at low energy can be dramatically enhanced by the kinematic factor.
 From the computational point of view this means that unphysical features may appear
 close to the origin if a Lorentzian broadening is used when calculating 
 the surface dielectric function of a system with a small band gap. We
 adopt therefore a tiny Lorentzian broadening of $\delta_L = 0.006$~eV
 when calculating the dielectric function, and afterwards convolute the 
 loss spectra with a Gaussian (FWHM = 0.3~eV) to approach the 
 experimental resolution.
\begin{figure}[!h]
 \centerline{\epsfig{file=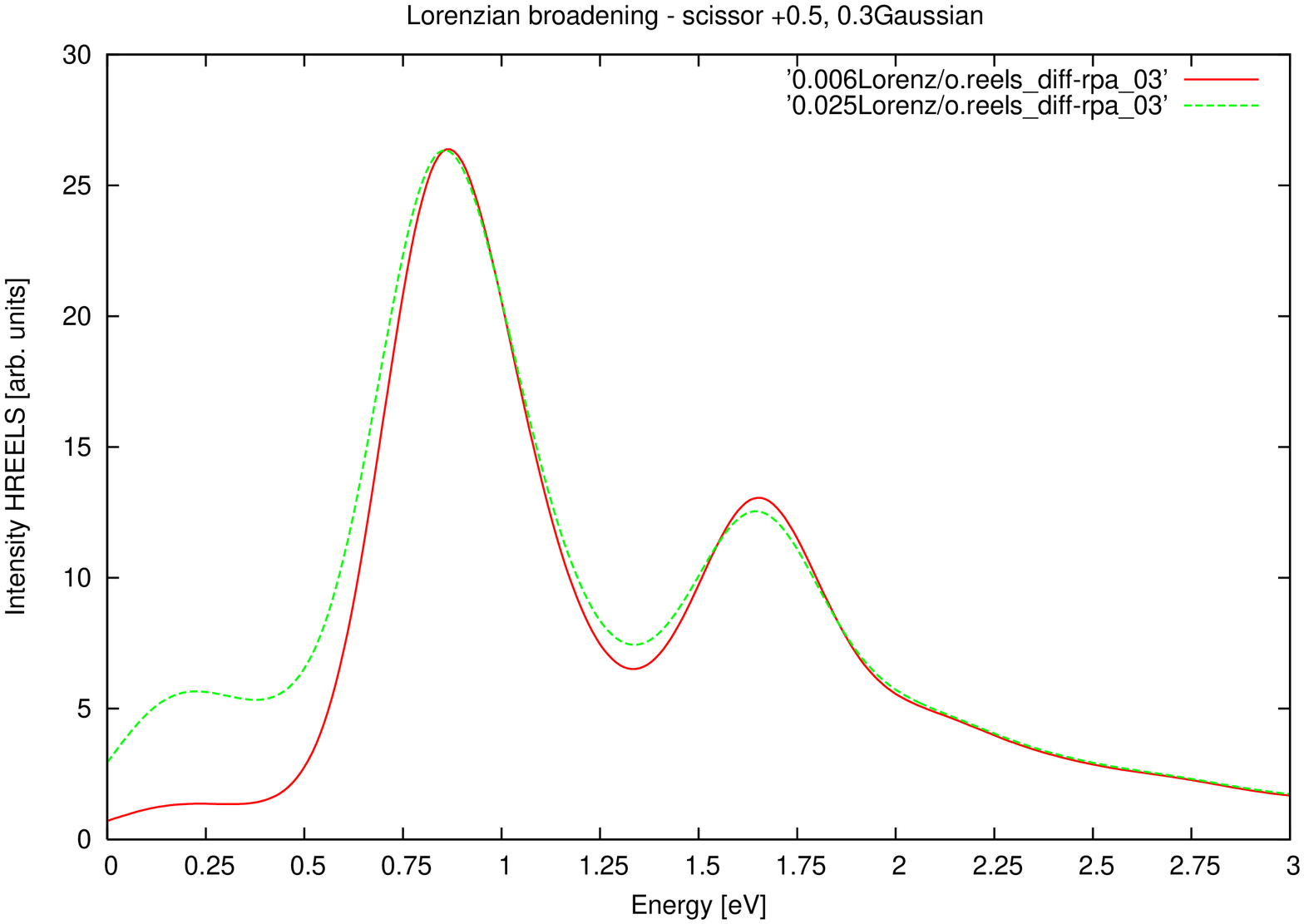,angle=270,clip=,width=10.0cm}}  
 \centerline{\epsfig{file=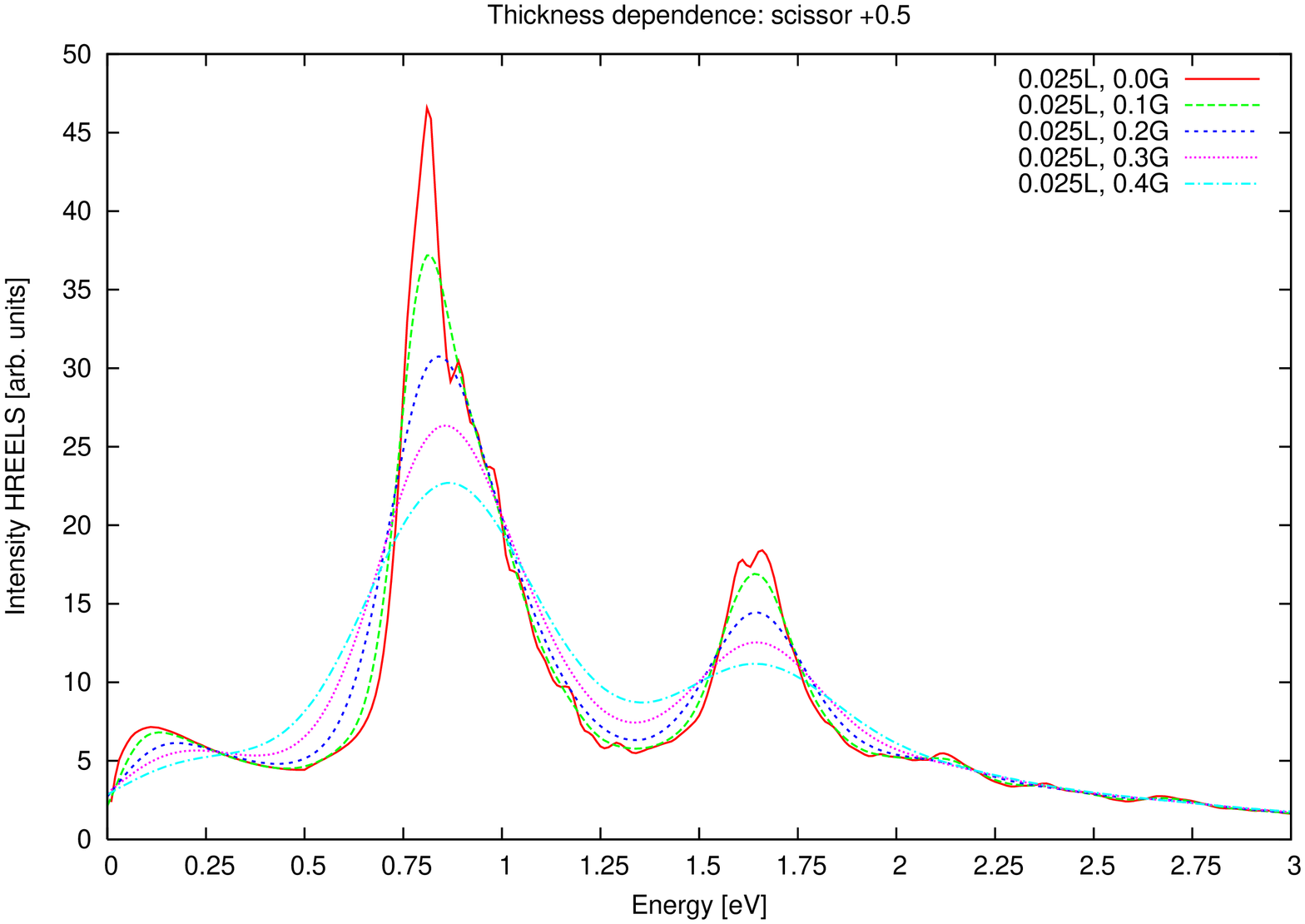,angle=270,clip=,width=10.0cm}}
  \caption[Broadening in the calculated EELS]{
          Effect of the Lorenzian broadening (top), 
          effect of the gaussian convolution (bottom)
          on the EEL spectra of a Si(100)-c(4$\times$2) 
          surface (E$_i$=7~eV and $\theta=$60).
         \label{fig:broadening}
          }
\end{figure}
\subsection{Slicing methods}
A crucial adjustable parameter present within the three-layer model of 
energy loss is the thickness $d$ of the surface layer, as used in Eq.~\ref{eq:eps_eff}.
According to the model, an electron impinging on the surface feels the potential
from this surface layer through its dielectric function $\varepsilon_s$ as well
as that of the bulk layer $\varepsilon_b$.
Obtaining $\varepsilon_s$ from the dielectric function of the slab or supercell
$\varepsilon_c$ is not trivial.
However early efforts used a simple subtraction of
the computed bulk dielectric function \cite{dels-moch-bar-91}:
\begin{equation}
  I(\omega) = (N_b -2N_s) d_l\varepsilon_b(\omega) + 2 N_s d_l\varepsilon_s(\omega)
\end{equation}
where I is the integral of the slab RPA dielectric susceptibility
$\varepsilon(\omega,z,z')$ over z and z', $d_l$ is the interlayer spacing, 
N$_s$ is the number of layers
at each surface with dielectric function $\epsilon_s(\omega)$ and N$_b$ is
the number of inner layers with bulk dielectric constant $\varepsilon_b(\omega)$.
Unfortunatly this approach cannot always guarantee perfect cancellation
of the bulklike layers in the supercell, and may lead
to unphysical negative loss features.
\begin{figure}[!h]
 \begin{center}
 \includegraphics[width=9.6cm]{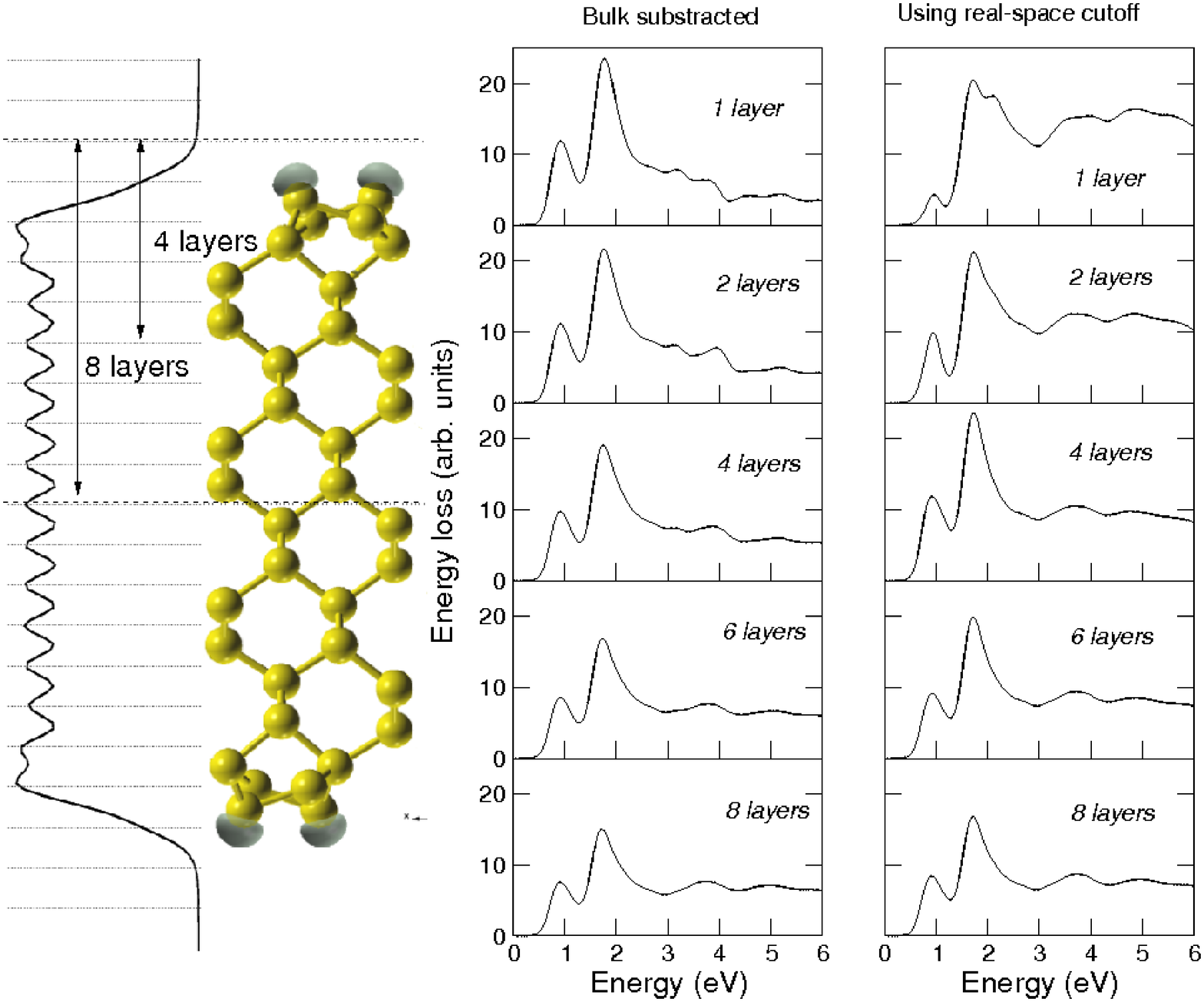}
 \end{center}
  \caption[Si(100)-$c(4\times2)$ slab]{
           Schematic diagram of the Si(100)-$c(4\times2)$ slab
           and calculated layer averaged charge density (left);
           atomic layers are marked with horizontal lines.
           Dependence of REELS spectrum on the chosen surface layer (right).
  \label{fig:slab_cutoff}}
\end{figure}
A better approach is to extract $\varepsilon_s$ directly using a ``cut off"
function as described in chapter \ref{ch:surface} or in 
Ref.~\cite{hogan_prb_y2003_v68_p035405}.
Nevertheless, the choice of $d$ remains somewhat arbitrary.
In this work we define the lower bound of the surface layer 
corresponding to the actual penetration depth of the electron 
for the chosen incident kinetic energy,
with the upper bound defined by the maximum extent of
the slab charge density or a typical surface state wavefunction
(see Fig.~\ref{fig:slab_cutoff}).
This turns out to be roughly one atomic layer.
We then checked that varying $d$ by an atomic layer did not change the results too much.
The dependence of the REEL spectra on the parameter $d$ 
is illustrated in Fig.~\ref{fig:slab_cutoff}.
\subsection{Detector integration}
In experimental EEL spectroscopy the detector
has a finite acceptance angle for collecting the scattered electrons.
Hence, in order to improve the quantitative estimation of the
EEL spectra at the surface, we implemented a method
to perform a numerical integration over the circular detector.
In particular we account better for the finite size of the detector window
through a random sampling of the circular area and in the
computation of the following integral:
 \begin{equation}
  \int A(k,k')\textrm{Im}g(q_{||},\omega)d\Omega
 \end{equation}
We implemented the method
in the Yambo code \cite{YAMBO}, the code we used for all further calculations.
\begin{figure}[h!]
 \begin{minipage}{5cm}
   \centerline{\epsfig{file=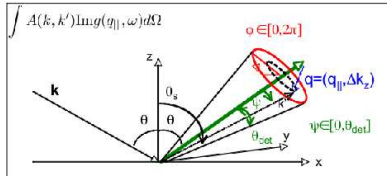,angle=0,clip=,width=5cm}} 
 \end{minipage}
  \hspace{1.2cm}
 \begin{minipage}{5cm}
  \caption[Integration geometry]{Integration geometry.}
 \end{minipage}
\end{figure}
In Fig.~\ref{fig:detector} an illustration of the importance of 
numerical detector integration scheme with
regard to relative intensity of peaks is shown.
\begin{figure}[!tbh]
  \centerline{\epsfig{file=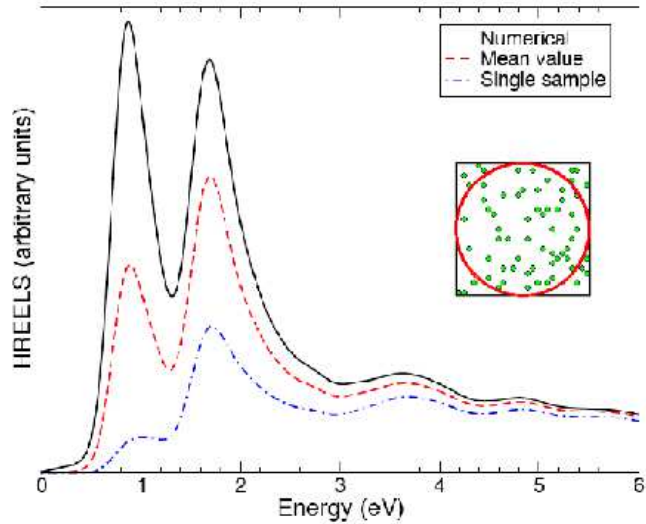,angle=270,clip=,width=8.5cm}} 
  \caption[Integration method]{
   Dependence of REEL spectra on detector integration method:
   numerical Monte Carlo, mean value scheme, and single point sampling.
   Example shown for $c(4\times2)$ surface,
   $E_0 =$ 40~eV; $\theta_0 = 60^\circ$; $\theta_\mathrm{det} = 1^\circ.$. 
   \label{fig:detector}
   }
\end{figure}
  \begin{figure}[!tbh]
 \begin{minipage}{5cm}
   \centerline{\epsfig{file=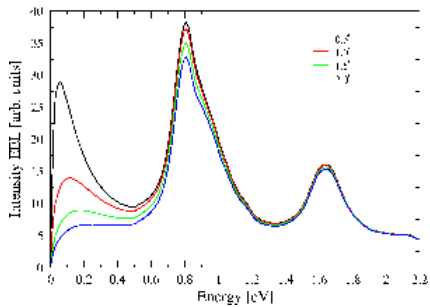,angle=0,clip=,width=5.5cm}} 
 \end{minipage}
  \hspace{1.2cm}
 \begin{minipage}{5cm}
 \caption[Detector size]{
   Dependence of REEL spectra on the detector size.
   \label{fig:detectorsize}
       }
 \end{minipage}
\end{figure}
A frequently used technique is to approximate the 
integral by an averaged value of the transferred momentum
${\bf q}$.
\begin{eqnarray}
 && \int d \Omega \rightarrow \int d{\bf q} \nonumber \nonumber \\
 && \int \mathit{f}(q)d{\bf q} \simeq \mathit{f}(\bar{q}) \nonumber
\end{eqnarray}
A comparaison between the two methods is shown in figure \ref{fig:detector}
where relative intensity of the low energy peaks is corrected
with converged numerical integration.
We also add a figure showing the effect of the detector size
on the EEL spectrum.

\section{Calculated spectra for Si(100)}
 A review of the current understanding of the structure of the clean Si(100) surface 
was given in the introduction and a schematic diagram of the three 
basic surface reconstructions (the \pone, \ptwo and \ctwo) is given 
in Fig.~\ref{fig:structures_clean}.
Throughout this work we will refer to the $[01\overline{1}]$ direction as $x$
and the $[011]$ direction as $y$, with $[100]$ being the surface normal $z$.
Here we present our results concerning optical and electronic spectroscopic study of
the clean Si(100) surface.

\subsection{Reflectivity anisotropy spectra}
Several theoretical studies of the RA spectra for the Si(100) surface have previously
been carried out, including tight binding calculations \cite{shkrekbtii_prl_y1993_v70_p2645},
discrete-dipole models \cite{mendoza_y2006_v74_p075318,hogan_prb_y1998_v57_p14843},
\textit{ab initio} calculations at the independent particle 
level \cite{kipp_prl_y1996_v76_p2810, kress_ss_y1997_v377_p398, gavrilenko_prb_y1998_v58_p12964, palummo_prb_y1999_v60_p2522,incze_prb_y2005_v71_035350, fuchs_prb_y2005_v72_p075353,fuchs_thesis_y2004}
as well as more recent studies including many-body effects~\cite{palummoBSE}.
Generally it was found that the best agreement with the experimental RA 
data \cite{shioda_prb_y1998_v57_R6823}
is obtained when the \ctwo\ or \ptwo\ models are used in the calculations,
while the \pone\ gives poor agreement.\\
Since we will use the same optical dielectric functions when computing the
energy loss spectra, we show for completeness (in Fig.~\ref{fig:RAS})
the results of our own supercell calculations
for the $c(4\times2)$, $p(2\times2)$ and $p(2\times1)$ reconstructions
at the independent particle level.
The reflectance anistropy (see chapter \ref{ch:surface}) is defined as
 \begin{equation} \textrm{RAS} = \frac{\Delta R_x}{R} - \frac{\Delta R_y}{R},
\end{equation}
where $\Delta R_i/R$ ($i=x,y$) is the normalized reflectivity
(i.e., relative to the Fresnel reflectivity).\\
As expected, we find that both $c(4\times2)$ and $p(2\times2)$ spectra
yield a good agreement with the experimental data.\\
The $p(2\times1)$ reproduces the low energy peak at 1.5~eV rather well,
but the comparison worsens at higher energy. Unfortunately, no experimental
data is available for the RAS of Si(100) in the near-IR range, and is hence limited
to $>$1.1~eV. 
\begin{figure}[!h]
   \centerline{\epsfig{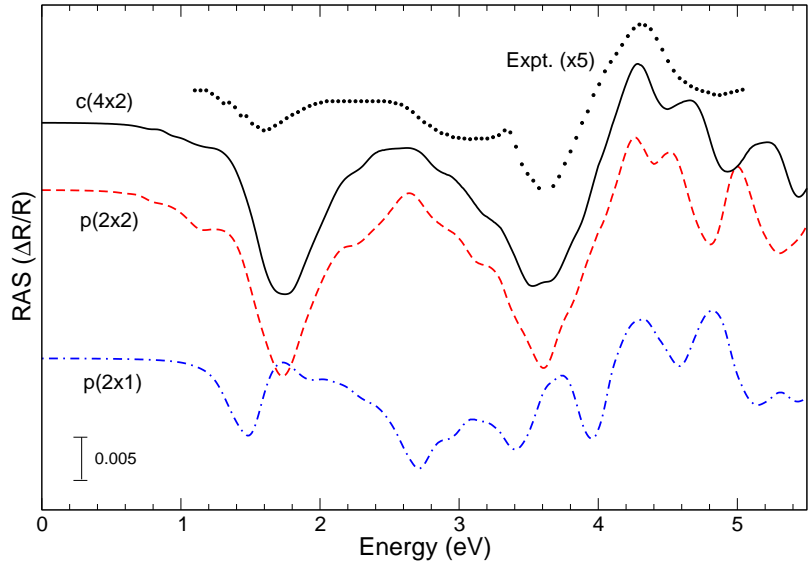}}
 \caption[Si(100)-$c(4\times2)$, $p(2\times2)$ and $p(2\times1)$: RAS]{
          RAS spectra of the $c(4\times2)$, $p(2\times2)$ and $p(2\times1)$
          reconstructions of clean Si(100), compared with the experimental
          spectrum of the nominal surface (scaled by a factor of 5).
          Experimental data are taken from Ref. \cite{shioda_prb_y1998_v57_R6823}.
          \label{fig:RAS}
          }
\end{figure}

 \subsection{Calculated REELS spectra at $E_0 =$ 40~eV}
\begin{figure}[!h]
 \begin{minipage}{8cm}
  \centerline{\epsfig{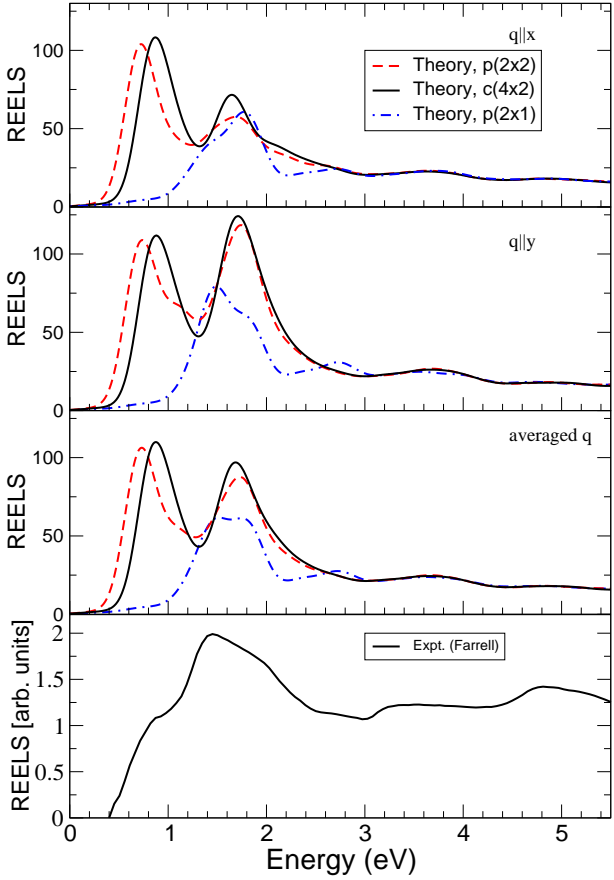}}
 \end{minipage}
 \hspace{1.2cm}
 \begin{minipage}{5cm}
   \caption[REELS $E_0 =$40~eV: Si(100)-$c(4\times2)$, $p(2\times2)$ and $p(2\times1)$]{
          REEL spectra of $c(4\times2)$, $p(2\times2)$ and $p(2\times1)$ 
          reconstructions of clean Si(100),
          and comparison with experiment (Farrell\ea\ \cite{farrell_prb_y1984_v30_p721}):
          $E_0 =$ 40~eV; $\theta_0 = 60^\circ$. The surface thickness is assumed to
          be $d = 8$ layers (plus one vacuum), i.e., the half slab. 
          A background signal has been subtracted from the experimental spectrum (see text). \label{fig:Farrell40eV}
}
  \end{minipage}
\end{figure}

  We now contrast the RAS results with the theoretical simulation of the HREELS experiment
of Farrell\ea\ \cite{farrell_prb_y1984_v30_p721} with incident
energy of $E_0 = 40$~eV, which roughly covers the same spectral range as 
the available RAS data.
Experimental results refers to specular geometry with incident angle of $\theta=60^\circ$. \\
\indent There exists a one-to-one correspondence between RAS and HREELS 
that can be justified with theoretical arguemnts when
we consider a spectral range below $E_1$.
In fact it can be written that RAS is proportional to 
Im$\varepsilon_s$ when Im$\varepsilon_b$ is close to zero.
On the other hand:
\begin{equation}
  \textrm{Im}g(q,\omega) \simeq \textrm{Im}\[\frac{1}{1+\varepsilon_s}\] 
       \simeq \frac{\textrm{Im} \varepsilon_s }
                   {(1 +\textrm{Re}\varepsilon_s)^2+\textrm{Im}\varepsilon^2_s}
\end{equation}
giving a direct proportionality of REEL respect to $Im \varepsilon_s$.
This correspondence is evident in Fig.~\ref{fig:eps_c4x2} where
HREEL peaks and Im$\varepsilon_s$ are represented.
The experimental one-to-one correspondance between RAS and HREELS was also 
previously illustrated by Arciprete\ea\ \cite{arciprete_prb_y2003_v68_p125328} 
for the case of GaAs(001)-$c(4\times4)$. \\
The experimental data, which are reproduced in Fig.~\ref{fig:Farrell40eV}, are 
characterized by surface-derived soulder at 0.9~eV ($S_0$) and a broad peak 
at 1.4-2.0~eV ($S_1$), 
and bulk-derived peaks at about 3.5~eV and 5~eV. 
The latter peaks have also been identified in second-derivative
off specular geometry spectra at $E_0$=100eV by 
Rowe and Ibach \cite{ibach_prl_y1973_v31_p102} 
as deriving from the bulk critical points, $E_1$ and $E_2$.
In the experimental spectrum of Fig.~\ref{fig:Farrell40eV} we have 
subtracted a background signal, taken to be that of the monohydride 
Si(100)-\pone:H surface, also reported in Ref.~\cite{farrell_prb_y1984_v30_p721}.

The results of our first principles calculations of HREELS are shown in Fig.~\ref{fig:Farrell40eV}
for the \pone, \ptwo\ and \ctwo\ reconstructions of Si(100).
From the comparison with experiment it is clear that the $p(2\times1)$ alone 
cannot reproduce the experimental signal since the shoulder at 0.8~eV ($S_0$) 
is missing from its theoretical spectrum.
A similar observation was made by Gavioli\ea\ \cite{gavioli_ss_y1997_v377_p360} 
based on tight binding calculations of the HREEL spectrum.\\
Moreover we observe that, as in the case of the RAS, it is difficult to distinguish
between the $c(4\times2)$ and $p(2\times2)$ calculations.
The $S_0$ peak appears at a slightly lower energy (by 0.1~eV) in the $p(2\times2)$
calculation; however, considering the approximations used in the present calculations 
(scissors shift), it is not sufficient to allow us to prefer the $c(4\times2)$ 
over the $p(2\times2)$.

As discussed in Section \ref{sec:th-eels}, the three-layer model of energy loss 
is based on the assumption that the electron does not penetrate the surface, 
so that losses occur from scattering off long range potentials above the surface.\\
In reality, electrons with a 40~eV kinetic energy actually penetrate the surface by
several atomic layers before elastic scattering occurs.
As a result, features in the loss which arise from scattering within the crystal itself
(as occurs naturally in transmission EELS) are missing from our theory.
Hence the calculated lineshape differs significantly from the experimental 
one above 2.5~eV.

\begin{figure}[!h]
 \centerline{\epsfig{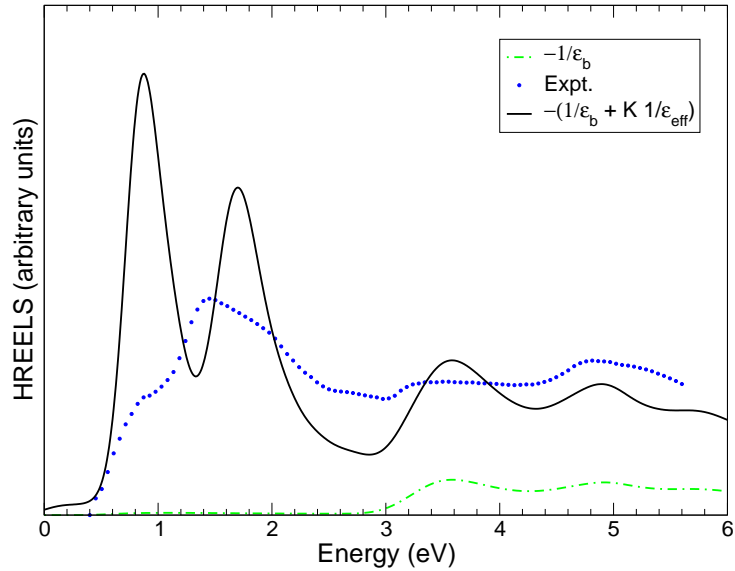}}
   \caption[REELS $E_0 =$40~eV with bulk contributions: Si(100)-c(4$\times$2)]{
     REELS spectra of $c(4\times2)$ calculated using a mixture of long range reflection 
     and short range transmittion loss functions, compared with experiment 
     (Farrell\ea\ \cite{farrell_prb_y1984_v30_p721}):
     $E_0 =$ 40~eV; $\theta_0 = 60^\circ$. 
    \label{fig:Farrell40eV_bulk}
     }
\end{figure}

To counteract this deficiency of the theory, we augment the reflection loss term 
with a second loss term that represents the trasmission loss, or ``bulk" loss 
within the subsurface layers:
\begin{equation}
  g(q,\omega) = g^\mathrm{3L}(q,\omega) + K|q| \frac{-1}{\varepsilon_b}
\end{equation}
The result is shown in Fig.~\ref{fig:Farrell40eV_bulk}.
Since the three-layer model already accounts well for losses in the surface layer
(the spectra presented here actually do not change much if a loss function
of the form $-1/\varepsilon_s$ is used),
we find we only need to increase the proportion of bulklike losses below
the surface to improve the agreement with experiment.

In spite of the agreement reached it is clear from the relative intensity
of the $S_0$ and $S_1$ peaks that one single reconstruction cannot reproduce
the experimental spectra.%

 \subsection{Calculated REELS spectra at low energy}
  Although many REEL spectra are present in the literature that study the
Si(100) surface, only a few high resolution spectra
that probe the interband transitions of Si(100)  are available. \\
\begin{figure}[!htb]
 \begin{minipage}{7cm}
  \centerline{\epsfig{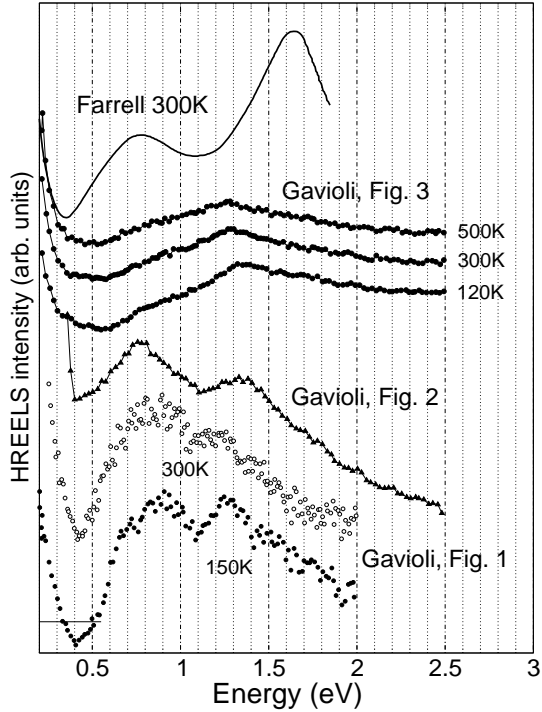}}
 \end{minipage}
\hspace{1.0cm}
 \begin{minipage}{5cm}
  \caption[HREELS experimental data]{ 
           HREELS experimental data from Farrell (solid line: $E_0 =$ 7~eV; $\theta_0 = 60^\circ$; 
           300K) and Gavioli (dots: $E_0 =$ 7.8~eV; $\theta_0 = 62.5^\circ$; 150K and 300K). 
           \label{fig:exp_reels}
           }
 \end{minipage}
\end{figure}
We reproduce in Fig.~\ref{fig:exp_reels} the HREELS data for low-energy incident
electrons reported by Farrell\ea\ \cite{farrell_prb_y1984_v30_p721} 
($E_0 =$ 7~eV; $\theta_0 = 60^\circ$ specular scattering; probably at T=300K) 
and Gavioli\ea\ \cite{gavioli_ss_y1997_v377_p360} 
(for a slightly different experimental setup: $E_0 =$ 6.8~eV; 
$\theta_0 = 62.5^\circ$; T=120--500K). \\
In both works, two main structures are identified in the spectra below 2~eV.
Farrell reports an adsorption edge at 0.4~eV with a peak at 0.75~eV 
(termed $S_0$), and a second edge at 1.1~eV with a peak at 1.65~eV 
(termed $S_1$).
We note that the $S_1$ peak was identified by Farrell as being the peak observed by
Ibach and Rowe \cite{ibach_prl_y1973_v31_p102,ibach_prb_y1974_v9_p1951} 
at $1.7\pm0.5$~eV and by Maruno\ea\ \cite{maruno_prb_y1983_v27_p4110} at about 2.0~eV.
In the more recent work by Gavioli\ea, various HREELS data were 
reported at temperatures ranging from 120 to 300~K, 
and at different analyzer focalizations (see Fig.~\ref{fig:exp_reels}).
At low temperature a shoulder is found at 0.68~eV,
as well as two main structures at 0.9~eV and 1.15--1.35~eV.
It is not clear, however, if the latter feature corresponds 
to the $S_1$ structures reported elsewhere, as any peaks appearing 
at around 1.7~eV in Gavioli's data seem to be obscured by the background noise.
Note that the main low energy peak occuring in the RAS is at 1.6~eV,
as seen in Fig.~\ref{fig:RAS}, which would appear to agree with 
the peak positions of Farrell\ea.
\begin{figure}[!h]
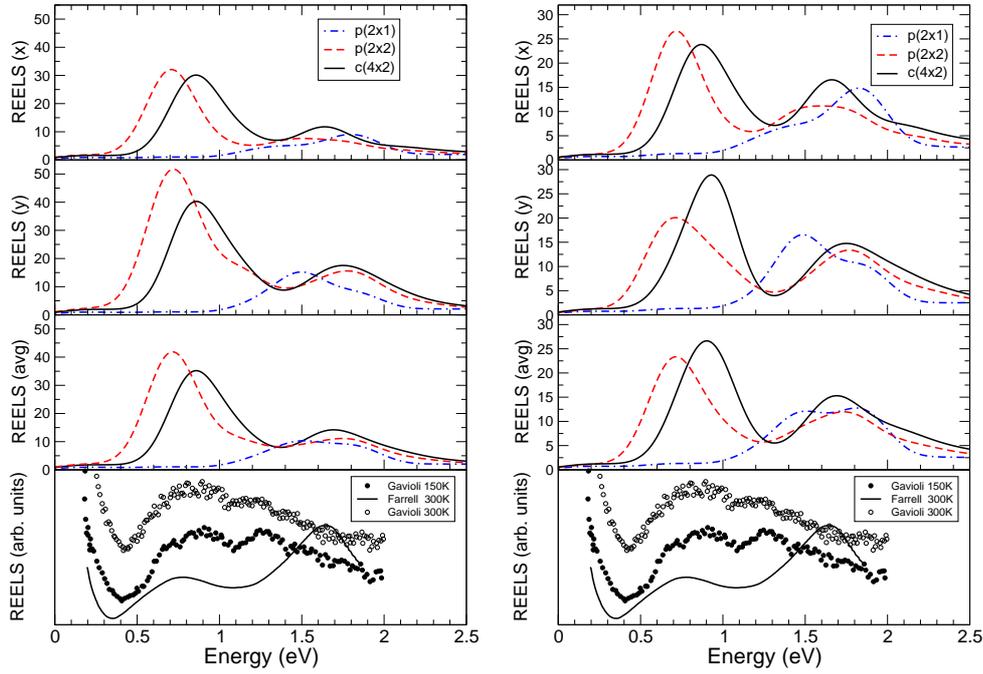

\begin{center}
 \begin{tabular}{c c}
  \epsfig{file=./i/as-Farrell_7eV_4L.eps,clip=,width=6.2cm} &
  \epsfig{file=./i/as-Farrell_7eV_2L.eps,clip=,width=6.2cm}
 \end{tabular}
 \end{center}
   \caption[REELS at $E_0 =$ 7~eV: Si(100)-\pone, \ptwo\ and \ctwo]{
   REEL spectra of \pone, \ptwo\ and \ctwo\ reconstructions of clean Si(100),
   for low energy incident beam ($E_0 =$ 7~eV; $\theta_0 = 60^\circ$).
   Left: Surface thickness $d = 4$ atomic layers;
   Lorenztian broadening $\delta_L = 0.006$~eV; Gaussian broadening of $\delta_G = 0.3$~eV.
   Right: Surface thickness $d = 2$ atomic layers.
   \label{fig:Farrell_7eV_4L}
   }
\end{figure}
In Fig.~\ref{fig:Farrell_7eV_4L}
we report the results of our \textit{ab initio}
simulation of this HREELS experiment.
As noted for the $E_0=40$~eV data, it is clear that the \pone\ model does not
yield the correct lineshape, as the $S_0$ peak is missing.
Both \ctwo\ and \ptwo\ structures succeed in reproducing
the double-peaked structure observed in the experiments.
In particular, the $S_0$ peak is well reproduced by the \ctwo\ model, and it
is possible that the shoulder observed at 0.68~eV points to the coexistence of
some \ptwo\ on the predominantly \ctwo\ surface.
The calculated energetic position of the $S_1$ peak is in reasonable
agreement with the data of Farrell, but not so much with that of Gavioli.
However, it is worth to mention that Farrell's data beyond 1.5~eV
can be affected to background and are not completly relible because 
some fit of experimental data has been performed and in the original 
paper the dashed line at that energies give us some doubts on 
the exact position of that peak. For this reason we feel 
more confident in the more recent Gavioli's data.

 \subsection{Analysis of spectra}
  In the following two sections we concern ourselves with interpreting 
the experimental peaks observed in the HREEL spectra below 3~eV for 
the kinematic setup of Farrell and Gavioli at $E_0 \approx 7$~eV.

\begin{figure}[!h]
 \centerline{\epsfig{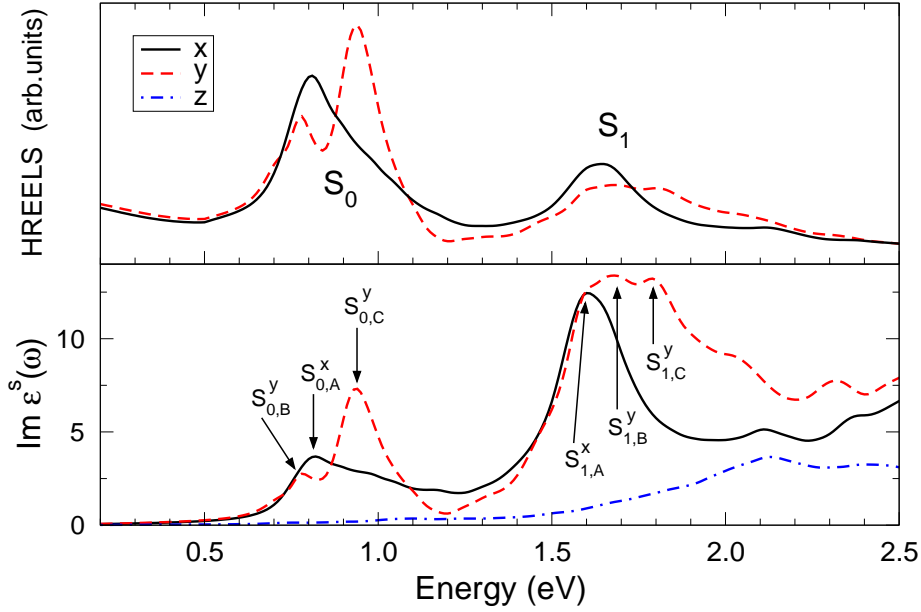}}
 \caption[Si(100) - HREELS and dielectric function]{
    HREEL (along $x,y$) spectra (top) 
    and surface dielectric functions ($x,y,z$)
    for $E_0 = 7$~eV, $\theta_0 = 60^\circ$.
 \label{fig:eps_c4x2}
          } 
\end{figure}

\begin{figure}[!h]
\begin{center}
 \begin{tabular}{c c c}
   \includegraphics[width=4.5cm]{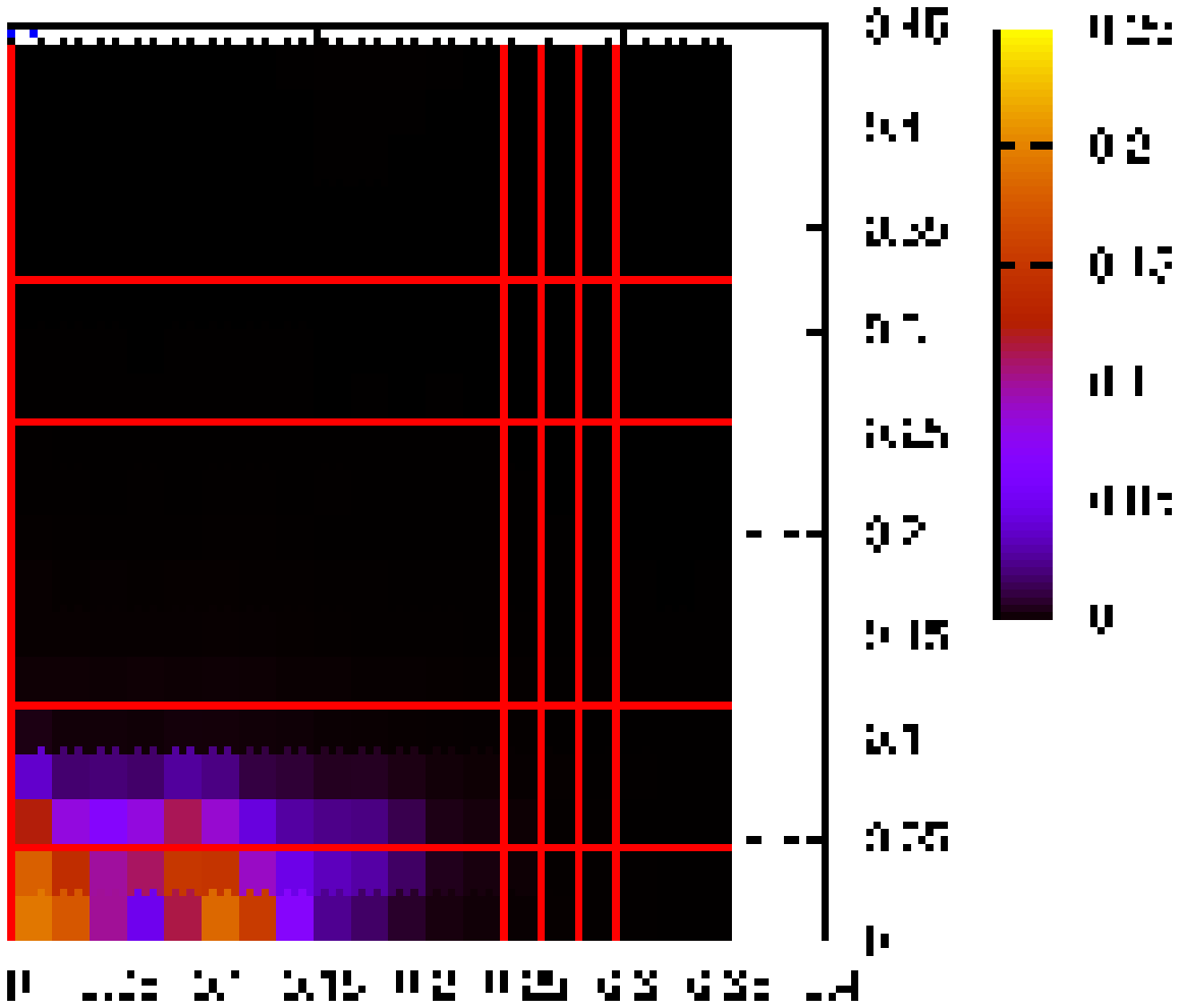} &
   \includegraphics[width=4.5cm]{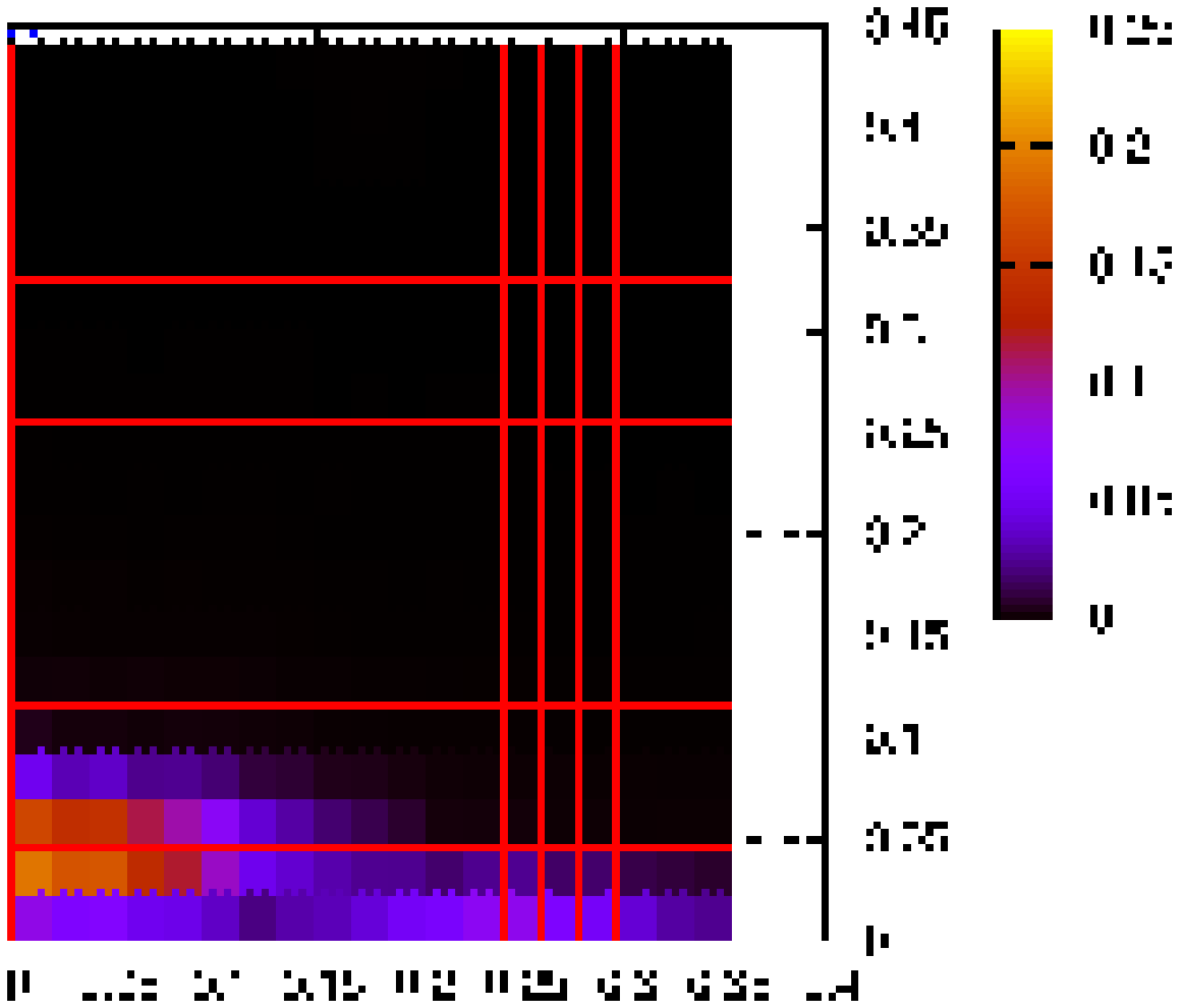} &
   \includegraphics[width=4.5cm]{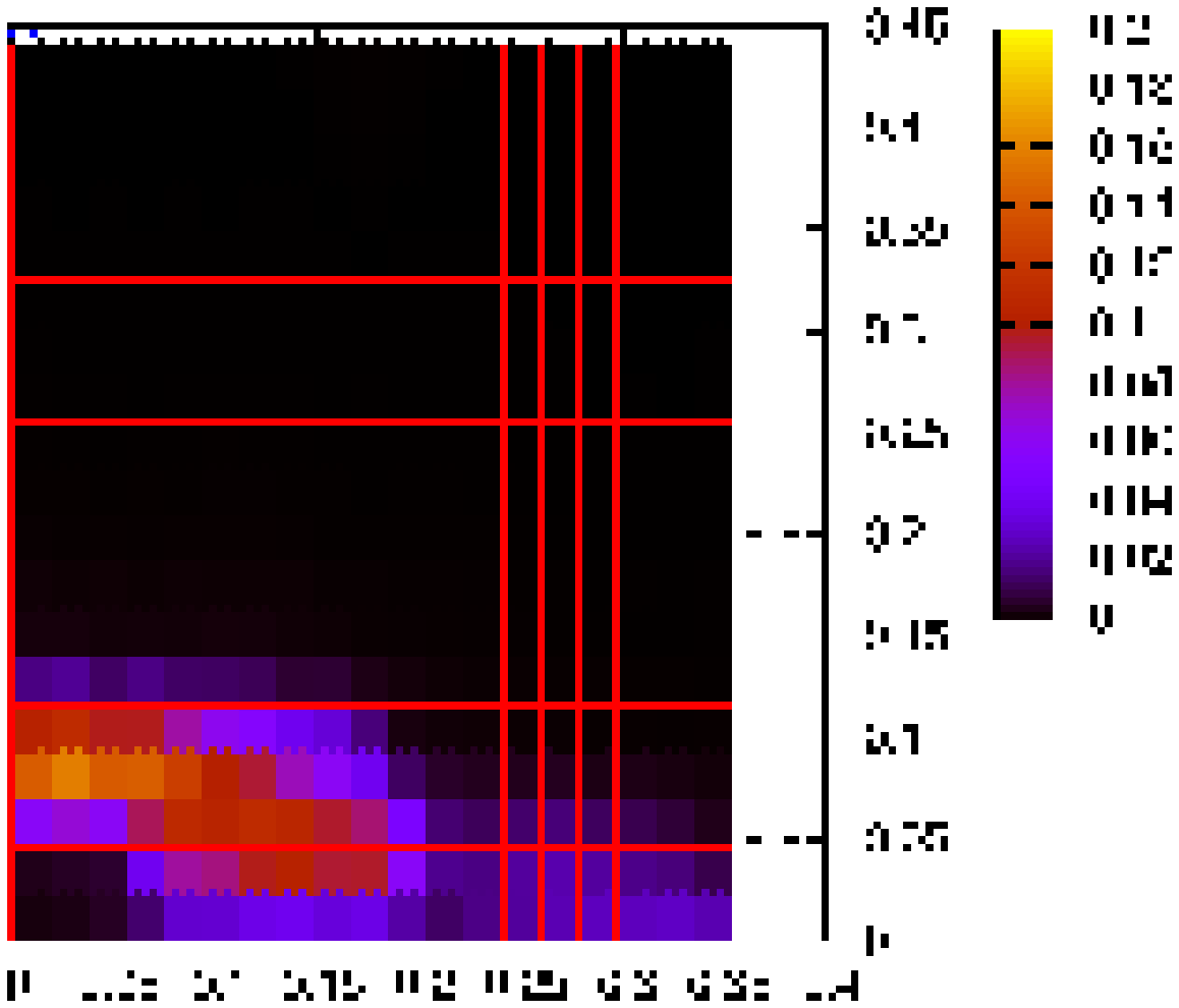}
  \end{tabular}
 \end{center}
\caption[Si(100)-c(4$\times$2) total oscillator strenght S$_2$]{ 
         Total oscillator strenght P$_E(k) (\delta E = 0.1eV)$
         as a function of $\bk$ contributing to $S_0$ peak.
         In particular we considered two perpendicular directions of polarization belonging 
         the surface plane: X and Y respectively and we analysed 3 energy 
         windows centred around the position
         of the peak in the surface epsilon (see Fig.\ref{fig:eps_c4x2}).
         In particular 0.32eV for x polarization (left) and 0.285eV (center) 
         and 0.44eV (right) for y polarization, namely S$_{0,A}$, S$_{0,B}$ 
         and S$_{0,C}$ respectively. 
       \label{fig:kspace_S0_x_y}
}
\end{figure}

\begin{figure}[!h]
 \begin{center}
  \begin{tabular}{c c c}
   \includegraphics[width=4.5cm]{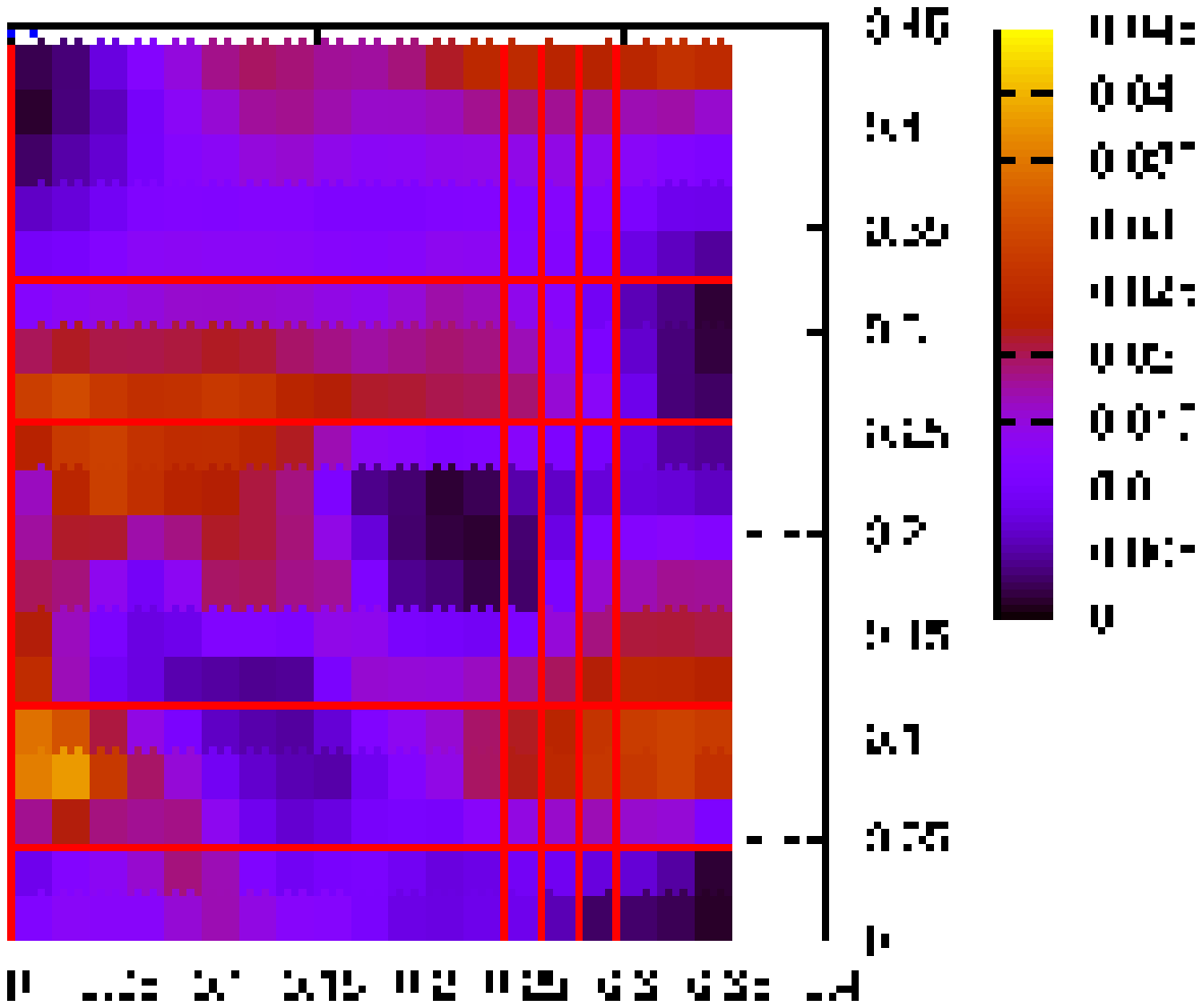} & &  \\
   \includegraphics[width=4.5cm]{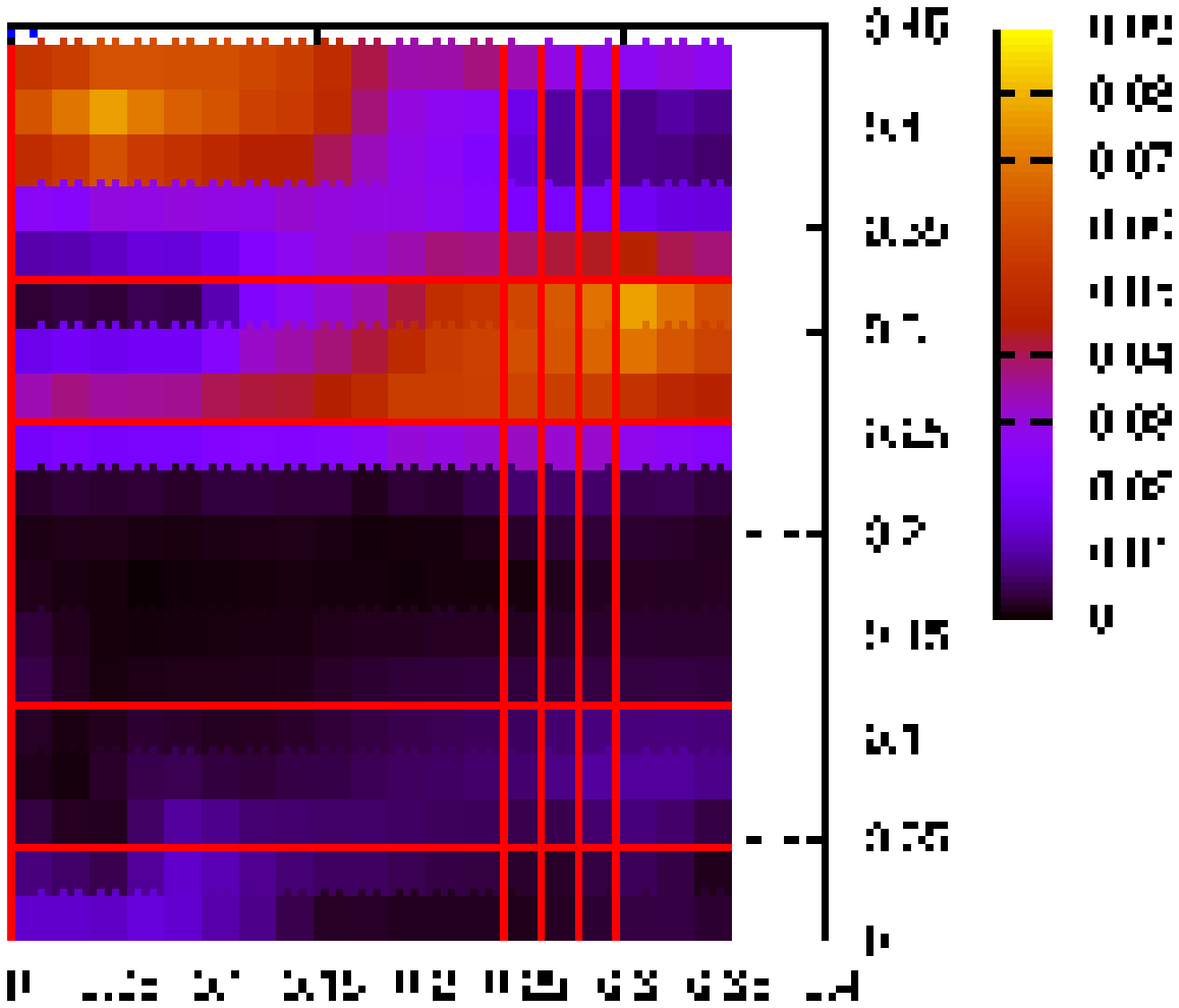} &
   \includegraphics[width=4.5cm]{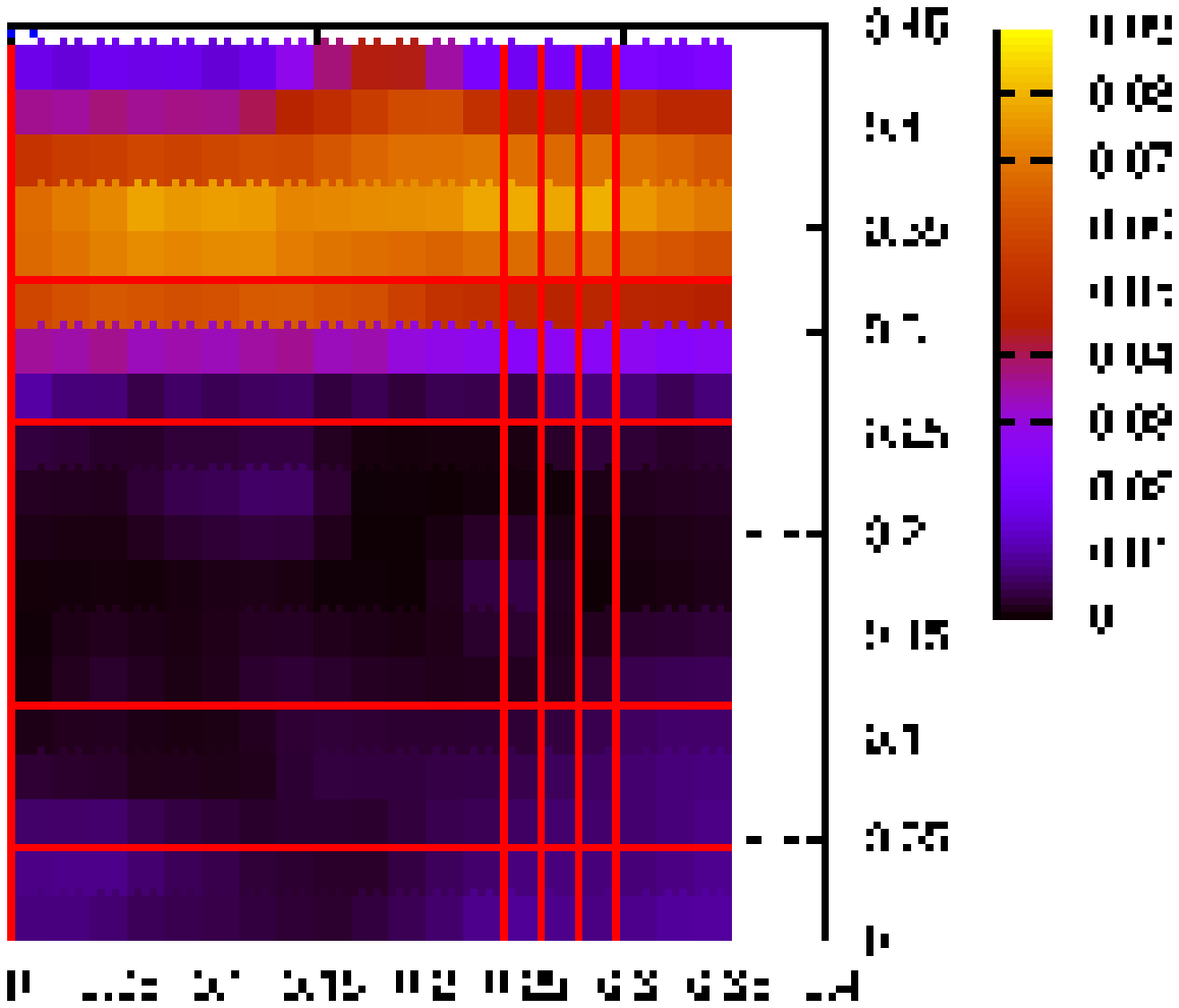} &
   \includegraphics[width=4.5cm]{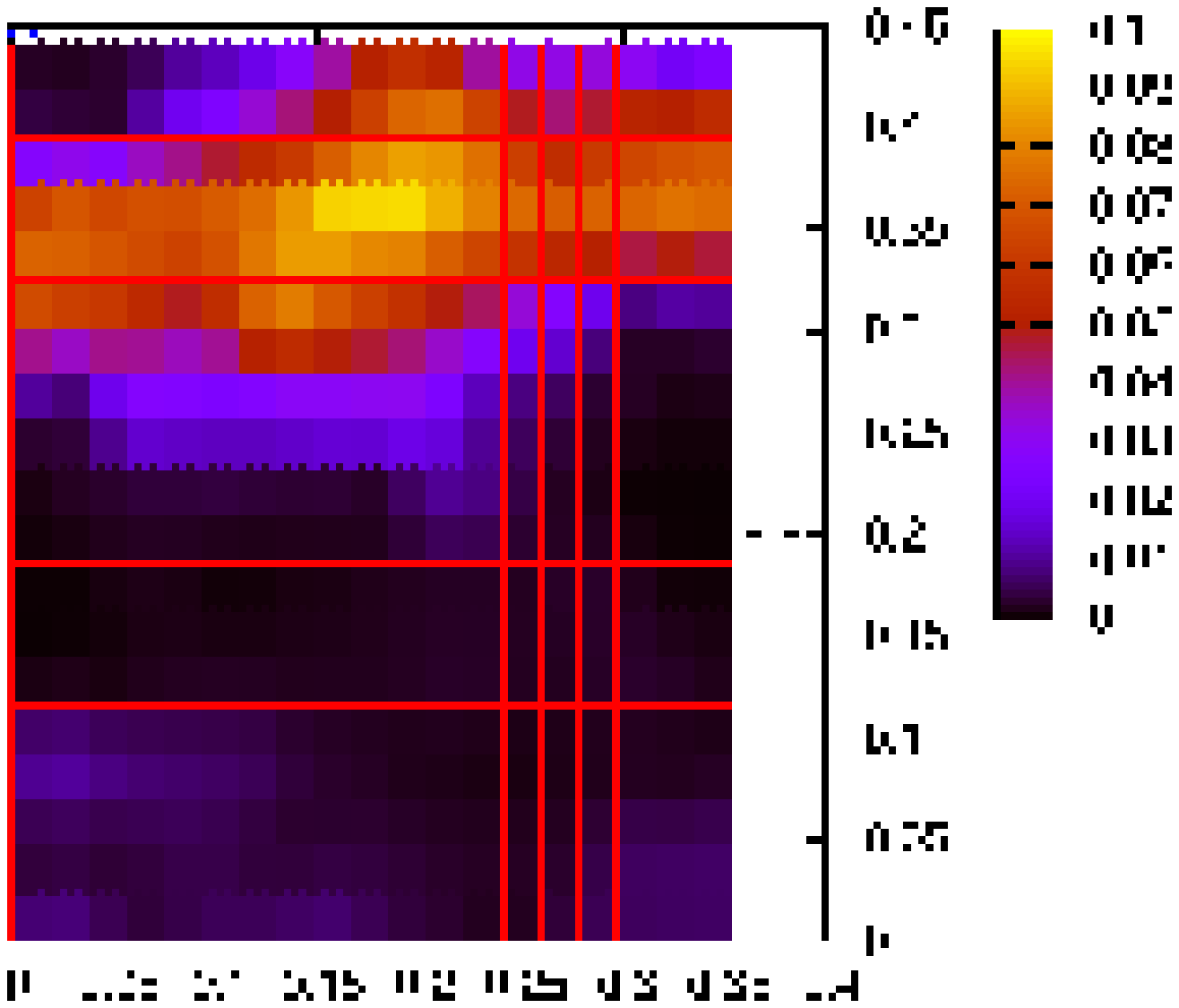}
   \end{tabular}
\end{center}
 \caption[Si(100)-c(4$\times$2) total oscillator strenght S$_1$]{
          Total oscillator strenght P$_E(k) (\delta E = 0.1eV)$
          as a function of $\bk$ contributing to $S_1$ peak.
          In particular we considered two perpendicular directions of 
          polarization belonging the surface plane:
          X (top) and Y (bottom) respectively and we analysed 3 energy 
          windows centred around the position
          of the peak in the surface epsilon (see Fig.\ref{fig:eps_c4x2}).
          We considered 3 energy windows centred at 1.11eV (top), for x (left) and y (right)
          polarization respectively, and at slightly highier energies (bottom)
          1.18eV (left) and 1.29eV (right) for y polarization, namely  S$_{1,A}$,S$_{1,B}$,
          S$_{1,C}$ and S$_{1,D}$ respectively.\label{fig:kspace_S1_x_y}
}
\end{figure}

Fig.~\ref{fig:eps_c4x2} compares the calculated surface dielectric function 
with the calculated HREEL spectra. \\
It is clear that there is a one-to-one correspondance between peaks 
in the energy loss and peaks in the imaginary part of the dielectric function.
Hence we can analyse $\varepsilon_2$ to characterize peaks in the HREEL spectra.
The experimental spectra feature two main peaks, termed $S_0$ and $S_1$ 
in the literature \cite{farrell_prb_y1984_v30_p721}.
The microscopic origins of these peaks is now analysed for the \ctwo\ reconstruction.
Figures~\ref{fig:kspace_S0_x_y} and ~\ref{fig:kspace_S1_x_y} show the total 
oscillator strength $\overline{P}_{E}(\bk)$,
as a function of $\bk$, corresponding to an energy window of width $2\delta$ centred around
a chosen peak energy $E$:
\begin{eqnarray}
  \overline{P_E}(\bk) &=& \sum_{v,c} |P_{v,c,\bk}|^2 \\
   &\textrm{ for }& E-\delta < E_{c\bk} - E_{v\bk} < E+\delta. \nonumber
  \label{eqn:transition}
\end{eqnarray}
 It is clear that $S_0$ arises from transitions located
 around the $\Gamma$ point for both polarizations 
 (see Fig.~\ref{fig:kspace_S0_x_y})
 On the contrary $S_1$ arises mostly from transitions along 
 $\bar{Y}$--$\bar{Y}'$ direction.

The location of these transitions with respect to the surface band structure
is shown in Fig.~\ref{fig:bands_c4x2_p2x1}.
Our bandstructure calculation for the clean \ctwo\ surface
 compares well with that previously published by Fuchs \cite{fuchs_thesis_y2004}.
In particular, a surface state is present at about 0.8~eV above the valence band maximum.
 
\begin{figure}[!h]
 \begin{center}
  \begin{tabular}{c c}
   \includegraphics[width=4.5cm]{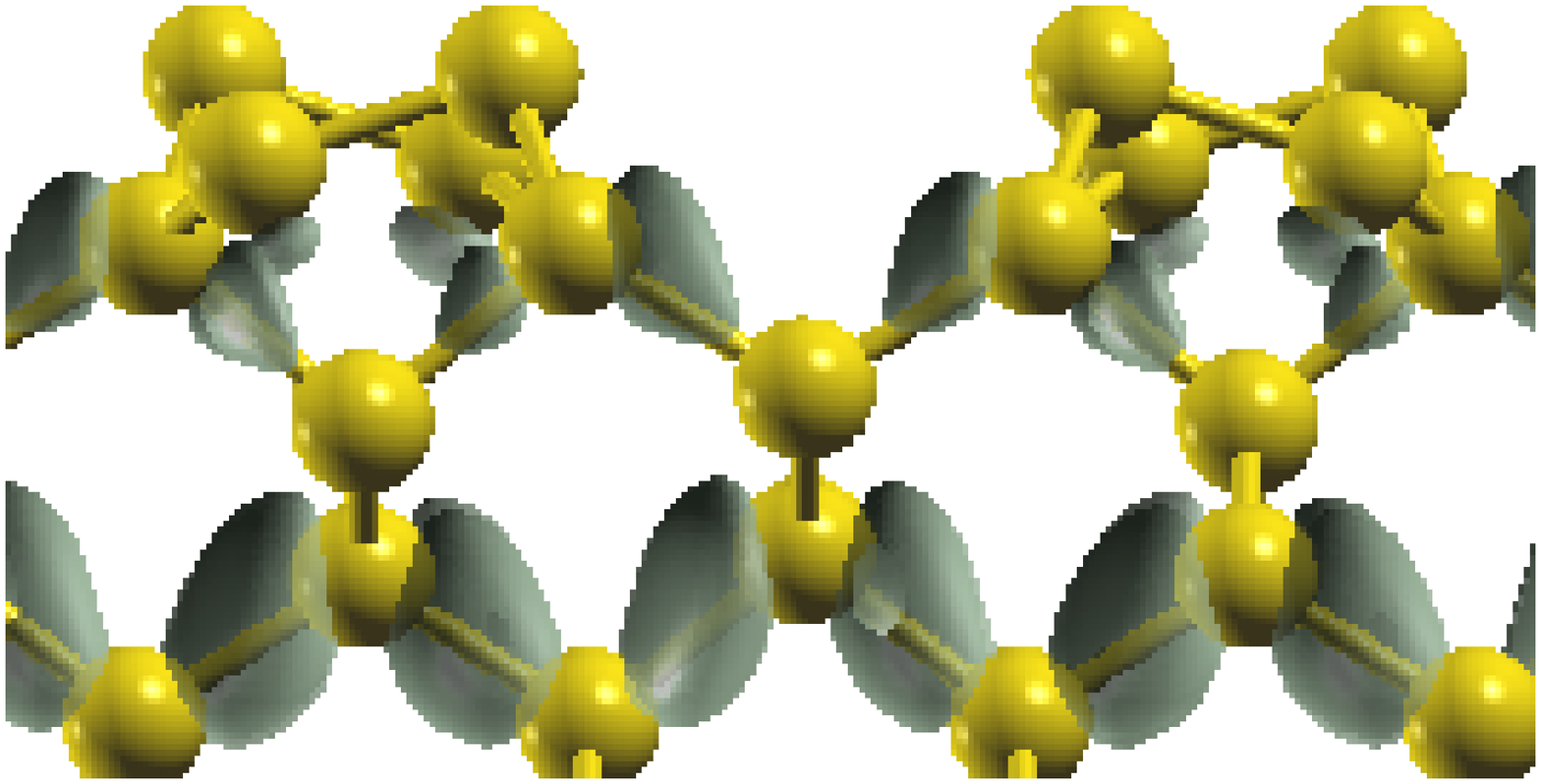} &
   \includegraphics[width=4.5cm]{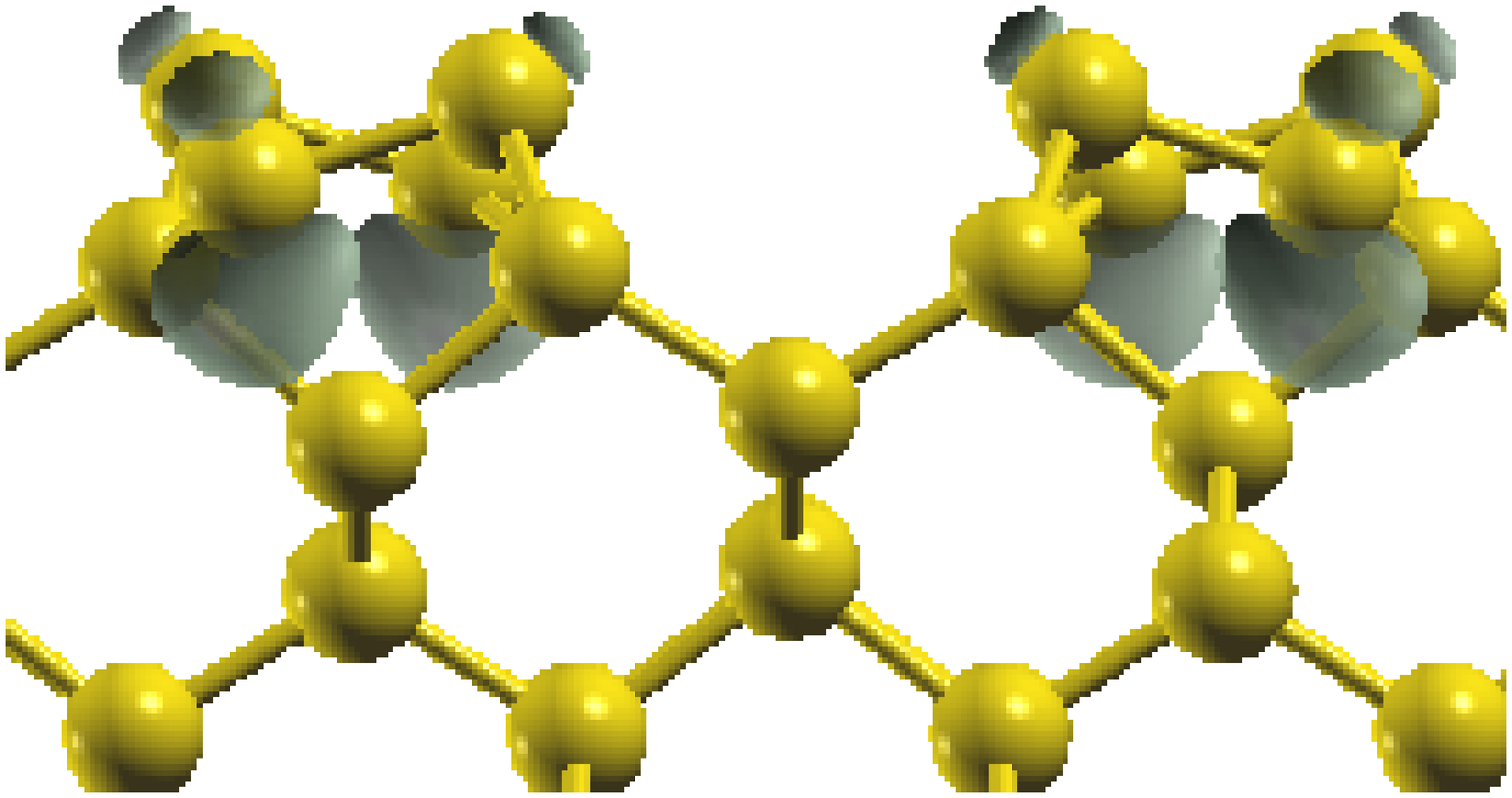} \\
   \includegraphics[width=4.5cm]{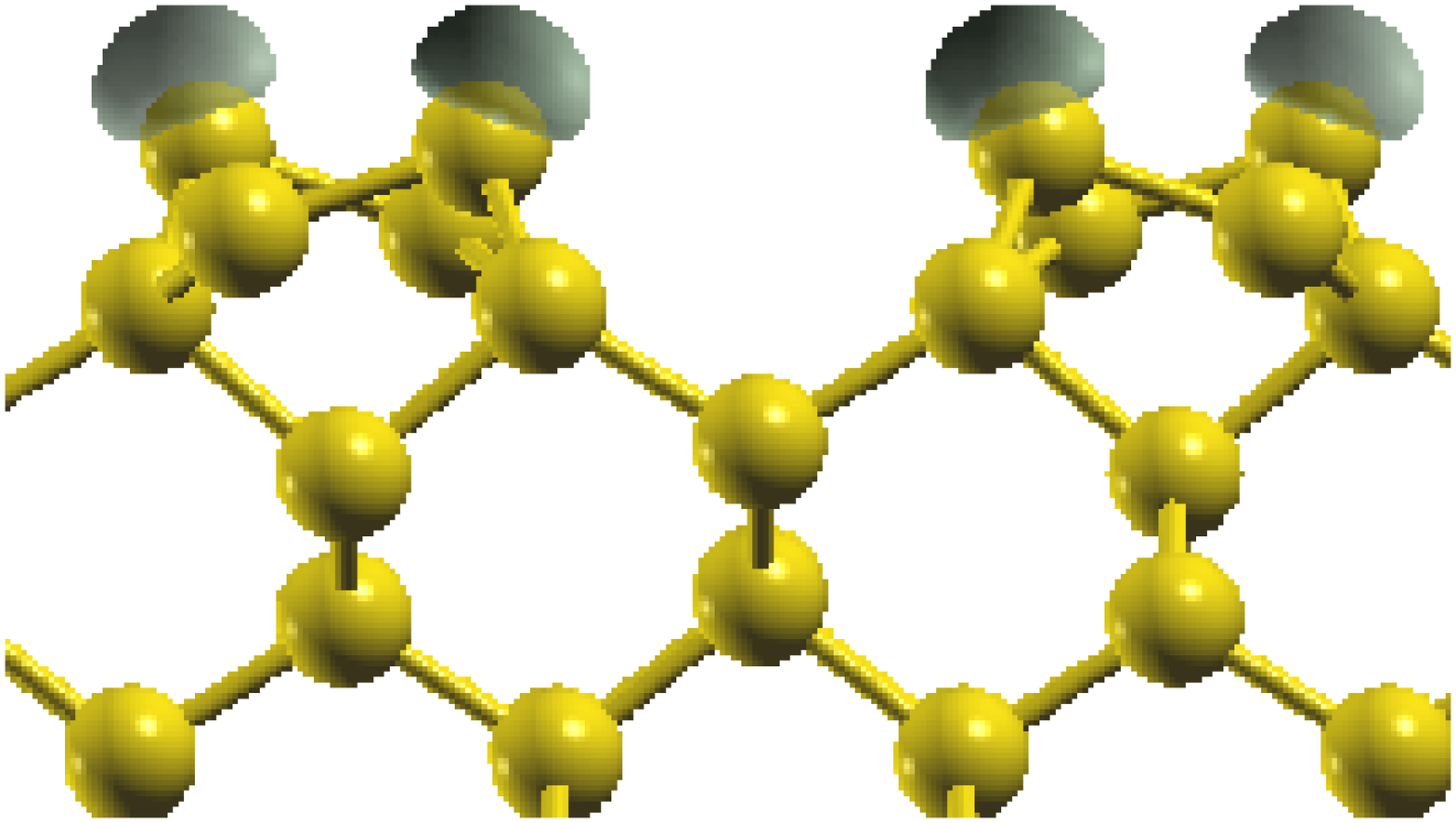} &
   \includegraphics[width=4.5cm]{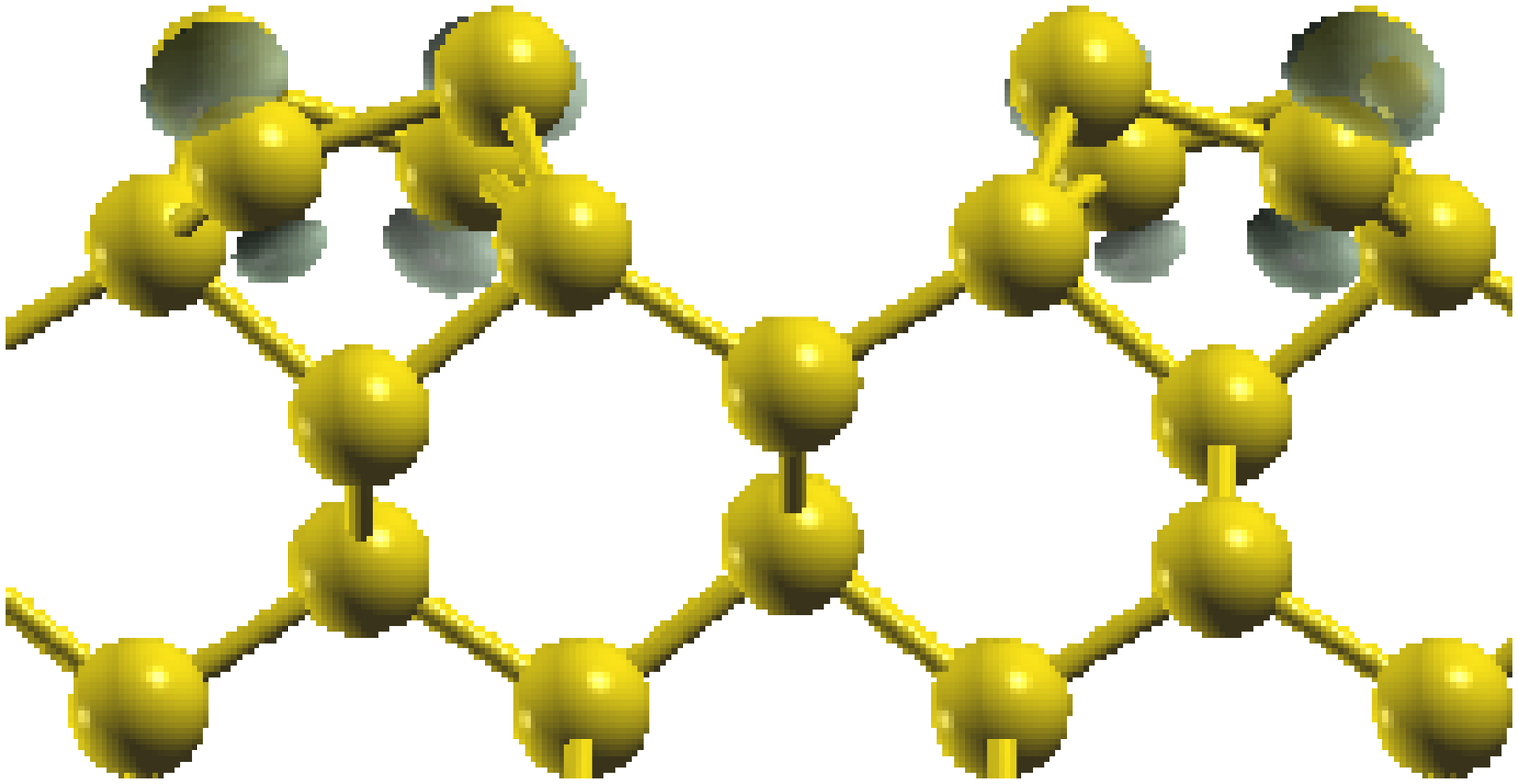}
  \end{tabular}
 \end{center}
\caption[Si(100)-c(4$\times$2) isosurface plots of $|\psi_{n\bk}(r)|^2$]{
         Isosurface plots of $|\psi_{n\bk}(r)|^2$ for represetative 
         states involved in transitions responsible for
         peaks in the low energy HREELS peaks $S_0$ at 0.8~eV (top) 
         and $S_1$ at 1.6~eV (bottom).
         Plots are obtained using Xcrysden \cite{xcrysden} package.
\label{fig:states_c4x2}
}
\end{figure}

\begin{figure}[!h]
 \centerline{\epsfig{file=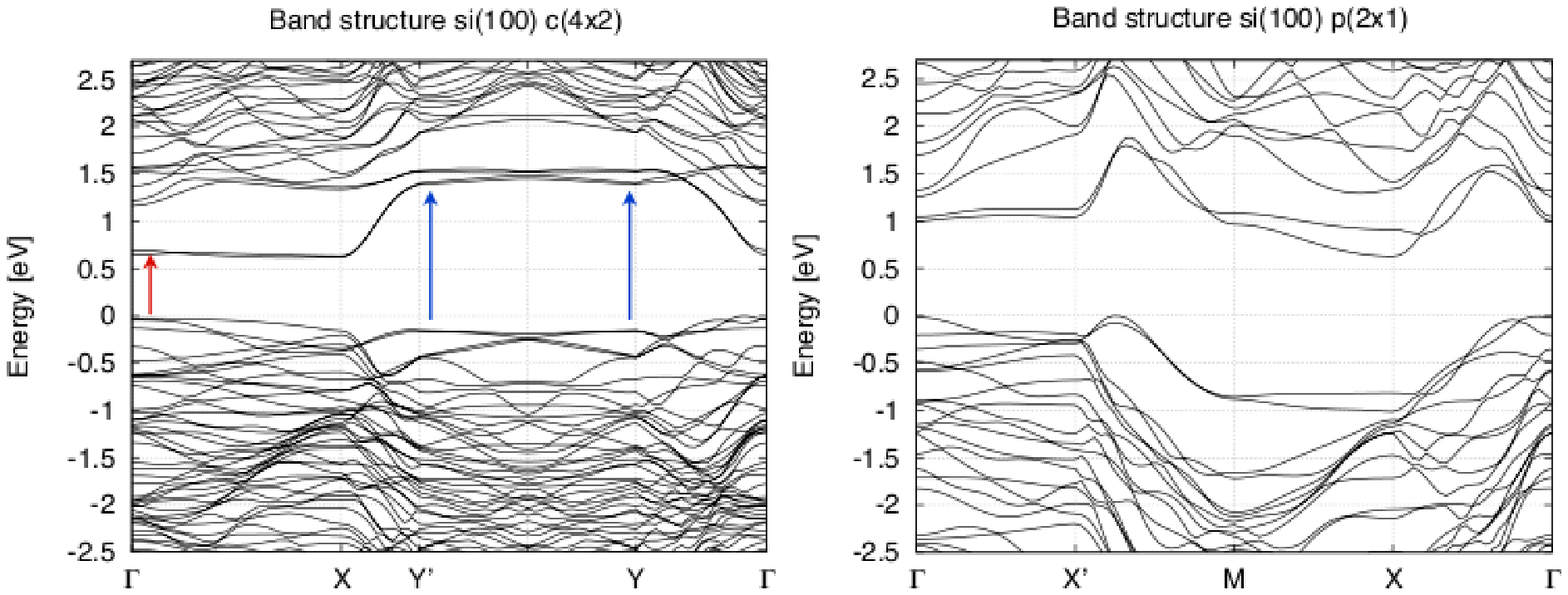,angle=0,clip=,width=13.5cm}}
   \caption[Si(100)­$c(4\times2)$ and -$p(2\times1)$: band structure]{
     Band structure of the $c(4\times2)$ (left) with indication of surface peaks.
     Compare with $p(2\times1)$ (right) to show absence of surface band.
     A rigid scissor shift of 0.5eV has been applied to unoccupied bands.
  \label{fig:bands_c4x2_p2x1}
 }
\end{figure}

Finally, we looked at $|\psi_{n,\bk}|^2$ for the
valence and conduction band states taking part in the strongest transitions.
These are plotted in Fig.~\ref{fig:states_c4x2} using 
Xcrysden package \cite{xcrysden}.
We found that $S_1$ derives from transitions between surface states:
dangling bonds and $sp^2$-hydridized $p_z$ orbitals,
or $\pi$ to $\pi^*$ orbitals as shown on Fig.~\ref{fig:states_c4x2}.
To have a confirmation of this statement an analysis of
Fig. \ref{fig:kspace_S1_x_y} and of the band structure
(top of Fig. \ref{fig:bands_c4x2_p2x1}) is required.
In fact, Fig.\ref{fig:kspace_S1_x_y} shows that the most part of
transitions contributing to $S_1$ comes from k points along $\bar{Y}$--$\bar{Y}'$ 
path in the SBZ and the band structure clarify which bands are involved.
In fact, looking at the band structure in Fig.~\ref{fig:bands_c4x2_p2x1}
it is clear that S$_1$ peak comes from transitions between the flat 
bands in the $\bar{Y}$--$\bar{Y}'$ direction (blue arrows).

On the other hand, $S_0$ is due to transitions between bulk states at the
valence band maximum (in $\Gamma$) and unoccupied surface states within 
the fundamental bulk bandgap.
A representative pair of states involved in this kind of excitation is
shown in Fig. \ref{fig:states_c4x2} (top).
Moreover, with a similar analysis as done for $S_1$, band structure and 
Fig. \ref{fig:kspace_S0_x_y} (red arrow) give a confirmation of the fact that $S_0$
is more symetric involving transitions around the $\Gamma$ point.

 \subsection{Discussion} \label{sec:si100-discussion}
   We now show how our HREELS calculations can be used 
 to better understand the structure of the Si(100) surface 
 as a function of temperature.
 Si(100) exhibits a \pone\ LEED pattern above the order--disorder
 transition temperature  of about 200K.
 Our findings (Fig.~\ref{fig:Farrell_7eV_4L})---that the \pone\ surface alone
 model cannot reproduce the experimental HREELS data---are consistent with current
 understanding of the surface at this temperature (rapid dimer flipping).
 At lower temperatures, the surface should be predominantly \ctwo.

 The experimental dependence of the HREEL spectra of Si(100) was studied by
 Gavioli\ea\ \cite{gavioli_ss_y1997_v377_p360}, and their results are reproduced
 in Fig.~\ref{fig:exp_reels}.
 In that work, the observed spectrum was explained as being due to a mixture of
 symmetric and asymmetric \pone\ structures. However, there is much
 evidence to show that symmetric dimers do not exist at 150K.
 Furthermore, they did not consider the \ctwo\ structural model, which we find
 is sufficient to produce a double-peaked HREEL spectrum (Fig.~\ref{fig:Farrell_7eV_4L}
 and ~\ref{fig:Farrell40eV}). \\
 Increasing the temperature, the $S_0$ peak is found to decrease by 5\% and 
 the $S_1$ peak by 40\%.
 As described previously, we took care to ensure that the calculated relative 
 intensities of the peaks are not artefacts of our calculation,
 as we checked their consistency with respect to the lorentzian and gaussian 
 broadenings, the surface layer thickness, detector integration settings, k-point 
 sampling.
 We can only conclude, therefore, that the $c(4\times2)$ is not the sole 
 reconstruction present at the surface. 

 Furthermore, we note that the $p(2\times1)$ spectrum does not contribute 
 to the $S_0$ peak, on the contrary \ctwo\ and \ptwo\ 
 have both some structure at that energies.
 Moreover, looking carefully at Fig.~\ref{fig:Farrell_7eV_4L}, 
 it is possible to link the 0.68~eV shoulder in Gavioli's data 
 at $T=150^\circ$ to the first peak of the \ptwo\ structure and
 the 0.9~eV peak, present also in the $T=300^\circ$ curve, 
 to the \ctwo\ structure.
 At the end we mention that the position of
 the S$_1$ peak needs a correction in order to
 predict Gavioli's experimental value between $1.15$~eV and $1.35$~eV
 for the \ptwo\, \ctwo\ and \pone\ structures.
 Spectra reported in this work, and the subsequent analysis, were
 carried out within the approximation of non-interacting 
 particles (RPA), using the DFT-LDA eigenvalues and wavefunctions.
 A straightforward scheme for incorporating many-body effects is to 
 apply a \emph{scissor} operator to the unoccupied states, following
 the recipe of Del Sole and Girlanda~\cite{dels-moch-bar-91}.
 In this way we compensate for the well known understimation of the 
 DFT-LDA band gap, and partially account for self energy and excitonic shifts
 in energy. A scissor shift of +0.5~eV has previously been determined 
 in other works on Si(100)~\cite{incze_prb_y2005_v71_035350} 
 as giving the best agreement with the experimental RA spectra. 
 We also confirm this result from a fit to the experimental data (see Fig.~\ref{fig:RAS}).
 Nevertheless, this value may not consistently describe the energetic positions 
 of all surface state features, which generally undergo many body corrections
 different from bulk ones.
 In order to determine the correct correspondance between surface-related experimental
 and theoretical energy loss peak, we performed some preliminary
 calculations including many body effects on a smaller (12 layers)
 \ctwo\ slab at a lower cutoff (12~Ry).
 Self energy corrections were computed within the so called GW approximation.
 Within this approach it is possible to solve self consistently
 a closed set of five equations (Hedin's equations \cite{hedin_pr_v139_y1965_pA796})
 connecting the Green function ($G$), the polarizability ($\Pi$), the screened Coulomb 
 and vertex interactions ($W$ and $\Gamma$) and the self energy ($\Sigma$).
 with the assumption $\Sigma = iGW$ and $\Gamma=1$.
 Excitonic and local field effects were accounted for by means of
 solving the Bethe Salpeter equation (BSE), connecting the full $\Pi$
 with the non interacting $\Pi^0$ via a 4-points kernel.\\
 Details of the approach are beyond the scope of this thesis,
 and can be found in Ref.~\cite{ORR02}.
 Our preliminary calculations on HREEL spectra below 2.5~eV
 show in fact that GW+BSE approach give a better agreement 
 with the experimental data of 
 Gavioli\ea~\cite{gavioli_ss_y1997_v377_p360} while 
 the RPA+scissor calculation give a misleadingly good comparison
 with the S$_1$ peak of Farrell\ea~\cite{farrell_prb_y1984_v30_p721}.
 Thus the combined GW+BSE approach is nowadays the state of the art
 for computing precise optical spectra, nevertheless it is very 
 expensive from the computational point.

\section{Conclusions}
   We studied the RA and REEL spectra of the Si(100) surface 
 modeling the surface with \pone\, \ptwo\ and \ctwo\ reconstructions.
 Our calculations of RA spectra are in agreement with previous works. 

 REEL spectra has been calculated for two experimental setup 
 according to the available experimental data from 
 Gavioli\ea\ \cite{gavioli_ss_y1997_v377_p360} 
 and Farrell\ea\ \cite{farrell_prb_y1984_v30_p721}.
 We confirmed that \pone\ cannot be the only reconstruction of the real surface 
 because S$_0$ peak is completely missing in spectra 
 and from a band structure analysis.\\
 The origin of the S$_0$ and S$_1$ peaks has been carefully 
 analysed considering the \ctwo\ model more rapresentative of 
 the surface. We have seen that S$_0$ arises from transitions
 involving bulk states around $\Gamma$ and surface states below 
 the bandgap. On the contrary S$_1$ involves only surface states.\\
 Moreover the 0.68~eV feature in Gavioli's data suggest some \ptwo\ present, 
 along with \ctwo\, but the experimental analysis also suggests that
 temperature can largely change the structural reconstruction
 because termal motion can easly induce a flip-flop of the
 dimers.

 In summary, we obtained several informations on the nature of the 
 low energy excitations of the Si(100) surface and the
 joint theoretical and experimental REEL spectroscopy 
 contributed to clarify the structural composition of this 
 still debated surface.

\chapter{The oxidized Si(100) surface} \label{ch:oxi}
  There are two reasons why the investigation of REEL spectra
of oxidized Si(100) is important.
First of all, studying the changes in the experimental spectrum that
occur after absorption of a foreign substance, such as oxygen or hydrogen,
is a widely used technique for elucidating the character of spectroscopic
features in the \emph{clean} surface. From a theoretical point of view,
it is not immediately apparent how spectral features related to surface
states are modified following atomic scale modifications.\\
Secondly, a thorough understanding of the oxidation process on Si(100)
at the atomic scale is of huge technological importance for the
development of electronic devices and nanodevices~\cite{baumvol_ssr_36_1_y1999}.
 
In spite of an extensive study over several decades, including electron 
energy loss~\cite{ibach_prb_y1974_v10_p710,ibach_prb_y1974_v9_p1951}
photoemission, RAS~\cite{borensztein_v95_p117402_y2005,katalin_submitted,katalin_inpreparation} 
and several theoretical 
investigations~\cite{incze_prb_y2005_v71_035350,fuchs_prb_y2005_v72_p075353,fuchs_thesis_y2004}, 
there remains some controversy about the reaction pathways, 
with different works suggesting dimer breaking, insertion of O into dimer backbonds, 
as well as silanone bound (O)Si=O formation~\cite{hemeryck_jcp_v126_p114707_y2007,chabal-prb-2002}.

In the present section  we aim to identify ``fingerprints'' in the REEL
spectrum that can help to distinguish between different adsorption sites
during the initial stages of oxidation, and to obtain further information 
about the surface states of the clean surface.

 \section{Atomic structure} \label{sec:oxi-atomic-structure}
  Among the vast experimental and theoretical work on oxidized Si(100),
a number of recent theoretical 
studies~\cite{incze_prb_y2005_v71_035350,fuchs_prb_y2005_v72_p075353,fuchs_thesis_y2004} 
proposed various possible adsorption sites at low and intermediate coverages, and
investigated the optical (RAS) and electronic properties of these systems.\\
\begin{figure}[!h]
   \centerline{\epsfig{file=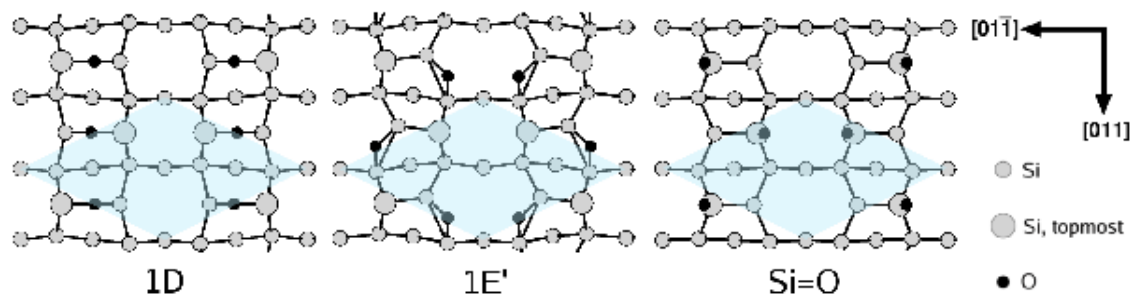,clip=,width=15.5cm}}
  \caption[Si(100):O - 1E$^\prime$, 1D and 1D Si=O models]{
           Schematic diagram of the 1E$^\prime$, 1D models of the oxidized Si(100)
           surface.
           Large circles indicate ``up'' silicon dimer atoms; small filled circles are oxygen atoms.
           The \ctwo\ unit cells is indicated by shaded regions.
 \label{fig:structure_ox}
 }
\end{figure}
\indent In this work we considered a wide range of oxygen adsorption sites that 
we can group according to the number of atomic oxygens 
involved: those with two atomic oxygens in the elementary cell
will be referred to as low-coverage (0.5ML) structures;
on the other hand, a coverage of 1ML will be described in our model by 
four oxygen atoms for cell.\\
%
%%%%%%%%%%%%%
\begin{table}
 \begin{center}
   \begin{tabular} {c c c c}
   \hline
 Structure  & dimer bridge [\AA] & dimer backbond [\AA] \\
  \hline
 \ctwo\ 1D  & 1.61  &  1.61 \\
 \ctwo\ 1E' & 1.78  &  1.75 \\
 \hline
 Structure & dimer bridge  [\AA] & Si=O  [\AA] \\
 \hline
  \ctwo\ 1D Si=O & 1.61 , 1.70 & 1.53 \\
  \ptwo\ 1D Si=O & 1.63 , 1.73 & 1.55 \\
  \hline
  \end{tabular}
 \end{center}
\caption[Structural lengths of Si(100):O]{
      Structural lengths of the oxidized Si(100) surface referring to the
      configurations studied. All calculations are performed in LDA.
      1D refers to oxygen in the bridge, 1E$^\prime$ refers to oxygen 
      in the backbond, Si=O refers to the silanone structure created 
      on the top silicon atom of the dimer.
      \label{tab:lattice-oxi}
               }
\end{table}
\indent Among the possible low-coverage structures, we choose
to compute the REELS for three characteristic local structural motifs 
(nomenclature follows that of Ref.~\cite{fuchs_prb_y2005_v72_p075353,fuchs_thesis_y2004}):
structure 1D, with an oxygen inserted into the surface Si dimer, and structure 1E',
in which oxygen inserts into the dimer backbond and a silanone structure, whereby 
oxygen is bonded via a double bond to the silicon dimer.
A schematic representation of these structures is given in Fig.~\ref{fig:structure_ox}.
Both surfaces have a low total energy (see Table~\ref{tbl:energies} and
Refs.~\cite{bechstedt-prb-2005,fuchs_thesis_y2004,yamasaki_prl_y2003_v91_146102})
and for this reason they are good candidates to represent the real system.
Regarding the silanone structure, we studied a configuration where oxygen is
bonded to one silicon atom, as proposed by Hemerick\ea\ in~\cite{hemeryck_jcp_v126_p114707_y2007}
showing a critical formation of silanone species (O)Si=O during initial oxidation.
Anyway, we found silanone structures being higher in energy when
only two atomic oxygens for cell are considered (see structure ``Si=O" with just
two oxygens added to the clean \ctwo\ base on top of a silicon atom of the
dimer in Table~\ref{tbl:energies}).\\
\indent Moreover, in the case of 1ML coverage, we considered
several structures appeared in litterature, 
starting from both 1D or 1E' base configuration.
First we studied the 1D base adding the oxygen on the backbond,
resulting a \ctwo\ 1D+1E' structure.
Furthemore we considered one oxygen in the bridge and one added in the
silanone bound because suggested in Refs.~\cite{hemeryck_jcp_v126_p114707_y2007,chabal-prb-2002}, 
finding a relaxed structure with two silanone species,
one free and the second connected to another silicon by a dative bond.
We analyzed this ``1D Si=O" structure built on both the \ctwo\ and
 \ptwo\ bases.
The latter case is considered because the two oxygens on
silanone bounds of adjacent dimers on the \ctwo\ reconstruction are very close
(see black circles in the central picture of Fig.~\ref{fig:structure_silanone}).
This is unlikely to occur in nature, because there would be extra strain
on the surface since steric interaction pushes away adjacent oxygens.
The \ptwo\ Si=O structure (see Fig.~\ref{fig:structure_silanone})
seems to be more reasonable for these reasons.\\
Table~\ref{tbl:energies} shows a summary of the studied structures with
corresponding total energies calculated. 
In order to compare the results, in case of the clean reconstructed surfaces we 
calculated the surface energy:
\begin{equation}
 E_{\textrm{surf}} = \frac{1}{2S}\left(E_{\textrm{tot}}(N) - NE_{\textrm{bulk}}\right)
\end{equation}
where N is the number of atoms in the slab, S is the surface cell area, $E_{\textrm{tot}}$ 
is the total energy and $E_{\textrm{bulk}}$ the total energy per atom of bulk Si.
The surface with the smallest surface energy is the most stable one.  \\
In case of the oxidized surface we reported the adsorption energy, 
i.e. the energy gain for adsorbing atom at the surface.
This quantity is calculated as:
\begin{equation}
 E_{\textrm{ads}} = -\frac{E_{\textrm{tot}} - E_{\textrm{clean}}-N_{\textrm{O}}E_{\textrm{free}}(O)}{N_{\textrm{O}}}
\end{equation}
where $E_{\textrm{tot}}$ is the total energy of the slab, $E_{\textrm{clean}}$ is the total energy of the corresponding clean surface, $E_{\textrm{free}}(O)$ is the energy of the free oxygen and $N_{\textrm{O}}$ is the number of oxygen atoms. \\
For all structures the inner four layers are fixed to the bulk positions,
assuming they are not influenced by the surface distortions.
Optimization is hence performed on the outer slab layers
using the Broyden-Fletcher-Goldfarb-Shanno minimization (BFGS)~\cite{B,F,G,S}.
Crosschecks of results have been performed with PWscf and ABINIT
codes~\cite{pwscf-long,abinit2}.
\begin{figure}[!h]
 \centerline{\epsfig{file=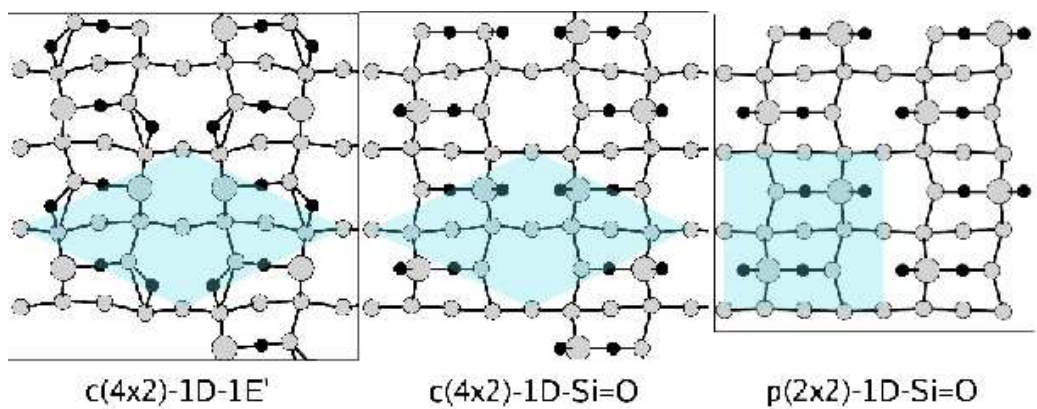,clip=,width=13.5cm}}
  \caption[Si--O related models of Si(100):O (1ML)]{
    Schematic diagram of the Si--O related models of the 1ML oxidized Si(100)
    considered in this work.
    Left: \ctwo\ with backbond and dative bonded silanone; Middle: \ctwo\ bridge bond
    and silanone; Right: \ptwo\ bridge bond and silanone.
    Large circles indicate ``up'' silicon dimer atoms; small filled circles are oxygen atoms.
    The \ctwo\ and \ptwo\ unit cells is indicated by shaded regions. \label{fig:structure_silanone}
  }
\end{figure}
\begin{table}[!h]
 \begin{center}
   \begin{tabular} {c c c c c c c}
   \hline
 base & structure & E$_{\textrm{surf}} $[eV/\AA$^2$]  & N$_{e^-}$ & N$_\textrm{O}$ & N$_k$\\
  \hline
  \ctwo\  & clean & 0.0903 & 256 & 0 & 6 \\
  \ptwo\  & clean & 0.0905 & 256 & 0 & 4 \\
   \hline
 base & structure & E$_{\textrm{ads}}$ [eV] & N$_{e^-}$ & N$_\textrm{O}$ & N$_k$\\
  \hline
  \ctwo\  & 1D   &-7.14   & 280 & 2 & 6 \\
  \ctwo\  & 1E'  &-7.22  & 280 & 2 & 8 \\
  \ctwo\  & Si=O &-5.73  & 280 & 2 & 2 \\
 \hline
  \ctwo\  & 1D+1E'  &-7.17  & 304 & 4 & 2 \\
  \ctwo\  & 1D Si=O &-6.76  & 304 & 4 & 2 \\
  \ptwo\  & 1D Si=O &-7.04  & 304 & 4 & 4 \\
 \hline
  \ctwo\  & ``A" in Fig.~\ref{fig:relax} & -6.74 & 304 & 4 & 2 \\
  \ctwo\  & ``B" in Fig.~\ref{fig:relax} & -6.82 & 304 & 4 & 2 \\
  \ctwo\  & ``C" in Fig.~\ref{fig:relax} & -7.45 & 304 & 4 & 2 \\
 \hline
  \end{tabular}
 \end{center}
\caption[Energetics of Si(100):O configurations studied]{
     A summary of energetics of all stable and metastable configurations 
     found after BFGS relaxation for the oxidized Si(100) surface.
     Structures are listed with their total energies, each force component
     is relaxed below the threshold of $0.13$~eV/\AA.
     \label{tbl:energies}
               }
\end{table}
Because of complexity, the case of one oxygen in the
backbond and one added in the silanone bond on top of the lower silicon
atom of the dimer, needs a separate discussion.
In spite of the fact that silanone structures have been suggested in the
litterature (see Refs.~\cite{chabal-prb-2002,hemeryck_jcp_v126_p114707_y2007}), 
both from theoretical and experimental grounds,  
in the present work the total energies founded are relatively high for the 
structures reported up to now.
Hence we tried to identify a low energy structure for a 1ML coverage that contains the
silanone motif. Although there are specific techniques for doing such simulations,
such as Car-Parrinello molecular dynamics, nudged elastic band simulations 
or potential energy surface mapping, we was able to identify one such structure 
by following a straightforward BFGS relaxation, starting from an undimerized
Si(100) surface with Si=O bonds in the vertical plane. The eventual
purpose is to identify any fingerprints in the REEL spectra that might correspond
to such Si=O bonds (see sec.~\ref{sec:exp-eels-oxi}).
In Fig.~\ref{fig:relax} the evolution of the total energy as this fictitious 
structure relaxes is shown.
We started from the undimerized Si(100) with Si-O-Si=O bonds staying
in the vertical plane (perpendicular to the surface).
After some BFGS steps, the Si-Si dimers are formed, but the Si-O-Si=O bonds are
still both present.
In configuration ``A" (see Fig.~\ref{fig:relax}), one of the Si=O starts
to bond the dimer until the metastable structure ``B" where there is still
the Si-O-Si=O bond in the plane (see blue arrows in Fig.~\ref{fig:relax}),
but the other oxygen is bonded in a Si-O-Si-O-Si chain, similar to 
a part of the SiO$_2$ crystal lattice.
Leaving this structure to relax we found the configuration ``C"
without any silanone bonds and characterized by
a silicon atom bonded to 3 oxygens.
In this last configuration Si(100):O (1ML) reaches the lowest total energy.
\begin{figure}[!h]
 \centerline{\epsfig{file=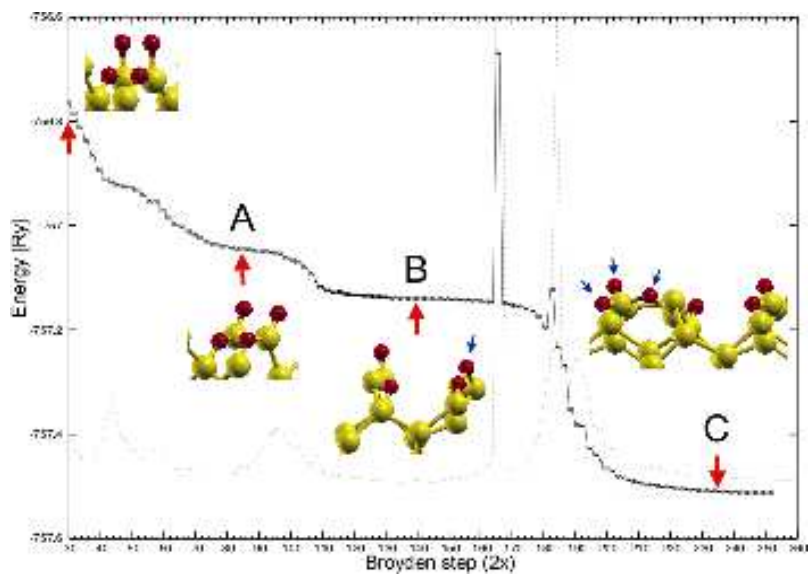,clip=,width=10.5cm}}
  \caption[Si(100):O - total energy configurations]{
  Total energy of configurations generated by relaxing Si(100)
  with one oxygen on the backbond and one oxygen on top of the
  higer silicon atom of the dimer.
 \label{fig:relax}
}
\end{figure}

\section{Electronic structure and the effect of oxidation}
\begin{figure}[!h]
  \begin{center}
   \begin{tabular}{c c}
    \includegraphics[angle=270,width=9.5cm]{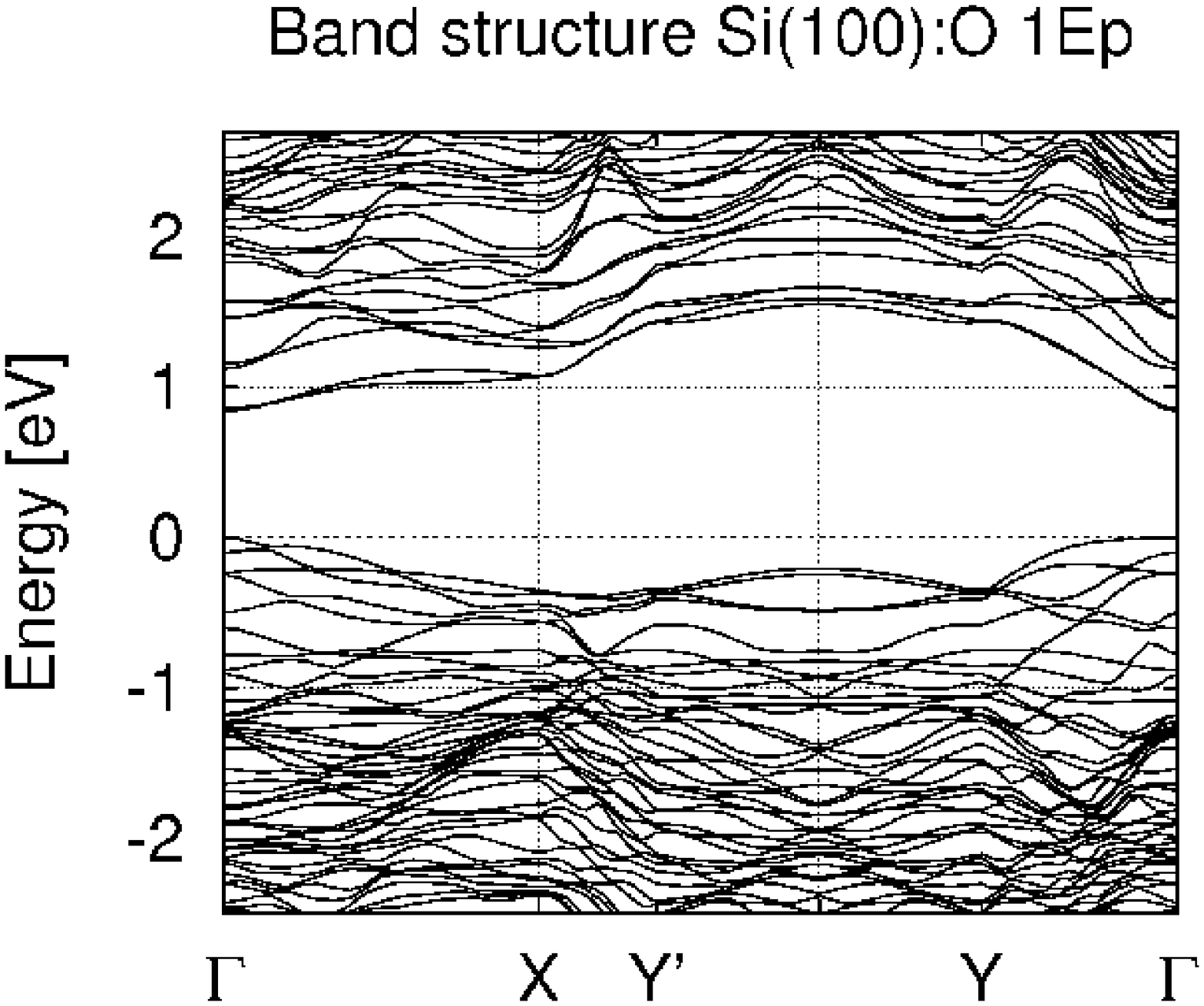} \\
    \includegraphics[angle=270,width=9.5cm]{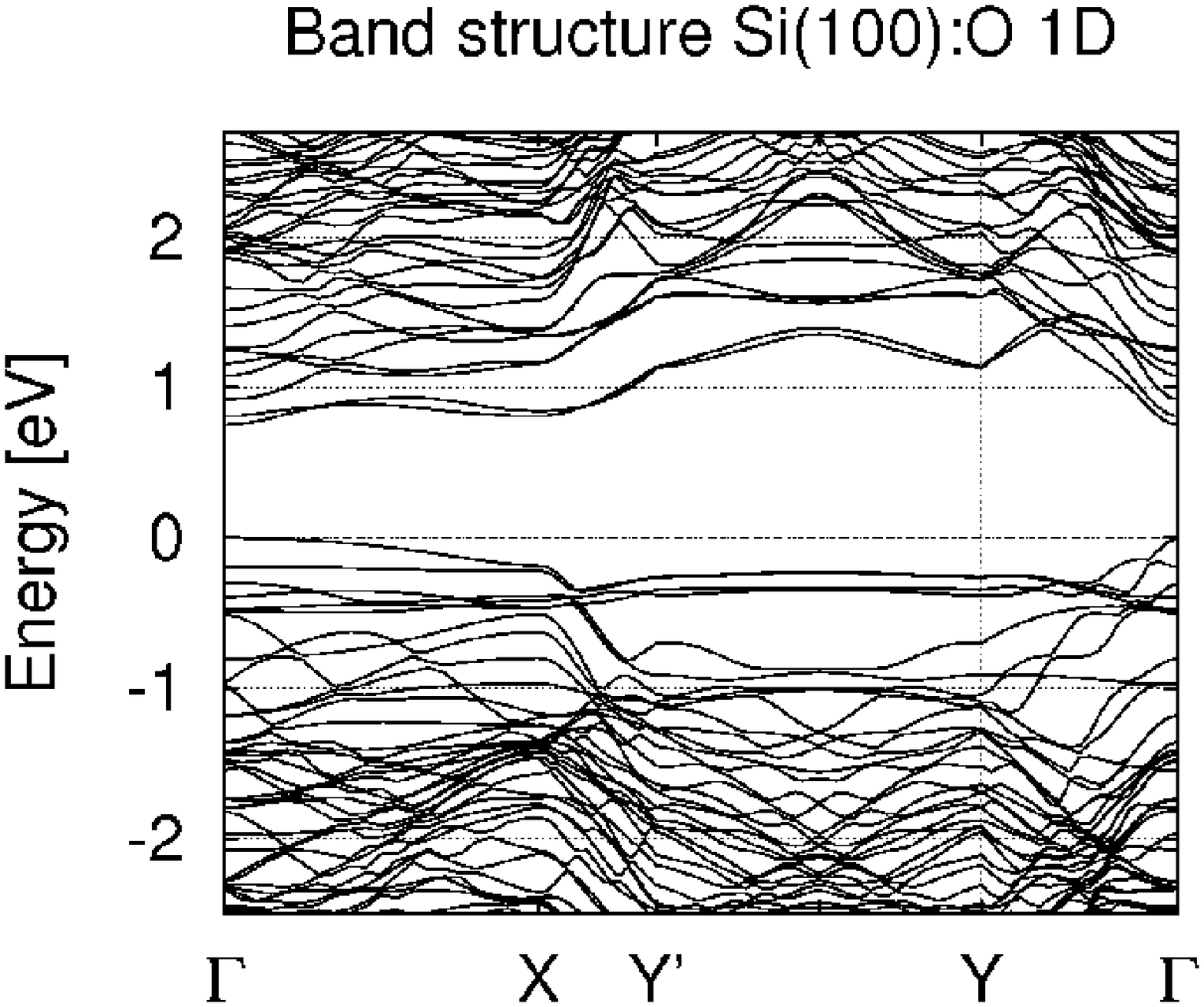}
   \end{tabular}
  \end{center}
\caption[Si(100):O 1E$^\prime$, 1D: band structure]{
    Band structures of the 1E$^\prime$ (top), 1D (bottom).
     models of the oxidized Si(100) surface. 
     A rigid upward shift +0.5eV (scissor shift)
     has been applied to the unoccupied bands.
     \label{fig:bands_oxidized}
         }
\end{figure}

\begin{figure}[!h]
  \begin{center}
   \includegraphics[angle=270,width=9.5cm]{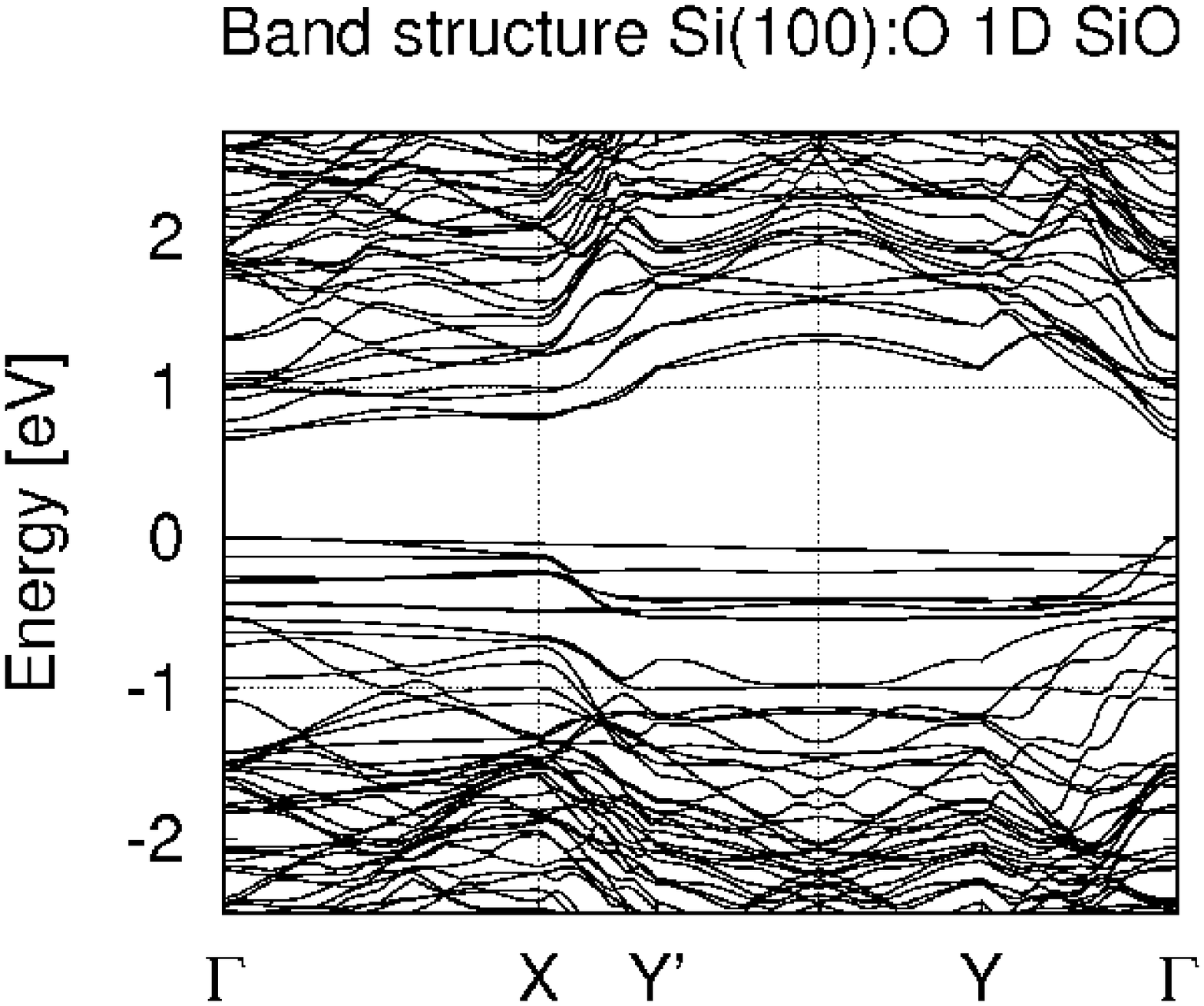}
  \end{center}
 \caption[Si(100):O 1D Si=O moled: band structure]{
       Bandstructure of the 1D Si=O model of the
       oxidized Si(100) surface. A rigid upward shift +0.5eV has been
       applied to the unoccupied bands. Flat bands along the
       $\bar{Y}$'-$\bar{Y}$ direction are due to the presence
       oxygen double bonded to the top silicon in the 
       dimer.
     \label{fig:bands_oxidized1}
         }
\end{figure}

\begin{figure}[!h]
 \begin{center}
  \includegraphics[angle=270,width=9.5cm]{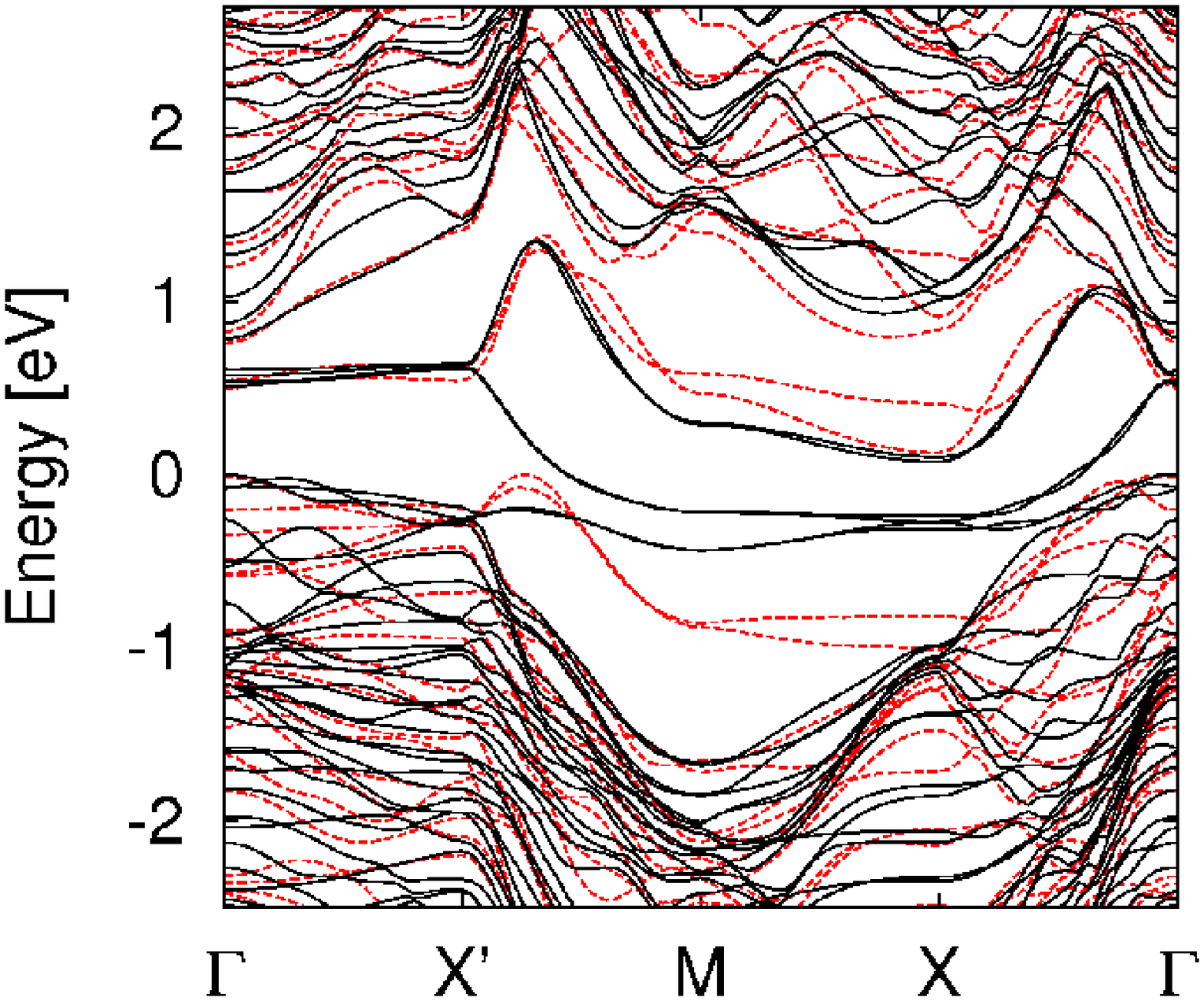}
 \end{center}
 \caption[Si(100)-p(2$\times$1) vs \pone\ Si=O band structure]{
     Bandstructure of the clean Si(100)-p(2$\times$1) (black line)
     compared with the ones of \pone\ Si=O (atomic oxygen)
     (red dashed line).
     \label{fig:bands_oxidized2}
         }
\end{figure}

  We present here the computed Kohn-Sham band structure of three models of oxidized
Si(100).
The electronic structure of the 1D and 1E' surfaces is shown in 
Fig.~\ref{fig:bands_oxidized} along a standard path
in the surface brillouin zone: $\Gamma$-$\bar{X}$-$\bar{Y}'$-$\bar{Y}$-$\Gamma$
(see Fig.~\ref{fig:th-SBZ} in chapter~\ref{ch:surface}).
Mostly, modifications respect to the clean
Si(100)--c(4$\times$2) concern a change of surface states inside the bandgap.
In fact, comparing Fig.~\ref{fig:bands_c4x2_p2x1} 
to Fig.~\ref{fig:bands_oxidized}, it is evident that oxygens move 
the surface states visible in the $\Gamma$-$\bar{X}$ direction, 
states that are responsible of the S$_0$ EEL peak of the clean surface, 
(see Fig.~\ref{fig:Farrell_7eV_4L} and Fig.~\ref{fig:exp_reels} in Chapter~\ref{ch:a-si100}).
The oxygen adsorbed modifies also bands in the $\bar{Y}'$-$\bar{Y}$ path.
Moreover, we present in Fig.~\ref{fig:bands_oxidized1} the computation of
the electronic states for the surface including a silanone (O)Si=O bound.
In Fig.~\ref{fig:bands_oxidized1} it is possible to see the flat bands in
the $\bar{Y}$'-$\bar{Y}$ path due to the presence of oxygen doubly bonded
to the top silicon of the dimer.
Furthermore we performed a test on the \pone\ structure to better analyse
the effect of the isolated Si=O bond. For this system we report
the bands compared to that of the clean \pone\ surface (see Fig.~\ref{fig:bands_oxidized2}).
The bandgap typical of a semiconductor is closed by the presence
of oxygen, hence the system is metallic.\\
\indent All calculations are performed within the DFT-LDA and 
a scissor shift of $+0.5$~eV (see sec.~\ref{sec:si100-discussion}) 
has been applied to the unoccupied bands in order to mimic the many body effects. 
Band structure calculations are performed with the ABINIT package~\cite{abinit2}.

\section{Experimental energy loss data} \label{sec:exp-eels-oxi}
  Experimental EEL spectra of the oxidized Si(100) surface
are not so numerous in litterature, at least in the range of energy
we are studying (i.e in general up to $\simeq$8-10~eV and in particular up to $\simeq$3~eV).
In this work we are refering to not-so-recent data from 
Ibach\ea\ ~\cite{ibach_prb_y1974_v10_p710,ibach_prb_y1974_v9_p1951} and at
the subsequent paper of Ludeke\ea\ ~\cite{ludeke_prl_y1975_v34_1170}
where experimental REELS are reported in terms of the second derivative 
of the spectrum respect to the energy in order to cut the large contribution 
of the elastic peak and evidence spectral features.\\
\begin{figure}[!h] 
 \centerline{\epsfig{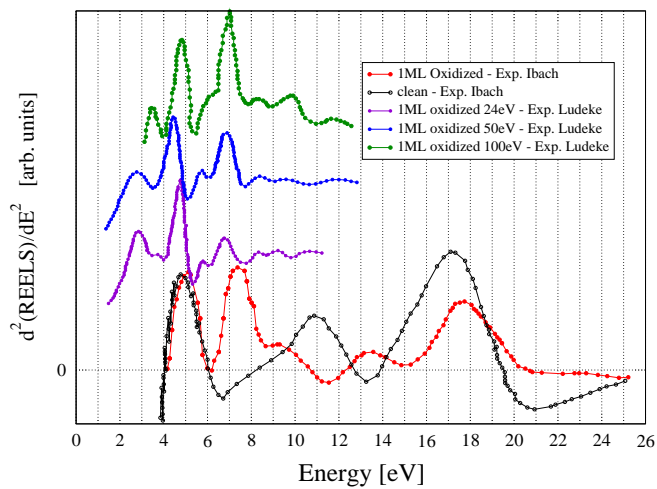}}
 \caption[Experiments: EEL second derivative spectra]{ 
   Experimental data from Ibach\ea\ and Ludeke\ea\ for the clean (black) and  
   oxidized (red and blue) Si(100) surface.\label{fig:all-exp-oxi}}
\end{figure}
In Fig.~\ref{fig:all-exp-oxi} all experimental data are collected.
The spectra from Ibach\ea\ are reported for the clean (black circles) 
and oxidized (red circles) surface.
In particular, Ibach used high incident energy of the electrons,
(E$_i$=100~eV) and an off specular geometry with
normal incidence and collecting electrons at 42.5$^\circ$ respect
to the perpendicular direction to the surface plane.\\
\indent From the analysis of Ibach's data we can conclude that there are two features
that seems to be due to the oxygen adsorption: 
a peak at around 7~eV, which appears in the oxidized surface, and the surface plasmon 
at around 12~eV which is splitted in two.
Moreover, the bulk plasmon, appearing at 17~eV in the clean surface,
is lowered and slightly blueshifted.\\ 
\indent The other experimental data reproduced in Figure~\ref{fig:all-exp-oxi}
are from Ludeke\ea\ ~\cite{ludeke_prl_y1975_v34_1170} (green, blue and violet 
circles in Fig.~\ref{fig:all-exp-oxi}) and refer to specular scattering geometry: 
backscattered electrons are detected and analyzed. The three sets refers to
different incident energies: 24~eV (violet), 50~eV (blue) and 100~eV (green).
Peaks at 3~eV and 5~eV (E$_1$ and E$_2$ respectively)
appear in all those sets of data. From the comparison of the three sets 
it is possible to conclude that the two main peaks around 5~eV and 7~eV are in 
agreement with Ibach's results.
Moreover the 7~eV peak increases with increasing incident electrons energy and
is interpreted by Ludeke\ea\ as an excitation of the SiO or SiO$_2$
molecules \cite{jevons_pps_y1935_v49_543,barrow_pps_y1954_v224_374}
in the Schumann region.
We will comment this assignement in the following
paragraphs, where we will show that our results do not support it.

\section{Theoretical energy loss spectra}
  In the following paragraphs we present REEL spectra calculations
of the oxidized Si(100) surface using the code YAMBO~\cite{YAMBO}.

\subsection{Clean versus oxidized Si(100): the low-energy part of the spectrum}
\begin{figure}[!h]
\begin{center}
 \includegraphics[width=8.5cm]{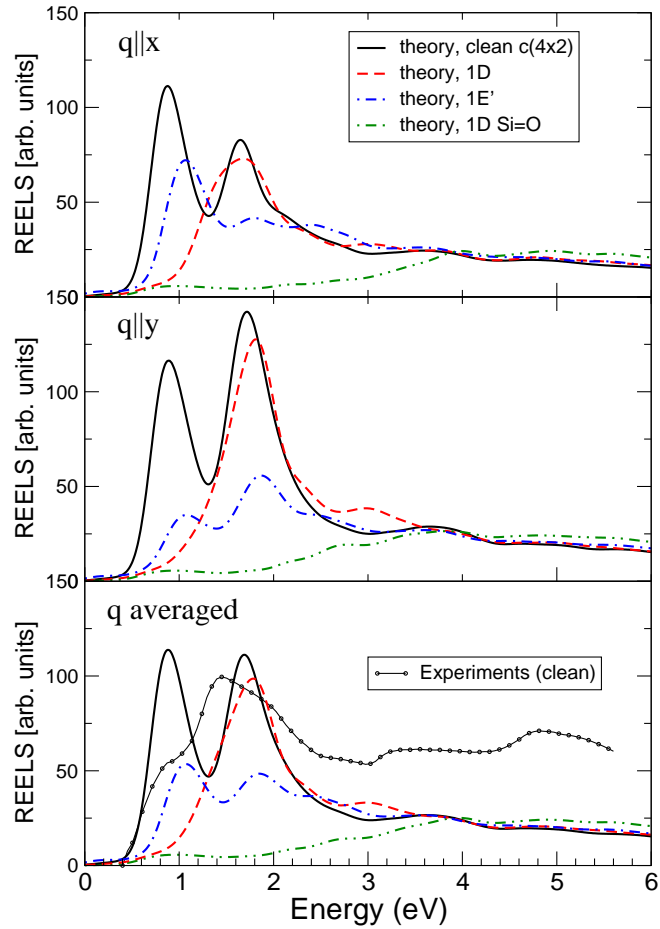}
\end{center}
\caption[ Si(100)-$c(4\times2)$ vs Si(100)-$c(4\times2)$:O 1D, 1E' and 1D Si=O: REELS at $E_0 = 40$~eV]{
         Experimental and computed REELS at $E_0 = 40$~eV 
         for clean Si(100)-$c(4\times2)$ is compared with the
         calculated spectra of 1D, 1E$^{\prime}$ and 1D Si=O 
         oxidized reconstructions. Experiments for the clean
         are taken from Farrell\ea\ in Ref.~\cite{farrell_prb_y1984_v30_p721}.
     \label{fig:reels_oxidized}
}
\end{figure}

  REEL spectra are often presented in literature in terms
of the second derivative of the energy loss. Moreover we start presenting
here our computed bare REEL spectra for the clean and oxidized Si(100).
In particular we considered the clean and the 0.5~ML oxygen-covered surfaces with 
oxygen inserted into the bridge (1D) or the backbond (1E$^\prime$).
We compare the REEL spectra computed for the experimental setup of
Farrell (Ref.~\cite{farrell_prb_y1984_v30_p721}: specular scattering at 
60$^{\circ}$ with incident energy E$_i$=40~eV) with measured spectra for the clean
Si(100). 
All results are shown in Fig.~\ref{fig:reels_oxidized} where theoretical spectra for
x and y polarization directions are also presented separately.
In the bottom panel of Fig.~\ref{fig:reels_oxidized} a comparison of the 
(unpolarized) experimental data with computed spectra averaged over the two 
polarization directions is shown.\\
It is evident that the two main peaks, namely the S$_0$ shoulder at 0.8~eV
and the peak S$_1$ at 1.3~eV in the experiments, appear also in the calculated
spectra for both the clean (black line) and the oxidized 1E$^\prime$ 
configurations (blue line).
On the contrary, oxygen in the bridge position (1D structure-­red line) 
completely kills the first peak of the spectrum.
Unfortunatly we did not find experimental data for this surface at low oxygen
coverage in the low energy range ($<3$~eV), but a similar behaviour is
shown in the case of a coverage of H$_2$O in Ref.~\cite{farrell_prb_y1984_v30_p721}.
In conclusion of this section we can say that REEL provide a useful tool
able to distinghuish between oxidized surface configurations, in particular
in the low energy spectral region.

\subsection{Clean versus oxidized Si(100): REEL spectra in a wider spectral reagion}
\begin{figure}[!h]
 \vspace{1cm}
\begin{minipage}{8cm}
 \includegraphics[width=8.5cm]{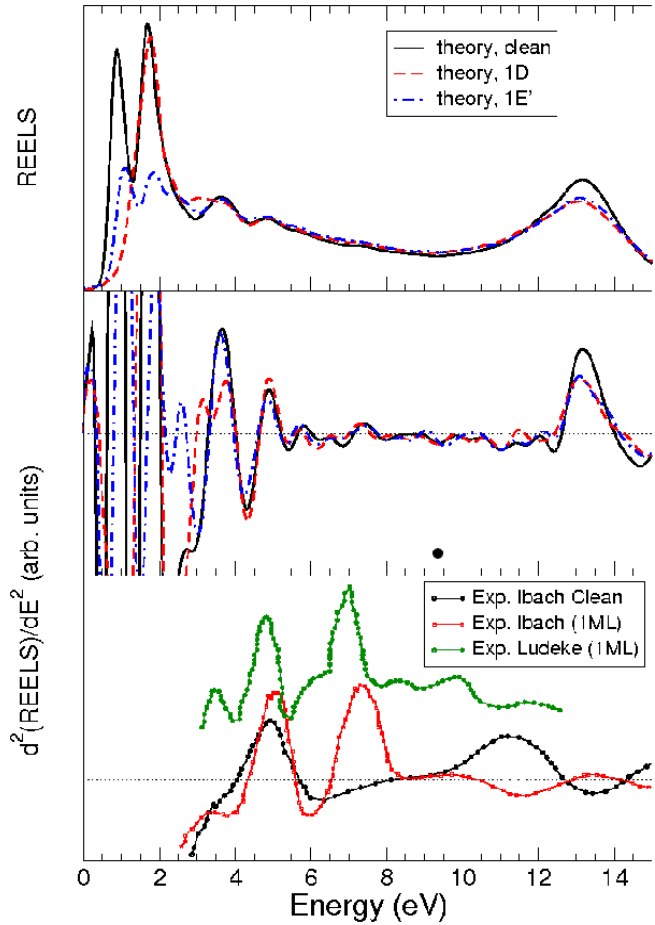}
\end{minipage}
\hspace{0.8cm}
 \begin{minipage}{5cm}
  \caption[Si(100)-$c(4\times2)$, 1D and 1E': REELS]{
           Computed bare REEL spectrum (top panel) for clean Si(100)-$c(4\times2)$ 
           compared with the 1D (oxygen in the bridge) and 1E' (oxygen in the
           backbond) oxidized reconstructions.       
           Calculated negative second derivative spectra of the same surfaces  
           (middle panel) shows slight differences in the spectral region
           above 3~eV. Experimental data (bottom panel) are also reported 
           from Refs.~\cite{ibach_prb_y1974_v9_p1951} and~\cite{ludeke_prl_y1975_v34_1170} 
           showing an important peak at around 7~eV, beyond the main peaks
           at 3~eV and 5~eV, interpreted as the E$_1$ and E$_2$ bulk silicon transitions.
           The 7~eV peak is absent in the experimental spectrum of the clean 
           (black circles) and is hence interpreted as due to oxidation 
           (red and green circles).
            \label{fig:ibach-1}
           }
 \end{minipage}
\end{figure}

\begin{figure}[!h]
\vspace{1cm}
\begin{center}
 \begin{tabular}{c c}
   \includegraphics[width=6.5cm]{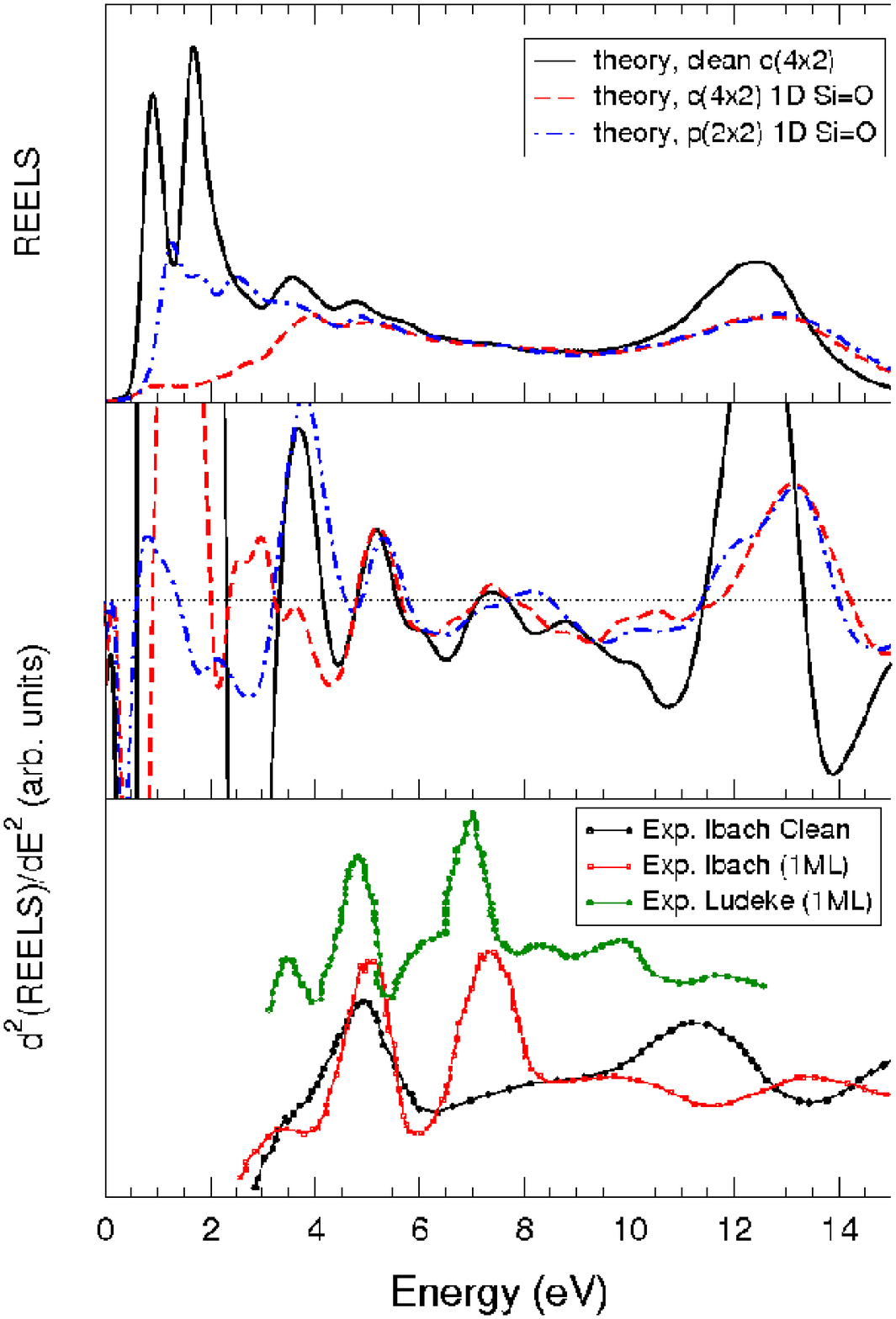}
 & \includegraphics[width=6.5cm]{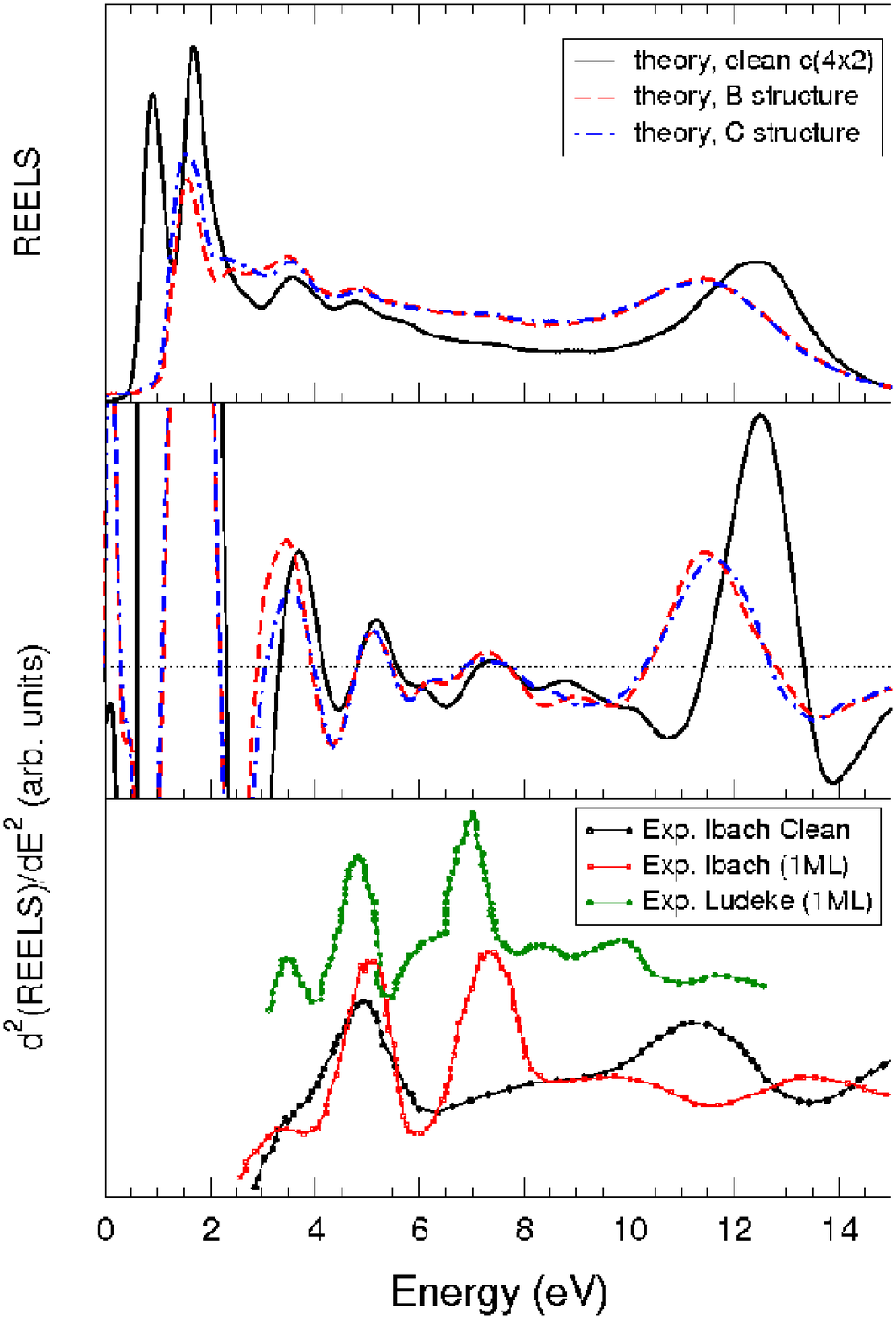}
 \end{tabular}
\end{center}
\caption[Si(100)-$c(4\times2)$, 1D Si=O, \ptwo\ 1D Si=O, B and C: REELS]{ Left:REELS for clean Si(100)-$c(4\times2)$ compared with the 1D Si=O 
            and the \ptwo\ 1D Si=O reconstructions. 
            Right:  REELS for clean Si(100)-$c(4\times2)$ compared with the B structure
            (metastable precursor) and C structure
            For comparison, experimental data from 
            Refs.~\cite{ludeke_prl_y1975_v34_1170,ibach_prb_y1974_v9_p1951}
            are reported.
\label{fig:ibach-2}
}
\end{figure}

  In this section we present computed REEL in terms of the second derivative spectrum.
In Fig.~\ref{fig:ibach-1} (top panel)
we show the bare REEL spectra for 
the clean, the 1D and 1E$^\prime$ structures (0.5~ML coverage)
computed for the experimental setup used by Ibach and 
Ludeke in their works~\cite{ibach_prb_y1974_v10_p710,ludeke_prl_y1975_v34_1170}
(i.e. incident energy E$_i$=100~eV with slightly different
geometry scattering\footnote{Ibach used off specular experimental 
geometry instead of backscattering geometry form a normal incident beam
of electrons used by Ludeke\ea\ in Ref.~\cite{ludeke_prl_y1975_v34_1170}.
However, at this level of approximation we did not find appreciable differences 
in case of spectra calculated with specular or off specular geometry.}).
In the middle panel of Fig.~\ref{fig:ibach-1} we show the computed 
second derivative spectrum.
Neglecting oscillations for energies below 3~eV, and except
for the intensity of the surface plasmon peak calculated around 15~eV,
the clean and the other two oxidized surfaces do not show any important differences. 
This is in contrast with experiments (bottom panel of Fig.~\ref{fig:ibach-1})
where a peak centred at around 7~eV appears after oxidation.
In order to understand the origin of this peak, we analysed the EEL spectra of all
configurations presented in Section~\ref{sec:oxi-atomic-structure}.
Following arguments presented in Ref.~\cite{ludeke_prl_y1975_v34_1170}
about the origin of the 7~eV peak, we show theoretical results for surfaces including 
a silanone bond on the \ctwo\ and \ptwo\
reconstructions of the clean Si(100) and an oxygen inserted in the
dimer bridge (Fig.~\ref{fig:ibach-2}).
We found slight differences with respect to the previous calculations:
the 7~eV peak still does not appear and the spectral features
are similar for the clean and the oxidized surfaces.
Unfortunatly, large differences between \ctwo\ and \ptwo\ structures appear
in the low region of the spectrum where experimental data are
not available.
However we can observe that the surface plasmon is
slightly lowered and blueshifted when a silanone bound is included in
the surface configuration.
Moreover we report in Fig.~\ref{fig:ibach-2} (left) the results concerning
the stable ``C" structure and his metastable precursor ``B" containing
one (O)Si=O bound depicted in Fig.~\ref{fig:relax}\footnote{
For \ctwo\ silanone structures we used 72k points in the IBZ and 450 bands,
corresponding to minimum $e$--$h$ transition energy of 13.27~eV
\ptwo\ required 64k points and 500 bands, corresponding to 15.09~eV, 
for the ``B" and ``C" structures 72k points and 500 bands, corresponding to 
15.09~eV, have been considered in order to fully converge.}.

Once more we can see that no features at 7~eV comparable
to the experimental peak appears in our calculations, and
even if Hemerick\ea\ ~\cite{hemeryck_jcp_v126_p114707_y2007}
suggests similar structures on small clusters as
energetically favorable, we can not conclude that silanone bond
or siloxane structures on Si(100) are representative of the
real surface.
Moreover, Ludeke\ea\ guessed that the 7~eV peak was related to molecular
excitations of Si=O bond or silicon monoxyde molecule.
Calculations performed in the present work, however, rule out
Si=O structures bonded to the surface.

It can be worth to mention that, taking as a reference
the surface plasmon peak of the clean Si(100) (represented
by the black continuous line in Figs.~\ref{fig:ibach-1},
~\ref{fig:ibach-2}), oxygen added 
into 1D and 1E$^\prime$ positions lowers the peak 
(see Fig.~\ref{fig:ibach-1}), while oxygen added into 
the silanone bonds lowers and redshifts the position 
(see Fig.~\ref{fig:ibach-2} left panel).
Conversely, calculated spectra for B or C configuration 
show a shift of the peak position to the opposite direction
(see Fig.~\ref{fig:ibach-2} right). \\
\indent At this point, several speculations can be done in order to 
explain the origin of the peak measured at 7.0~eV, which does
not appear in theoretical spectra for the considered structures.
From the point of view of calculations we must underline 
that they are performed only at the RPA level.
A further analysis could call for the use of approximations 
beyond RPA in order to describe eventually strong excitonic effects.
Moreovoer, we mention that the oxidation process of Si(100) is still under debate,
for example, we can not exclude the formation of 
clusters of SiO$_2$ during oxidation. 
In addition, looking at the intensity dependence of the peak, increasing with the 
energy of the incident electrons (see Fig.~\ref{fig:all-exp-oxi}),
we could interpret the peak as due to
transitions from states originating from structures below the surface level,
in fact, higher energy electrons are expected to penetrate
deeper in the sample. 
Within these hypotesis the present theory could not be adeguate to treat, 
for example, multiscattering processes.
In addition we note that the experimental data are old and it is not clear
how well characterized and clean the surfaces are. 
However we conclude that our calculations does not 
support Ludeke's interpretation about the origin of the 7~eV peak,
because does not appear in all the considered model surfaces 
(including Si=O or O-Si-O bonds). 

\section{Conclusions}
  In the present section we draw a summary of the previous analysis,
and some conclusions.
We calculated the relaxed atomic position of several oxidized si(100)
surface reconstructions at 0.5ML (1D and 1E') of coverage.
Moreover we analyzed 1ML coverage structures with a silanone bund
and an oxygen in the bridge of the dimer (1D Si=O), with a \ptwo\
and a \ctwo\ base.
Finally we found a structure ``B" characterized by a silanone bond and
Si-O-Si-O-Si  chains, to be the metastable precursore of the ``C" structure,
the most stable, with one silicon atom bonded to three oxygens 
(see Fig.~\ref{fig:relax}).\\
\indent We calculated the bare REEL spectrum in the cases
of the oxydized Si(100) previously considered.
We have been able to relate the change in the spectra with the changes in the 
 band structure in the case of both prototypical configuration (oxygen in the bridge
 and oxygen in the backbond). 
 Furthermore we calculated the second derivative
 spectra thanks to the implementation of the SG method in the YAMBO code \cite{YAMBO}.
This allowed us to compare EELS results with the experimental data.
We have shown that the excitation at 7~eV can not be related to
 a molecular excitation of Si=O or O-Si-O, as suggested by Ludeke\ea, because
the configuration with an oxygen on top of the higher silicon atom in the
 dimer does not reproduce that structure.
At the moment we are not able to give an alternative explanation for the 
origin of this spectral feature, but we can confirm that the reconstruction 
studied in this paper (including the most stable structures 1D, 1E$^\prime$, 
 1D Si=O \ptwo\ and \ctwo\, ``C" and the metastable precursor with Si--O and Si$=$O molecule)
 do not reproduce this peak. Still, recent works by Hemerick\ea\ and Chabal\ea\
 have pointed out the critic presence of silanone bonds during the oxidation
 process.
We can only guess that including many body effects in the calculations could
provide more accurate theoretical results, taking into account self-energy and excitonic
effects, in order to help in shedding light on that problem.

\chapter{Subtleties in electronic excitations of open shell molecules} \label{ch:a-molecules}
   In this chapter we present a theoretical study of BeH,
a simple heteronuclear diatomic molecule with an unpaired
electron.
We considered BeH molecule as the simplest exemple
of an isolated open shell system for which the ground state
is expected to be spin polarized.
We present calculations of energy levels and density of states.
Convergence studies and problems arising up in order to
correctly describe excitation spectra are also underlined.
At the end some excitation energies calculated for this system
are compared to available experimental data and the agreement 
is discussed.

  \section{Brief review of TDDFT for isolated systems}
In this section we briefly review the Casida's approach to treat the molecular 
excitation of isolated systems. 
This framework is a reformulation of the TDDFT in the configuration space
and the details can be found in Ref.~\cite{casida,casida_book}.
In fact all the observables are represented using
the DFT--KS eigenvectors in such a way that the non--interacting 
response function results to be diagonal. \\
We note that in case of isolated systems the physical
quantity we deal with is the polarizability of the system $\alpha(\omega)$
that is defined when an external perturbation $\delta V^{ext}(t) = z\delta E_{z}(t)$
is applied to the system.
In this case the x component of the dipole momentum is defined by: 
$\delta d_x = -q\delta x$ where q is the charge of the electron and 
$\delta x = \int d^3x \delta \rho(x,t)x$. \\
Hence, the polarizability is defined by:
\begin{equation}
 \alpha_{xz}(\omega) = -\int d^3 x \frac{q\delta x }{E_z(x)}
\end{equation}
and using the Lehmann representation it can be written by:
\begin{equation}
 \alpha_{xz}(\omega) = 
   \sum_{I} 2(E_I - E_0)\frac{\langle \Psi_0| \widehat{x}|\Psi_I\rangle 
        \langle \Psi_I|\widehat{z} |\Psi_0  \rangle}{(E_I - E_0)^2-\omega^2} 
 \label{eq:alpha1}
\end{equation}
Where $E_I, E_0,\Psi_I,\Psi_0$ are the energies and the many body wavefunctions
of the excited and ground state respectively.
Using the basis set of operators $\{\widehat{a}_{n\sigma}^{\dagger},{\widehat{a}}_{m\sigma} \}$ 
we can define the quantity:
\begin{equation}
 P_{ij\sigma} (t) = \langle \Psi^{KS} |{\widehat{a}}_{i\sigma}^{\dagger}(t){\widehat{a}}_{j\sigma}(t) | \Psi^{KS}\rangle
\end{equation}
corresponding to the density in the configuration space, for which the following 
relations held:
\begin{eqnarray}
  \delta P_{ij\sigma} (t_1) =  \int dt' \chi^{KS}_{ij\sigma,hk\tau}(t_1-t_2)\delta V_{hk\tau}(t_2) \\
  \delta P_{ij\sigma} (t_1) =  \int dt' \chi_{ij\sigma,hk\tau}(t_1-t_2)\delta V^{ext}_{hk\tau}(t_2) 
\end{eqnarray}
where the KS response function is written over this basis set:
\begin{eqnarray}
 \chi^{KS}_{ij\sigma,hk\tau}(\omega) &=& \sum_{I} 
               \frac{\langle \Psi^{KS}_0|{\widehat{a}}_{j\sigma}^{\dagger}{\widehat{a}}_{i\sigma}| \Psi^{KS}_I\rangle
                     \langle \Psi^{KS}_I|{\widehat{a}}_{h\tau}^{\dagger}{\widehat{a}}_{k\tau}| \Psi^{KS}_0\rangle}
                    {\omega -(E_I-E_0)+i\eta}\nonumber\\
           &-&\frac{\langle \Psi^{KS}_0|{\widehat{a}}_{h\tau}^{\dagger}{\widehat{a}}_{k\tau}| \Psi^{KS}_I\rangle
                    \langle \Psi^{KS}_I|{\widehat{a}}_{j\sigma}^{\dagger}{\widehat{a}}_{i\sigma}| \Psi^{KS}_0\rangle}
                    {\omega +(E_I-E_0)+i\eta}
\end{eqnarray}
In case of independent particles and applying the $\widehat{a}$ and $a^\dagger$ operators we obtain the reduced 
expression for the response function:
\begin{equation}
 \chi^{KS}_{ij\sigma,hk\tau}(\omega) = \delta_{\sigma \tau}\delta_{j k}\delta_{i h} \frac{f_{j\sigma}-f_{i\sigma}}{\omega-(\varepsilon_i^{KS}-\varepsilon_j^{KS})}
\end{equation} 
Hence, in the configuration space the TDDFT equation held the form (see Ref.~\cite{casida,casida_book}):
\begin{equation}
 \sum_{kl\tau}^{f_{k\tau}-f_{h\tau}\ne0} \left[\delta_{\sigma \tau}\delta_{j h}\delta_{i k} 
   \frac{\omega -(\varepsilon_{h\tau} - \varepsilon_{k\tau}) }{ f_{k\tau} - f_{h\tau} } - K_{ij\sigma,hk\tau}(\omega)
   \right]\delta P_{hk\tau}(\omega) = \delta V^{ext}_{ij\sigma}(\omega) 
\end{equation}
This equation can be reorganized and written as a matrix equation 
(see Ref.~\cite{casida,casida_book} for mathematical details) exploiting the hermitian 
nature of the kernel $K_{ij\sigma,hk\tau}$. 
Moreover assuming that the KS orbitals and 
the kernel are real quantities (infinite lifetime of the excitations) and after some algebra 
we can write:
$$
\left[\left(
 \begin{array}{cc} 
  A(\omega) + B(\omega)  &         0             \\
  0                      &  A(\omega) - B(\omega)\end{array}\right) 
 - \omega \left(  
\begin{array}{cc} 0 &  -C \\ 
                 -C  & 0 
 \end{array}\right) 
\right]  
$$
$$
\left( 
\begin{array}{c} 
  \textrm{Re}\, \delta P (\omega)  \\ 
  -i\textrm{Im} \, \delta P (\omega)  
 \end{array}\right) = 
\left(\begin{array}{c} 
  \textrm{Re} \,\delta V^{ext}(\omega)  \\ 
  -i\textrm{Im} \, \delta V^{ext} (\omega)  
 \end{array}
\right) 
$$
where:
\begin{eqnarray}
  A_{ij\sigma,hk\tau}(\omega) &=& \delta_{\sigma\tau}\delta_{ih}\delta_{jk} 
        \frac{\varepsilon_{h\tau}-\varepsilon_{k\tau}}{f_{hk}-f_{k\tau}} -K_{ij\sigma,hk\tau}(\omega) \\
  B_{ij\sigma,hk\tau}(\omega) &=& -K_{ij\sigma,kh\tau}(\omega) \\ 
  C_{ij\sigma,hk\tau}(\omega) &=& \frac{\delta_{\sigma\tau}\delta_{ih}\delta_{jk}}{f_{h\tau}-f_{k\tau}}
\end{eqnarray}
and the polarizability as:
\begin{equation}
 \alpha_{xz}(\omega) = 2\vec{x}^{\dagger}S^{-1/2}(\Omega(\omega)-\omega^2)^{-1}S^{-\/2}\vec{z}
 \label{eq:alpha2}
\end{equation}
where $S(\omega) = C(A-B)^{-1}C$ and $\Omega(\omega) = S^{-1/2}(A+B)S^{-1/2}$.
Finally, comparing this last expression~(\ref{eq:alpha2}) with eq.~(\ref{eq:alpha1}) 
it is possible to write the eigenvalue equation:
\begin{equation}
 \Omega(\omega_I)F_I =\omega_I^2 F_I
 \label{eq:casida}
\end{equation}
where:
\begin{eqnarray}
 \Omega(\omega_I)_{ij\sigma,hk\tau}(\omega) &=& \delta_{i,h}\delta_{j,k}\delta_{\sigma,\tau}(\varepsilon_{k,\sigma} 
         - \varepsilon_{h\sigma})^2 +\\
           && + 2\sqrt{(f_{i\sigma}-f_{j\sigma})(\varepsilon_{j\sigma} - \varepsilon_{i\sigma})} K_{ij\sigma,hk\tau} 
                \sqrt{(f_{h\tau}-f_{k\tau})(\varepsilon_{k\tau} - \varepsilon_{h\tau})} \nonumber
\end{eqnarray}
Eq.~(\ref{eq:casida}) is called Casida equation (see Ref.~\cite{casida,casida_book}) and allows 
to map the problem of finding the excitation spectrum of a system too an eigenvalue problem
for the matrix $\Omega(\omega)$.

  \section{Closed and open shell systems: the problem of the correct counting of excitations}
   We divide now the discussion to the case of closed and open shell systems.
In the first case the approach presented in the previous section predicts,
in principle, the correct excitation energies.
However, in the second case some problem limits 
its straightforward applicability. 
In the next paragraphs we present the problem with an example.
\subsection{Closed shell systems}
Let us start with the case of the closed shell systems,
i.e. systems having the same wavefunction for the
two spin channels.
In this case it is possible to diagonalize the $\Omega$ matrix, 
presented in the previous section, exploiting the symmetries
resulting from the adiabatic approximation.
\begin{figure}[h!]
 \begin{center}
 \includegraphics[width=8.5cm,clip,]{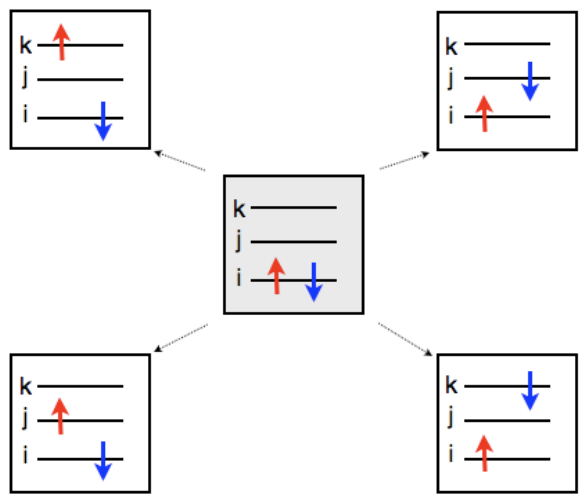}
\end{center}
\caption[Excitations for a closed shell system]{Schematic representation of the possible
excitations generated in a 3 levels system with 2 electrons with opposite
spins.
 \label{fig:closed-shell}}
\end{figure}
Let us consider a three levels system consisting of two electrons (see Fig.~\ref{fig:closed-shell}),
the ground state is characterized by two coupled electrons 
both occuping the $i$ level, two additional unoccupied levels 
$j$ and $k$ are available to the excitations.
A spin--unpolarized ground state is expected and the $\Omega$ matrix has 
important symmetry properties, so that its eigenvectors
will be either symmetric or anti-symmetric with rispect to spin flip.
Assuming the adiabatic approximation, the $\Omega$ matrix, can be written as:
$$
\left( \begin{array}{cc}
\Omega_{\uparrow\uparrow} & \Omega_{\uparrow\downarrow} \\
\Omega_{\uparrow\downarrow} & \Omega_{\downarrow\downarrow}
\end{array} \right)
\label{eq:omega}
$$
such that :
$$
\Omega_{\uparrow\uparrow} = \Omega_{\downarrow\downarrow} \quad \textrm{and} \quad \Omega_{\uparrow\downarrow} = \Omega_{\downarrow\uparrow}
$$
The eq.~(\ref{eq:casida}) is decoupled in:
\begin{eqnarray}
  &&  F_{I\uparrow}=  F_{I\downarrow} \quad, \quad (\Omega_{\uparrow\uparrow}+\Omega_{\uparrow\downarrow}) F_{I\uparrow}   = \omega_I^2 F_{I\uparrow}
  \label{eq:singlet}\\
  &&  F_{I\uparrow} = -  F_{I\downarrow}\quad, \quad  (\Omega_{\uparrow\uparrow}-\Omega_{\uparrow\downarrow}) F_{I\uparrow} = \omega_I^2 F_{I\uparrow}
  \label{eq:triplet}
\end{eqnarray}
In case of eq.~(\ref{eq:singlet}) for a given electron--hole pair,
both spin channels interfere constructively,
TD total charge changes, but not the TD magnetization
(Spin-singlet excited state). 
Conversely in case of eq.~(\ref{eq:triplet})
both spin channels interfere destructively,
no TD total charge change, TD magnetization change
(Spin-triplet excited state).
In this case the resulting eigenvectors are written in the form: 
$$
\left( \begin{array}{c}
1 \\
1 \end{array} \right)
\quad , \quad 
 \left( \begin{array}{c}
1 \\
-1 \end{array} \right)
$$
Hence, summarizing, it is possible to group the results in two kind of 
excitations:
\begin{eqnarray}
 \phi^{1/2}_0  = \frac{1}{\sqrt{2}}\left(|j\uparrow i\downarrow \rangle +|i\uparrow j\downarrow \rangle \right)  \quad  \quad  \quad \quad\textrm{singlet}
\end{eqnarray}
and 
\begin{eqnarray}
 \phi^{1}_{-1} &=& |i\uparrow j\downarrow \rangle \\
 \phi^{1}_0    &=& \frac{1}{\sqrt{2}}\left(|j\uparrow i\downarrow \rangle - |i\uparrow j\downarrow \rangle \right) \quad \quad  \quad   \quad \textrm{triplet}\\
 \phi^{1}_1    &=& |i\downarrow j\downarrow \rangle
\end{eqnarray}
This states are simultaneous eigenvectors of H$^\textrm{KS}$, S$^2$ and S$_z$.
\subsection{The problem of the open shell systems}
\begin{figure}[h!]
 \begin{center}
  \includegraphics[width=8.5cm,clip]{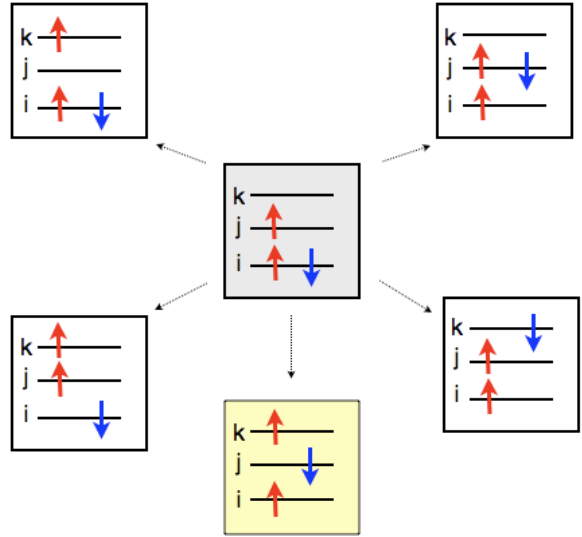}
\end{center}
\caption[Excitations for an open shell system]{
  Schematic representation of the possible
  excitations generated in a 3 levels system 
  consisting of 3 electrons. The yellow square represents the 
  double excitation missed by the casida's framework. \label{fig:open-shell}
  }
\end{figure}

The simplest case of a system with spin--polarized ground state consists
of three electrons that can be arranged within three levels.
Fig.~(\ref{fig:open-shell}) describe the ground state configuration and the possible 
spin--collinear excitations.
In this case the $\Omega$ matrix does not have the same symmetries
presented in the previous section. This time it contains the 
elements related to the excitations $jk\uparrow$, $ij\downarrow$, $ik\uparrow$ and $ik\downarrow$
resulting with $4\times4$ dimension. Moreover any symmetry does not 
allow to divide by blocks the whole $\Omega$. However the selection rules
allow to group the excitations in two sets: conserving the spin $\Delta S =0$ and 
those for which $\Delta S = 1$. \\
\indent Diagonalizing H$^{\textrm{KS}}$, S$^2$ and S$_z$ simultaneously we
obtain the following doublet for the ground state:
$$
 \phi^{2}_0  = |\bar{i}ij\rangle 
$$
and the excited states (doublet):
\begin{eqnarray}
 \phi^{2}_{1/2}  &=& | i \bar{i} k \rangle \\
 \phi^{2}_{1/2}  &=& | i \bar{j} j \rangle\\
 \phi^{2}_{1/2}  &=&  \frac{1}{\sqrt{2}} \left(|\bar{i}jk \rangle +|ij\bar{k}\rangle \right)  \\
 \phi^{2}_{1/2}  &=&  \frac{1}{\sqrt{6}} \left(|\bar{i}jk \rangle +|ij\bar{k}\rangle -2 |i\bar{j}k\rangle \right)
\end{eqnarray}
and the quadruplet:
\begin{eqnarray} 
 \phi^{4}_{1/2} = \frac{1}{\sqrt{3}} \left(|\bar{i} jk \rangle + |ij\bar{k} \rangle + | i\bar{j}k \rangle \right)
\end{eqnarray}
where we adopted the notation $\bar{i}$ to indicate the $\downarrow$ spin occupation of the $i$ state.
The double excitation $|i\bar{j}k\rangle$, represented by the yellow square in fig.~\ref{fig:open-shell},
is present because the eigenvectors
must be simultaneous solutions of the eigenvalue problem for H$^{\textrm{KS}}$ and S$^2$.
The excitations of this system present states with three half occupied levels 
(see Fig.~\ref{fig:open-shell}) that are mixed by the S$^2$ operator
with similar states with inverted spins. \\
It is worth to note that this kind of excitations are missing when 
we attempt to build the eigenstates of H$^{\textrm{KS}}$, S$^2$ and S$_z$
starting from the single particle excitations. In this case we obtain three doublet:
\begin{eqnarray}
 \phi^{2}_{1/2}  &=& | i \bar{i} k \rangle \\
 \phi^{2}_{1/2}  &=& | i \bar{j} j \rangle \\
 \phi^{2}_{1/2}  &=&  \frac{1}{\sqrt{2}} \left(|\bar{i}jk \rangle +|ij\bar{k}\rangle \right) 
\end{eqnarray}
and a triplet:
\begin{eqnarray} 
 \phi^{3}_{1/2} = \frac{1}{\sqrt{2}} \left(|\bar{i} jk \rangle + |ij\bar{k} \rangle \right)
\end{eqnarray}
As the theory is exact and the sole quantity requiring an approximation is the kernel
the lack of the excitations description is due to the adiabatic assumption.
The research of an improved kernel able to correctly describe the missed excitations is
beyond the scope of this thesis. \\
However, in the following paragraphs an application to the case of the 
calculations of the excitation energies of the 
BeH molecule is presented.
 
  \section{A simple open shell molecule: BeH}
\begin{figure}
 \begin{center}
  \includegraphics[width=8.5cm,clip=]{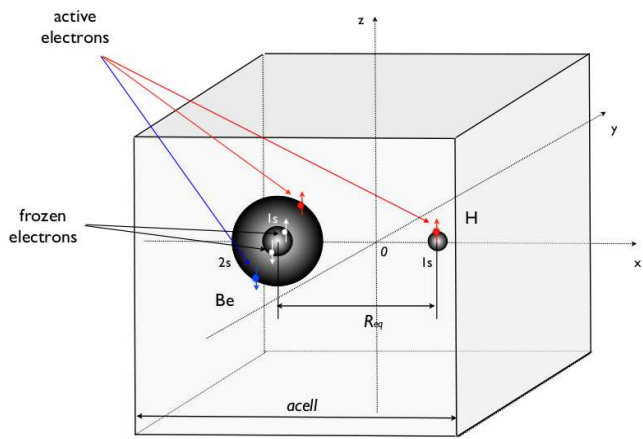}
 \end{center}
 \caption[BeH molecule]{ 
           Schematic view of the atoms in the cell 
           used to represent the BeH molecule (acell is not on scale). 
           Red and blue circles represent active electrons, 
           white circles represent Beryllium 1$s$
           electrons. The effect of these core electrons is described by 
           a suitable pseudopotential (see text).
  \label{fig:BeH-geo} 
  }
\end{figure}

   A BeH molecule consists of five electrons, four due to 
Beryllium and one to Hydrogen. 
Configuration of the Beryllium atom ($^4$Be: $1s^2 2s^2$) suggests
his divalent nature, in fact usually the two electrons in the $1s$ level 
are not involved in chemical bonds. 
Therefore we describe the joint effect of the core and the 
inner electrons of Beryllium with a suitable pseudopotential.
In particular we used a Troullier--Martins  LDA pseudopotential
available in the on line repository of the ABINIT code~\cite{ABINIT,abinit2}.
Following this recipe, only three valence electrons
of the molecule are involved in DFT calculations and are expected to determine the 
main properties of BeH, including the spin polarization of its ground state.\\
\indent Bond length between Beryllium and Hydrogen is assumed to
be R$_{eq}=1.3426$\AA\ (see Ref.~\cite{casida_jms_v527_p229_y2000})
and the cubic supercell is centred in the middle point of
the line joining the centers of the two atoms.
A schematic representation of the molecule is depicted in 
Figure~\ref{fig:BeH-geo} where the active electrons (blue and red circles) 
are underlined and distinguished respect to the spectator electrons
(white circles).
Spin variable is also shown in order to evidence that spin 
is not compensated.
In the following paragraphs we will show our results about
spin resolved calculations performed with the ABINIT package~\cite{ABINIT}.

  \section{Energy levels and density of states}
\begin{figure}
 \begin{center}
  \includegraphics[angle=0,width=8.5cm,clip=]{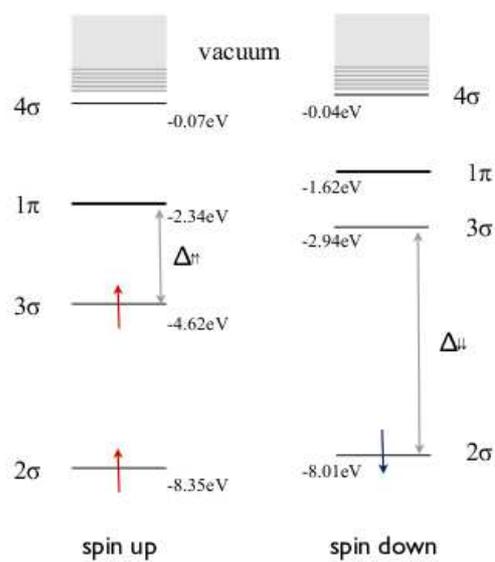}
 \caption[BeH: spin resolved KS energy levels]{ 
          Spin resolved Kohn-Sham energy levels for the BeH molecule
          calculated within DFT--LSDA. Energy levels
          are calculated for a 55~Bohr cubic unit supercell with a 
          plane--wave cutoff energy E$_{cut}$=18~Ha and 20 Kohn--Sham states..
  \label{fig:levels-BeH}
         }
 \end{center}
\end{figure}

   In order to obtain converged energy levels and 
the spin resolved density of states of BeH, we 
performed total energy minimization within
DFT in local spin density approximation. \\
\indent We obtained the theoretical energy levels 
(see Fig.~\ref{fig:levels-BeH}) imposing a convergence 
tolerance of $10^{15}~$Ha$^2$ for the largest squared residual defined by:
$  \langle nk|(H-E)^2|nk \rangle, \,\, \textrm{where} \,\,\,\, E = \langle nk|H|nk\rangle$.
In order to establish reliable results we performed
convergence tests on the supercell size
(see Sec.~\ref{sec:BeH-convergence}).
Here we comment the results summarized in Table~\ref{tbl:levels}.
We compared our results with theoretical calculations
reported in Ref.~\cite{casida_jms_v527_p229_y2000} and  
experiments from Ref.~\cite{herzberg_book}.
We obtained a satisfactory agreement of the
Kohn and Sham energy of the highest occupied orbital
with respect to calculations performed in Ref.~\cite{casida_jms_v527_p229_y2000} 
with localized basis orbital.\\
\indent Even if there is not analogous of the Koopmans theorem, providing a physical 
interpretation of the Hartree Fock eigenenergies, for the KS
eigenvalues, it is possible to interpret the highest occupied molecular orbital (HOMO) 
energy calculated within DFT as the ionization energy of the system 
(see ref.~\cite{olof}). \\
However in the present case of the BeH molecule, if we assume that the E$_{3\sigma_{\uparrow}}$ 
energy referred to the vacuum should be the ionization energy of the system, 
we face with a discrepancy with respect to the experimental value reported in Ref.~\cite{herzberg_book}.
In fact, the experimental value of 8.21~eV disagrees with 
$-\epsilon^{\textrm{HOMO}}=4.62$~eV, calculated within DFT-LSDA, 
and also with 4.60~eV obtained in Ref.~\cite{casida_jms_v527_p229_y2000}.
\begin{figure}
 \begin{center}
  \includegraphics[angle=0,width=8.5cm,clip=]{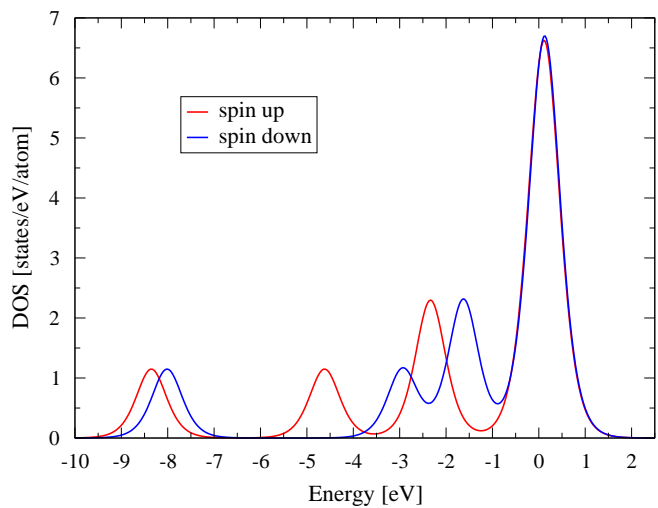}
 \end{center}
 \caption[BeH: spin resolved DOS]{ 
          Spin resolved density of states for BeH
          calculated within DFT--LSDA with the code ABINIT~\cite{ABINIT}.
          Red and blue lines represent the spin up and down components. 
          A smearing of $0.008$~Ha is applied and a 55~Bohr length 
          unit cell is considered.
  \label{fig:dos-BeH}
         }
\end{figure}
This discrepancy is due to the inadequacy of the 
LDA exchange and correlation potential
(see Ref.~\cite{myrta,myrta_jcp_y2002})
that does not have an asymptotic $-1/r$ required behaviour.
This fact lead to the lack in the description of the levels staying close 
to the vacuum region and the excitation energies 
of the molecule involving these levels. \\
\indent Figure~\ref{fig:dos-BeH} represents the distribution of
the density of states obtained after convergence tests,
discussed in detail in Sec.~\ref{sec:BeH-convergence} and, 
 in Figure~\ref{fig:dos-BeH}, we report the
density of states (DOS) calculated within the same approximations.
When spin symmetry is broken, BeH molecule shows an
asymmetric distribution of DOS (see Fig.~\ref{fig:dos-BeH})
with respect to the two spin channels.
We considered a cutoff energy E$_{\textrm{cutoff}}=18$~Ha and a cell size
a$_0= 55$~Ha as converged values.
The first three peaks, centred at $-8.35$~eV, $-8.01$~eV and $-4.62$~eV
represent the occupied states, also depicted in
Fig.~\ref{fig:levels-BeH}, moreover $3\sigma_{\downarrow}$ 
and $1\pi$ orbitals are shown, in particular 
the height of peaks related to $1\pi$ orbitals is doubled
with respect to the $\sigma$ orbitals bacause of their
degenerancy giving contributions to the DOS peak coming 
from a double number of states.
A smearing of 0.008~Ha is applied to both distributions.
\subsection{Convergence issues} \label{sec:BeH-convergence}
Convergence tests have been performed respect to two
quantities: (i) the cutoff energy $E_{\textrm{cut}}$ determining the number of 
plane wave of the basis set and (ii) the supercell dimension $acell$.

In order to determine a suitable value for $E_{\textrm{cut}}$ 
we checked the values of the KS-eigenvalues: 
E$_{2\sigma_{\uparrow}}$, E$_{2\sigma_{\downarrow}}$ 
and E$_{3\sigma_{\uparrow}}$, corresponding to the occupied
states (see Figure~\ref{fig:levels-BeH} and Table~\ref{tbl:levels}).
Since we used a Troullier--Martin (TM) pseudopotential for Be, 
we found a resonable converged value of E$_{cut}=18$Ha (corresponding 
to 36Ry) see Table~\ref{tbl:levels} and Fig.~\ref{fig:levels-BeH}.
However we verified that E$_{cut}=10$~Ha can be considered a converged value
in order to calculate the first excited states of the molecule,
in fact, the HOMO--LUMO distance for each spin channel 
does not change significantly.
 Table~\ref{tbl:levels} summarize the converged values for 
$\Delta_{\uparrow\uparrow}=1\pi_{\uparrow}-3\sigma_{\uparrow}$ and
$\Delta_{\downarrow\downarrow}=3\sigma_{\downarrow}-2\sigma_{\downarrow}$,
that distances are also evidenced in Fig.~\ref{fig:levels-BeH}.
\begin{figure}
\begin{center}
  \includegraphics[width=8.5cm, clip=]{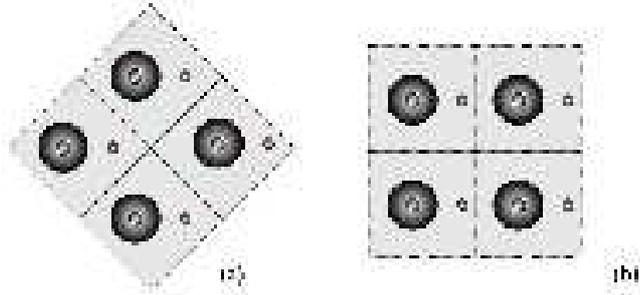}
\end{center}
 \caption[BeH: supercell]{ 
          Top view of the BeH molecule (black circles) 
          inside a cubic cell, grey zone represents the vacuum
          region around the molecule (cell is not on scale).
          As an exemple we report a comparison between two possible
          choices of the molecule placement inside the cell: distance
          between atoms of the replicas is larger in case (a) than (b).
         \label{fig:cell-shape}
         }
\end{figure}
Moreover, a set of tests with respect to the cell size have been performed
in order to find the best compromise between accuracy and
practical feasibility of calculations.
In fact, a large vacuum region around the molecule is required 
in order to obtain the correct behaviour of continuum states above
the vacuum level (see Fig.~\ref{fig:convergence}) 
and to avoid spurious interactions between replicas,
however this request can increase significantly the 
computational workload. \\
It is worth to mention that a good choice of the cell shape 
(see Fig.~\ref{fig:cell-shape}) can slightly optimize the 
numerical convergence.
We considered the molecule placed in the middle of a cube 
of length $a_0$ ($acell$) with the straight line joining
Beryllium and Hydrogen atoms parallel to
one face of the cube (Fig.~\ref{fig:cell-shape}(b)).

We performed several tests increasing $a_0$ for a
fixed cutoff energy and number of bands.
In Fig.~\ref{fig:convergence} a summary of the
results is depicted: in the actual range of $a_0$ considered,
occupied states slightly change because of the
dimension of the supercell, on the contrary the unoccupied
levels are strongly influenced and require a large
cell size ($a_0>50$~Bohr). 
This fact can be important if we want to calcuate
excited energies involving higher energy levels.
In Fig.~\ref{fig:levels-BeH} we can distinguish discrete unoccupied 
states ($1\pi,3\sigma,4\sigma$) and a thick serie of 
states above the vacuum level (grey region).
In fact, increasing the cell size, 
the upper levels become closer and closer 
with each other (see Fig.~\ref{fig:convergence}) 
and a continuum of states, in the limit case of an 
infinite supercell.
\begin{table}
 \begin{center}
  \begin{tabular}{ccccc}
\hline
     Level & present work & Ref.~\cite{sangalli_thesis_y2007} & Ref.~\cite{casida_jms_v527_p229_y2000}&Exp. Ref.~\cite{herzberg_book} \\
\hline
     $2\sigma_{\uparrow}  $ & -8.35  & -8.35  & - & - \\
     $2\sigma_{\downarrow}$ & -8.01  & -7.99  & - & -\\
     $3\sigma_{\uparrow}  $ & -4.62  & -4.63  & -4.60 & 8.21 \\
\hline
     HOMO--LUMO & present work & Ref~\cite{sangalli_thesis_y2007} & & \\
\hline
     $ \Delta_{\uparrow\uparrow}$    & 2.28  & 2.29  & &\\
     $ \Delta_{\downarrow\downarrow}$& 5.07  & 5.16  & &\\
\hline
 \end{tabular}
 \end{center}
\caption[Energy levels of BeH]{
         Theoretical spin resolved energy levels and HOMO-LUMO distance 
         for BeH molecule calculated within DFT-LDA using the code 
         ABINIT~\cite{ABINIT}.
         A 55~Bohr cell, $E_{\textrm{cut}}=$18~Ha and 20 bands
         have been used and considered converged values.
         \label{tbl:levels}
         }
\end{table}
\begin{figure}
\begin{center}
  \includegraphics[width=8.5cm, clip=]{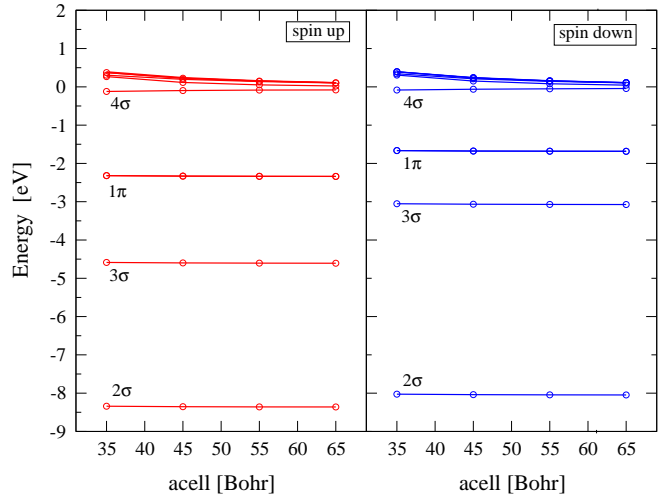} 
\end{center}
 \caption[BeH: convergence]{
         Energies of the molecular orbitals of
         BeH as a function of the cell size. 
         Spin up (right panel, red circles) and down (left panel,
         blue circle) are distinguished. 
         Unoccupied states are strongly influenced to the 
         cell size and require a$_0 > 50$~Bohr. 
         Tests are performed with a cutoff energy of 10~Ha.
         \label{fig:convergence}
         }
\end{figure}
In conclusion we assumed that a cell size of 55~Bohr is 
required in order to correctly calculate the three lowest 
eigenvalues. 
On the contrary, for excited states calculations, involving 
higher energy levels in the vacuum, a larger cell 
is recommended.
Our conclusions are in agreement with a slightly different treatment
discussed in Ref.~\cite{sangalli_thesis_y2007}.

  \section{Excitation energies}\label{sec:exc}
   %Discrepanza rispetto al valore sperimentale :
%4.63eV (Casida 4.60eV) vs 8.21eV. 
%ERRORE DOVUTO ALL'INADEGUATEZZA DEL POTENZIALE DI SCAMBIO E CORRELAZIONE LDA
%==> ECCITAZIONI MANCANTI.
%3) Calcolo delle Eccitazioni con ABINIT TDLSDA
In this section we report results concerning excited state
calculations within the TDLSDA formalism.
For that we used a recent implementation of the method in 
the ABINIT software~\cite{ABINIT,abinit2,sangalli_thesis_y2007}.
In Table~\ref{tbl:exc} we summarize the results about 
the main excitation energies calculated. 
We did not consider excitations involving continuum states because
they are extremely influenced by the countour conditions.
However, in order to have more reliable results we used a larger 
supercell with respect to the case of ground state calculations, 
in particular a 70$\times$70$\times$70~Bohr cubic cell is considered 
with a cutoff energy of 10~Ha and 50 bands.

Even if the ionization energy disagrees with respect to
experimental data making a shorter distance between occupied states 
and vacuum level, our first calculated excitation 
energy $\Pi$ connecting the spin-collinear states 
$3\sigma \rightarrow 1\pi$ is comparable 
with experimental value reported in Ref.~\cite{colin_cjp_v53_p2142_y1975}.
The TDLSDA theoretical value ($2.39$~eV) understimates the experimental
measure ($2.48$~eV) but improves the DFT-LSDA HOMO-LUMO distance ($2.27$~eV)
calculated as a bare difference between Kohn-Sham eigenvalues 
(see Table~\ref{tbl:levels} and~\ref{tbl:exc}).

However we observed that every excitation involving states
higher than the $4\sigma$ level are strongly influence by the cell size.
For this reason we evidenced with bold carachter in Table~\ref{tbl:exc}
the more reliable results, i.e. excitations 
excluding states above $4\sigma$.
\begin{table}
 \begin{center}
  \begin{tabular}{c c c c c c c }
\hline
  Excitation & elec. conf. & this work & Ref.~\cite{casida_jms_v527_p229_y2000} & Ref.~\cite{ipatov_preprint} & Exp.$^a$& Exp.$^b$\\
%~\cite{colin_cjp_v53_p2142_y1975} \\
\hline
  $\Pi   : 3\sigma_\uparrow \rightarrow 1\pi_\uparrow        $               & $2\sigma^2 1\pi^1$               & \bf{2.37} & 2.391  & 2.2479 & 2.56  & 2.48  \\%2.395
  $\Sigma: 3\sigma_\uparrow \rightarrow 4\sigma_\uparrow     $               & $2\sigma^2 4\pi^1$               & 4.51      & 4.593  & 4.5103 & 5.51  &   -   \\ %54.519 
\hline                                                                                                                                                        
  $\Sigma: 3\sigma_\uparrow \rightarrow 5\sigma_\uparrow     $               & $2\sigma^2 5\sigma^1 $           & 4.62      &  -     & 4.6300 & 5.61 &   -   \\%4.693
  $\Pi: 3\sigma_\uparrow \rightarrow 2\pi_\uparrow           $               & $2\sigma^2 2\pi^1    $           & 4.70      &  -     & 4.7047 & 6.31 & 6.317 \\%4.798
  $\Sigma: 2\sigma_\downarrow \rightarrow 3\sigma_\downarrow $               & $2\sigma^1 3\sigma^2 $           & \bf{5.41} &  5.129 & 4.8049 &  -   &   -   \\% 5.536  
  $\Pi: 2\sigma_{\uparrow\downarrow} \rightarrow 1\pi_{\uparrow\downarrow} $ & $2\sigma^1 3\sigma^1 1\pi^1$     & 5.66      &  -     & 5.1685 &  -   &   -   \\%5.654 
  $\Pi: 2\sigma_{\uparrow\downarrow} \rightarrow 1\pi_{\uparrow\downarrow} $ & $2\sigma^1 3\sigma^1 1\pi^1$     & \bf{7.15} &  -     &    -   &  -   & 7.46  \\%7.145 
  $\Sigma: 2\sigma_\downarrow \rightarrow 4\sigma_\downarrow $               & $2\sigma^1 3\sigma^1 4\sigma^1$  & 8.01      &  -     &    -   &  -   &   -   \\%7.962
\hline
  \end{tabular}
 \end{center}
\caption[Excitation energies of BeH]{
         Theoretical excitation energies for BeH molecule calculated within
         the TDDFT framework and ALDA approximation using the code ABINIT~\cite{ABINIT}.
         Energy values are reported in eV. Converged results are obtained
         with a cutoff energy of E$_{cut}$=10~Ha, 70$\times$70$\times$70
         Bohr and 50 bands. We compare our results with calculated values 
         within two TDDFT implementations in the 
         code DeMon2K (see Ref.~\cite{casida_jms_v527_p229_y2000,ipatov_preprint}).
         Exp.$^a$ are reproduced from Tab. I of
         Ref.~\cite{petsalakis_jcp_v97_p7623_y1992} and Exp.$^b$
         from Ref.~\cite{colin_cjp_v53_p2142_y1975}.
         \label{tbl:exc}
         }
\end{table}
Moreover we underline that the $\Pi$ excitations,
involving three half occupied states
($2\sigma^1 3\sigma^1 1\pi^1$),
appear at two excitation energies (5.66~eV and 7.15~eV)
and with the same orbital configuration.
These $\Pi$-excitations are distinguished by their total spin, i.e.
 S$=1/2$ in one case and higer total spin in the other. 
Thanks to the Hund's rule we can identify the excitation with S$=1/2$ with 
higer energy ($7.15$~eV), and this lead us to discard the 5.66~eV 
excitation.
A different argument leads Casida\ea\ to reject the same 
excitation identifying the same excitation we discarded as 
\emph{spin contaminated}.
Comparison with experiments available in 
Ref.~\cite{colin_cjp_v53_p2142_y1975,petsalakis_jcp_v97_p7623_y1992}
are satisfactory, in particular with respect to the 
case of $\Pi : 3\sigma_\uparrow \rightarrow 1\pi_\uparrow $
and $\Pi: 2\sigma_{\uparrow\downarrow} \rightarrow 1\pi_{\uparrow\downarrow}$ 
excitations.
Discrepancy about $\Pi: 3\sigma_\uparrow \rightarrow 2\pi_\uparrow $ excitation 
is due to the problem previously underlined: in fact $2\pi$ states are higer 
than $4\sigma$ level. \\
Comparison with results obtained in Ref.~\cite{casida_jms_v527_p229_y2000,ipatov_preprint}
are also aligned with our conclusions.

  \section{Conclusions}
   We studied the BeH molecule as the simplest exemple of open shell system.
In our pseudopotential approach only three valence electrons
are involved in the eigenvalue problem, for this reason
total spin is not compensate and the electronic ground state 
is spin polarized.
 
Spin resolved density of states and energy levels have been 
computed within DFT--LDA approach. 
Convergence tests with respect to the supercell size have shown the 
sensitivity of the unoccupied states above the vacuum level
to the cell dimension.
Moreover we found a discrepancy between the
energy of the HOMO orbital calcualted (in agreement with 
others theoretical works, see ref.~\cite{casida_jms_v527_p229_y2000}
and~\cite{ipatov_preprint}) and the ionization energy measured. 
This is due to the well known inadequacy of the
asimptotic behaviour of the LDA exchange and correlations potential.

In conclusion we calculated the first excitation energies 
of the molecule using TDDFT within the adiabatic approximation.
We compared our reliable results with
experimental data finding a satisfactory agreement even 
if a part of the excitations can not be reproduced by 
frequency--independent kernels such as the ALDA one.

\chapter{A paradigmatic case for semicore and spin-polarization effects \\
          in electronic spectra of solids: \\bulk iron} \label{ch:a-iron}
  
Semicore states of transition metals such as iron are outside the reach of
``standard'' pseudopotential of DFT which, even when \emph{non linear core
corrections} are adopted, include them in the frozen atomic core.
In this work we present an analysis of several pseudopotentials for iron generated
in the Troullier Martins and Hamann scheme assuming
both local density and generalized gradient approximations (LDA, GGA)
for the exchange--correlation functional and considering core-valence
partitions with and without including  semicore orbitals among valence states.
Non linear core corrections are considered and the pseudopotential
transferability has been checked.
We calculate structural and electronic properties of the $\alpha$ phase of iron
and we present a comparison between calculated optical conductivity and
experimental data.

  \section{Pseudopotentials for iron}
   Pseudopotentials (PP) are a well-established tool in
\emph{ab initio} structure calculations of solids.
A review of the topic can be found in literature
ranging from the most influential works~\cite{phillips_pr_v116_y1959_p287, HSC79,K80,BHS82,
kleinman_prl_v48_y1982_p1425,vanderbildt_prb_v41_y1990_p7892,
blochl_prb_v41_y1990_p5414,gonze_prb_v44_y1991_p8503}
to other important but less fundamental papers~\cite{kleinman_prb_v21_y1980_p2630,louie_prb_v26_y1982_p1738,
troullier_prb_v43_y1991_p1993,hamann_prb_v40_y1989_p2980}.
Important advantages of the pseudopotential approach can be summarized
in the following two points:
first, by replacing the atom by a pseudoatom, the number of orbitals
which have to be calculated is reduced, and,
second, using plane waves, the size of the basis set can be substantially reduced,
because the pseudo-wavefunctions are smoother than their all-electron
counterparts.

\begin{figure}[ht!]
 \begin{center}
 \includegraphics[width=8.5cm,angle=0,clip=]{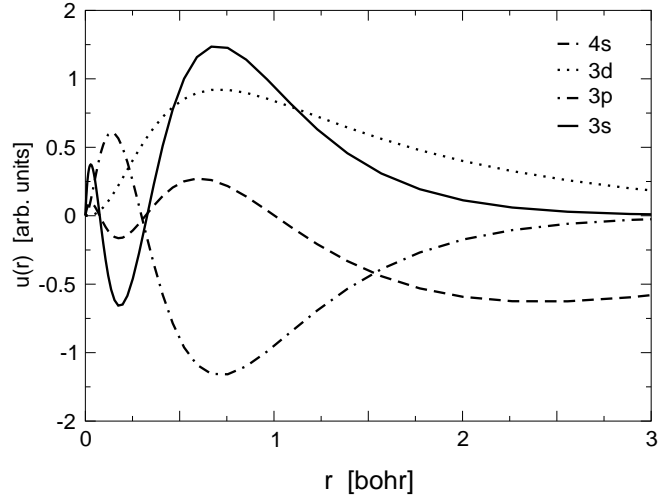}
  \caption[Iron: wave functions]{
           Wave functions of $3s3p3d$ and $4s$ level of iron can be distinguished
           by their different localization. The not-negligible overlap between
           $3d$ and $3s$ leads to an ambiguity in the definition of core and valence states
            and to the need of corrections to the frozen core approximation.
             }
  \end{center}
 \label{fig:core-valence}
\end{figure}

However in all cases where an overlap between valence and core wave functions
exists, the \emph{frozen core} approximation underlying the construction of
all pseudopotentials is not well satisfied.
One way to overcome this problem is the inclusion of a core correction considering
the non linear contribution of the core charge to the
exchange-correlation potential (NLCC).
Another more straightforward solution is
the explicit inclusion of the semicore electrons into the valence shell. \\
In this work we considered both these approaches in order to build
pseudopotentials for the specific case of iron.
Such pseudopotentials are used to calculate the structural and electronic
properties of the $\alpha$ phase of bulk iron. In conclusion we report our results
for the optical conductivity compared with experimental data.
The $3d$ wavefunctions of iron are strongly localized and show a significant overlap with
the $3s3p$ orbitals, although the latter are much lower
in energy.
Looking at Fig.~\ref{fig:core-valence}, it is evident that
the separation of the electronic system into well-isolated core and
valence shells is ambiguous, because of the large overlap between $3s$
and $3d$ states.\\
For this reason, we considered two electronic configurations:
one with 8 ($3d^64s^2$ states) and the other with 16 ($3s^23p^63d^64s^2$ states) valence electrons.
We used two schemes for pseudopotential generation, referring to
Hamann~\cite{hamann_prb_v40_y1989_p2980} and
Troullier--Martins~\cite{troullier_prb_v43_y1991_p1993} works respectively.
In the former scheme a fixed cutoff radius r$_{\mathrm{cl}}$ by an exponential function
$\exp[-(r/r_{\mathrm{cl}})\lambda]$, in the latter scheme wavefunctions 
are built with a parametric form $r^{\mathrm{l}}\exp[\rho(r)]$
below r$_{\mathrm{cl}}$, where $\rho(r)$ is a polynomial of order six in r$^{2}$. \\
In the case of a standard pseudoatom with 8 valence electrons we verified
the importance of including non linear core corrections (NLCC) to correct
GGA pseudopotentials presenting fake wiggles in the r $\to$ 0 limit due to the
gradient dependence of the exchange-correlation approximation. \\
In the case of a 16 electrons pseudoatom we could
not generate a transferable pseudopotential in
generalized gradient approximation, neither in
the Troullier-Martins nor in the Hamann scheme.
The only pseudopotential with good transferability was
generated within the local density approximation following 
the Hamann scheme. \\
\newline
\newline
 \begin{figure}[h!]
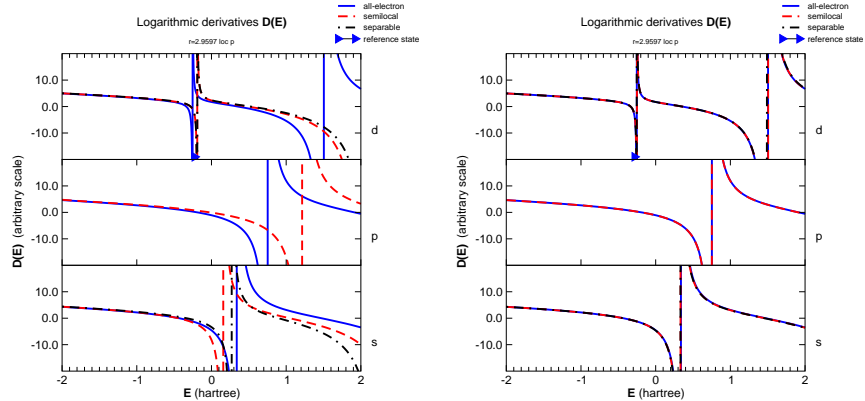

  \begin{center}
     \begin{tabular}{cc}
  \includegraphics[angle=0,clip=,scale=0.26]{./i/afe-Dtmlder.eps} &
  \includegraphics[angle=0,clip=,scale=0.26]{./i/afe-Dhlder.eps}
     \end{tabular}
   \end{center}
    \caption[Iron: PP transferability]{
     Logarithmic derivatives in the case of a 16 electrons pseudoatom.
     Left: Pseudopotentials generated according to the Troullier-Martins
     scheme show discrepancies between
     the differents curves, and are less transferable.
     Right: Pseudopotentials generated within the Hamann scheme are more
     transferables and consequently more
     accurate in reproducing scattering properties.
   \label{fig:log}
                  }
\end{figure}

In order to check the pseudopotential transferability
we studied logarithmic derivatives and compared them to the
their all-electrons counterpart. \\
Figure~\ref{fig:log} shows logarithmic derivatives in case
of Troullier--Martins (left) and Hamann (right) pseudopotentials
respectively.
Hamann pseudopotential are clearly more transferable, and therefore
more accurate in describing scattering properties.
However we should be aware that Hamann pseudopotentials are
usually much more expensive in terms of plane waves required for convergence
than the ones generated within the Troullier Martins scheme.
As a general rule, depending on the specific system and properties we want to describe, a compromise between accuracy and efficiency should guide the choice of the suitable pseudopotential.
 \label{sec:a-iron-pseudo}
 \section{Properties of bulk iron}
   In this section we report the results of DFT calculations of 
structural and electronic properties of bulk iron.
We also show a Time Dependent DFT calculation of the optical
conductivity, and we compare our results with experimental data.
\subsection{Structural properties}
An useful comparison between the
generated PP can be based on the evaluation
of physical quantities. In particular we calculated
the lattice parameter $a_0$ of iron.
We considered iron in his $\alpha$ phase,
the most stable at normal temperatures,
where the metal presents a body centred cubic (BCC)
crystal structure.

\begin{figure}[ht!]
  \centerline{\epsfig{file=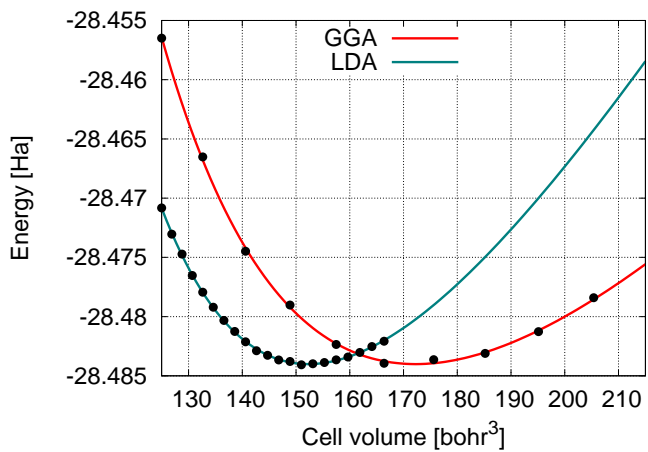,width=2.5in,angle=270,clip=}}
 \caption[Birch--Murnaghan EOS for iron]{
  Birch--Murnaghan fit of DFT total energy calculations for iron (8 valence electrons)
  performed within local density (continuous line) and generalized gradient (dotted line)
  approximations. Lattice parameter calculated within LDA are lower than the
  experimental value, and GGA partially correct this understimation.
}
 \label{fig:lattice}
\end{figure}

The optimized value of $a_0$ has been obtained by an estimation 
of the minimum of the Birch--Murnaghan~\cite{B52,M44} equation of
state for solids:
\begin{equation}
  E(V) = \frac{B_0 V}{B_0'(B_0' - 1)}\left[B_0' \left( 1 - \frac{V_0}{V} \right)
                                 + \left(\frac{V_0}{V}\right)^{B_0'} - 1 \right]
 \label{eq:BM}
\end{equation}
where $B_0$ is the bulk modulus and $V_0$ is the cell
volume at the minimum.
We used the ABINIT package~\cite{ABINIT,abinit2} to compute
converged total energy for different values of the cell
parameter a$_0$, and we fitted the obtainted points
with Eq.~\ref{eq:BM} (see Fig.~\ref{fig:lattice}).
%%%%%%%%%%
% Tabular
%%%%%%%%%%
\begin{table}[h!]
 \begin{center}
  \begin{tabular}{cccc}
  \hline
 Pseudopotential & a$_{0}$ & B$_{0}$   & $  \mu  $     \\
                 & [Bohr]  &  [GPa]    & $[\mu_{B}]$   \\
  \hline
  \hphantom{0} 8e- TM GGA NLCC \hphantom{00} & \hphantom{0}5.56 & \hphantom{00} 67.86  & 2.47\hphantom{0} \\
  \hphantom{0} 8e- TM LDA NLCC \hphantom{00} & \hphantom{0}5.34 & \hphantom{0} 106.31  & 2.11\hphantom{0} \\
              16e- H LDA NLCC \hphantom{0}   & \hphantom{0}5.27 & \hphantom{0} 101.23  & 2.19\hphantom{0} \\
  \hline
  Exp.$^{(a)}$ \hphantom{00}             & \hphantom{0}5.42 & \hphantom{0}   168   & 2.22\hphantom{0} \\
  \hline
  \end{tabular}
 \end{center}
 \caption[Iron: lattice parameter]{
        Comparison of lattice parameter for iron calculated with different pseudopotentials.
        Exp.$^{(a)}$ are experimental data reproduced from Ref.~\cite{kittel,acker_prb_v37_p6827_y1988}.
        \label{tbl:lattice}
        }
\end{table}
In Tab.~\ref{tbl:lattice} we report a summary of the lattice parameters
and bulk moduli obtained with the fitting procedure.
Moreover Fig.~\ref{fig:lattice} shows a comparison between LDA and GGA
results, confirming a well known understimation of the experimental value
by LDA, that can be partially corrected using GGA.

 \section{Electronic properties}
   \subsection{Electronic properties}  \label{sec:elec-prop}
\begin{figure}[h!]
  \begin{center}
    \begin{tabular}{c}
  \includegraphics[angle=0, scale=0.4,clip=]{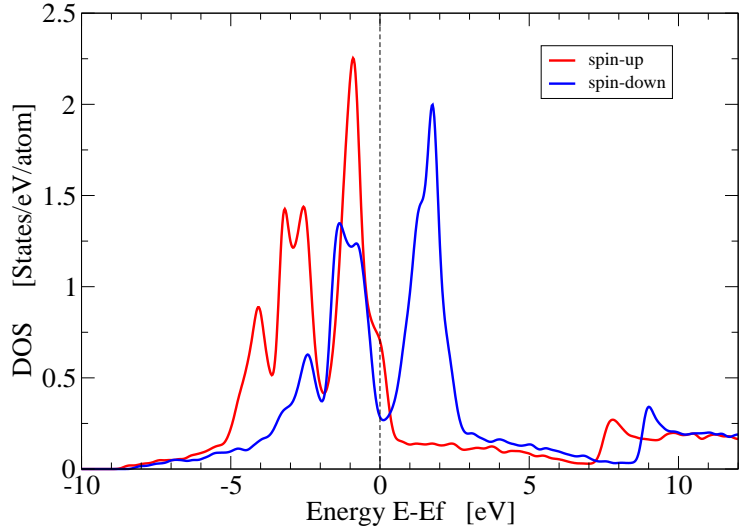} \\
    \end{tabular}
    \caption[Spin resolved DOS of BCC iron]{
     Spin resolved density of states of BCC iron calculated within DFT--LDA
     with ABINIT~\cite{ABINIT}. Majority (red line) and minority (blue line)
     spin channels are presented. The Fermi energy is set to zero for clarity.
                  }
  \label{fig:dosfe}
 \end{center}
\end{figure}

\begin{figure}[h!]
 \begin{center}
  \includegraphics[angle=270, scale=0.6]{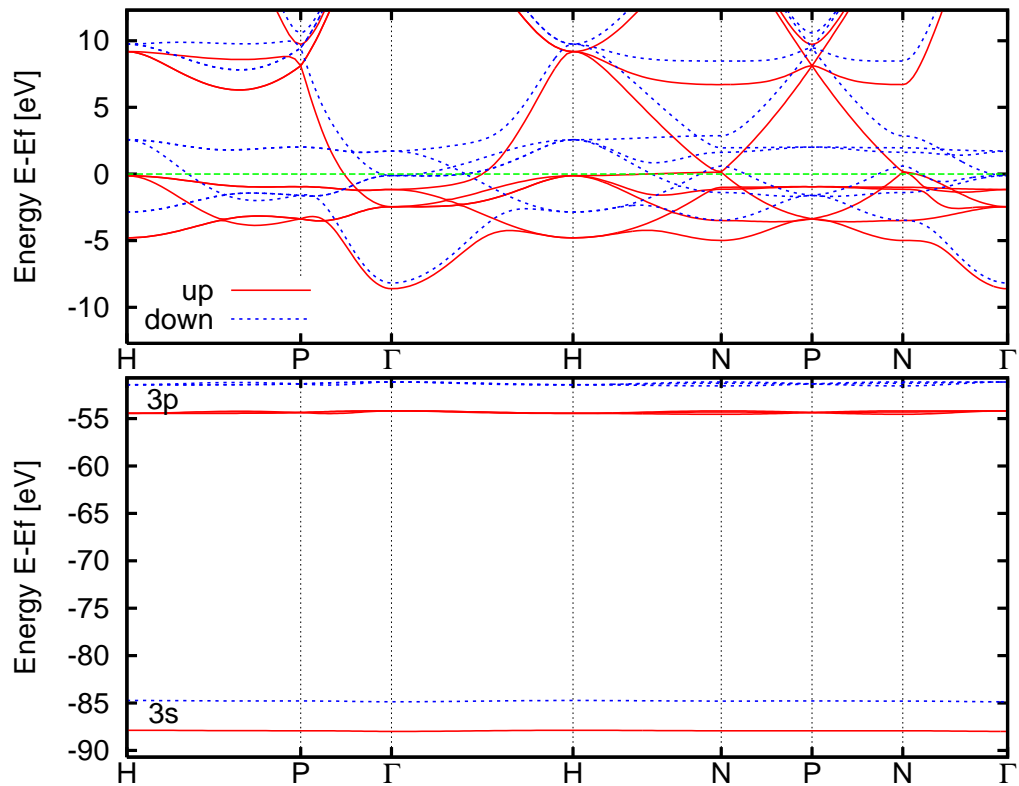}
    \caption[Spin resolved band structure of BCC iron]{
     Spin resolved band structure of BCC iron
     along standard high simmetry directions of the Brillouin zone,
     calculated in LDA approximation with the ABINIT~\cite{ABINIT} code.
     Valence (top) and semicore (bottom) states are distinguished in different panels
     in order to evidence energy scales.
            }
  \label{fig:bands}
   \end{center}
\end{figure}

\begin{figure}[h!]
  \begin{center}
    \begin{tabular}{cc}
  \includegraphics[angle=270, scale=0.5,clip=]{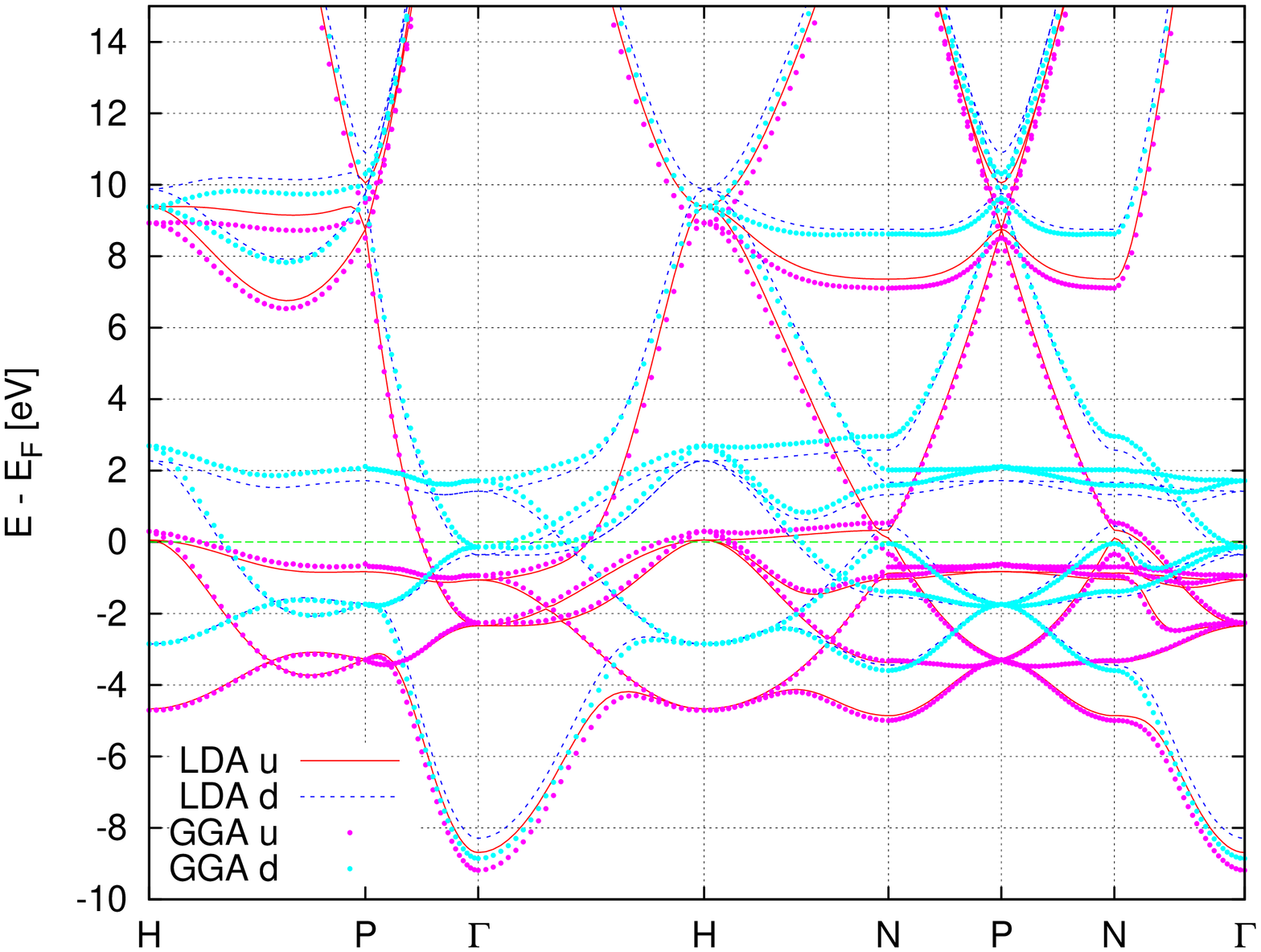} \\
  \includegraphics[angle=270, scale=0.5,clip=]{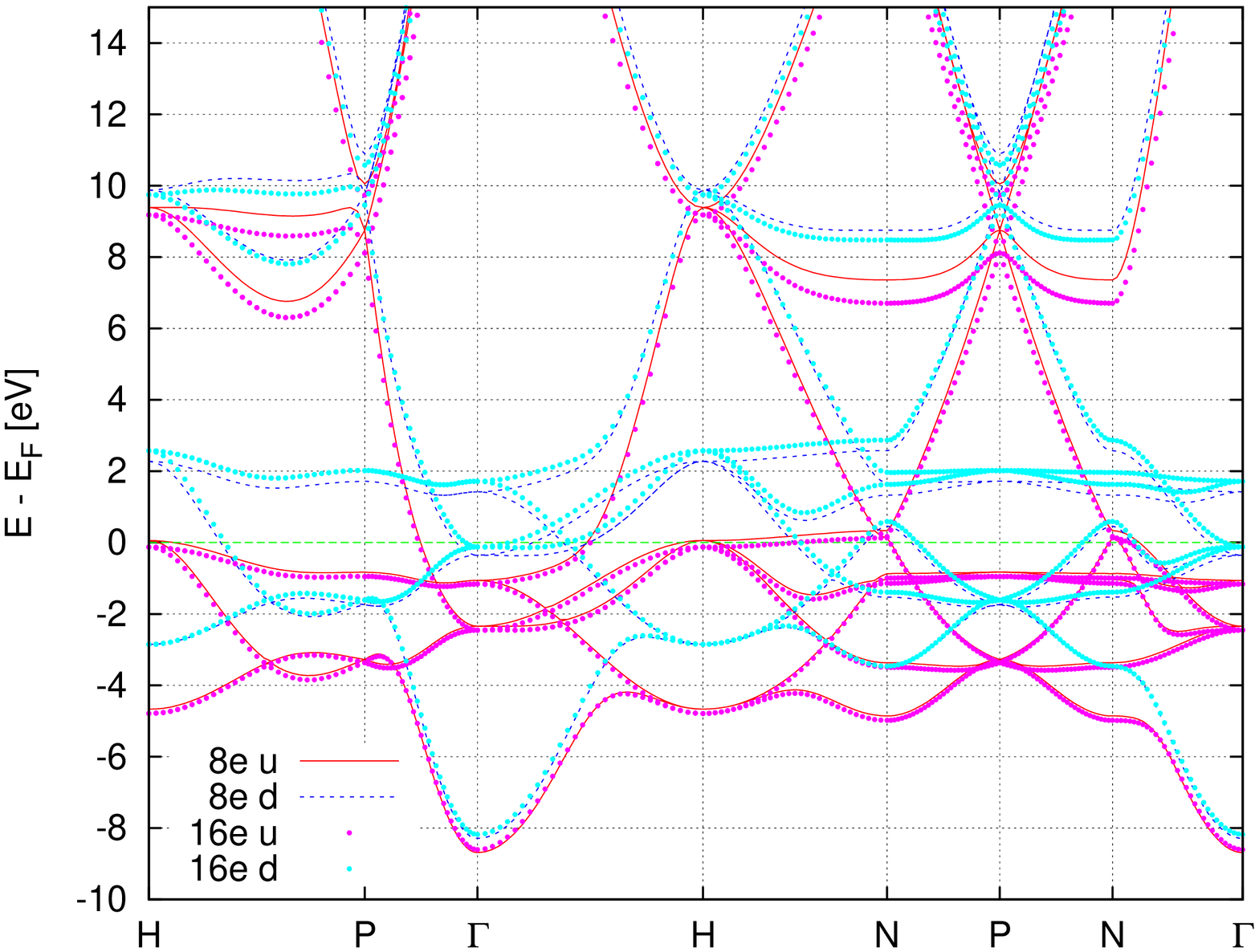}
    \end{tabular}
    \caption[Valence bands of BCC iron]{
     Spin resolved band structure (valence states) of BCC iron calculated 
     at the experimental lattice paramter.
     Top: comparaison between GGA and LDA calculation;
     Bottom: comparaison between LDA calculations with and without semicore states.
                  }
  \label{fig:bands-confronto}
 \end{center}
\end{figure}

 We present here our results for the density of states (DOS)
 around the Fermi energy and the electronic band structure
 including semicore states.
 These calculations have been performed with the ABINIT
 package~\cite{ABINIT} in local spin density approximation (LSDA).
We verified that a random sampling of {\bf k} points 
in the Brillouin Zone (BZ) is more efficient than
the use of a regular grid, in fact with 5000 {\bf k} DOS 
is converged. In Fig.~\ref{fig:dosfe} two curves 
represent converged DOS where
the two spin channels show an asymmetric 
distribution allowing to identify the majority and minority 
component.
The corresponding band structure is presented in Fig.~\ref{fig:bands},
where high symmetry crystallographic directions are chosen
in agreement with previous 
literature~\cite{callaway_pr_y1955_v99_500,singh_prb_y1975_v11_287,cococcioni_thesis}.
We used Hamann LDA pseudopotentials generated including semicore
states in order to describe the core bands $3p$ and $3s$ bands.
We obtain band energies of Fe-3s (85~eV) and Fe-3p (52~eV)
as depicted in Fig.~\ref{fig:bands}.
In the same figure we show the effect of the spin on the electronic
properties, removing most of the accidental degenerancies
of band structure and splitting the flat core states.
In addition we report in Fig.~\ref{fig:bands-confronto} (top panel) a comparison
between bands calculated within LDA or GGA, concluding that
the GGA corrections do not change dramatically the electronic 
structure making LDA a satisfactory approximation for our scope. 
Even the use of semicore PP (bottom panel in Fig.~\ref{fig:bands-confronto})
do not influence considerably the estimation of valence states,
therefore it is possible to conclude that our approximations 
are reasonable in order to predict correctly the optical properties
of bulk iron.\\
On the contrary, the case of $3p$ levels is more subtle.
In fact, because the angular momentum $\vec{l}$ is not zero
for $p$ levels (i.e. $l$=1), the total quantum number
$\vec{j} = \vec{s}+\vec{l}$ must be considered, and the spin orbit
coupling can influence the exact degenerancy and position of the bands.
Even if this case will not be treated in the present work, because
beyond the aim of the thesis, it is worth to mention that a fully
relativistic approach is required in order to obtain reliable
results for this kind of states.
Conversely, the $3s$ states are correctly predicted within 
the DFT--LSDA approach and the up--down splitting 
$\Delta_{\uparrow\downarrow} = 3s_{\uparrow} - 3s_{\downarrow}$ 
is comparable with the available experimental 
data~\cite{acker_prb_v37_p6827_y1988}.
For completeness, we report in Tab.~\ref{tbl:dispersione}
the calculated splitting and dispersion of both the $3s$ and $3p$ 
levels. 
\begin{table}
 \begin{center}
 \begin{tabular}{cccc}
 \hline
  state & $\Delta E_{\uparrow\downarrow}$ & Dispersion & $\Delta E_{\uparrow\downarrow}^{exp.}$ \\      
        &   [eV]     &   [meV]    &   [eV]   \\
 \hline
  $3p$ & 3.03 & 360 (at N) &         - \\   
  $3s$ & 3.15 & 120 (at $\Gamma$) & 4.9\\
\hline
 \end{tabular}
 \end{center}
 \caption[Core states splitting of iron]{
      Calculated splitting of the core states compared with available experimental data 
      and calculated dispersion. \label{tbl:dispersione}
      }
\end{table}

\begin{figure}[h!]
  \begin{center}
    \begin{tabular}{c}
  \includegraphics[angle=0, width=10.5cm,clip=]{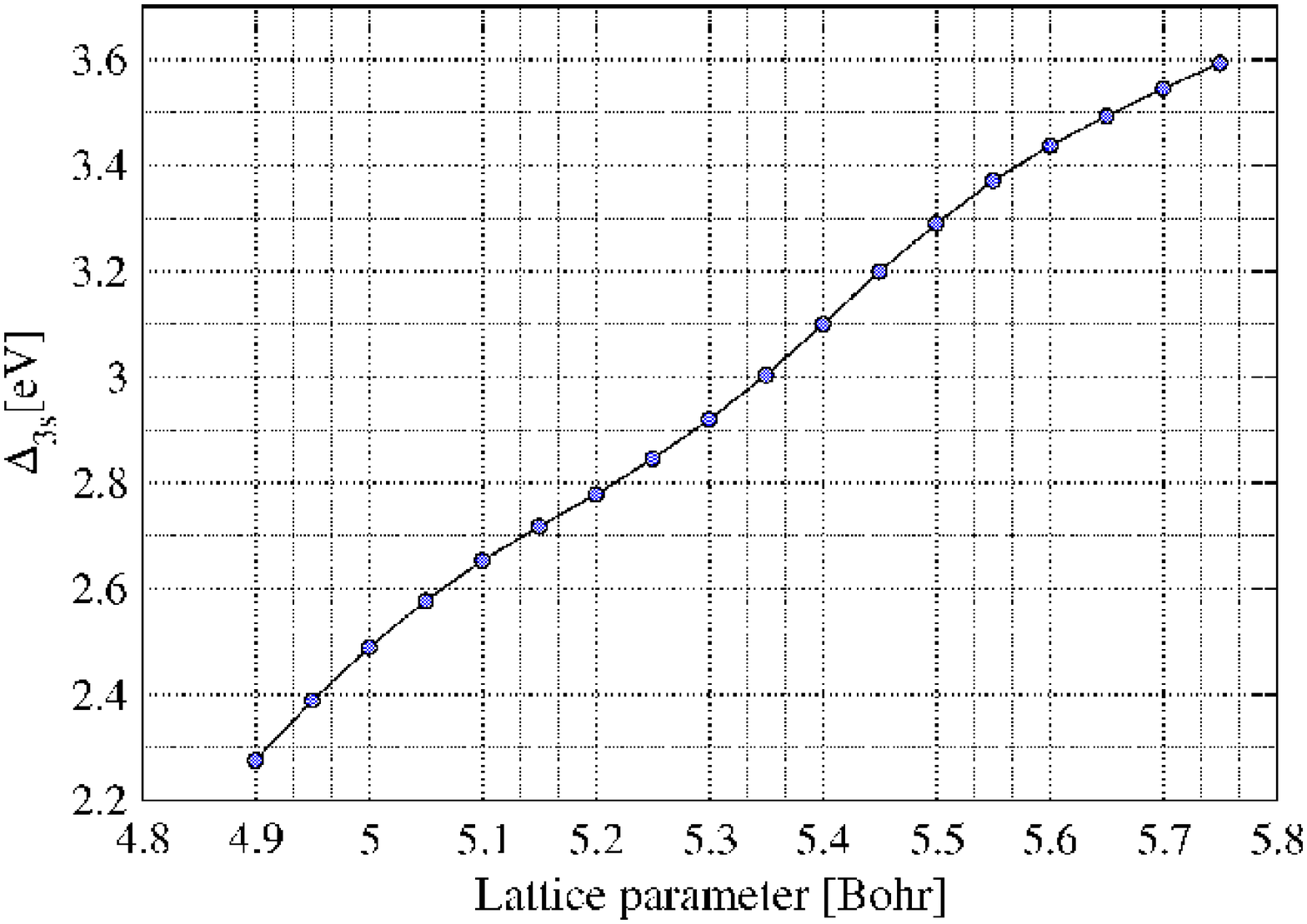} \\
  \includegraphics[angle=0, width=10.5cm,clip=]{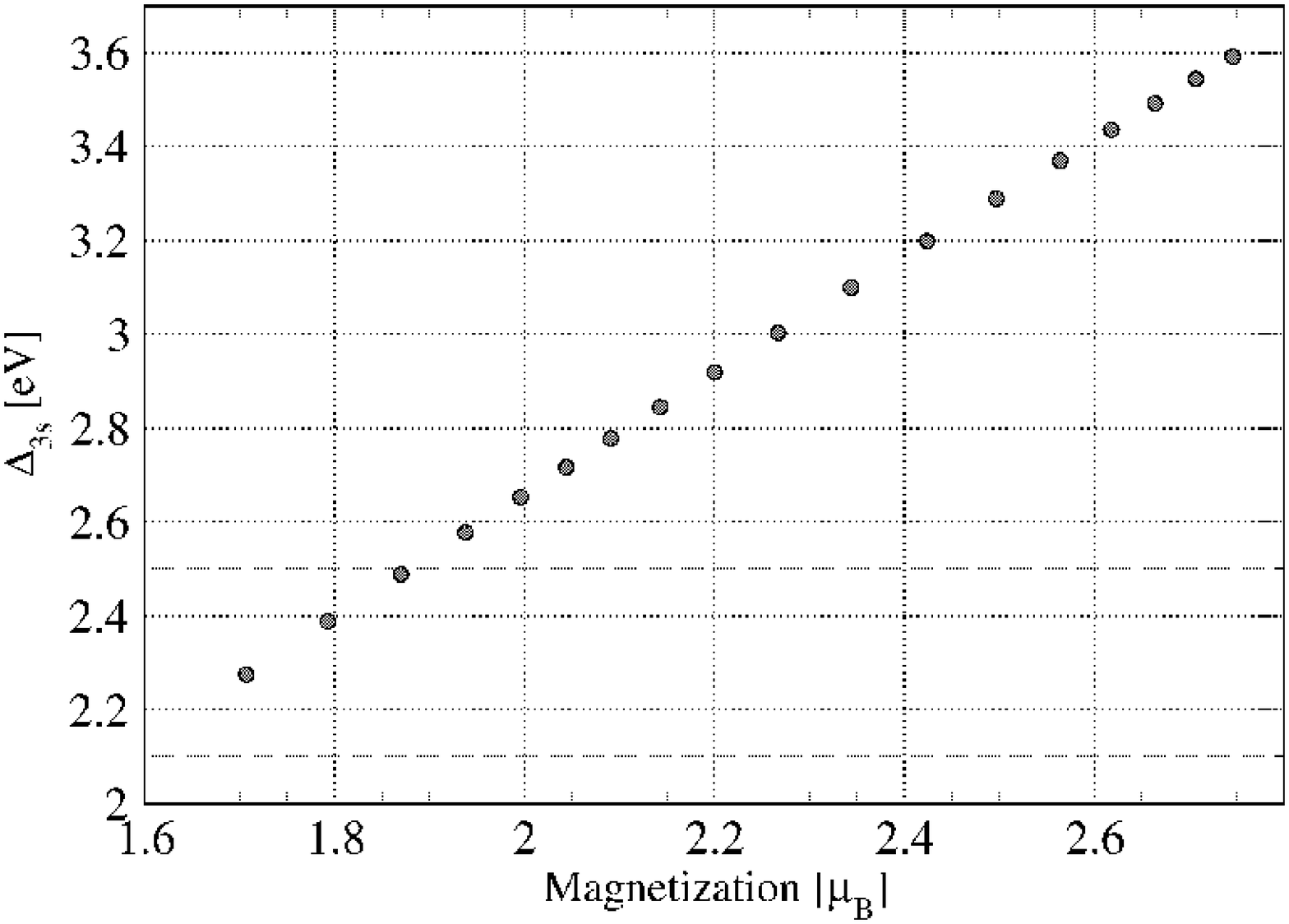}
    \end{tabular}
    \caption[Fe-3s splitting]{Splitting of 3s levels 
     as a function of the lattice parameter (top panel)
     and the magnetization (bottom panel).
                  }
  \label{fig:splitting}
 \end{center}
\end{figure}

In conclusion we also report an analysis of the
relation between total magnetization $\mu$ of iron and 
the 3$s$ splitting.
It was suggested in the past~\cite{hufner-prb-v7-p2333-y1973,fadley-prl-v23-p1397-y1969} 
that $s$ core-level splitting could 
be used to monitor the magnetic moment or the hyperfine field,
because the splitting should vary linearly with the spin 
state of the unfilled inner shell.
X-ray photoelectron spectroscopy (XPS) measurements for nonmetallic
transition metal compounds, on rare-earth metals
and ionic compounds~\cite{hufner-prb-v7-p2333-y1973,fadley-prl-v23-p1397-y1969,wertheim-prb-v7-p556-y1973} 
are compatible with this scheme. \\
However, Fe-3$s$ XPS splitting studied in crystalline and amorphous alloys
have shown a poor correlation between the 3$s$ splitting 
and the magnetic moment of the solid
(see Fig. 3 in Ref.~\cite{acker_prb_v37_p6827_y1988}).
Our calculations show a linear relation of such a splitting
with respect to the total magnetization of bulk iron 
(see Fig.~\ref{fig:splitting} top panel), giving some 
basis to better investigate the apparent disagreement 
with experimental data. \\
We calculated the total magnetization defined as the 
difference between the majority and minority spin density 
integrated over the unit cell $\mu = \int_{\Omega} \rho_{\uparrow}(r) - \rho_{\downarrow}(r)$.
This quantities has been evaluated for several lattice 
parameter values of bulk Fe, centred around the 
experimental one (5.42~Bohr).
Changes of the lattice parameter influence
the local distance between atoms, reproducing the effect of 
a change of pressure. Moreover changes on lattice geometry affect 
total magnetization.
The dependence of the splitting with respect to the 
pressure is shown in Fig.~\ref{fig:splitting} bottom panel.

  \section{Optical conductivity}
\begin{figure}[h!]
 \begin{center} 
   \epsfig{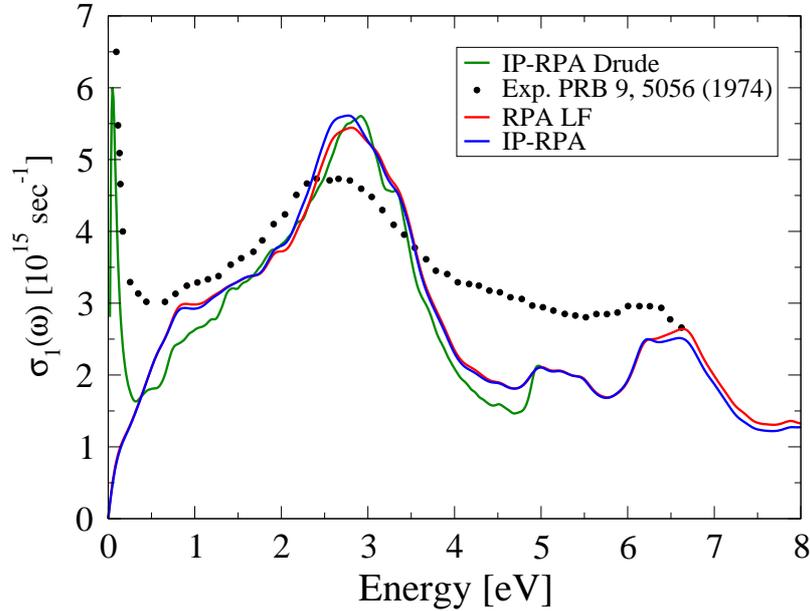}
 \end{center}
 \caption[Iron: optical conductivity]{
          Comparison between optical conductivity experimental 
          data from Refs.~\cite{johnson_pr_y1974_v9_5056}
          and RPA calculation performed with the DP code~\cite{DP}.
          Results with and without the inclusion of crystal 
          local field effects (LF) are shown by the red and blue lines, 
          respectively.
          }
 \label{fig:sigma}
\end{figure}

   In this last section we report the calculations
of the dielectric function ($\varepsilon = \varepsilon_1 + i\varepsilon_2$)
obtained within the  indipendent particle random phase approximation
(IP-RPA) and including crystal local fields effects (RPA-LF).
This quantity is closely related to the electronic structure
and follows straightforward from theory, however
experiments are often presented in terms of the
optical conductivity ($\sigma=\sigma_1+i\sigma_2$,
see Refs.~\cite{johnson_pr_y1974_v9_5056}).
However, $\varepsilon$ and $\sigma$ are connected by the well known
relations:
\begin{eqnarray}
 \sigma_1 &=& \frac{\varepsilon_2 \omega}{4 \pi} \\
 \sigma_2 &=& \frac{(1-\varepsilon_1)\omega}{4 \pi}
\end{eqnarray}
In Fig.~\ref{fig:sigma} we show the optical conductivity of bulk iron
calculated with the DP code~\cite{DP}.
There is a nice agreement between the calculated curves
and the experimental data reported in Ref.~\cite{johnson_pr_y1974_v9_5056} 
with respect to the general shape and the position of the maxima.
We have not included any intraband contribution in our calculation and
this is the reason why the sharp structure in the low--energy region
of the experimental spectra is not reproduced (Fig.~\ref{fig:sigma}).
We report, for comparison, a recent calculation of the 
optical conductivity including the Drude peak
(see Ref.~\cite{cazzaniga_thesis_y2008}
for details about the treatment of the intraband transitions), in 
order to show that only the low energy part of the spectum
is affected by this approximation. \\
Spin flip is also not allowed and this bring us to conclude that
the maximum of the computed conductivity near 2.5~eV results
from transitions between collinear spin states, in particular
between states of the majority spin channel.
It is easy to identify these excitations looking at Fig.~\ref{fig:dosfe}:
transitions connect occupied states and the unoccupied
states just above the Fermi energy for the same spin channel (dashed line).
Crystal local fields have negligible effects, and this is
in agreement with the cases of others metals.
As shown in Fig.~\ref{fig:sigma} by the red and blue lines.
%%%%%%

 \section{Conclusions}
   We succeeded in generating a set of pseudopotentials for
iron, a prototypical transition metal for which an obvious
separation between core and valence orbitals does not
exist.

Semicore states have been either explicitly included in the valence
set, or treated within the usual \emph{non linear core corrections}
scheme.
Our set of PP has been tested against structural and electronic properties
of BCC bulk iron, including the lattice constant, spin resolved
density of states, band structure.

We also reported an analysis of the Fe-3s splitting 
with respect to the total magnetization.
We found a linear dependence between these two 
quantities and between splitting and the lattice parameter.
This results support the use of the s core-level 
splitting as a monitor of magnetization, and are
compatible with a similar behaviour presented by 
non metallic transition metal compounds, rare earth metals
and ionic compounds.
Moreover we found that the 3s splitting calculated at the experimental 
value of the lattice parameter substantially agrees with XPS core level 
measurements reported in Ref.~\cite{acker_prb_v37_p6827_y1988}.
However the linear trend theoretically predicted is in apparent 
disagreement with experimental results where 3s XPS peak splittings 
do not correlate with the Fe magnetic moments.

Concerning the optical conductivity good agreement between RPA results
and the experimental data is found. This allows one to conclude that the main
spectral feature are well reproduecd by spin-collinear excitations.

\chapter{Iron, cobalt and nickel pyrites} \label{a:sulfur}
 Transition-metal pyrites form a series of compounds with a large variety 
in electrical, magnetic and optical properties.
In particular, recently there has been a renewed interest in
iron, cobalt and nickel disulfides because of their potential
in future technological applications.
Iron and nickel-controlled doping of CoS$_2$ gives rise to a tunable 
source of highly spin-polarized electrons~\cite{wang_prb_y2004_v69_p094412}. 
This property is extremely interesting to design new devices exploiting
the spin character of the electrons in addition to their charge.
In fact the essence of the current focus area termed \emph{spintronic},
or \emph{spin-electronics}, is to use the electron's spin, as well
as its charge in creating new devices or enhancing the functionality
of the existing ones~\cite{wolf_science_v294_p1488_y2001,appelbaum_nature_v447_p295_y2007}. \\
\indent In this section we will present structural, magnetic 
and electronic properties of FeS$_2$, CoS$_2$ and NiS$_2$
calculated within the density functional theory framework. A 
comparison of our results with respect to experimental data and to previous 
calculations is also discussed.

  \section{Motivations}
\begin{figure}
 \begin{center}
  \includegraphics[width=10.5cm,clip=,angle=0]{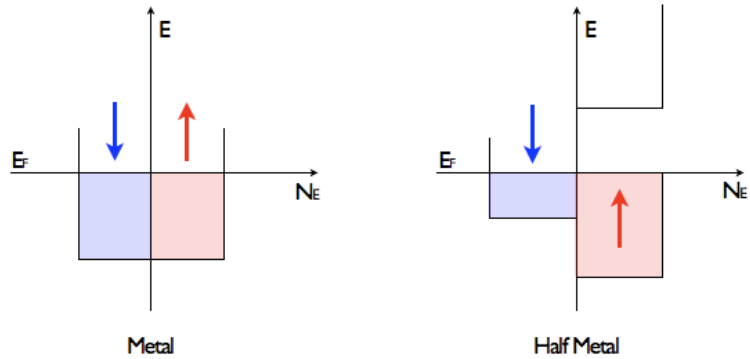}
 \end{center}
  \caption[Schematic representation of metal and half metal]{
     Schematic representation of the density of states in case of
     a metal (left) and an half metal (right).
 \label{fig:hm}}
\end{figure}

\begin{figure}[h!]
 \begin{minipage}{7cm}
  \includegraphics[width=8.0cm,clip=,angle=0]{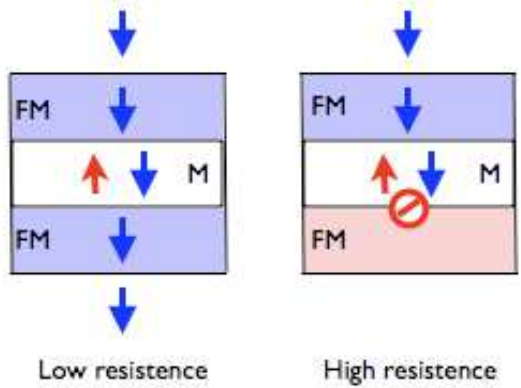}
 \end{minipage}
\hspace{1.2cm}
  \begin{minipage}{5cm}
  \caption[Schematic representation of spin transport]{
     Schematic representation of spin polarized transport
     from a FM material to a metal and a FM material again.
     According to the reciprocal configuration of the spins,
     transport is allowed or forbidden.
 \label{fig:ic}
          }
  \end{minipage}
\end{figure}

 The rapid developement of spin valve-based magnetic read heads 
and the emergence of spintronic~\cite{wolf_science_v294_p1488_y2001,appelbaum_nature_v447_p295_y2007}
has thrown up a need for a better understanding of
spin-polarized materials~\cite{prinz_science_v282_p1660_y1998}. \\
Spintronic is based on the up or down spin of the carriers
rather than on electrons or holes as in traditional semiconductors
electronics.
In particular spin-polarized transport will occur naturally in any
material for which there is an imbalance of the spin populations
at the Fermi level.
This imbalance is present in ferromagnetic metals where the
density of states for spin up and down electrons are
shifted in energy with respect to each other.
Commonly in these materials there is an unequal filling of the bands,
which is the source of the net magnetic moment,
and causing the up and down carriers at the Fermi level to be
different in number, character, and mobility.
This inequality produces a net spin polarization
but the sign and magnitude of that polarization depends on
the specific material.
Materials having at the Fermi level only one occupied spin band 
are usually called \emph{half metals} (see Fig.~\ref{fig:hm}). \\
\indent A fundamental component in any spintronic device is a ferromagnetic
electrode which is used as a source of polarized electrons, and 
an high value of spin polarization 
P=(N$_{\uparrow}$-N$_{\downarrow}$)/(N$_{\uparrow}$+N$_{\downarrow}$) at the Fermi level 
can provide significant benefits. 
 When the spin polarized electron current crosses
a sample having a non zero total magnetization
the only states that are available to the carriers are those 
for which the spins of the carriers are parallel to the spin 
direction of those states at the Fermi level.
If the magnetization of the material is reversed,
the spin direction of those states also reverses. 
Thus, depending on the direction of magnetization of 
a material relative to the spin polarization of the current, 
a material can be either a conductor or an insulator for electrons of a
specific spin polarization (see Fig.~\ref{fig:ic}). \\
\indent The largest effect is generally seen for the 
most highly polarized currents, therefore, there are continuing 
efforts to find 100\% spin-polarized conducting materials.
However, partially polarized materials
(such as Fe, Co, Ni and their alloys), are adequate 
to develop technologically useful devices and can show 
unexpected properties.
For example, transition metal compounds have beed attracting 
extensive attention for that reason. Among them, FeS$_2$, CoS$_2$ and NiS$_2$
have been studied from the 
experimental~\cite{jarret_prl_y1968_v21_p617,folkerts_jpcssp_y1987_v20_p4135,ferrer_ssc_y1990_v74_p913,wang_prb_y2006_v73_p144402} 
and theoretical~\cite{bullett_jpcssp_y1982_v15_p6163,zhao_prb_y1993_v48_p15781,Shishidou_prb_y2001_v64_pR180401} 
point of view.
Moreover these compounds have been used to build and to 
study Co$_{1-x}$Fe$_x$S$_2$ and Ni$_{1-x}$Co$_x$S$_2$ compounds,
recently predicted to be tunable half metals (see 
Refs.~\cite{mazin_apl_y2000_v77_p3000,wang_prb_y2006_v73_p144402,wang_prl_y2005_v94_p056602,ramesha_prb_y2004_v70_p214409,cheng_jap_y2003_v93_p6847} 
for a review of works on Co$_{1-x}$Fe$_x$S$_2$ and 
Ref.~\cite{mabatah_prl_y1977_v39_p494} for Ni$_{1-x}$Co$_x$S$_2$). \\
\indent In the following sections we will focus on structural and magnetic
properties of FeS$_2$ and CoS$_2$, as a first step to be able in the 
future to describe more complex compounds.
 
  \section{Structural and magnetic properties}\label{sec:stru-sulf}
\begin{figure}
 \begin{center}
  \includegraphics[angle=0,width=15.0cm]{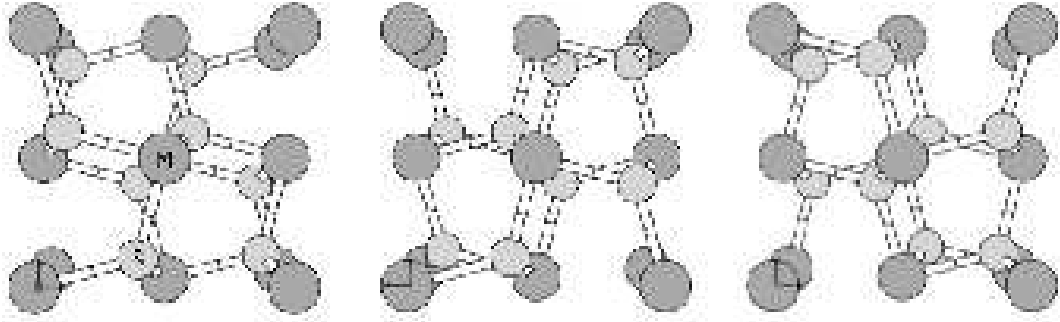}
  \caption[Pyrite structure]{
           Atoms rearrangement in pyrite structure along three orthogonal directions.
           Dark grey circles represent metal atoms, smaller light grey circles depict
           sulfur atoms. Graphs are generated with Xcrysden package~\cite{xcrysden}.
  \label{fig:pirite-structure}
           }
 \end{center}
\end{figure}

\begin{figure}
\begin{minipage}{5cm}
  \includegraphics[angle=0,width=9.5cm,clip=]{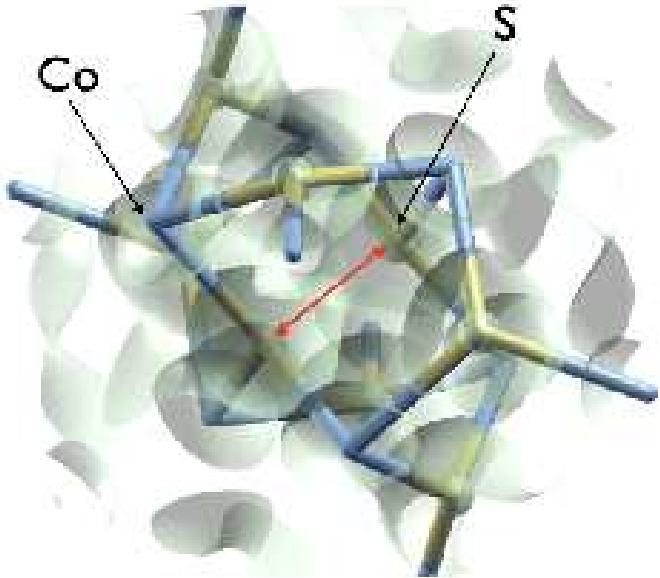}
\end{minipage}
\hspace{4.4cm}
\begin{minipage}{4cm}
  \caption[Isosurfaces of $|\psi_{n\bk}(\br)|^2$ for CoS$_2$]{
           Isosurfaces of electronic density (grey regions) around the 
           atoms of CoS$_2$. Cobalt atoms (light blue)
           and sulfur atoms (yellow) are represented 
           inside the crystal structure. The point of view 
           is chosen in order to evidence the characteristic S--S
           bond in the center of the crystal (red arrows). 
           Graphs are generated with the Xcrysden package~\cite{xcrysden}.
  \label{fig:ss-sulfur}}
 \end{minipage}
\end{figure}

\begin{figure}[hb!]
 \begin{center}
  \includegraphics[angle=0,width=14.0cm,clip=]{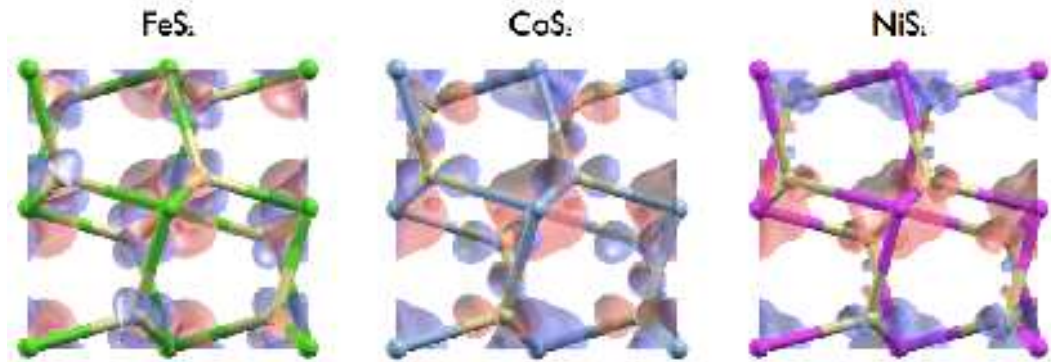}
\caption[ Isosurfaces of $|\psi_{n\bk\uparrow}(\br)|^2$ and 
          $|\psi_{n\bk\downarrow}(\br)|^2$ for FeS$_2$, CoS$_2$ and NiS$_2$.]{
          Electronic density isosurfaces for the up (red) and down (blue)
          spin component for FeS$_2$ (left), CoS$_2$ (center) and NiS$_2$ (right). 
          Graphs are generated with the Xcrysden package~\cite{xcrysden}.
  \label{fig:isosurfaces-sulfur}
         }
 \end{center}
\end{figure}

 All transition metal disulfides (MS$_2$ with M a 3$d$--transition metal atom),
crystallize in a cubic pyrite structure of space group
$T^6_h(Pa3)$ in which metal atoms are located 
in face-centred postions. Structure can be 
considered as an NaCl-like grouping of metal and chalcogen 
atom pairs (sulfurs).
In Fig.~\ref{fig:pirite-structure} the atoms arrangement is shown from three
perpendicular views. Every metal atoms (dark grey circles) is surrounded 
by six nearest-neighbour sulfurs in a distorted octahedral environment, 
while each sulfur (light grey circles) bonds to one sulfur 
(S-S bond) and three metals in a distorted tetrahedral arrangement.
Distance of sulfur--sulfur pairs (S--S) is short because of a
covalent bond. 
The formation of S--S pairs is characteristic feature of 
these structures. \\
In particular metal atoms are located at the positions
(0,0,0), (0,1/2,1/2), (1/2,0,1/2) and (1/2,1/2,0), 
the eight sulfur atoms instead 
are located at position 
$\pm(u,u,u)$, $\pm(u+1/2,1/2-u,\bar{u})$, 
$\pm(\bar{u},u+1/2,1/2-u)$, $\pm(1/2-u,\bar{u},u+1/2)$.
The values of $u$ and $a$ (the lattice parameter)
are taken from Wyckoff~\cite{wickoff}, in particular we used 
$u=0.386$ and $a=5.407$~\AA\ in case of FeS$_2$, 
$u=0.389$ and $a=5.524$~\AA\ in case of CoS$_2$ and 
$u=0.395$ and $a=5.677$~\AA\ in case of NiS$_2$. \\
\indent We performed a geometric optimization using the BFGS algorithm~\cite{B,F,G,S}
with the ABINIT~\cite{ABINIT} code.
The optimized S--S distances of FeS$_2$, CoS$_2$ and NiS$_2$ 
are reported in Tab.~\ref{tbl:SS}, and compared with 
values reported in Ref.~\cite{bullett_jpcssp_y1982_v15_p6163}.
All the ground state calculations have been performed within the DFT-GGA framework
(Perdew-Burke-Ernzerhof (PBE) parametrization~\cite{PBE}) with a cutoff 
energy E$_{\textrm{cut}}=34$~Ha. 
The geometry relaxation is performed by setting a tolerance of 
$0.02$ on the ratio of differences of forces to maximum force,
reached twice successively, will cause a self consistent cycle to stop.
 \begin{table}
 \begin{center}
  \begin{tabular}{c c c c c c}
   \hline 
    System  & $a_0$ [\AA] & S--S [\AA] & S--S [\AA] & $\mu$/at. [$\mu_B$] & $\mu$/at. [$\mu_B$] \\ 
            & Ref.~\cite{wickoff} &  this work  & from Ref.~\cite{bullett_jpcssp_y1982_v15_p6163} & this work &  Ref.~\cite{jarret_prl_y1968_v21_p617,mazin_apl_y2000_v77_p3000} \\ 
  \hline  
     FeS$_2$ & 5.407 & 2.18  & 2.14  & $\simeq 0$  & $\simeq 0$ \\
     CoS$_2$ & 5.524 & 2.29  & 2.12  &   0.98      &    0.9     \\
     NiS$_2$ & 5.677 & 2.06  & 2.06  &    0        &     -      \\
   \hline
  \end{tabular}
 \end{center}                                        
 \caption[Optimized S--S distances FeS$_2$, CoS$_2$ and NiS$_2$]{ 
           Optimized S--S distances calcluated for FeS$_2$, CoS$_2$ and NiS$_2$ 
           are compared with values reported in Ref.~\cite{bullett_jpcssp_y1982_v15_p6163}. 
           The calculated ground state magnetization per metal atom is compared
           to experimental values reproduced from 
           Ref.~\cite{jarret_prl_y1968_v21_p617,mazin_apl_y2000_v77_p3000}
     \label{tbl:SS}
      }
 \end{table}
The agreement of the computed S--S distance with respect to 
previous calculations is reasonable, as listed in Tab.~\ref{tbl:SS}.
Moreover, Figure~\ref{fig:ss-sulfur} shows the electronic density distribution
for a suitable value of the chosen isosurfaces in order to evidence
the formation of the characteristic S--S bond
in the center of the crystal structure. \\
\indent The magnetic character of the electronic ground state 
of all the compounds has been also analyzed.
The values of total magnetization per metal atom are reported in Tab.~\ref{tbl:SS}
with experimental reference values reproduced from literature. \\
The three compounds display different behaviours:
FeS$_2$ presents a total magnetization close to zero,
CoS$_2$ has a ferromagnetic ground state with
total magnetization per cobalt atom close to 1~$\mu_B$,
and NiS$_2$ is non--magnetic.
Experimental studies on transition metal disulfides
(see Ref.~\cite{jarret_prl_y1968_v21_p617}) revealed the non magnetic
nature of all this class of compounds, except for the case of CoS$_2$.
Experiments indicate that FeS$_2$ as well as NiS$_2$ are
paramagnetic semiconductors. In addition, at low temperature
NiS$_2$ presents a transition to an antiferromagnetic phase.\\
Our calculations predict correctly this behaviour.
In particular for CoS$_2$ we find a ground state electronic configuration
with a total magnetic moment $\mu=$0.98~$\mu_B$ per
cobalt atom at T=0$^{\circ}$~K, which is close to the experimental value
$\mu=$0.9~$\mu_B$ (see Ref.~\cite{jarret_prl_y1968_v21_p617}). \\
\indent In addition we report in Fig.~\ref{fig:isosurfaces-sulfur}
the electronic density isosurfaces that evidences
the distribution of spin up and spin down components.
In the case of FeS$_2$ the spin up and down
charge distributions are located at the same positions without any
motif distinguishig between the two components.
On the contrary, in case of CoS$_2$ it is possible to appreciate a regular
ordering of the electronic density distribution according to the two spin
components (red and blue respectively in Fig.~\ref{fig:isosurfaces-sulfur}).
The maxima of the distributions are located at the metal sites of the
crystal and the up-down spin density is alternated in adiacent sites. \\
NiS$_2$ still presents a regular ordering of the spin
density distributions but, as the total magnetization is found to be
zero, suggests an antiferromagnetic character.
Even if the study of NiS$_2$ is beyond the scope of this thesis, we notice
that this result is in agreement with Ref.~\cite{jarret_prl_y1968_v21_p617}.

  \section{Electronic properties}
\begin{figure}[h!]
 \begin{center}
  \includegraphics[angle=0,width=12.5cm,clip=]{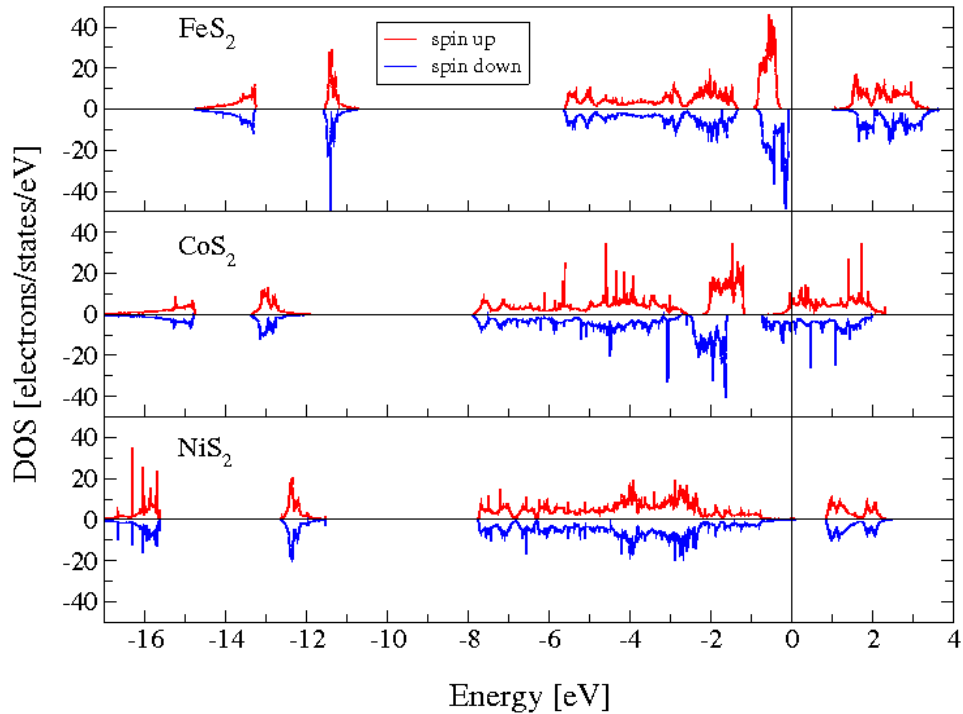}
 \end{center}
  \caption[Spin resolved DOS of FeS$_2$, CoS$_2$, NiS$_2$]{
   Spin resolved density of states of FeS$_2$, CoS$_2$, NiS$_2$
   calculated with tethraedron method with the software ABINIT~\cite{ABINIT}.
  \label{fig:dos-sulfur}
          }
\end{figure}

 In order to analyze the electronic properties of FeS$_2$, CoS$_2$ and NiS$_2$
we calculated the spin resolved density of state (DOS) within DFT-GGA (PBE parametrization).
The software ABINIT~\cite{ABINIT} has been used in order to perform convergence
study and calculate the final distribution.
For FeS$_2$ we used a Monkorst Pack~\cite{MP76} grid of $12\times12\times12$ 
{\bf k} points in the Brillouin zone (BZ), corresponding to 76~{\bf k}, applying
a rigid shift of $0.5$ in the three spatial directions. 
For CoS$_2$ the grid consisted of $10\times10\times10$ {\bf k}
points and $6\times6\times6$ for NiS$_2$.
A thermal broadening (cold smearing) $tsmear=0.007$~Ha is applied to simulate
the metallic occupation of levels following the recipe of
Ref.~\cite{marzari_thesis}\footnote{The smeared delta function is
defined by: $\delta_s = \frac{1}{\sqrt{\pi}}e^{-x^2}\left(ax^3-x^2-\frac{3}{2}ax +\frac{3}{2}\right)$
where $a=-0.5634$, this choice minimizes the bump.}.\\
Density of states are calculated using the tetrahedron method giving a faster 
convergence over the number of {\bf k} poins required with respect
to the standard integration over the Brillouin zone.\\
\indent In Fig.~\ref{fig:dos-sulfur} the DOS of FeS$_2$, CoS$_2$ and NiS$_2$
are shown. The three compounds present a similar density of states 
but different electronic behaviour.
We discuss now our results that are in general agreement 
with experimental 
works~\cite{folkerts_jpcssp_y1987_v20_p4135,ferrer_ssc_y1990_v74_p913,bither_ic_2208_1968} 
and theoretical calculations based on linear combination of atomic orbital
(LCAO) or semiempirical and self-consistent tight-binding (TB)
approach~\cite{bullett_jpcssp_y1982_v15_p6163,zhao_prb_y1993_v48_p15781,lauer_prb_29_6774_1984}. \\
The first two lowest energy bands are associated with bonding and
antibonding pairs of orbitals of the sulfur atom dimers
(s$\sigma$ and s$\sigma^*$ orbitals). 
The following complex structure is due to sulfur 3$p$ and 
the metal 3$d$ orbitals. 
Then the crystal field of the disulfide anions (sulfurs)
splits the cation (metal) 3$d$ non--bonding
orbitals into three low-lying 3$d(t_{2j})$ and two 
higher-energy 3$d(e_{g})$ levels (see Fig.~\ref{fig:dos-sulfur}). \\
All the disulfides studied have completely filled 
3$d(t_{2j})$ orbitals, and as the atomic number increases
the additional electrons (none for iron, one for cobalt 
and two for nickel) fill in the 3$d(e_{g})$ orbitals.
The different filling of these bands determine the
electronic nature of the three compounds studied. \\
\indent In particular, our calculations predict FeS$_2$ as 
a small band gap semiconductor, in agreement with 
experiments (optical and conductivity
measurements~\cite{ferrer_ssc_y1990_v74_p913} and X--ray
photoemission spectoscopy~\cite{folkerts_jpcssp_y1987_v20_p4135,heide_jssc_33_17_1980})
and theoretical works~\cite{bullett_jpcssp_y1982_v15_p6163,zhao_prb_y1993_v48_p15781}.
From our calculations the value of the FeS$_2$ band gap turns out
to be 1.2~eV and 0.98~eV for the spin-up and spin-down components respectively
(Fig.~\ref{fig:dos-sulfur}, top panel).
Even if a more detailed analysis including 
band structure calculations is required, 
we can already conclude that the calculated values 
are close to the experimental ones ranging from 
0.9~eV to 1.2~eV (see Ref.~\cite{ferrer_ssc_y1990_v74_p913})
and for which there is no general consensus.
Difficulties in the exact experimental determination of the band gap
of FeS$_2$ could rise because of a large difference between 
the indirect and direct gap as pointed out by 
Zhao\ea\ \cite{zhao_prb_y1993_v48_p15781}. 
In fact in that work all calculations in the local
density approximation lead to a minimum indirect gap
E$_{ig}=0.59$~eV and to a smallest direct gap E$_{dg}=0.74$~eV.\\
\indent Concerning CoS$_2$, we found similar structures 
for low-energy states with respect to FeS$_2$, but shifted 
to lower energies (see Fig.~\ref{fig:dos-sulfur}, central panel).
Moreover our DFT-GGA calculations confirm the half metallic nature 
of CoS$_2$, where the two spin components present
density of states characteristic of a semiconductor (red line)
or a metal (blue line).
In Fig.~\ref{fig:dos-sulfur} the Fermi energy 
is set to zero for clarity. \\
It is worth to mention that half--metallicity of CoS$_2$ 
has been recently discussed in literature from both the
experimental and theoretical points of view.
Direct measurements of spin polarization and 
magnetotransport~\cite{wang_prb_y2004_v69_p094412} 
show that CoS$_2$ is not completely polarized
(i.e. $P = \frac{N_{\uparrow}(E_F)+N_{\downarrow}(E_F)}{N_{\uparrow}(E_F)-N_{\downarrow}(E_F)}$ 
is  not large).
On the contrary, the temperature variation of the electronic 
structure, studied by means of optical reflectivity measurements confirms 
the half metallicity of the compound.
From the theoretical point of view 
Shishidou\ea\ \cite{Shishidou_prb_y2001_v64_pR180401}
presented a detailed comparison between LDA and GGA 
electronic structure calculations within 
density functional, full potential linearized augmented 
plane waves calculations. 
In that work the authors conclude that GGA greatly 
modifies the LDA band structure 
from metallic to half metallic.
These conclusions are in agreement with our finding 
displayed in Fig.~\ref{fig:dos-sulfur}. \\
\indent In the case of NiS$_2$ the spin density of states 
presents a symmetric behaviour that confirm the
apparently non magnetic nature of that compound and the 
zero total magnetization obtained from our calculations
(see sec.~\ref{sec:stru-sulf}).
However, as underlined preivously, NiS$_2$ presents a local regular ordering 
of the spin density distribution justifying its antiferromagnetic nature
(see Refs.~\cite{kikuchi_jpsj_v45_y1978}).
Among the transition metal disulfides, NiS$_2$ is considered
to be a Mott-Hubbard insulator~\cite{wilson_pm_v23_y1971}.
For that reason more specific theories
are required in order to correctly describe NiS$_2$ properties. 

 \section{Conclusions}
  We presented a systematic study of structural, magnetic and
electronic properties of FeS$_2$, CoS$_2$ and NiS$_2$,
three examples of 3d-transition metal sulphides with cubic
pyrite structure.
As anticipated in the introduction of this chapter,
the interest on these systems is recenlty renewed, due to the
great potential of such systems for spintronic applications. \\
\indent From our calculations we obtained the S--S relaxed distance,
characteristic of these compounds, 
the ground state total magnetization and the spin resolved density of
states.
We correctly predict that CoS$_2$ is the sole magnetic system of this class
of compounds, with a total magnetization per metal atom $\mu=$0.98$\mu_B$, 
in agreement with experimental values.
On the contrary FeS$_2$ and  NiS$_2$ do not show any total magnetization,
the former being non--magnetic and the latter presenting locally a regular
arrangement of the two spin components of the electronic 
density distribution, typical of an antiferromagnetic material. 
Moreover DOS calculated for CoS$_2$ is typical of an half metal, while 
FeS$_2$ and NiS$_2$ typical of small band gap semiconductors. \\
All these results are in nice agreement with experimental data
and previous theoretical calculations. \\
\indent These results now open the way to study also the optical properties
of such compounds.\\
In conclusion, as anticipated in the introduction of this chapter,
the interest on these systems is recenlty renewed due to the
great  potential for the emerging spintronic technological applications.
In fact, when CoS$_2$ is doped with iron or nickel
in a solid solution like Co$_{1-x}$Fe$_x$S$_2$ or Co$_{1-x}$Ni$_x$S$_2$.
the metal to semiconductor transition can be tuned increasing
the iron or nichel concentration.
Moreover the spin polarization can be controlled generating 
an highly polarized electron 
source~\cite{wang_prb_y2006_v73_p144402,mazin_apl_y2000_v77_p3000,wang_prl_y2005_v94_p056602,ramesha_prb_y2004_v70_p214409,cheng_jap_y2003_v93_p6847,mabatah_prl_y1977_v39_p494}.

\chapter*{Conclusions}
\markboth{Conclusions}{Conclusions}
 This thesis is devoted to \emph{ab initio} calculations
of ground state and excited state properties of realistic systems
within the density functional theory (DFT) and its Time Dependent 
generalisation (TDDFT). \\
\indent We used \emph{theoretical spectroscopy} tools in order to study 
several systems with different dimensionality (surfaces, molecules, bulk crystals).
Starting with the dielectric function $\varepsilon(\omega)$, 
obtained by the response function in linear regime, 
we calculated the anisotropy reflectivity 
(RA) spectra and the reflectance electron energy loss (REEL) spectra for 
the Si(100) clean and oxidized surfaces.
In the case of the clean surface, we considered three surface reconstructions
\pone, \ptwo\ and \ctwo.
Thanks to the comparison between experiments and numerical simulation,
we were able to rule out the \pone\ reconstruction and to define the 
origin of the REEL peaks without ambiguity.\\
In the second part of the work, we evidenced the problem of the correct 
description of excitation spectra of open shell molecules within DFT--LDA.
We calculated the energy levels and the first excitation energies
for the BeH molecule in the TDDFT framework. 
These calculations have been a first step in order to approach the case
of magnetic metals and half metals.
In fact, this part of the work was dedicated to the study of 
optical properties of magnetic iron alloys, that are very interesting 
materials for new applications in the spintronic domain. \\
In this framework we evaluated ground state properties and conductivity
of the bcc crystal phase of iron in order to validate 
the theoretical approach comparing the results with 
experimental data. \\
Finally we studied the ground state properties of some 3$d$--transition
metal disulfides, i.e. FeS$_2$, CoS$_2$ and NiS$_2$, having a commun 
cubic pyrite structure but different electronic and magnetic properties.
Starting from these compounds it is possible to create more complex 
systems such as iron 
or nickel doped CoS$_2$ alloys, that are interesting for the design of new 
technological devices based on spin electronics. \\
\indent From the numerical point of view, we implemented an original 
method in the large scale ab initio code DP in order to calculate 
the indipendent 
particle dynamical response function $\chi^{0}({\bf r},{\bf r}',\omega)$,
built from the eigenvalues and eigenvectors of the Kohn and Sham 
hamiltonian.
We have demostrated that the method, based on the Hilbert transform, 
is efficient for large size systems, as in the case of surfaces.
Moreover, we have generalized the code to spin variable in order to study 
magnetic properties of realistic applications. \\
\\
\\
Milano, December 2008

\chapter*{Scientific contributions}
\addcontentsline{toc}{chapter}{Scientific contributions}
\markboth{Scientific contributions}{Scientific contributions}
 \section*{List of papers}
 \begin{itemize}
 \item \underline{L. Caramella}, G. Onida, F. Finocchi, L. Reining and F. Sottile
      \emph{Optical properties of real surfaces: Local-field effects at oxidized Si(100)(2$\times$2) computed with an efficient numerical scheme}, Phys. Rev. B {\bf 75}, 205405 (2007). \\
(See chapter~\ref{ch:hilbert}) \\
 It has been published also in ``Virtual Journal of Nanoscale Science \& Technology'', 14 May 2007,Volume 15 issue 19.
  \item \underline{L. Caramella}, G. Onida and C. Hogan, \emph{Dielectric response and electron energy loss spectra of an oxidized Si(100)(2$\times$2) surface}, in Epioptics-9, The Science and culture, ed. by A. Cricenti, World Scientific, Singapore 2008, Vol. 29, p. 62. \\
 (See chapter~\ref{ch:a-si100})
  \item \underline{L. Caramella}, D. Tavella and G. Onida
    \emph{Norm conserving pseudopotentials for iron with semicore states} 
    sumbitted (2008). \\
  (See chapter~\ref{ch:a-iron})
  \item \underline{L. Caramella}, C. Hogan, G. Onida and R. Del Sole
    \emph{Calculation of High Resolution EEL Spectra of reconstructed Si(100) surfaces}, to be submitted (2008). \\
  (See chapter~\ref{ch:a-si100})
  \item C. Hogan, \underline{L. Caramella}, G. Onida and R. Del Sole
        \emph{Electron energy loss of the oxidized Si(100) surface}, in preparation (2008). \\
  (See chapter~\ref{ch:oxi})
  \item \underline{L. Caramella}, D. Sangalli, G. Onida \emph{et al.}
    \emph{Subtleties in electronic excitations of open shell molecules}, in preparation (2008). \\
  (See chapter~\ref{ch:a-molecules})
  \item \underline{L. Caramella} \emph{et al.} 
    \emph{Optical properties of FeS$_2$ and CoS$_2$}, in preparation (2008). \\
  (See chapter~\ref{a:sulfur})
\end{itemize}

\section*{Selected oral contributions to conferences}
 \begin{itemize}
  \item International School of Solid State Physics - Epioptics-10, 20--27, June 2008, Erice (Italy): \emph{Ab initio calculation of loss function of clean 
and oxidized silicon surfaces}.
  \item 5$^{th}$ Nanoquanta Young Researchers' Meeting, 20--23, May 2008, Modena (Italy): \emph{Quantitative predictions of Electron Energy Loss Spectra at the Si(100) surface}
  \item 4$^{th}$ Nanoquanta Young Researchers' Meeting, 15--18, May 2007, San Sebastian (Spain): \emph{Optical properties of real surfaces:
local field effects at oxidized Si(100)(2$\times$2)
computed with an efficient numerical scheme}
  \item R\'eunion G\'en\'eral du GDR DFT ++, 27--30, March 2007, L'Escandille, Autrans (France):\emph{Optical properties of real surfaces:
local field effects at oxidized Si(100)(2$\times$2)
computed with an efficient numerical scheme}.
  \item APS March Meeting, 5--9, March 2007, Denver (Colorado):\emph{Optical properties of real surfaces: local field effects at
oxidized Si(100)(2$\times$2) computed with an efficient numerical scheme}.
  \item International School of Solid State Physics - Epioptics-9, 20--26, July 2006, Erice (Italy):\emph{Application of an efficient numerical scheme for the 
computation of response functions to study the optical properties of the 
oxidized Si(100)(2$\times$2) surface}.
  \item 10$^{th}$ Nanoquanta General Meeting, 12--15, September 2005, Bad Honnef (Germany):\emph{A Hilbert transform--based scheme for the efficient computation
of response functions and its application to study the optical properties
of the oxidized Si(100)(2$\times$2) surface}.
\end{itemize}

\newpage

\appendix
\chapter{Determination of second derivative spectra}
\begin{figure}[!tbh]
 \centerline{\epsfig{file=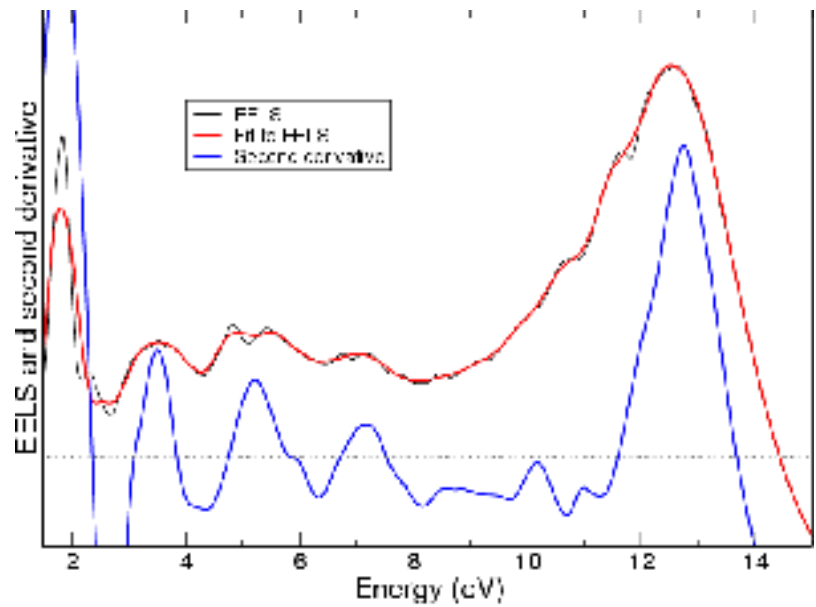,angle=270,clip=,width=10.5cm}}
 \caption[Second derivative spectra of EEL]{
      Computation of the second derivative spectra (blue line) of a typical
      EEL spectrum. The bare EEL spectrum (black line) and the resulting
      fit with polynomials (red line) are also shown.
  \label{fig:SG}
   }
\end{figure}

\markboth{Appendix}{Appendix}
 We implemented a Savitski--Golay (SG) smoothing algorithm 
\cite{savitzky_ac_y1964_v36_p1627} in order to compute the derivative spectra. \\
The SG algorithm fits the EEL curve with polynomials
preserving spectral features:
\begin{equation}
  g_i = \sum_{n=-n_L}^{n_R} c_N f_{i+N}
\label{eq:SG}
\end{equation}
where $g_i$ represents the smoothed function, and $c_N$
the coefficients obtained by the coefficient matrix of polynomial of
degree M by the relation:
\begin{equation}
 c_N = (A^TA)^{-1}(A^T e_n)_0
\end{equation}
where $A_{ij}=i^j$ is the coefficient matrix of polynomial of degree M
and $e_n$ is the unit vector. 
Then SG algorithm returns the derivative of $k$-degree of 
the smoothed curve. 
When $k>1$ the coefficients in equation \ref{eq:SG} have 
to be normalized replacing  $c_N$ with $c_Nk!$. \\
Smoothing step is required to correctly identify the physical
EEL peaks, in fact, derivative can dramatically enhance
the noise of the EEL spectra due to the discrete k point sampling. 
See Fig.~\ref{fig:SG} for an example of the method.

\bibliographystyle{ieeetr}
\addcontentsline{toc}{chapter}{Bibliography}
 \bibliography{tesi_biblio}
\end{document}